\documentclass[a4paper,12pt]{article}
\usepackage{authblk}
\usepackage{amsfonts}
\usepackage{amssymb}
\usepackage{amsthm}
\usepackage{amsmath}
\usepackage{array}
\usepackage{booktabs}
\usepackage{cases}
\usepackage{cite}
\usepackage{color}
\usepackage{dsfont}
\usepackage{enumerate}
\usepackage{epsfig}
\usepackage{multirow}
\usepackage{graphicx}
\usepackage{geometry}%
\usepackage[colorlinks,linkcolor=red]{hyperref}
\usepackage{pdflscape}
\usepackage{longtable}
\usepackage{latexsym}
\usepackage{longtable}
\usepackage{lscape}
\usepackage{multirow}
\usepackage{mathrsfs}
\usepackage[numbers]{natbib}
\usepackage{rotating}
\usepackage{setspace}
\usepackage{pst-node}
\usepackage{pst-grad}
\usepackage{pst-xkey}
\usepackage{pstricks}
\usepackage{pst-gantt}
\usepackage[section]{placeins}
\usepackage{tipa}
\usepackage{titlesec}
\usepackage{titletoc}
\usepackage{times}
\usepackage{tikz}
\usepackage[T1]{fontenc}
\usepackage[utf8]{inputenc}
\usepackage{url}
\usepackage{subfigure}
\usepackage{diagbox}
\usepackage{color}
\usepackage{supertabular}
\usepackage{tabularx}
\usepackage{lscape}
\usepackage{amssymb}
\usepackage{bbding}
\usepackage{hyperref}
\usepackage{verbatim}

\title{Digital Twin-Oriented Complex Networked Systems based on Heterogeneous Node Features and Interaction Rules}%

\author[1]{Jiaqi Wen, Bogdan Gabrys, Katarzyna Musial \\ Complex Adaptive Systems Lab, Data Science Institute, University of Technology Sydney \\ jiaqi.wen@student.uts.edu.au, bogdan.gabrys@uts.edu.au, katarzyna.musial-gabrys@uts.edu.au}

\begin{document}
\setlength{\baselineskip}{18pt}%

\date{~}
\maketitle
\vspace{-1cm}

\begin{abstract}

This study proposes an extendable modelling framework for Digital Twin-Oriented Complex Networked Systems (DT-CNSs) with a goal of generating networks that faithfully represent real systems.  Modelling process focuses on (i) features of nodes and (ii) interaction rules for creating connections that are built based on individual node's preferences. We conduct experiments on simulation-based DT-CNSs that incorporate various features and rules about network growth and different transmissibilities related to an epidemic spread on these networks. We present a case study on disaster resilience of social networks given an epidemic outbreak by investigating the infection occurrence within specific time and social distance. The experimental results show how different levels of the structural and dynamics complexities, concerned with feature diversity and flexibility of interaction rules respectively, influence network growth and epidemic spread. The analysis revealed that, to achieve maximum disaster resilience, mitigation policies should be targeted at nodes with preferred features as they have higher infection risks and should be the focus of the epidemic control. 


\vspace{1ex}
{\noindent{\bf Keywords:}
Social Networks; Network Dynamics; Complex Network Simulations; Heterogeneous Nodes Attributes and Features; Dynamic Processes on Networks}
\vspace{1ex}


\vspace{0.2cm}

\end{abstract}
\newpage
\tableofcontents
\newpage
\newpage

\section{Introduction}

Over the years, Complex Networked Systems (CNSs) have been modelled with increasing structural complexity levels that resulted in models with higher and higher accuracy. The ultimate goal of these research efforts is to build models that are an accurate reflection and, sometimes, extension of reality. Family of models that perfectly reflect real world networked systems can be seen as Digital Twin Oriented Complex Networked Systems (DT-CNSs) \citep{wen2022towards}. There are no current realisations of these conceptual ideas and the goal of the research presented in this paper is to bring us closer to the ultimate goal of DT-CNSs.

The complexity of CNSs results from the heterogeneity of CNSs' components and interactions between them as well as dynamic processes (e.g. epidemic spread) on those systems. Current studies on CNSs focus on the least complex scenario where a single dynamic process takes place on a static network without changes of the network components (nodes and edges) or the process parameters \citep{zhang2021vulnerability,zhang2018influence,wang2019simulation,wang2021multi,pastor2001epidemic,ganesh2005effect}. To increase complexity of modelling, and in this way to approach Digital Twin, a natural way forward is to increase the heterogeneity of nodes' features (characteristics of each specific node) and their preferences to create relationships as well as allow the system to evolve over time, both from the perspective of structure and process.

The complexity of modelling CNS is constrained by the system's observability. The observability is concerned with the ability to faithfully reconstruct the state of a system from a limited set of measured variables in finite time \citep{aguirre2018structural,wen2022towards}.
It not only determines the available ground-truth information to be represented and modelled but also the information to be simulated towards achieving the ultimate goal of a Digital Twin (DT). With different levels of observability, we categorise the CNS components as data-driven, simulation-based and hybrid \citep{wen2022towards}, each composed of real data \citep{budka2013molecular,musial2013creation,skarding2020foundations}, purely simulated data \citep{ashraf2019simulation} and both real and simulated data respectively. Current studies generally build CNSs with hybrid components (e.g. real networks and simulated dynamic processes) due to the partial observability of a real-world scenario \citep{zhang2021vulnerability,zhang2018influence,wang2019simulation,wang2021multi,pastor2001epidemic,ganesh2005effect,jovanovski2021modeling,carchiolo2021mutual,zhang2018convergence}.

In this study, we propose an extendable 
modelling framework for DT-CNSs, composed of (i) a flexible network model that allows for static network formation based on heterogeneous node features and interaction rules related to connection preferences, and (ii) a process model for the Susceptible-Infected (SI) spread with a predetermined seed selection strategy and an infection rate. We propose a set of evaluation metrics to validate the faithfulness of DT-CNSs, considering their similarity with real system, efficiency and reproducibility of the network patterns and the infection occurrence on such networks. We present a case study on the disaster resilience of social networks considering their infection occurrence within specific time and social distance given an epidemic outbreak. Our experiments focus on various simulated networks, where nodes have heterogeneous attributes and the interactions between the nodes are driven by various rules (e.g. social networks driven by age features and an interaction rule -- preference for a similar age when nodes interact). 
The results show that the heterogeneous features and interaction rules (related to preference for features) influence the network structure and can enhance epidemic outbreak. Nodes with preferred by others features have more connections and, as a result, get exposed to higher infection risk within the same proximity to the initial infection node. 

The main contributions of this study are:
\begin{itemize}
 \item Constructing an extendable DT-CNS modelling framework composed of a network model based on heterogeneous nodes' features and rules driving the interactions between nodes and a process model spreading over the network;
 \item Creating an evaluation protocol for faithfulness, concerned with similarity, efficiency and reproducibility;
 \item Validating the influence of heterogeneous nodes' features and interaction rules on network growth and level of epidemic spread;
 \item Suggesting disaster mitigation policies dependent on age diversity and social preference.
\end{itemize}

The rest of this study is structured as follows. Section \ref{Rep1-1section2} provides background in the space of dynamics of and on the networks. Section \ref{Rep1-1section3} presents the methodology of building and driving a CNS towards a DT. Following this, section \ref{Rep1-1section4} builds and evaluates the simulation-based CNSs. Finally, in section \ref{Rep1-1section5}, we conclude the study and outline future directions. 

\section{Background}
\label{Rep1-1section2}
This section presents the network and process models developed and employed in the current studies. It discusses (i) interaction rules that drive the network growth through preference for node features and (ii) transmission rules governing the node adaptability to an epidemic spread in different conditions.

\paragraph{Network model} aims at providing a faithful representation of a network that minimises information loss \cite{wen2022towards}. 
There are two widely employed principles for network formation and growth: preferential attachment \cite{jeong2003measuring,abbasi2012betweenness,barabasi1999emergence,tsiotas2020preferential,abbasi2012betweenness} and homophily \cite{mcpherson2001birds,kossinets2009origins,boucher2015structural,ertug2022does}. By incorporating these two principles, this study starts to build DT-CNSs based on observable networked information

Preferential attachment describes the "rich get richer" principle \citep{abbasi2012betweenness}, where highly connected nodes increase their connectivity faster than their less-connected peers \citep{jeong2003measuring}. A scale-free network model based on the preferential attachment to higher node degrees was first proposed in \citep{barabasi1999emergence}. With time, more and more studies in the network science space started looking into other topological features, such as clustering coefficient, closeness centrality and eigenvector centrality \citep{tsiotas2020preferential,abbasi2012betweenness}. In our study, we allow preferential attachment to the nodes with specific feature values concerned with topology and attributes.

Tendency to associate ourselves with other similar to us is called homophily \citep{mcpherson2001birds}. It can either be choice homophily or induced homophily. The first one is a preference for similar attributes, e.g. people tend to interact with those similar interests. Induced homophily is a result of the interaction opportunities due to structural proximity \citep{kossinets2009origins,boucher2015structural,ertug2022does} (e.g. people who interact because they work in the same building). In the network embedding paradigms, the induced homophily generally suggests that highly connected nodes should be embedded closely in the latent representation space \citep{zhou2018dynamic}. In contrast, the choice homophily effect leads to the consideration of attribute proximity \citep{liao2018attributed}. In our study, we account for both choice homophily and induced homophily based on the proximity of structures and attributes.

\paragraph{Process model} for the spreading phenomenon on networks has two components: 
(i) seed selection strategy for the first contagious nodes, 
and (ii) the transmissibility that describes the spread conditional on node features \citep{brodka2020interacting,krol2015propagation,sadaf2022maximising}.

Seed selection strategies identify single/multiple seeds (source nodes of the spreading process) in the initial/subsequent steps (single-stage seeding/sequential seeding) from which the spread starts. Current studies generally select seeds based on the centrality measures, such as degree centrality, betweenness centrality, closeness centrality and eigenvector centrality \citep{comin2011identifying,karczmarczyk2018influencing} or identify driver nodes and using them as seeds \citep{sadaf2022maximising}. 

The transmissibility involved in the contagion models depends on many, different elements. For example, node adoptability, concerning the probability to adopt a spread, rises when there are multiple exposures for multiple contagions/simplicial contagion, compared with a simple contagion \citep{min2018competing,jovanovski2021modeling}. In addition, the transmissibility can also vary with specific interactions (e.g. physical contact, airborne spread, etc.) among node pairs or groups \citep{liu2020using}. Particularly, \cite{min2018competing} explains the current contagion models with conditional transmissibility, such as the threshold model, stochastic contagion model, diffusion percolation model, independent cascade model and Reed–Frost model. 

Overall, current network models are built on interaction rules, either related to preference for topological features (preferential attachment) or preference for similar node attributes (homophily). Current rule-based process models focus on seed selection strategies and transmission rules. However, none of the studies combines heterogeneous features and interaction rules and investigates their influence on the network growth and the corresponding epidemic spread. Therefore, this study proposes an extendable modelling framework for DT-CNSs based on heterogeneous node features and diverse interaction rules.

\section{Digital Twin Orientated Complex Networked System}
\label{Rep1-1section3}
In this part, we provide a description of an extendable CNS simulator in section \ref{Rep1-1section31} and propose an evaluation protocol for the faithfulness of CNSs in section \ref{Rep1-1section32}. In section \ref{Rep1-1section33} we describe the optimisation process to obtain CNSs model that faithfully represent reality and approach DT. 

\subsection{An extendable DT-CNS framework}
\label{Rep1-1section31}
The extendable CNS simulator enables to model the dynamics of and on the networks. It is devised with (i) a network model based on the interaction rules of homophily and preferential attachment, as well as (ii) a process model (transmission mechanism) that uses different seed selection strategies and transmission rules dependent on varying conditions. 

\subsubsection{Network and Network Dynamics} 

In this study, we focus on social networks composed of nodes (people), edges (people's interactions) and the attributes of nodes and edges. We model the interactions between nodes with network dynamics governed by the interaction rules.

Network at time $t$ can be represented as $G_{t}=\{V_t,E_t,A_t,Z_t\}$ based on the observable and simulated information of the changeable network components: nodes $V_t=\{v_{1,t},\cdots, v_{n_t,t}\}$, edges $E_t=\{e(v_{i,t},v_{j,t})|v_{i,t},v_{i,t}\in V, i \neq j\}$, node attributes $A_t = \{a(v_{1,t}),\cdots,a(v_{N_t,t})\}$ and the edge attributes $Z_t = \{z(v_{i,t},v_{j,t})|v_{i,t},v_{i,t}\in V, i \neq j\}$. Network dynamics drives network formation and growth over time.

The node attribute vector $a(v_{i,t})$ for node $v_i$, referring to \cite{ashraf2019simulation}, is defined as 
\begin{equation}
 a(v_{i,t}) = \{\mathbf{f}(v_{i,t}), \mathbf{p}(v_{t,i}), \mathbf{w}^p(v_{t,i}), \mathbf{h}(v_{i,t}), \mathbf{w}^h(v_{t,i})\} \quad v_i\in V
\end{equation}
which includes an $l$-length feature vector $\mathbf{f}(v_{i,t})$ ($l$ is the number of features) and a social-DNA (sDNA) defined with another four $l$-length vectors. Features presented in $\mathbf{f}(v_{i,t})$ are characteristics that describe each individual $v_{i,t}$, such as age, gender and interests. $\mathbf{p}(v_{t,i})$ determines the preference for features with a trinary value: $0$, $1$, or $-1$, which each, respectively, indicates: $0$ -- no preferential attachment to the corresponding node feature in $\mathbf{f}(v_{i,t})$, $1$ -- preferential attachment to a higher feature value or $-1$ -- to a lower feature value. For example, the scale-free network model, where nodes connect with other popular nodes, uses pure preferential attachment (a favourable preference) to node degrees. Similarly, $\mathbf{h}(v_{t,i})$ determines the preference for similar features with a trinary value: $0$, $1$, $-1$, each, respectively, representing no homophily ($0$), heterophily ($1$) and homophily ($-1$). For example, in a social network where people like to interact with people of similar ages, we would use pure homophily to describe the negative preference for age differences. Both of these preferences are followed with a weighting factor $\mathbf{w}^p(v_{t,i})$ and $\mathbf{w}^h(v_{t,i})$ within the range of $(0,1]$. 

The feature $\mathbf{f}(v_{i,t})$ and the sDNA $\{\mathbf{p}(v_{t,i}), \mathbf{w}^p(v_{t,i}), \mathbf{h}(v_{i,t}), \mathbf{w}^h(v_{t,i})\}$ can vary with nodes in a frozen time scale and/or mutate over time. They can also co-evolve with the network topology based on the rules of network growth:

\begin{equation}
 \pi(v_{i,t},v_{j,t}) = [\frac{1}{2}\pi^p(v_{i,t},v_{j,t}) + \frac{1}{2}\pi^h(v_{i,t},v_{j,t})+\epsilon_{ij,t}]*I_{ij,t}
 \quad v_i,v_j\in V, i\neq j
 \label{sumScore}
\end{equation}
where the network growth is driven by 
ranks of score $\pi(v_{i,t},v_{j,t})$ of any node pair through mutual evaluation concerned about their preferences. It is calculated as the sum of preferential attachment score $\pi^p(v_{i,t},v_{j,t})$, the homophily score $\pi^h(v_{i,t},v_{j,t})$ and the random interference $\epsilon_{ij,t} \sim\mathcal{N}(0,\sigma^2)$ that follows a random normal distribution. $I_{ij,t}(\eta)$ is a binary vector that denotes the an encounters or zero encounters of this node pair with $1$ or $0$ dependent on the encounter rate $\eta\in[0,1]$.

The preferential attachment score $\pi^p(v_{i,t},v_{j,t})$ incorporates the preference $\mathbf{p}(v_{i,t})$ for the other nodes with higher/lower feature values for $\mathbf{f}(v_{i,t})$:

\begin{equation}
\begin{aligned}
 \pi^p(v_{i,t},v_{j,t}) = \frac{1}{2l}\mathbf{f}(v_{j,t})^\tau (\mathbf{p}(v_{i,t}) \odot \mathbf{w}^p(v_{i,t}))+\frac{1}{2l}\mathbf{f}(v_{i,t})^\tau (\mathbf{p}(v_{j,t}) \odot \mathbf{w}^p(v_{j,t}))+1 
\end{aligned}
\end{equation}

The homophily score $\pi^h(v_{i,t},v_{j,t})$ incorporates the preference $p(v_{i,t})$ for the other nodes with similar/dissimilar feature values $f(v_{i,t})$:
\begin{equation}
\begin{aligned}
 \pi^h(v_{i,t},v_{j,t}) &=& \frac{1}{2l}|\mathbf{f}(v_{i,t})-\mathbf{f}(v_{j,t})|^\tau (\mathbf{h}(v_{i,t}) \odot \mathbf{w}^h(v_{i,t}))\\
 & + &\frac{1}{2l}|\mathbf{f}(v_{i,t})-\mathbf{f}(v_{j,t})|^\tau (\mathbf{h}(v_{j,t}) \odot \mathbf{w}^h(v_{j,t}))+1
\end{aligned}
\end{equation}

The score list $\Pi_t = \{\pi(v_{i,t},v_(j,t))|v_{i,t},v_{i,t}\in V, i \neq j\}$ stays frozen with fixed nodes and node attributes (features and sDNA). The CNS simulator connects $\lambda_{e,t}$ pairs of nodes based on the score ranks and generates a static network to represent discrete interactions at this frozen time point, with an edge intensity of

\begin{eqnarray}
\gamma(v_{i,t},v_{j,t}) = \left\{
\begin{array}{ll}
\frac{\pi(v_{i,t},v_j,t)+2l}{4l} & \textrm{if $\lambda_{e,t}$ is reached;}\\
0 & \textrm{if else.}\\
\end{array}
\right.
\label{rearrange}
\end{eqnarray}

The score list $\Pi_t$ changes with any node feature change/preference mutation. This may result in changes of node pair ranks and their connections, finally leading to network evolution. 

\subsubsection{Process and Process Dynamics} 

In this study, we also investigate the epidemic spreading process on the social networks. 
Process Dynamics drives the spreading process $\mathrm{P} = \{\Omega,Pr_t,\mathbf{B}_t,\mathbf{R}_t\}$ on the networks and is defined by the following elements: seed selection strategy $\Omega$, transmission probability $Pr_t$ given others' adoptability $\mathbf{B}_t=\{\mathbf{b}(v_{i,t})|v_{i,t}\in V_t\}$ and the resulting infection status $\mathbf{R}_t = \{\mathbf{r}(v_{i,t})|v_{i,t}\in V_t\}$.

The seed selection strategy identifies the first contagious nodes at the beginning of epidemic spread. The seed selection strategy $\Omega=\{\omega_{1},\omega_{2},\cdots\}$ for single/multiple seeds in the initial/sequential stages is defined with a set of rules $\omega_t$:

\begin{equation}
 \omega_t(\mathbf{s}_t,\mathbf{w}^s_t,k_t)
\end{equation}
where $k_t$ represents the number of seeds selected at time $t$. $\mathbf{s}_t$ is a $l$-length vector that determines the seed preference for node features $\mathbf{f}(v_{i,t})$ with values: $0$, $1$ or $-1$, each representing no preference, preference for higher/lower feature values, respectively. This is accompanied by a $l$-length weighting vector $\mathbf{w}^s_t$, including weights of preference within range $(0,1]$. The first $k_t$ seeds are selected at time $t$, dependent on the descending ranks of node score $\epsilon(v_{i,t})$:
\begin{equation}
 \epsilon(v_{i,t}) = \mathbf{f}(v_{i,t})^\tau(\mathbf{s}\odot\mathbf{w}^s)
\end{equation}
For example, we can build a 
seed selection strategy based on node popularity preference. In this context, popular nodes, featured with significant node degrees, are preferred and assigned high score ranks, finally selected as the contagious seeds. 

The seed selection results in the nodes' infection status $\mathbf{r}(v_{i,t})$, a $m$-length vector that determines whether the node is infected or not 
with binary values of $0$ or $1$. Its transition from the previous infection status $\mathbf{r}(v_{i,t-1})$ to the current status $\mathbf{r}(v_{i,t})$ is dependent on the node adoptability $\alpha(v_{i,t})$, also termed as the Process DNA (pDNA):

\begin{equation}
 \mathbf{b}(v_{i,t}) = \{\mathbf{c}(v_{i,t}),\Theta(v_{i,t}),\mathbf{w}^c(v_{i,t})\}
\end{equation}

The pDNA is represented through three $q$-length vectors: $\mathbf{c}(v_{i,t})=[c_1(v_{i,t}),\cdots,c_q(v_{i,t})]$ specifies the $q$ independent conditions of node $v_{i,t}$ considering the node infection status $\mathbf{r}(v_{i,t})$ and the node features $\mathbf{f}(v_{i,t})$. The independent conditions presented in $\mathbf{c}(v_{i,t})$ identify various individual characteristics that pose each individual at infection risk and determine how the epidemic spreads. As an example, there is an epidemic spread between males with physical contact. In this context, the independent conditions include gender (male/female) and exposure (with/without physical contact).

$\Theta(v_{i,t})=[\theta_1(v_{i,t}),\cdots,\theta_q(v_{i,t})]$ represents the adoptability-thresholds to be considered when evaluating each node condition; and $\mathbf{w}^c(v_{i,t})=[w_1^c(v_{i,t}),\cdots,w_q^c(v_{i,t})]$ includes the multiplier effect of node adoptability on state transition, within the value range $(0,1]$. %

With the pDNA $\mathbf{b}(v_{i,t})$, the transition probability $Pr(\mathbf{r}(v_{i,t})|\mathbf{r}(v_{i,t+\Delta t}))$ can be calculated as:
\begin{equation}
log(Pr(\mathbf{r}(v_{i,t})|\mathbf{r}(v_{i,t-\Delta t}))) = \sum\limits_{j=1}^{q}\delta_{c_j(v_{i,t}),\theta_j(v_{i,t})}*log(w_j^c(v_{i,t})) 
\end{equation}
where $\delta$ is a Kronecker function that determines, with binary values between $0$ and $1$, whether a specific condition $c_j(v_{i,t})$ for node $v_{i,t}$ meets the threshold $\theta_j(v_{i,t})$. To be more specific, $\delta_{c_j(v_{i,t}),\theta_j(v_{i,t})}=0$ indicates the elimination of the multiplier effect $w_j^c(v_{i,t})$, while given $\delta_{c_j(v_{i,t}),\theta_j(v_{i,t})}=1$, $w_j^c(v_{i,t})\longrightarrow0$ indicates full resilience of the node.

\subsection{An Evaluation Protocol on Faithfulness}
\label{Rep1-1section32}

Below, we propose an evaluation protocol to formally capture different interpretations of faithfulness under the constraints of observability that vary with the no/complete/partial ground-truth scenarios.


\subsubsection{Complexity}
Emergent patterns observed in CNSs and their complexity are the starting point of evaluation. 
The evaluation metrics of the emergent patterns existing in CNSs in the Network dimension and the Process dimension (See Tab.~\ref{tab1}) respectively focus on the characteristics of network topology, network attributes, the transmissibility of the spread and the infection occurrence.

\begin{table}[h]
\centering
\small
\caption{Measures of complex patterns emerging from CNS representation.}
\label{tab1}
\setlength{\tabcolsep}{3pt}
\renewcommand{\arraystretch}{1.5}
\begin{tabular}{|c|c|p{85pt}|p{100pt}|p{100pt}|}
\hline
\multicolumn{2}{|c|}{CNS components} & Pinpoint & Local & Global \\
\hline
\multirow{6}{*}{Network} & \multirow{4}{*}{Topology}& Node \citep{kendrick2018change}, Edge \citep{wahid2019predict} & Triad \citep{musial2012triad}, Quadrangle \citep{jia2021measuring}, Community \citep{qin2018adaptive,jin2018robust,wang2016autonomous}, Clustering coefficient \citep{jia2021directed}, Closure coefficient \citep{jia2021directed} &Degree distribution \citep{musial2013kind}, Shortest path length \citep{musial2013kind}, Modularity \citep{musial2013kind,jin2020modmrf}, Assortativity, Betweenness, Closeness. \\
\cline{2-5}
 & \multirow{2}{*}{Attribute}& Node attribute \cite{liu2021block,wang2019community}, Edge attribute \cite{liu2020semi,wang2019community} & & Feature distribution, Hill numbers \citep{chao2014rarefaction}\\
\hline
\multirow{3}{*}{Process} & \multirow{2}{*}{Spread}& Transmissibility to nodes
& Transmissibility to groups & Transmissibility to the population\\
\cline{2-5}
& Infection & Infected nodes & Infected groups & Infected population \\
\hline
\end{tabular}
\end{table}

These measures range from (i) the pinpoint level connected to the individual component of the network/process (e.g. a node/edge added/removed over time; transmissibility to a node,
together with its resulting infection status); (ii)  the local level concerned with the local interactions among grouped individuals (e.g. clustering coefficient employed to assess local structure); and (iii) the global level concerned with the emergent global characteristics resulting from the interactions within the population (e.g. degree distribution based on the number of edges connected to each node; Hill numbers that describe the diversity of the population (nodes) based on node attributes).

\subsubsection{Performance}
Performance of modelling CNS dynamics, given the 
quantified complexity of CNS, can be evaluated in respect to different requirements in terms of the CNS patterns' 
reproducibility, the corresponding similarity between the CNS patterns with a target state (e.g. the target network topology required in link prediction tasks; etc.) as well as efficiency.

\begin{table}[htp]
\centering
\small
\caption{Measures used to assess performance of simulation, prediction and/or control models of CNS dynamics.}
\label{tab2}
\setlength{\tabcolsep}{3pt}
\renewcommand{\arraystretch}{1.5}
\begin{tabular}{|c|c|p{265pt}|}
\hline
 Requirement & Level & Measures \\
\hline
\multirow{10}{*}{Similarity} & \multirow{4}{*}{Pinpoint}&Precision \cite{wahid2019predict,gao2017community}, Recall \cite{gao2016hybrid}, Area Under the Precision–Recall (AUPR) curve \cite{dong2015coupledlp}, Receiver Operating Characteristic (ROC) curves, Area Under the ROC (AUC)\cite{lu2011link}, Geometric Mean of AUC and PRAUC (GMAUC) \cite{junuthula2016evaluating}, Error Rate \cite{chen2018gc}, SumD \cite{li2014deep}, Kendall's Tau Coefficient (KTC) \cite{bu2019link}, Micro/Macro/Weighted Average Precision/Recall/F1 Score \cite{patel2021graph,chen2020multi} \\
\cline{2-3}
& \multirow{2}{*}{Local} & 2-sample Kullback-Leibler divergence \cite{della1997inducing}, Manhattan distance \cite{faisal2020comparative}, Canberra Distance \cite{faisal2020comparative}, Euclidean distance \cite{koutra2013deltacon}, Matusita distance \cite{koutra2013deltacon} \\
\cline{2-3}
& \multirow{4}{*}{Global}& 2-sample Kullback-Leibler divergence \cite{della1997inducing}, Manhattan distance \cite{faisal2020comparative}, Canberra Distance \cite{faisal2020comparative}, Euclidean distance \cite{koutra2013deltacon}, Matusita distance \cite{koutra2013deltacon}, Earth Mover’s Distance \cite{nikolentzos2017matching}, Similarity metrics respectively based on entropy distance, spectral distance, modality distance, cosine of the angle between two graphs \cite{wang2018graph}. \\
\hline
 \multirow{2}{*}{Efficiency}& Data processing & Time delays compared with the real time data flow \cite{wang2018real} \\
 \cline{2-3}
& Simulation/Modelling & Runtime of the simulation/modelling \cite{kim2019advancing,ashraf2019simulation}\\
\hline
\multirow{3}{*}{Reproducibility} & Same data & Yes/No \\
\cline{2-3}
& Same statistics & 2-sample Kullback-Leibler divergence \cite{della1997inducing} \\
\cline{2-3}
& Same phenomena & $\omega$-index \cite{telesford2011ubiquity}, City organization index \cite{courtat2011mathematics} \\
\hline
\end{tabular}
\end{table}

Based on the measures describing complex emerging patterns in a CNS, similarity between the created and the target CNS patterns also ranges from the pinpoint level, the local level, to the global level.

The pinpoint level similarity is generally considered in prediction tasks, like the link prediction and its evaluation with precision \cite{lu2011link} where the similarity is interpreted in the context of the links predicted vs links actually appearing. In contrast, the local and the global similarity are generally employed in the 
simulation optimisation of CNSs toward the target network topological features.

Efficiency required for the CNS dynamics involves the discussion from two perspectives: the efficiency of data processing, concerned with the time delays compared with the real-time data flow~\cite{wang2018real}, as well as the efficiency of modelling, connected to the runtime of CNSs~\cite{kim2019advancing,ashraf2019simulation}.

Reproducibility requires the equivalent results of the same tasks \cite{ivie2018reproducibility}. It varies with the increasingly demanding levels of equivalence in the recreated data, ranging from the same phenomena and statistic distribution to the same data. Current CNSs generally fall into the category of the same statistics. 
For example, the scale-free network model generates networks that share the same node degree distributions based on a given parameter set.

\subsubsection{Faithfulness}
Faithfulness of CNS representation and modelling is defined and evaluated based on complexity and model performance. Simulation-based, data-driven, and hybrid CNSs each poses different requirements of efficiency, reproducibility and similarity, which results in different meanings of faithfulness (See Tab.~\ref{tab3}). 

\begin{table}[htp]
\centering
\small
\caption{Requirement of a faithful CNS representation and modelling under observability constraints.}
\label{tab3}
\setlength{\tabcolsep}{3pt}
\renewcommand{\arraystretch}{1.5}
\begin{tabular}{|c|c|c|c|c|}
\hline
 Requirement & Level & Simulation-based & Data-driven & Hybrid\\
\hline
\multirow{3}{*}{Similarity} & \multirow{1}{*}{Pinpoint}& & \FiveStar&\FiveStarOpen \\
\cline{2-5}
& \multirow{1}{*}{Local} &\FiveStar & & \FiveStarOpen\\
\cline{2-5}
& \multirow{1}{*}{Global}&\FiveStar & & \FiveStarOpen\\
\hline
 \multirow{2}{*}{Efficiency}& Data processing & &\FiveStar & \FiveStarOpen \\ 
 \cline{2-5}
& Simulation/Modelling & \FiveStar & \FiveStar & \FiveStar\\
\hline
\multirow{3}{*}{Reproducibility} & Same data & & \FiveStar& \\
\cline{2-5}
& Same statistics & \FiveStar & & \FiveStar\\
\cline{2-5}
& Same phenomena & \FiveStar & & \FiveStar \\
\hline
\end{tabular}
\end{table}


As is shown in Tab.~\ref{tab3}, the solid and the hollow five-pointed star 
each represents the fulfilment or partial fulfilment
of the corresponding requirement for each type of modelling. The data-driven CNSs, as well as the data-driven components of hybrid CNSs, generally require pinpoint similarity with the real data and efficiency in data processing and modelling. The simulation-based CNSs and the simulation components of Hybrid CNSs need global and local similarity, as well as efficiency of a simulation process. In terms of reproducibility, data-driven CNSs, in the ideal case, are required to generate the same data, mimicking the reality. In contrast, for the simulation-based CNSs and the hybrid CNSs, the same statistics and phenomena are needed and sufficient. For example, some data-driven CNSs employ network embedding methods in node classification tasks \citep{hong2019deep,abu2020n,ashraf2019simulation}. They require a pinpoint level of similarity with ground-truth information about network topology and attributes, data processing and modelling efficiency, and reproducible classification results for validation and comparative analysis. The simulation-based CNSs, such as the scale-free network models, simulate realistic networks based on the assumptions of interaction rules -- the preferential attachment to popular nodes \citep{fortunato2006scale,sohn2017small}. They require a global level similarity and the reproducibility of the statistics that measure the required network characteristics, such as the degree distributions reproduced by the same scale-free network model. Some hybrid CNSs employ real social networks and simulate the spreading processes on these networks \citep{qiu2016effects}. They first represents the real networks with pinpoint level similarity and then, based on these networks, simulate spreading processes with reproducible spreading results.

\subsection{Optimisation towards a Digital Twin}
\label{Rep1-1section33}

\begin{figure}
 \centering
 \includegraphics[width=0.99\textwidth]{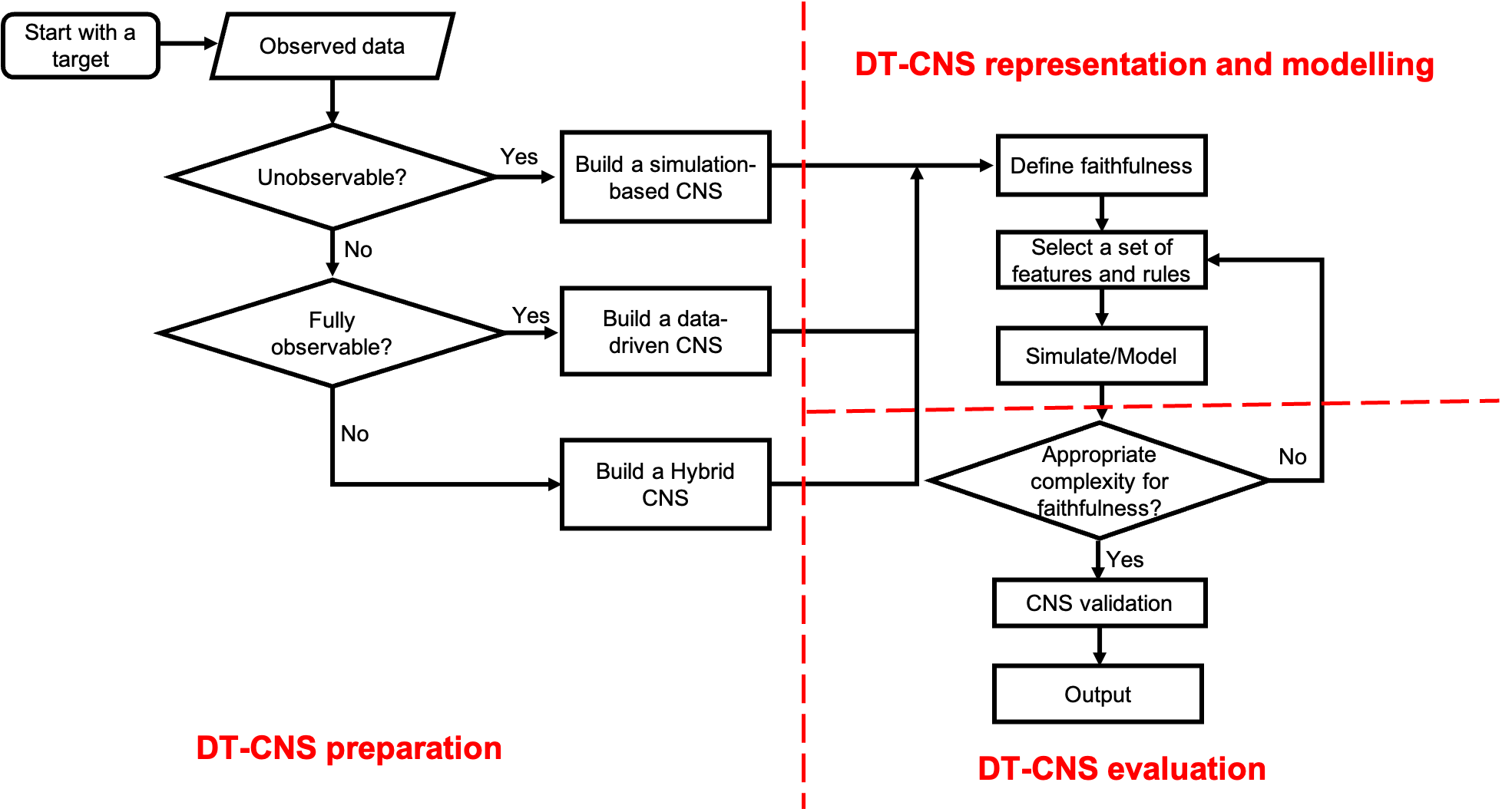}
 \caption{The procedures of building Digital Twin-Oriented Complex Networked Systems.}
 \label{flow}
\end{figure}

Under observability constraints, we optimise the DT-CNSs for an appropriate level of complexity, which enables a minimised information loss with a limited number of variables. We also create a set of modelling procedures to drive the optimisation of DT-CNSs considering different CNS types and their required faithfulness given the level of observability (See Fig.~\ref{flow}).

As shown in Figure~\ref{flow}, the devise and optimisation of the CNS towards a DT go through the DT-CNS preparation, DT-CNS representation and modelling, and the DT-CNS evaluation procedures. We prepare a DT-CNS by targeting a specific state of real systems and, under the observability constraints, deciding the CNS type (See Section \ref{Rep1-1section31}). After that, we determine the faithfulness needed for a DT-CNS under the evaluation protocol proposed in Section \ref{Rep1-1section32}. In addition, we devise and optimise an extendable CNS towards a DT through an appropriate selection of features (See Tab.~\ref{feat}) and rules (See Tab.~\ref{rule}). 

\subsubsection{Features}
Features in the DT-CNS context represent the attributes and the topological characteristics of nodes and edges in CNSs. As is shown in Tab.~\ref{feat}, the network dimension has the ascribed and topological features, each resulting in and from the network growth. The process dimension focuses on the ascribed and the resulting conditions that change transmissibility and influence the spreading process. They are ascribed or resulting conditions dependent on whether they exist before or after the spread. Nodes with specific ascribed features can have different susceptibilities given an epidemic exposure, such as gender and age. The epidemic spread can also result in conditions such as the states of infected, immune and susceptible. These resulting conditions also change the susceptibility of nodes in the next round of epidemic spread.

\begin{table}[htp]
\centering
\small
\caption{Examples related to features and the conditions in the DT-CNS representation and modelling.}
\label{feat}
\setlength{\tabcolsep}{3pt}
\renewcommand{\arraystretch}{1.5}
\begin{tabular}{|l|l|l|l|p{100pt}|p{100pt}|}
\hline
\multicolumn{4}{|c|}{Features\&Conditions} & Data-driven & Simulation-based \\
\hline
\multirow{1}{*}{Network}&\multirow{1}{*}{Node}&\multirow{1}{*}{Ascribed features} & \multirow{1}{*}{Continuous}& Height, weight, etc. & Normal distribution, uniform distribution, etc. \\
\cline{4-6}
& & & \multirow{1}{*}{Discrete}&Age, etc. &Poisson distribution, etc. \\
\cline{4-6}
& & & \multirow{1}{*}{Categorical}& Gender, nationality, group, etc & Bernoulli distribution \\
\cline{3-6}
& & \multirow{1}{*}{Topological features} & \multirow{1}{*}{Continuous}& \multicolumn{2}{|p{200pt}|}{Node clustering coefficient, centrality, etc.} \\
\cline{4-6}
& & & \multirow{1}{*}{Discrete}&\multicolumn{2}{|p{200pt}|}{Node degree, shortest path lengths, etc.}\\
\cline{2-6}
&\multirow{1}{*}{Edge}&\multirow{1}{*}{Ascribed features} & \multirow{1}{*}{Continuous}& Intensity of contact/relationship, etc.
& Normal distribution, uniform distribution, etc. \\
\cline{4-6}
& & & \multirow{1}{*}{Discrete}& &Poisson distribution, etc. \\
\cline{4-6}
& & & \multirow{1}{*}{Categorical}& Relations, Types of contact, etc. & Bernoulli distribution \\
\cline{3-6}
& &\multirow{1}{*}{Topological features} & \multirow{1}{*}{Continuous}& \multicolumn{2}{|p{200pt}|}{Duration of contact/relationships, etc.} \\
\cline{4-6}
& & & \multirow{1}{*}{Discrete}&\multicolumn{2}{|p{200pt}|}{Frequency of contact/Rating of relationships, etc.}\\
\hline
\multirow{1}{*}{Process}
&\multirow{1}{*}{Seed}&\multirow{1}{*}{Resulting conditions} & \multirow{1}{*}{Categorical}& \multicolumn{2}{|p{200pt}|}{Infection status.} \\
\cline{2-6}
&\multirow{1}{*}{Spread}&\multirow{1}{*}{Ascribed conditions} & \multirow{1}{*}{Categorical}& \multicolumn{2}{|p{200pt}|}{Immunity, susceptibility, etc.} \\
\cline{3-6}
& &\multirow{1}{*}{Resulting conditions} & \multirow{1}{*}{Categorical}& \multicolumn{2}{|p{200pt}|}{Infection status, an exposure, multiple exposure, etc.} \\
\hline
\end{tabular}
\end{table}

\subsubsection{Rules}
Rules are used to govern the behaviours of nodes in the DT-CNSs and they can be prescribed or learned from data. Those rules drive the network growth and the propagation of the spreading process based on various CNS features. Rules that are considered in the framework are presented in Tab.~\ref{rule}. We parameterise the network growth with the social DNA vector based on preferential attachment and homophily rules. In addition, we model the spreading process through seed selection and the process DNA vector for node adoptability. The experiments on DT-CNSs in this study investigate the network growth based on the interaction rules presented in Tab.~\ref{rule}. The seed selection strategy and the transmission rules in Tab.~\ref{rule} are part of the design but will be developed in the future research.

\begin{table}[htp]
\centering
\small
\caption{The examples of possible rules governing the interactions, seed selection and epidemic transmission in DT-CNS representation and modelling.}
\label{rule}
\setlength{\tabcolsep}{3pt}
\renewcommand{\arraystretch}{1.5}
\begin{tabular}{|p{50pt}|p{55pt}|l|l|p{100pt}|l|}
\hline
\multicolumn{2}{|p{105pt}|}{Rule} & Parameters & Value range &Description & Terminology \\
\hline
\multirow{1}{*}{Network}&Preferential attachment& $\mathbf{p}(v_{i,t})$ & $\{-1,0,1\}$& Preference for larger/smaller feature values & \multirow{1}{*}{social DNA} \\
\cline{3-5}
& &$\mathbf{w}^p(v_{i,t})$ & $(0,1]$ &Corresponding weight pf preference& \\
\cline{2-5}
& Homophily & $\mathbf{h}(v_{i,t})$ & $\{-1,0,1\}$ &Preference for similar/dissimilar features& \\
\cline{3-5}
& &$\mathbf{w}^h(v_{i,t})$ & $(0,1]$& Corresponding weight of preference& \\
\hline
\multirow{1}{*}{Process}& Node adoptability&
$\Theta(v_{i,t})$ & $\{0,1\}$ & Adoptability threshold for each condition. & process DNA \\
\cline{3-5}
& &$\mathbf{w}^c(v_{i,t})$ & $(0,1]$ & The multiplier effect on transmissibility when meeting the threshold&\\
\cline{2-6}
& Seed selection &$\mathbf{s}_t$ & $\{-1,0,1\}$& seed preference for node features& seed strategy \\
\cline{3-5}
& &$\mathbf{w}^s_t$ & $(0,1]$& Corresponding weight of preference &\\
\cline{3-5}
& &$k_t$ & $(0,1]$ & Number of seeds &\\
\hline
\end{tabular}
\end{table}

\subsubsection{Trade-offs}
Trade-offs between different 
requirements occur over the entire modelling and evaluation process to achieve highest possible faithfulness. These requirements involve the evaluation of CNSs with different measures, respectively, considering the three aspects: similarity, efficiency and reproducibility. However, constrained by the available CNS measures and modelling techniques for multi-objective optimisation, we make trade-offs in the modelling and evaluation process (See Section \ref{Rep1-1section32}) including:


\begin{itemize}
\item single-objective optimisation that combines different measures of one of these aspects: similarity, efficiency or reproducibility
\item multi-objective optimisation considering all three aspects: similarity, efficiency and reproducibility
\item minimised information loss for faithfulness and minimum number of variables under the constraint of observability
\end{itemize}

Current studies combine different similarity measures to calculate the loss function for an optimised CNS \citep{zhou2018dynamic}. Exceptionally, \citep{tsiotas2020preferential} proposes a comparative directed graph to compare the similarity of CNSs based on the sum of different ranks considering different similarity measures. 

The integration and trade-offs between similarity, efficiency and reproducibility emerge as a research gap under the proposed evaluation criteria.
The data-driven scenarios have focused on minimised information loss and a minimum number of variables for years, which involves the discussion on feature selection/extraction and model validation. However, the features in the context of DT-CNS, simulation-based, data-driven or hybrid, differ between those scenarios due to the obseravability challenges and the simulation of informative and realistic features for DT-CNSs call for in-depth study in the future. 

Finally, to validate the performance of a DT-CNS under the evaluation criteria, data-driven CNSs, with more concerns related to the minimised information loss and the reproducible representation accuracy, can employ the cross-validation approaches through data split and repeated experiments. In contrast, for the simulation-based and hybrid CNSs with partial observability, CNS ensembles that create a series of CNS representations with quantifiable, reproducible and realistic characteristics
are expected to prove the validity of DT-CNSs.

\section{Results}
\label{Rep1-1section4}

In this section, we conduct experiments on simulation-based DT-CNSs to investigate the disaster resilience of different social networks considering the infection occurrences on these networks in an epidemic outbreak. We conduct a comparative analysis of simulation-based DT-CNSs driven by selected rules (preferential attachment and homophily) and distributions of the ascribed feature -- age.
We take the degree distribution of a scale-free network (composed of $90$ nodes and $1400$ edges, built with Barabasi-Albert network model \cite{barabasi1999emergence}) as the target for the simulation of the static networks. To simulate the social networks, we assume $90$ nodes, $1400$ edges and an encounter rate of $0.8$ for the respective node pairs. A simulation-based dynamic process takes place on the networks without changing its parameters over the temporal scale. We evaluate these models in terms of complexity and their faithfulness, considering the global level similarity (measured by the difference between degree distributions), simulation efficiency, and reproducibility of the same statistics.

We respectively term the involved simulation-based DT-CNSs as $DT-CNS_{U}$, $DT-CNS_{B}$, $DT-CNS_{I}$, $DT-CNS_{L}$, $DT-CNS_{R}$ and $DT-CNS^{H+}$, $DT-CNS^{H-}$ $DT-CNS^{P+}$, $DT-CNS^{P-}$ and $DT-CNS^{HP}$, to characterise the DT-CNS paradigms with uniform ($U$), bell ($B$), inverse bell ($I$), left-skewed ($L$) and right-skewed ($R$) age distributions, respectively driven by preferential attachment to larger feature values ($P+$), preferential attachment to smaller feature values ($P-$), heterophily ($H+$), homophily ($H-$), and rules that optimally incorporate the above principles for a target degree distribution.

\subsection{Simulation-based Networks}
In this section, we build the simulation-based networks with $90$ nodes, $1400$ edges and a single feature -- age, ascribed to each node. We presume that the age involved in disaster resilience range between $[0,90]$ while categorising the age groups every ten years \cite{russell2020estimating}. 

\subsubsection{Feature}
\begin{figure*}[htp] 
	\centering
	\includegraphics[width=1\linewidth]{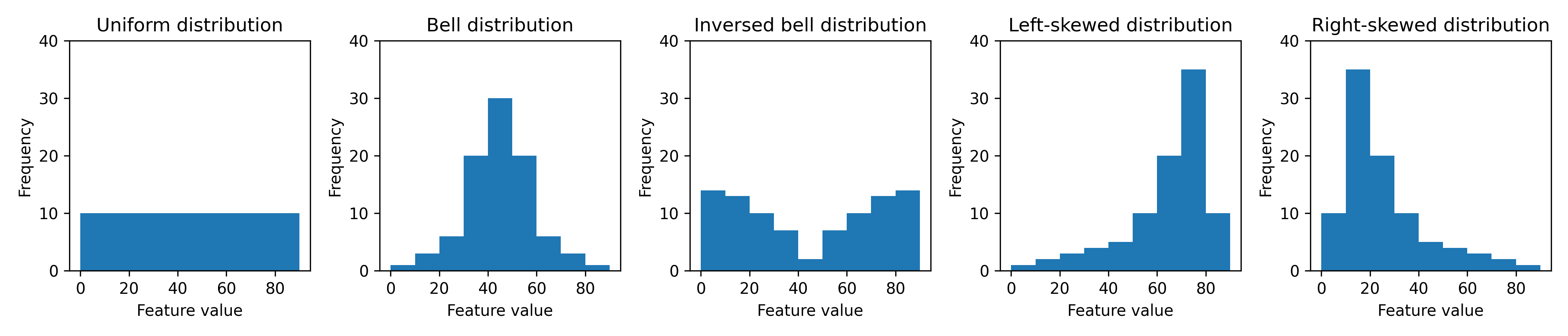}
	\caption{The age distributions used in the experiments.}
\label{featfig}
\end{figure*}

We simulate the uniform, the bell, the inverse bell, the left-skewed and the right-skewed shape distributions for a fixed number of nodes allocated in each age group. We also introduce uncertainty through a randomly generated age value within the range required by each age group (See Fig.~\ref{featfig}).

In Fig.~\ref{featfig}, the numbers of nodes in each age group vary depending on  distribution. There are $10$ nodes evenly allocated for each age group in a uniform distribution. In contrast, more/fewer nodes are in their 40s in a bell/inverse bell distribution than in other age groups. In a left-skewed distribution, most nodes are over 60, and vice versa for a right-skewed distribution. We evaluate the age diversity with the Hill number, which measures the effective number of equally abundant species (age groups)~\cite{chao2014unifying,alberdi2019guide}. The varying diversity levels differentiate the respective age distributions, indicating the population's age differences and resulting in different connection patterns. 

\begin{figure*}[htp] 
	\centering
	\includegraphics[width=0.75\linewidth]{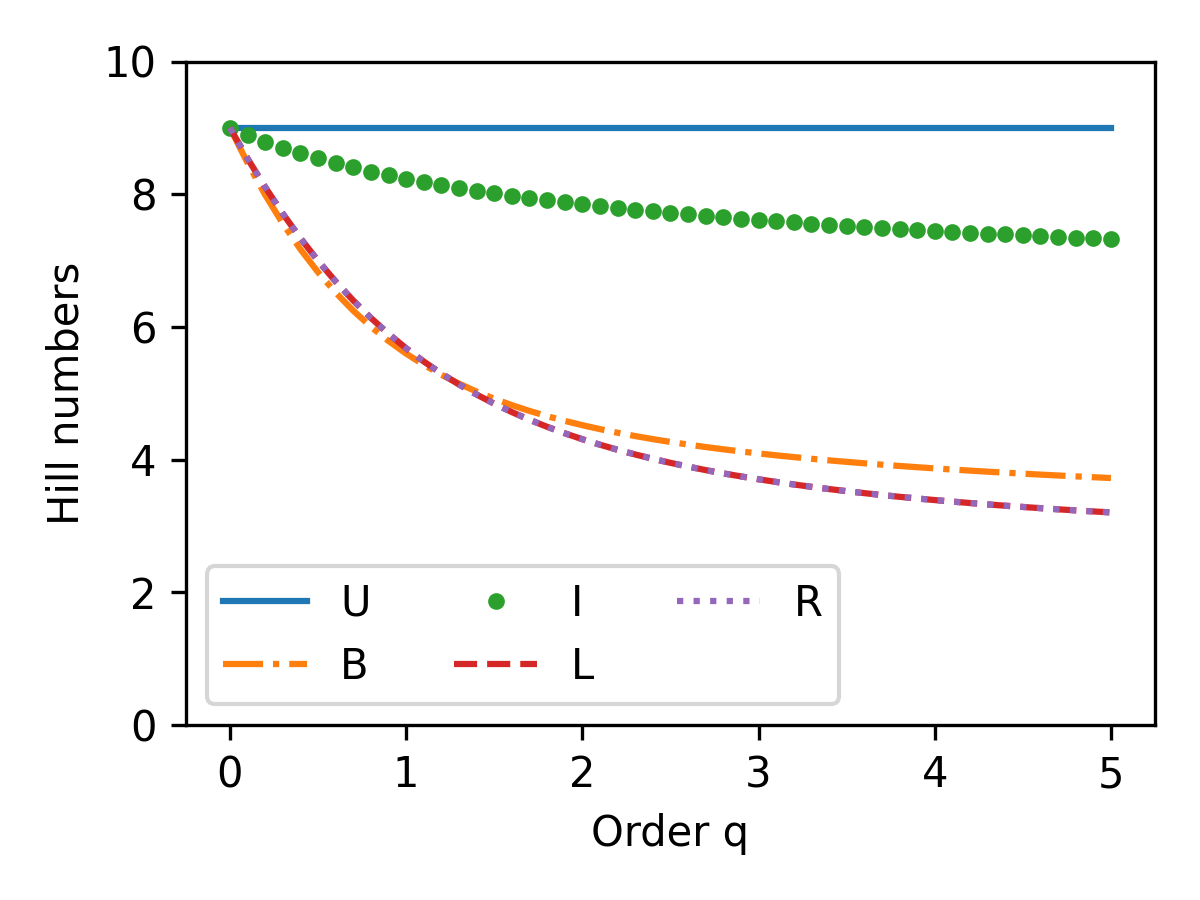}
	\caption{The Hill number for each age distribution.}
\label{hill}
\end{figure*}

In Fig.~\ref{hill}, we calculate the Hill numbers with an order value that determines the sensitivity to the relative frequencies of the species (age groups), valuing between $0$ and $5$. A higher order value indicates a higher sensitivity level to the relative number of nodes allocated in each age group. With an order value of $0$, the Hill numbers get insensitive to the abundance of the respective species (age groups) and keep at $9$ the number of age groups for all the age distributions \citep{qiu2016effects}. With the same order value, the higher Hill number indicates higher diversity. As the age distributions transform from the uniform, inverse bell, bell to skewed age distributions, their Hill numbers decrease, indicating less age diversity. 

\subsubsection{Preference}
Given the age feature and the different rules of network formation, including the preferential attachment to larger feature values (P+), the preferential attachment to smaller feature values (P-), the heterophily (H+), the homophily (H-) as well as the optimised combination of both the preferential attachment and the homophily principles (PH), the scores of node pairs vary a lot and indicate different strengths of relationships. The parameter set-ups of preferences and weights of preferences is included in Table \ref{params} in the appendix \ref{app}.

\subsubsection{Network Topology}

\begin{figure*}[htp] 
	\centering
	\subfigure[$DT-CNS^{P+}_{U}$]{
		\begin{minipage}[b]{0.185\linewidth}
			\includegraphics[width=1\linewidth]{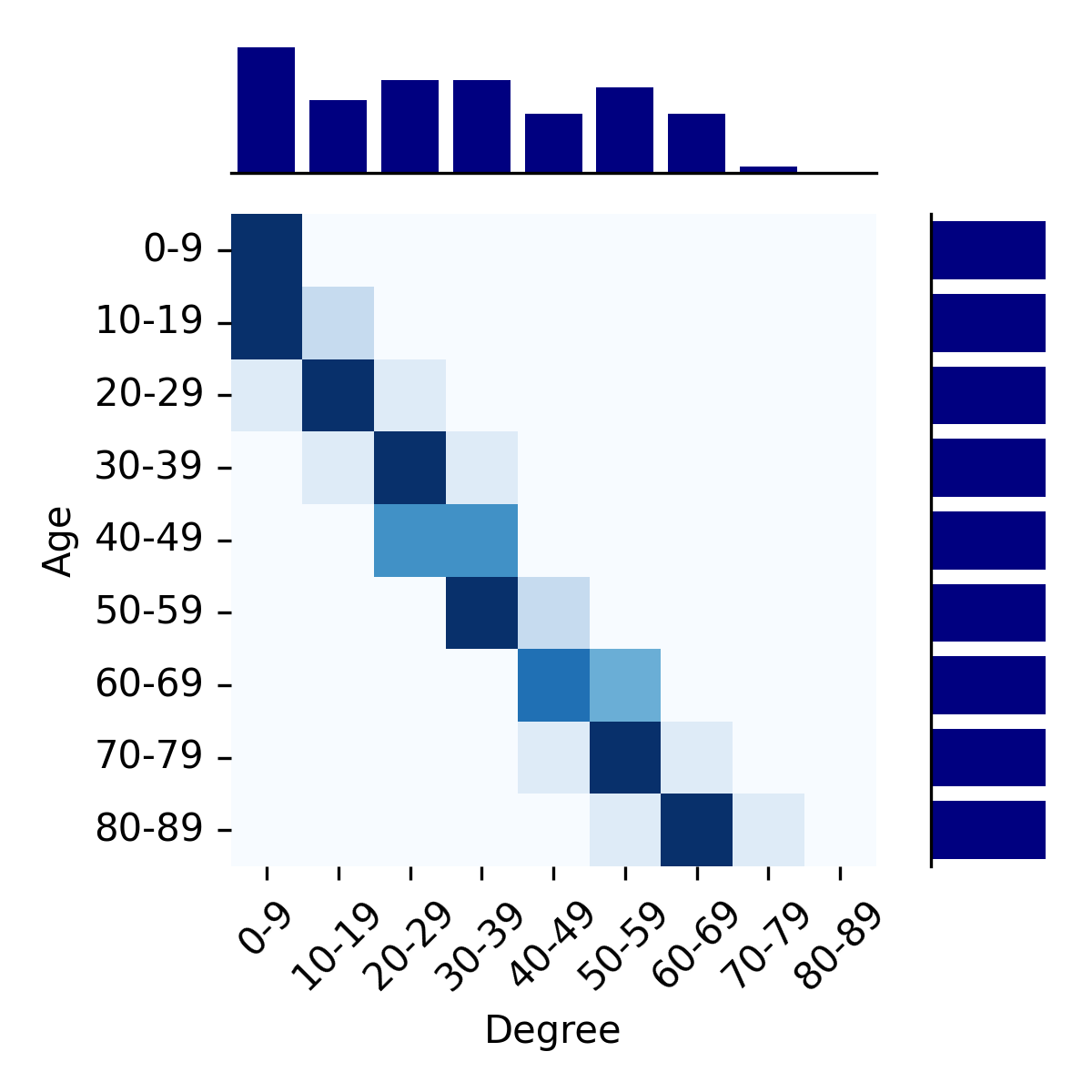}
	\end{minipage}}	
 	\subfigure[$DT-CNS^{P+}_{B}$]{
		\begin{minipage}[b]{0.185\linewidth}
			\includegraphics[width=1\linewidth]{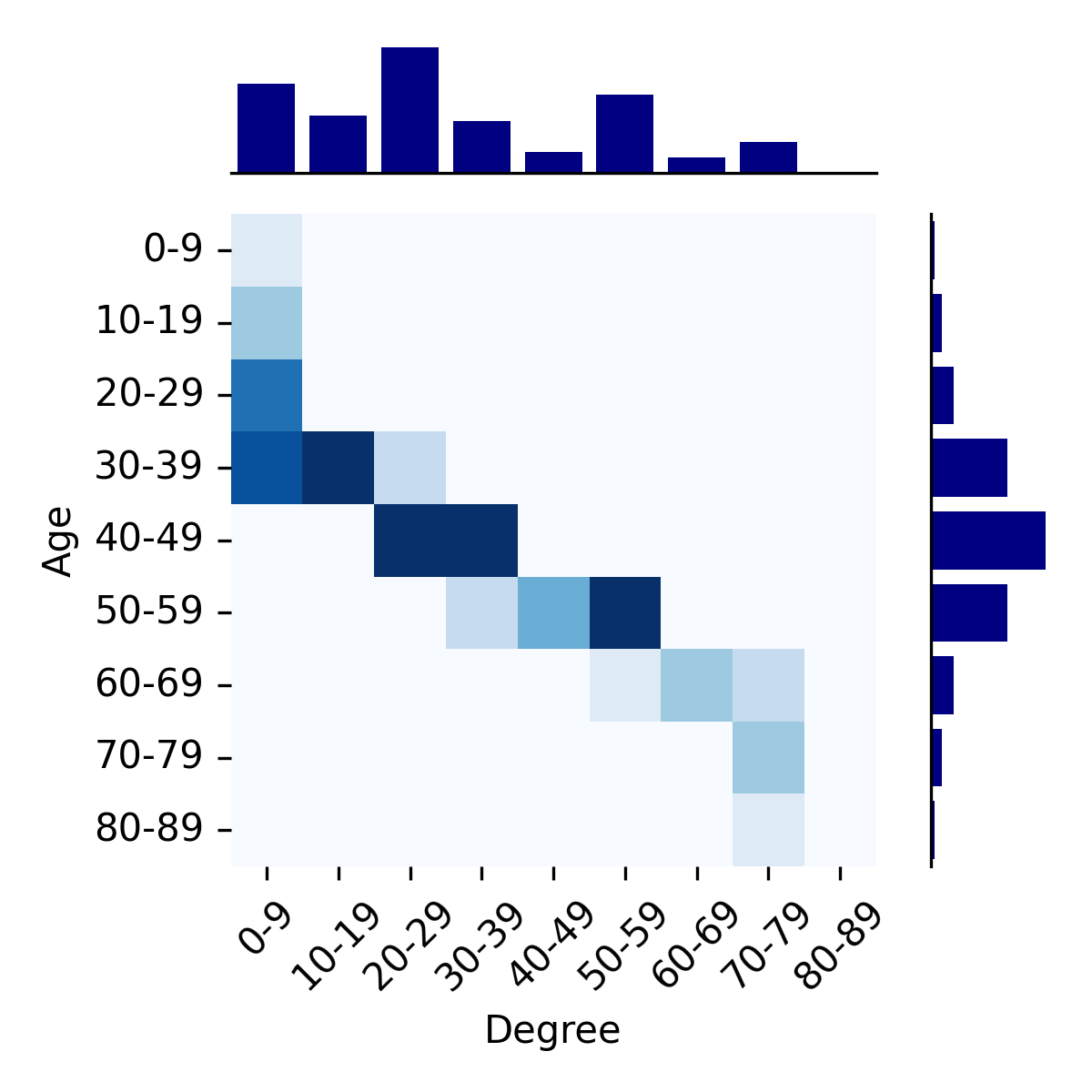}
	\end{minipage}}	
 	\subfigure[$DT-CNS^{P+}_{I}$]{
		\begin{minipage}[b]{0.185\linewidth}
			\includegraphics[width=1\linewidth]{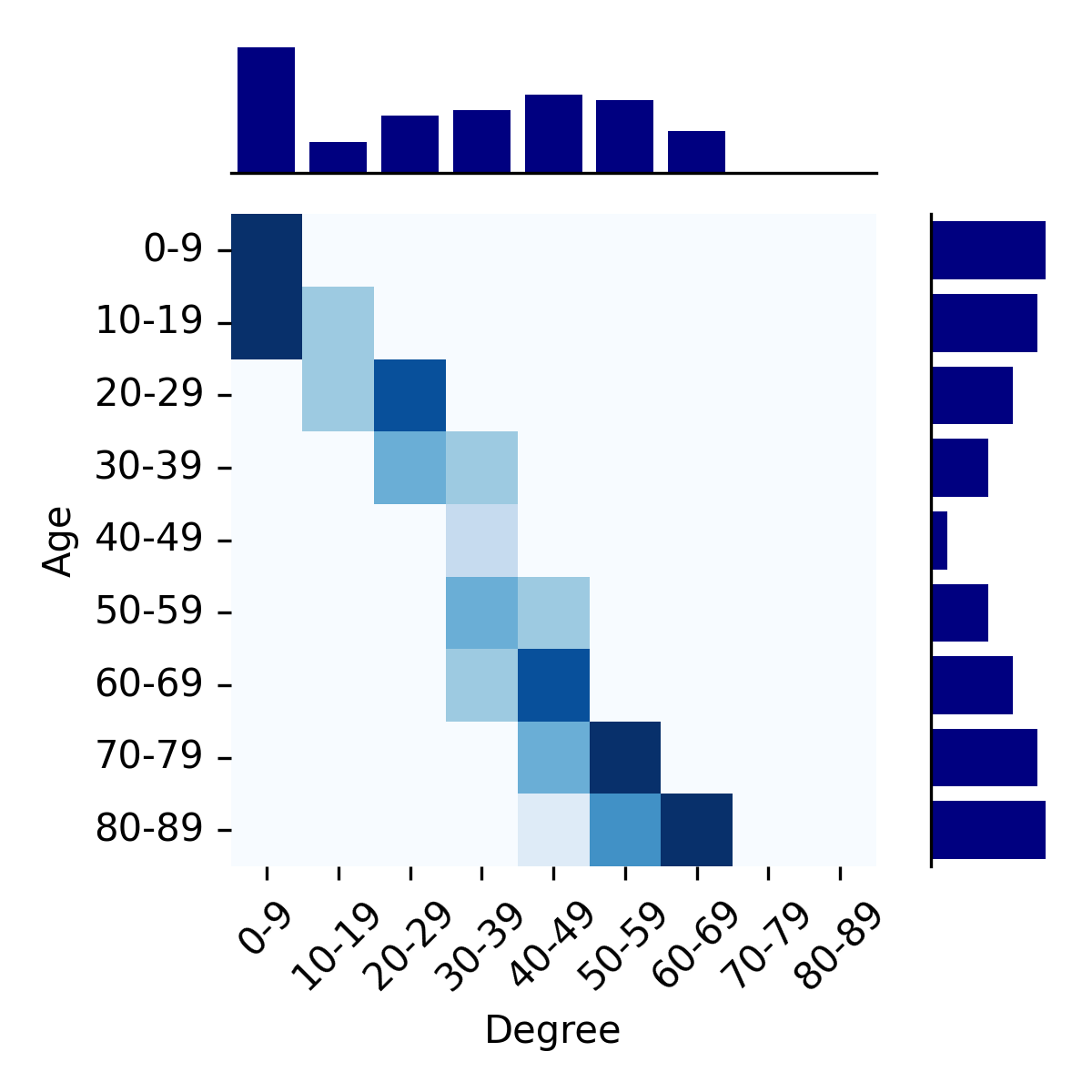}
	\end{minipage}}	
  	\subfigure[$DT-CNS^{P+}_{L}$]{
		\begin{minipage}[b]{0.185\linewidth}
			\includegraphics[width=1\linewidth]{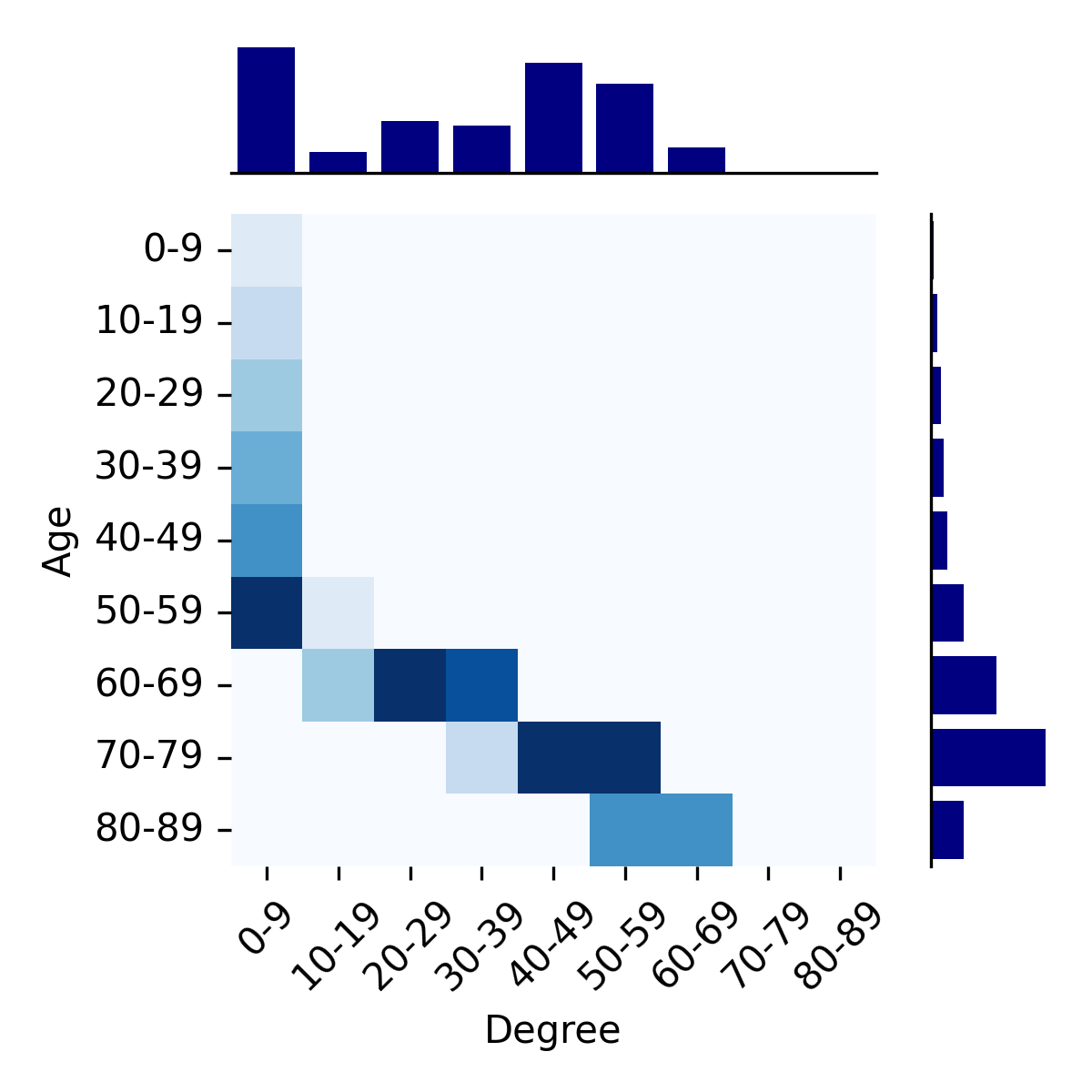}
	\end{minipage}}	
  	\subfigure[$DT-CNS^{P+}_{R}$]{
		\begin{minipage}[b]{0.185\linewidth}
			\includegraphics[width=1\linewidth]{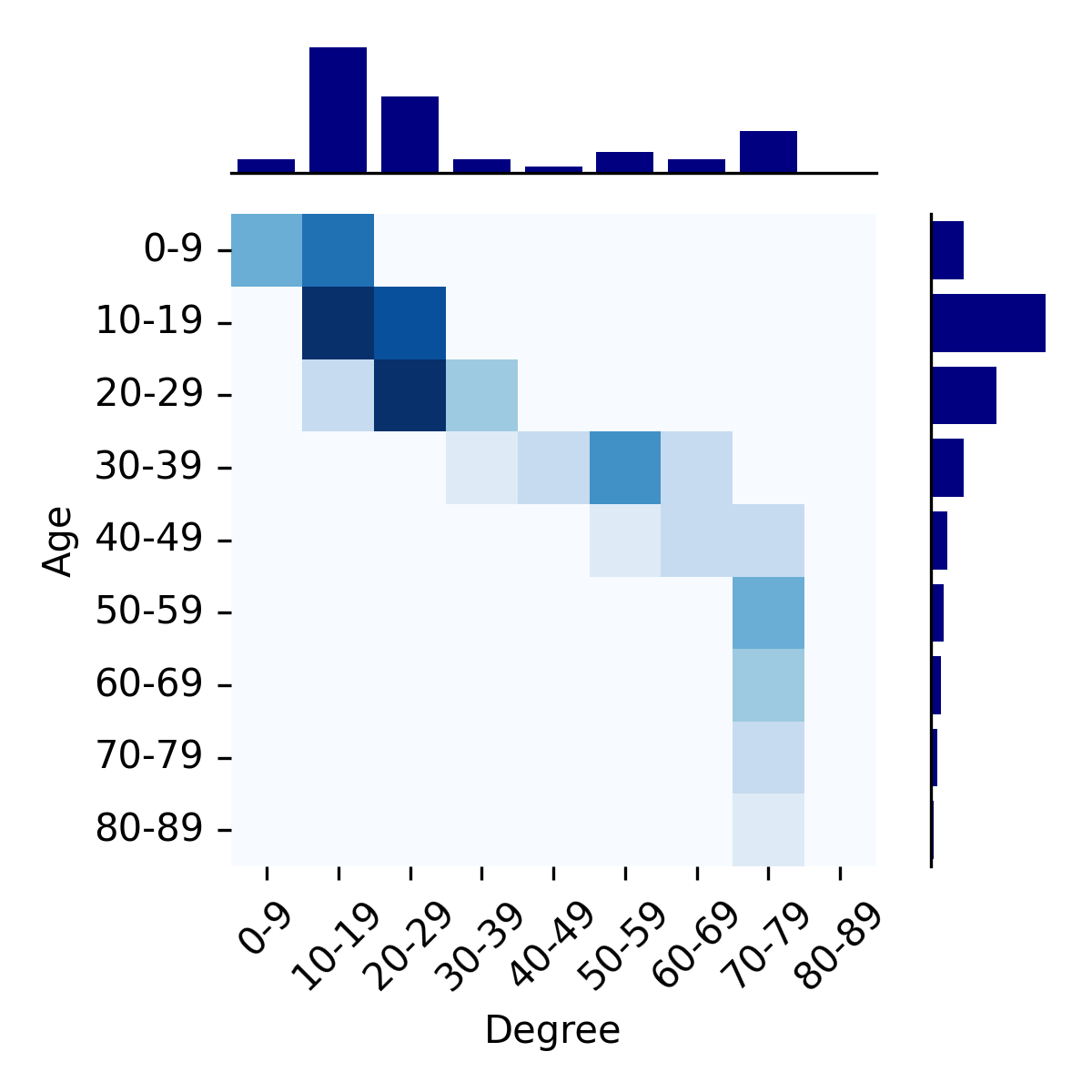}
	\end{minipage}}	\\
 	\subfigure[$DT-CNS^{P-}_{U}$]{
		\begin{minipage}[b]{0.185\linewidth}
			\includegraphics[width=1\linewidth]{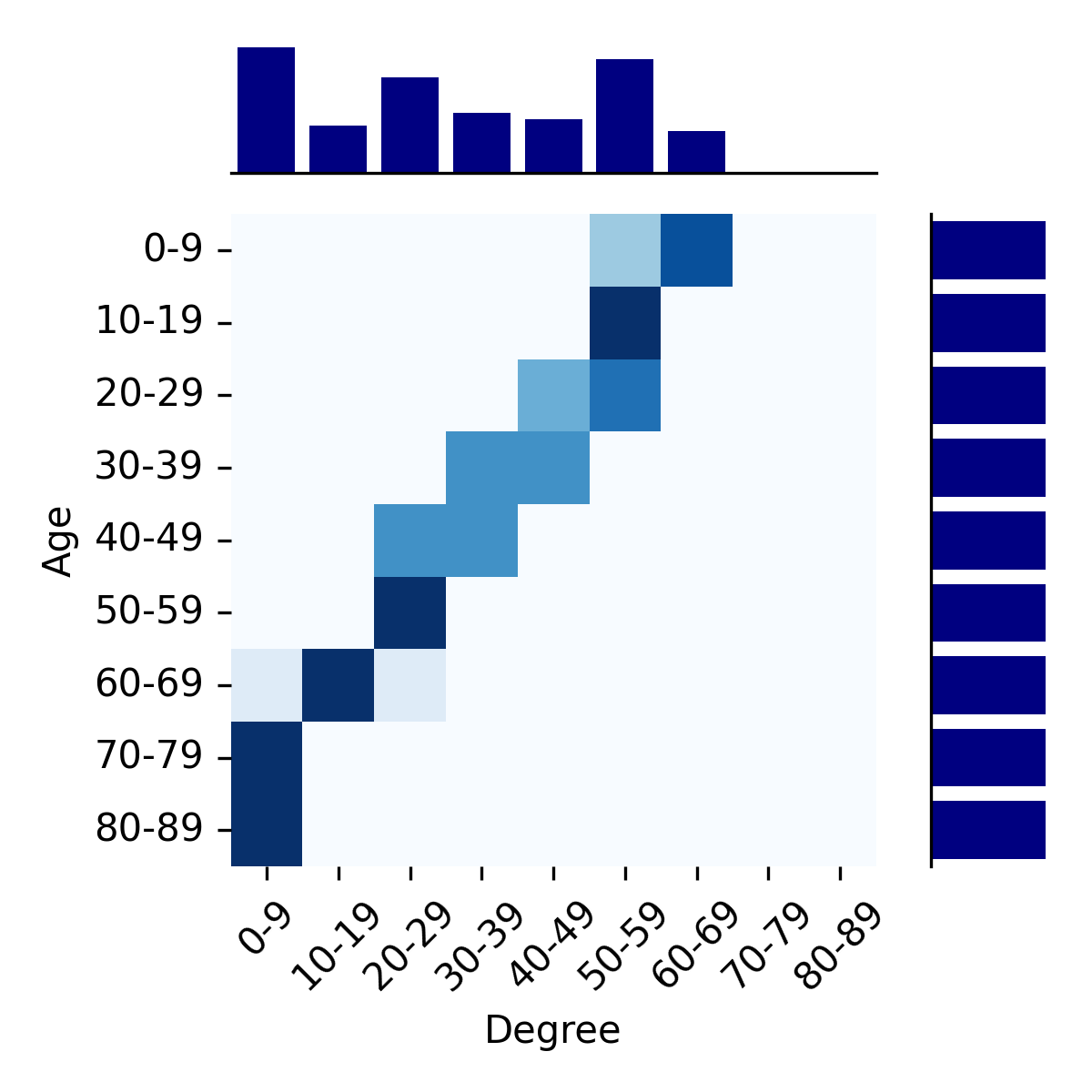}
	\end{minipage}}	
 	\subfigure[$DT-CNS^{P-}_{B}$]{
		\begin{minipage}[b]{0.185\linewidth}
			\includegraphics[width=1\linewidth]{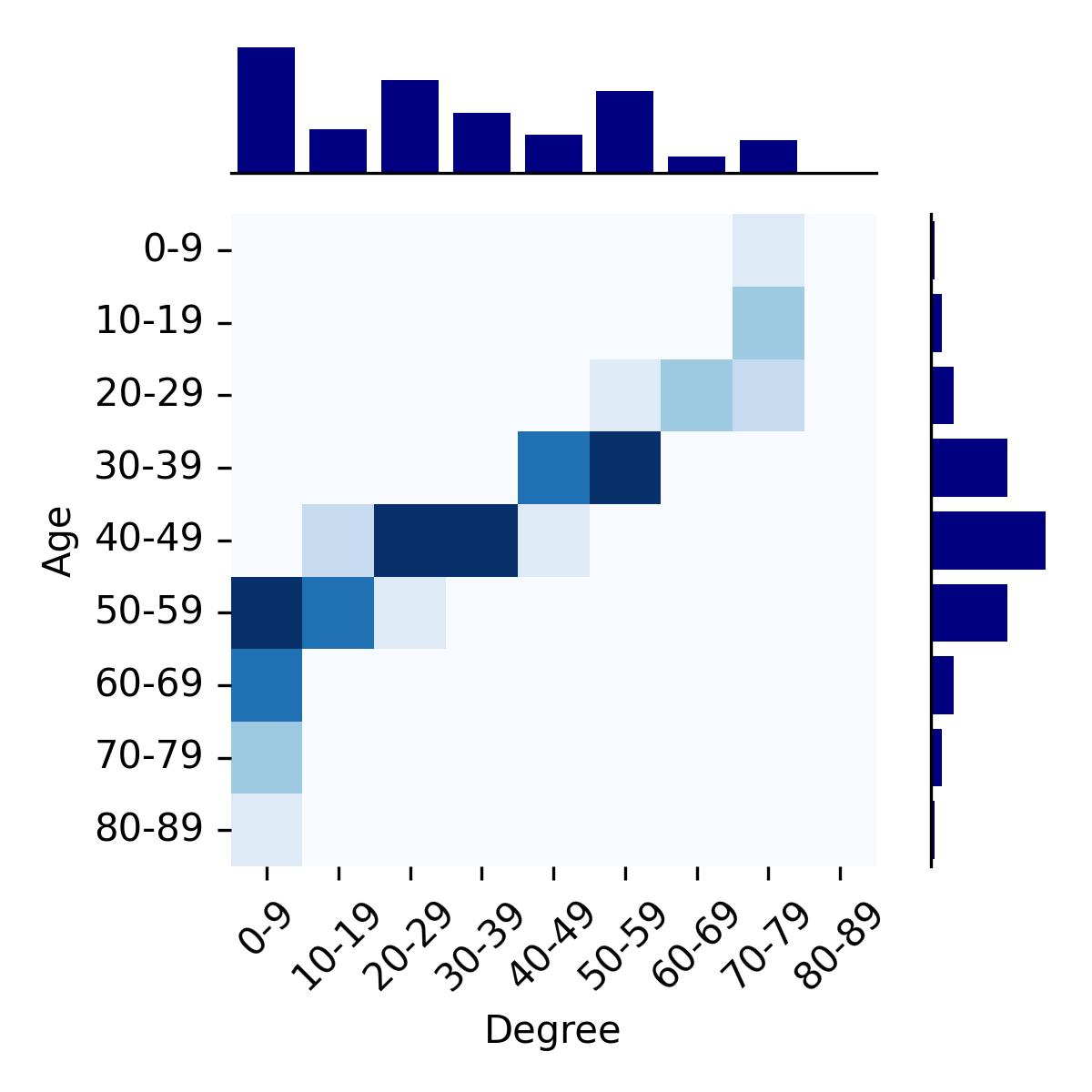}
	\end{minipage}}	
 	\subfigure[$DT-CNS^{P-}_{I}$]{
		\begin{minipage}[b]{0.185\linewidth}
			\includegraphics[width=1\linewidth]{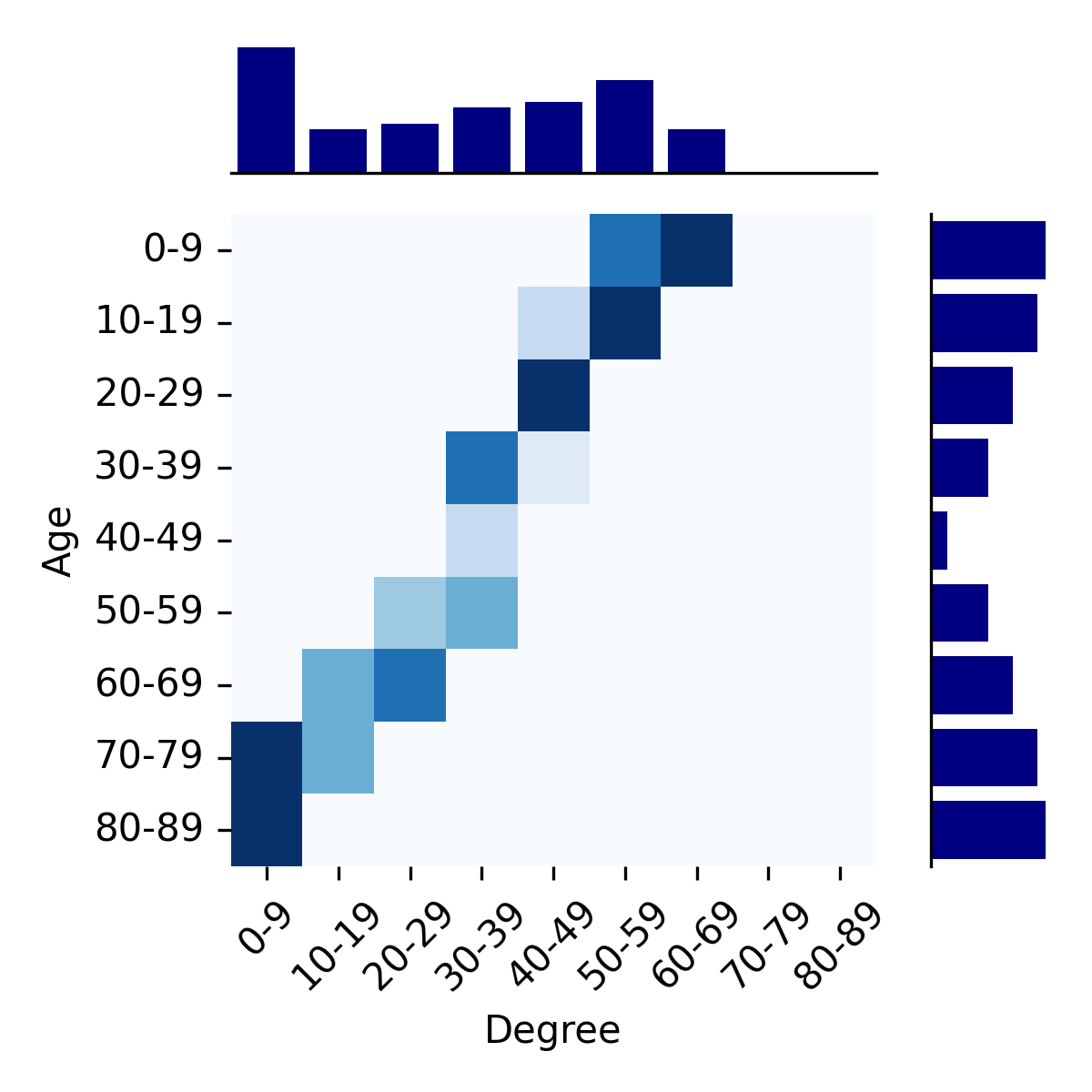}
	\end{minipage}}	
  	\subfigure[$DT-CNS^{P-}_{L}$]{
		\begin{minipage}[b]{0.185\linewidth}
			\includegraphics[width=1\linewidth]{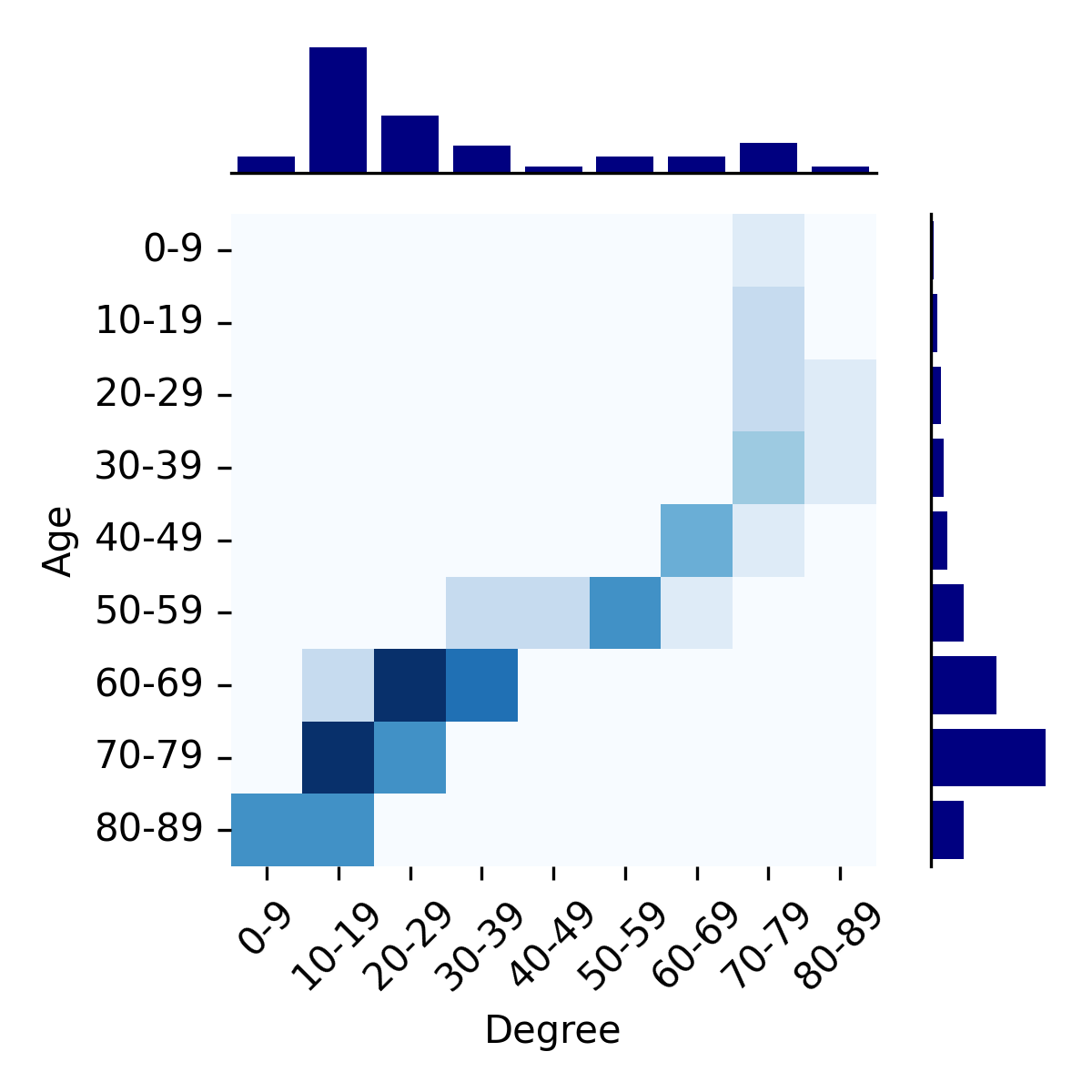}
	\end{minipage}}	
  	\subfigure[$DT-CNS^{P-}_{R}$]{
		\begin{minipage}[b]{0.185\linewidth}
			\includegraphics[width=1\linewidth]{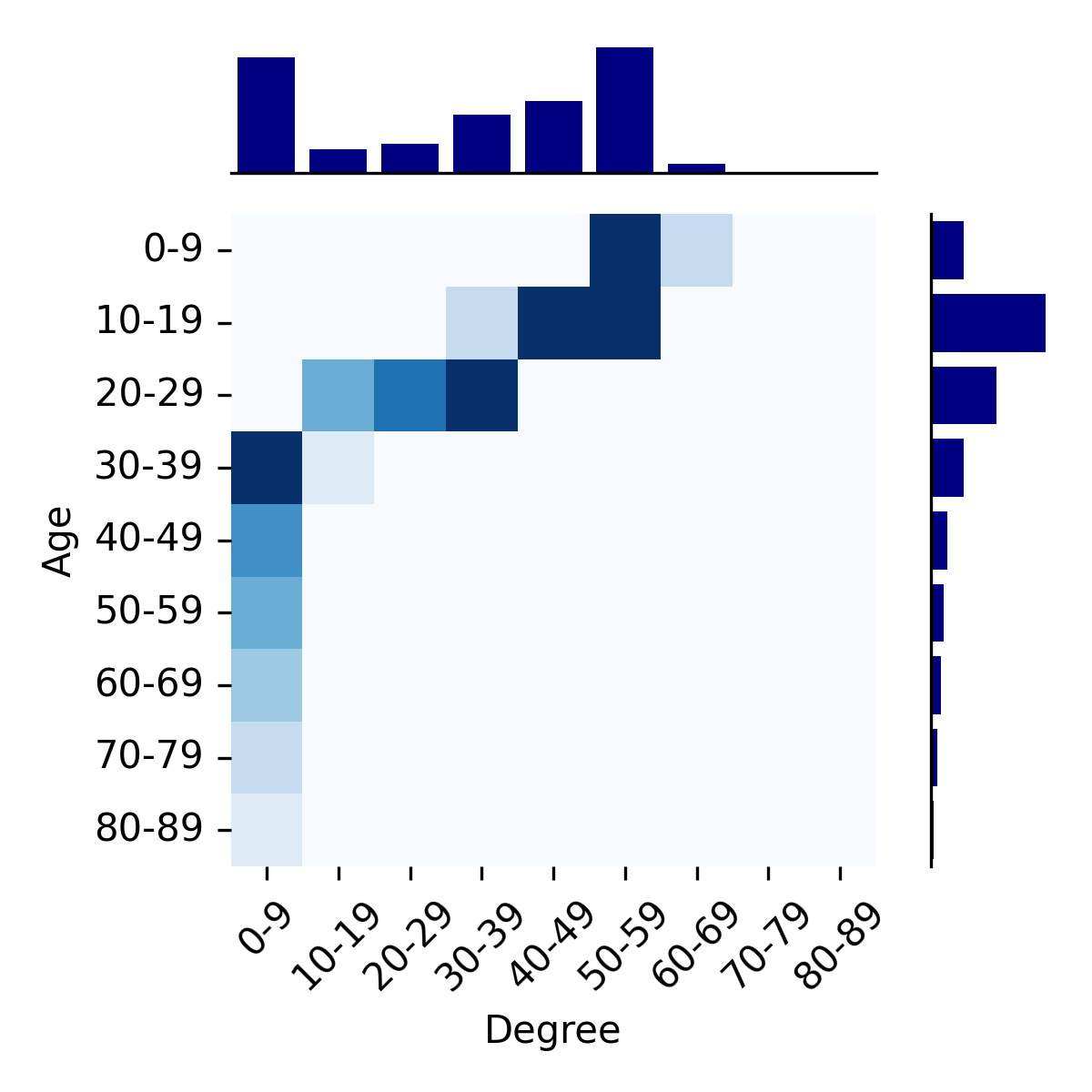}
	\end{minipage}}	\\
 	\subfigure[$DT-CNS^{H+}_{U}$]{
		\begin{minipage}[b]{0.185\linewidth}
			\includegraphics[width=1\linewidth]{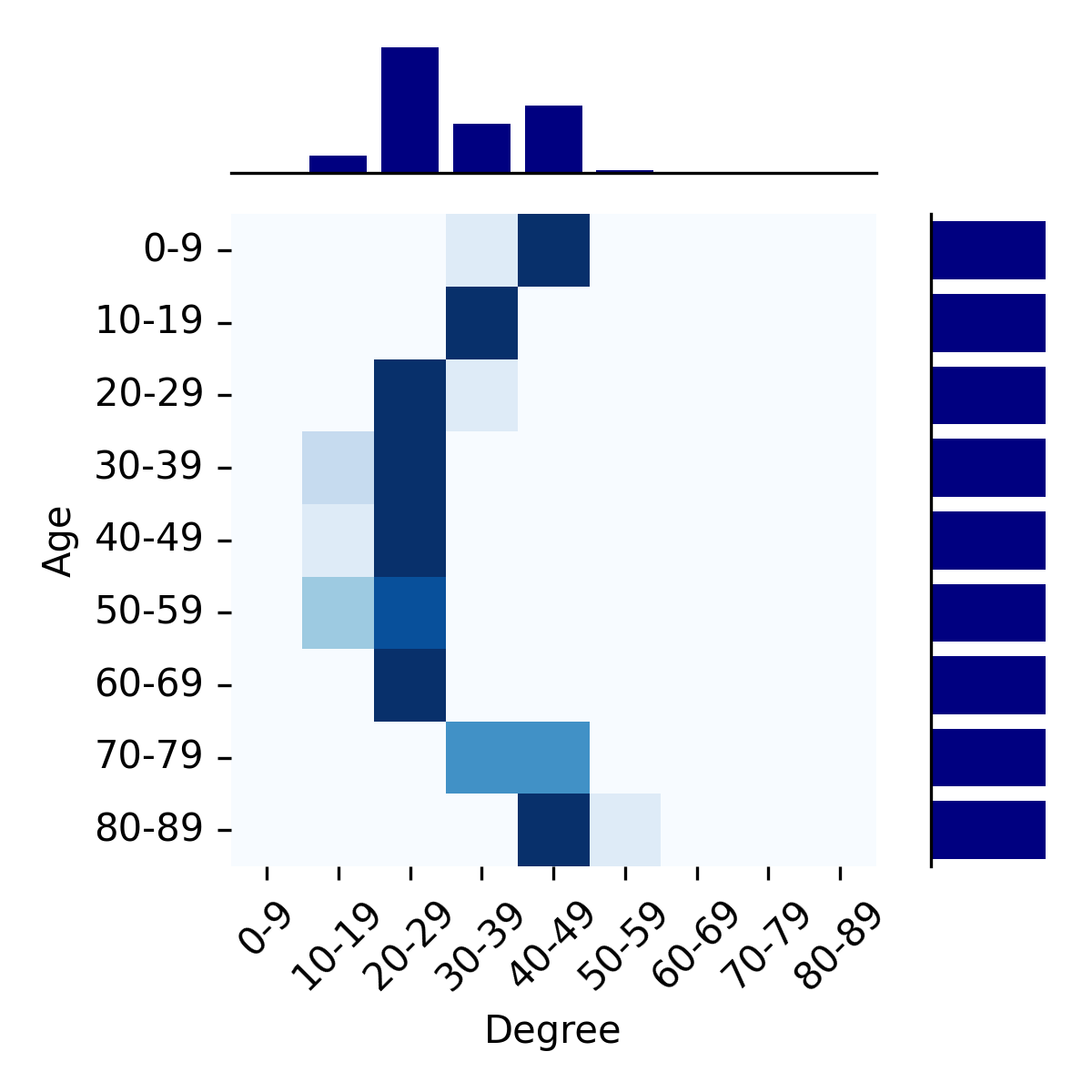}
	\end{minipage}}	
 	\subfigure[$DT-CNS^{H+}_{B}$]{
		\begin{minipage}[b]{0.185\linewidth}
			\includegraphics[width=1\linewidth]{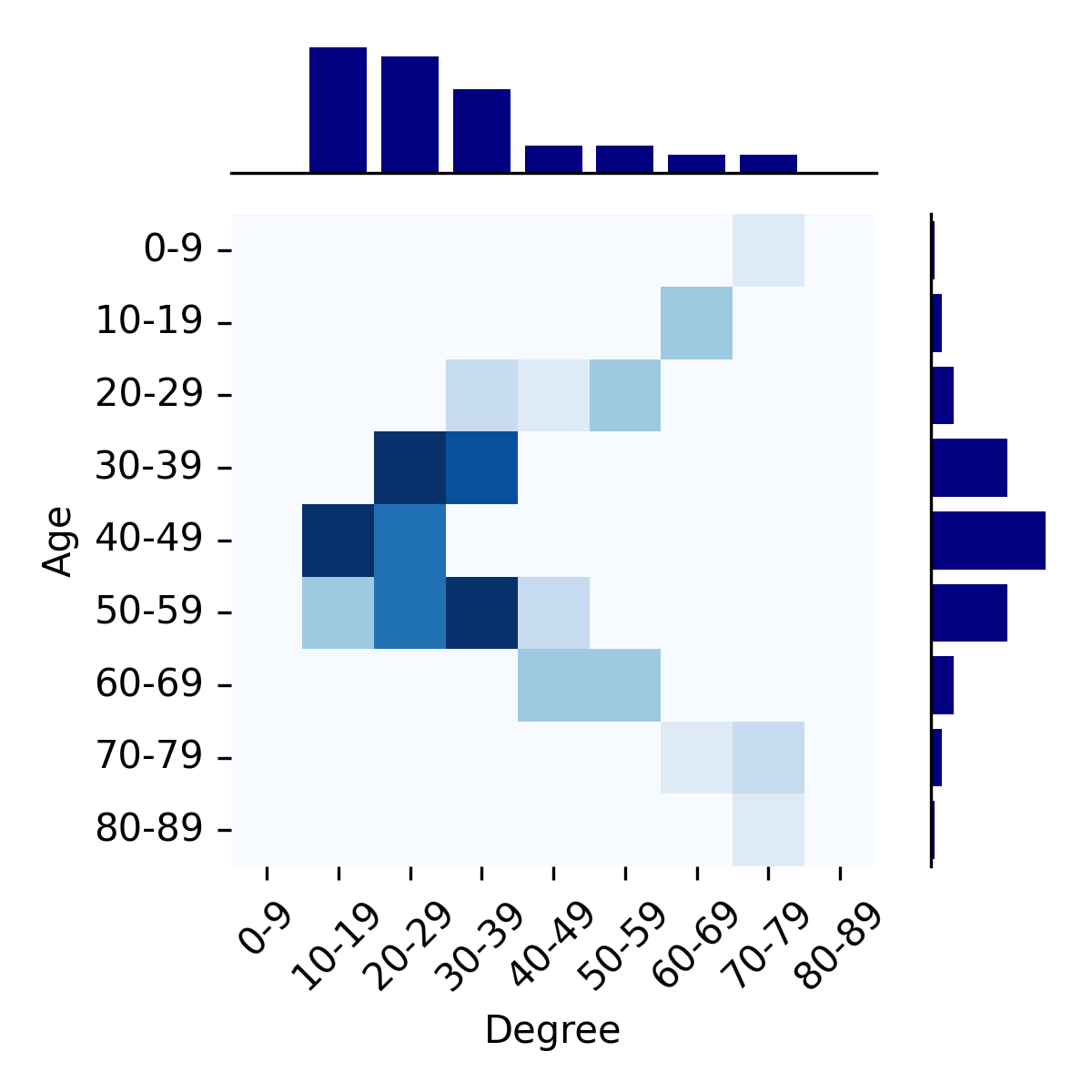}
	\end{minipage}}	
 	\subfigure[$DT-CNS^{H+}_{I}$]{
		\begin{minipage}[b]{0.185\linewidth}
			\includegraphics[width=1\linewidth]{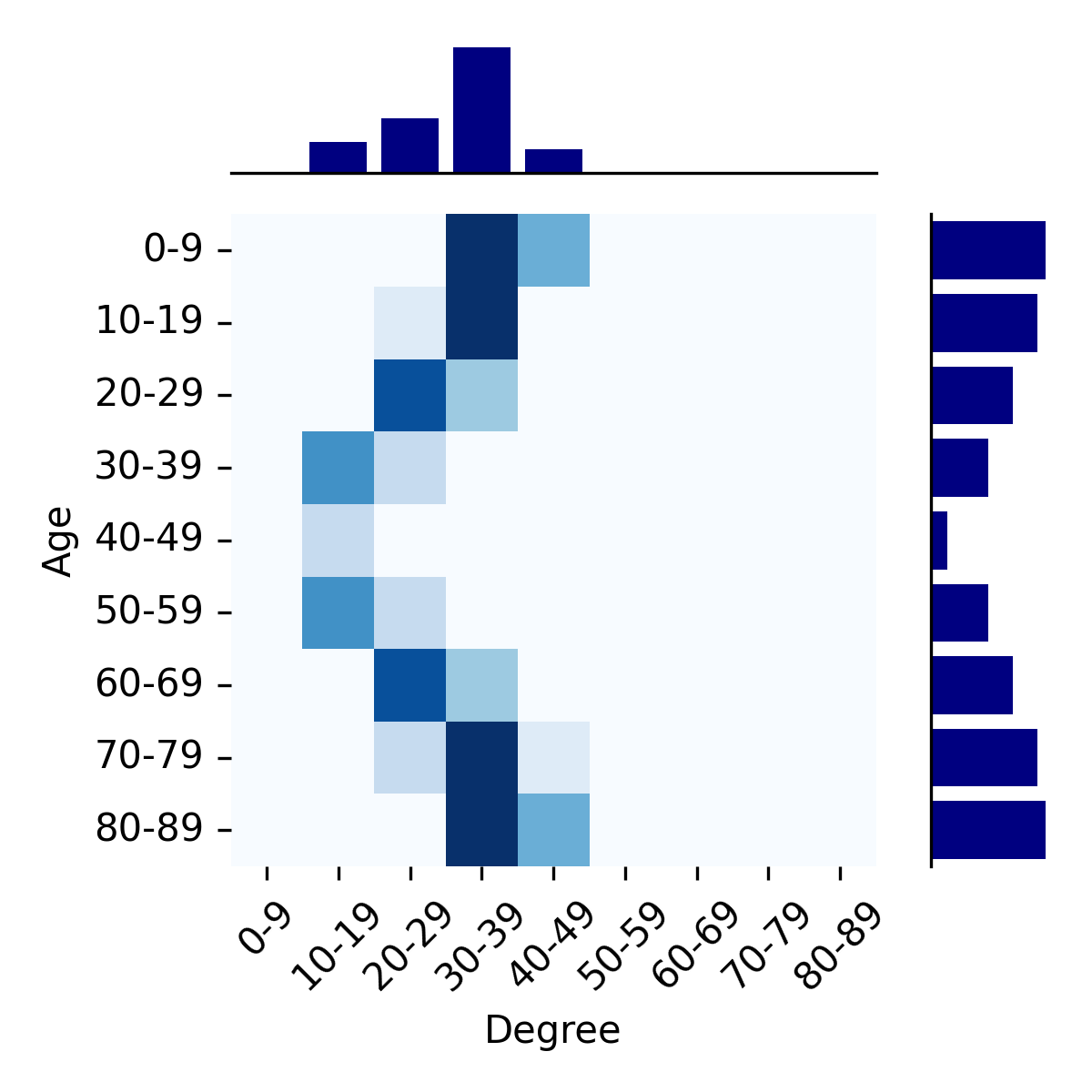}
	\end{minipage}}	
  	\subfigure[$DT-CNS^{H+}_{L}$]{
		\begin{minipage}[b]{0.185\linewidth}
			\includegraphics[width=1\linewidth]{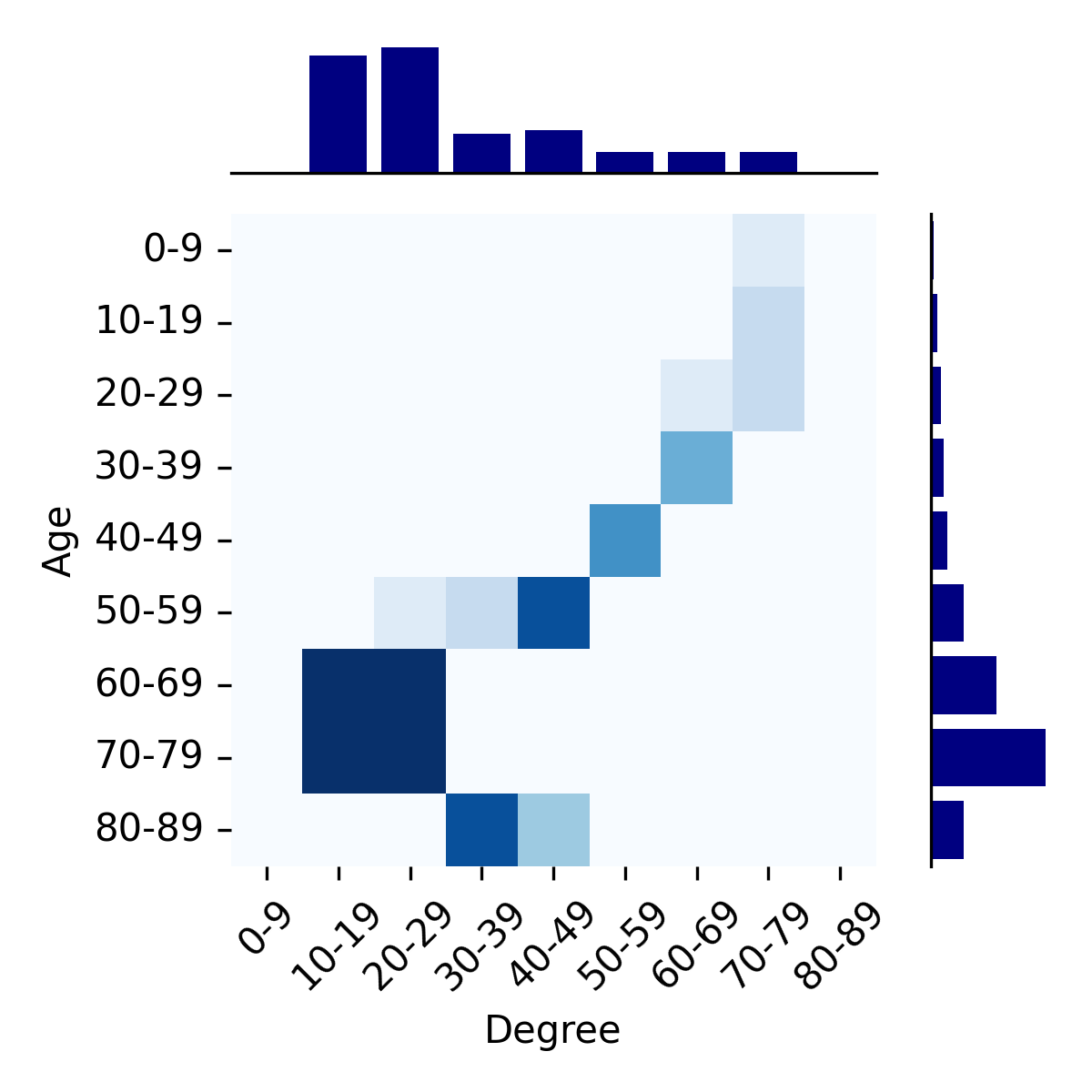}
	\end{minipage}}	
  	\subfigure[$DT-CNS^{H+}_{R}$]{
		\begin{minipage}[b]{0.185\linewidth}
			\includegraphics[width=1\linewidth]{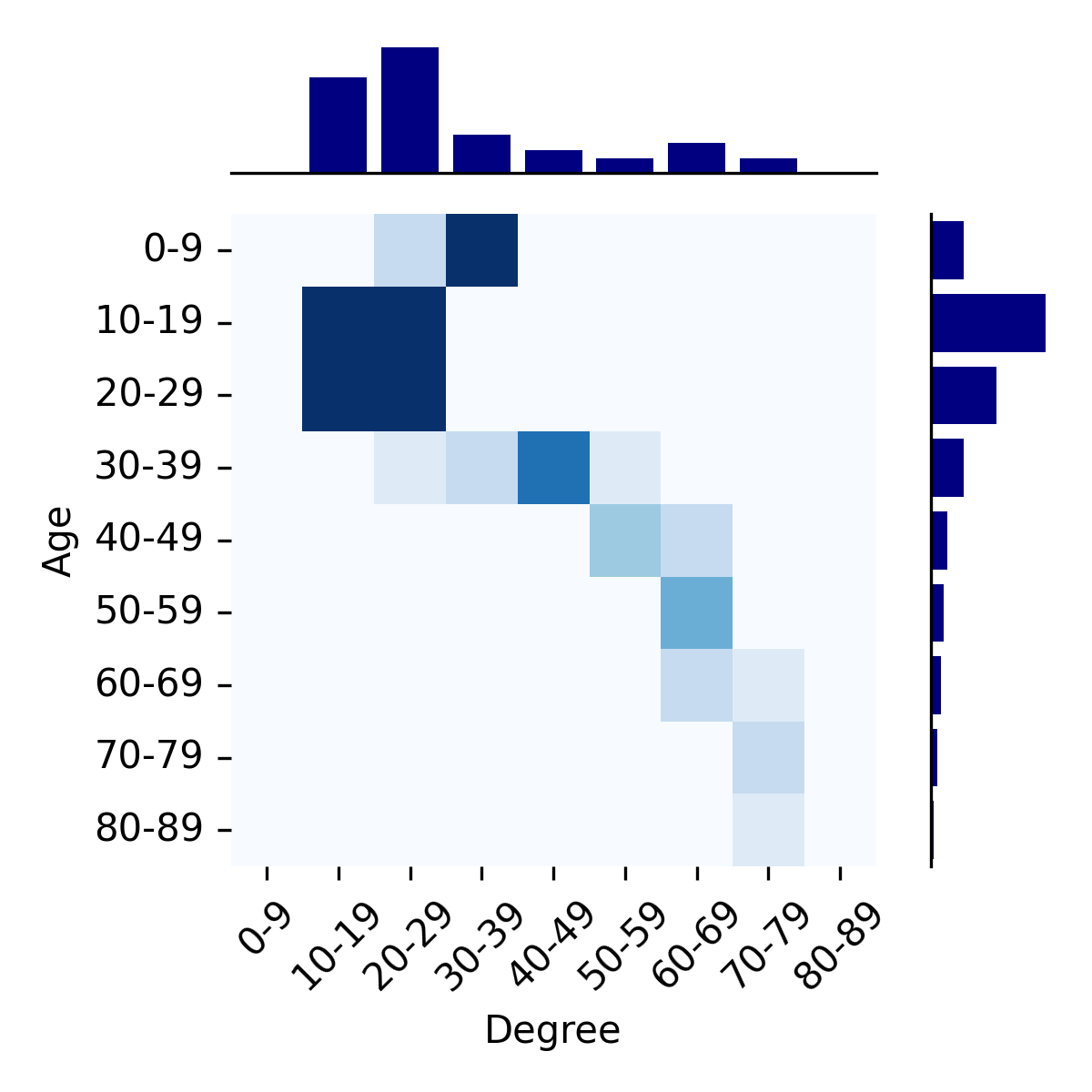}
	\end{minipage}}	\\
 	\subfigure[$DT-CNS^{H-}_{U}$]{
		\begin{minipage}[b]{0.185\linewidth}
			\includegraphics[width=1\linewidth]{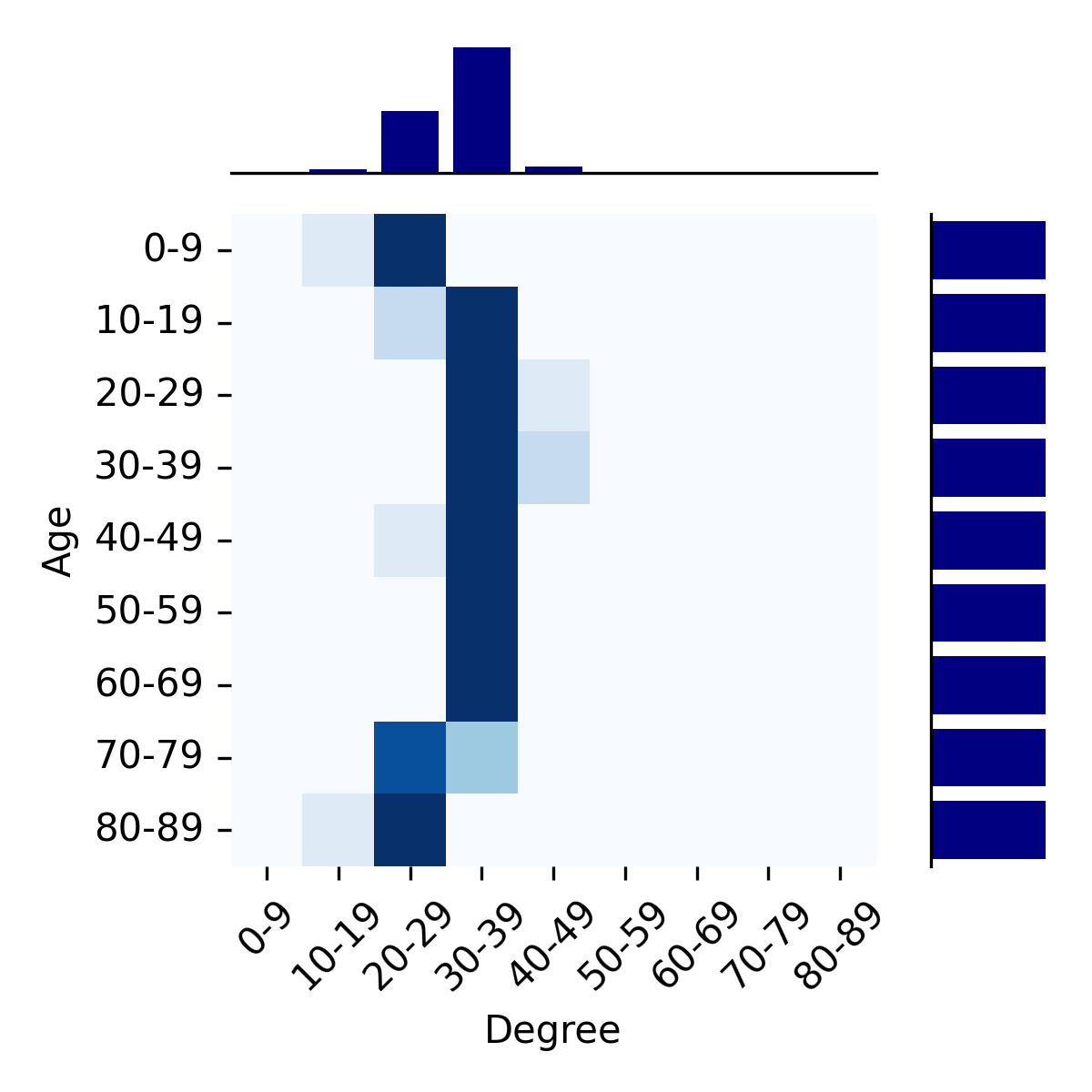}
	\end{minipage}}	
 	\subfigure[$DT-CNS^{H-}_{B}$]{
		\begin{minipage}[b]{0.185\linewidth}
			\includegraphics[width=1\linewidth]{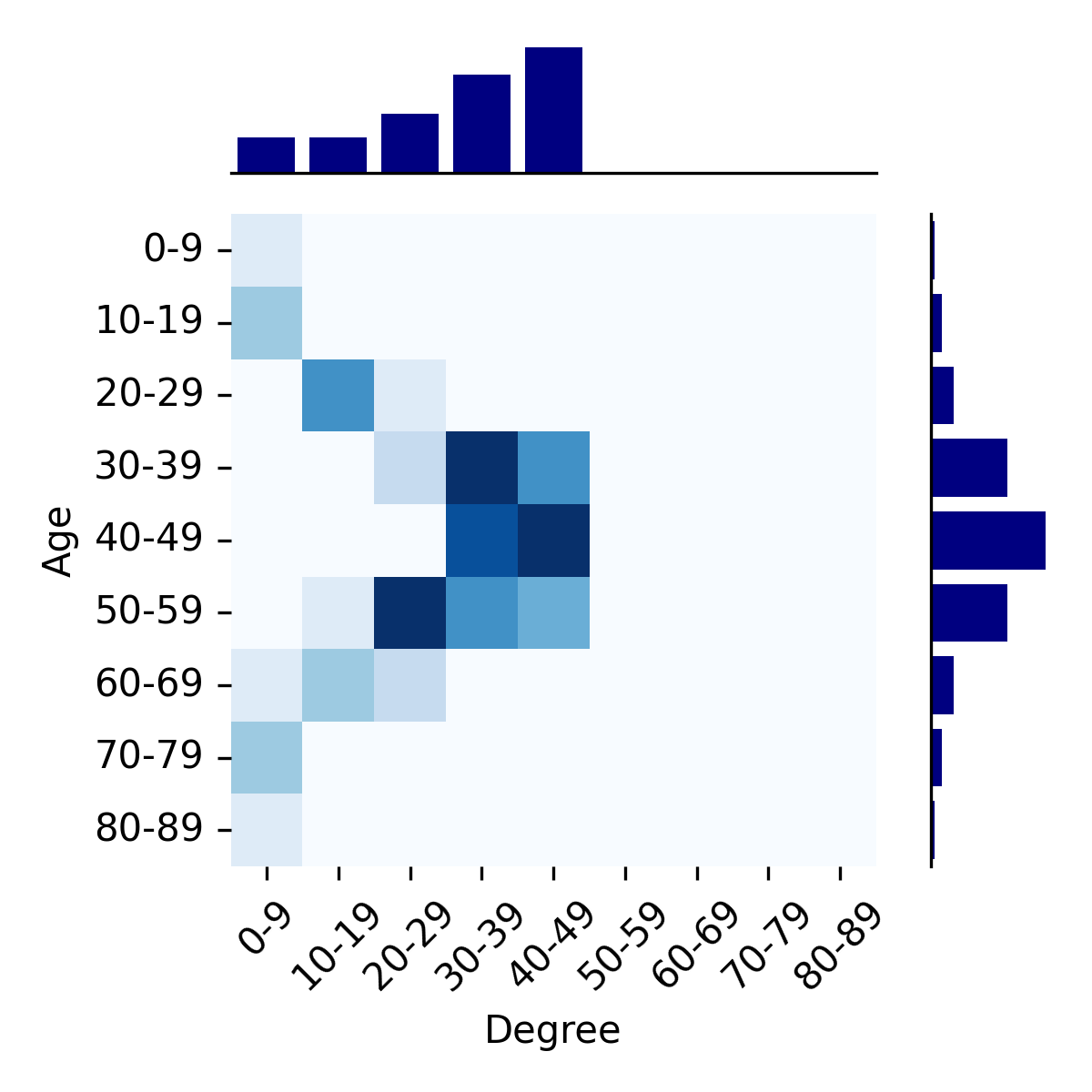}
	\end{minipage}}	
 	\subfigure[$DT-CNS^{H-}_{I}$]{
		\begin{minipage}[b]{0.185\linewidth}
			\includegraphics[width=1\linewidth]{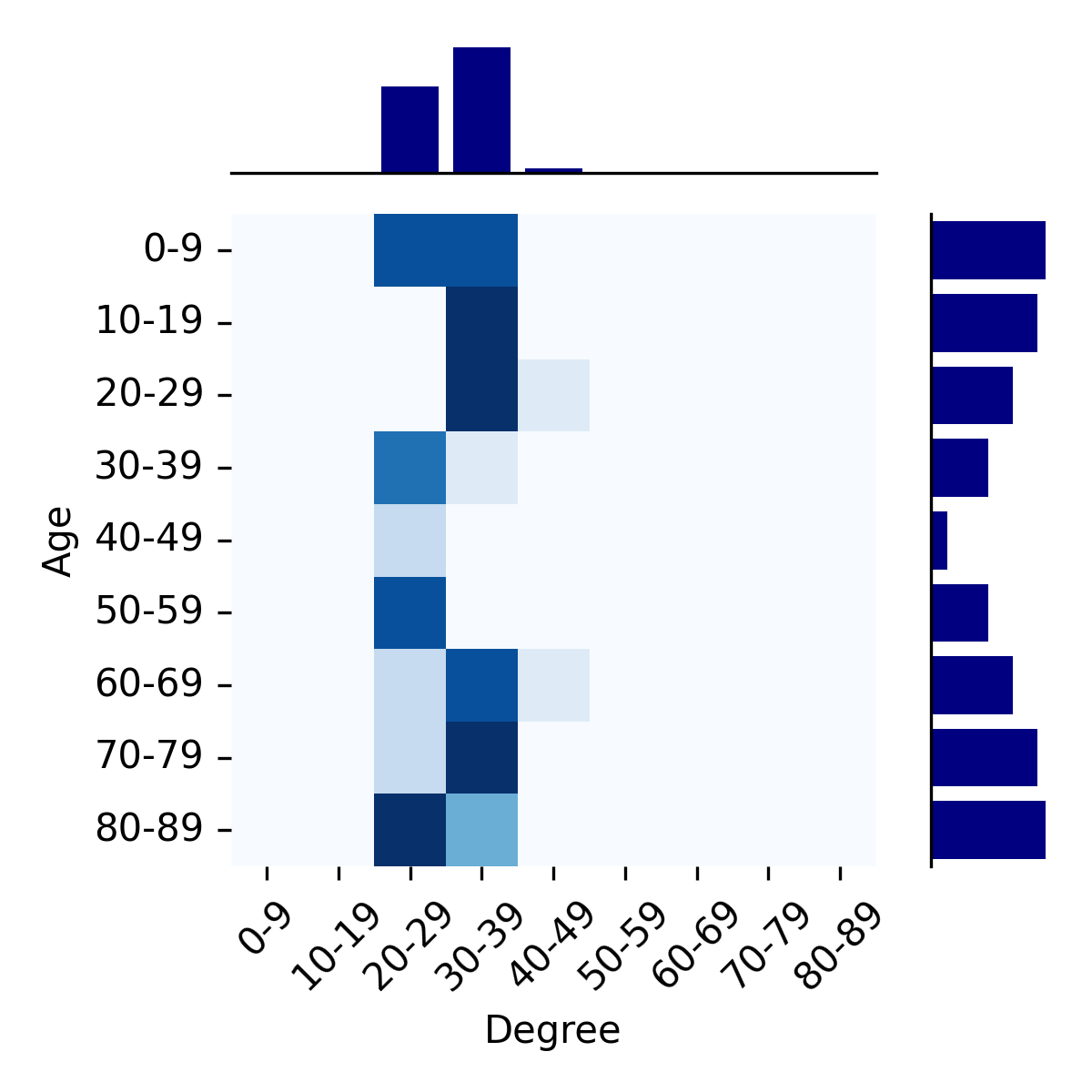}
	\end{minipage}}	
  	\subfigure[$DT-CNS^{H-}_{L}$]{
		\begin{minipage}[b]{0.185\linewidth}
			\includegraphics[width=1\linewidth]{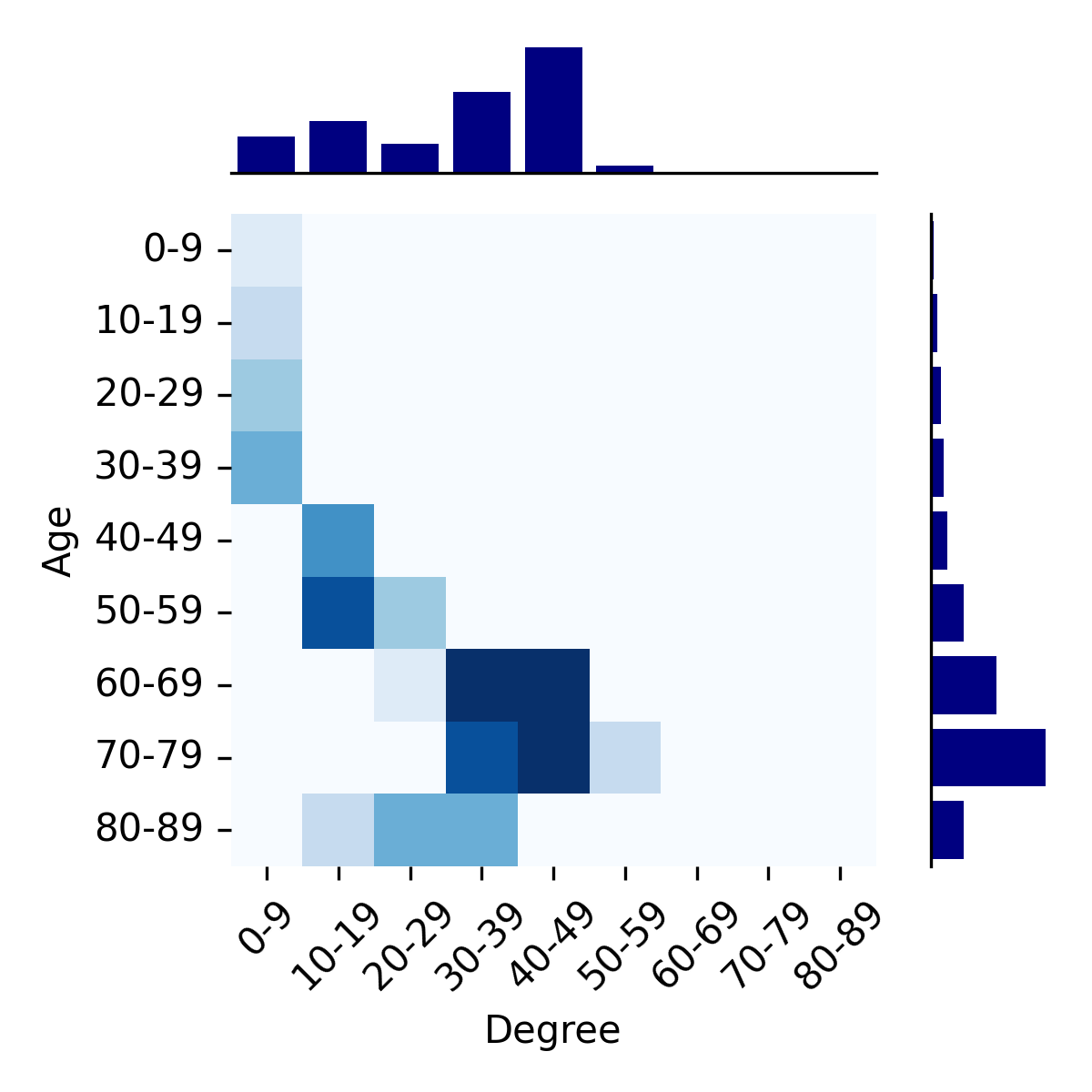}
	\end{minipage}}	
  	\subfigure[$DT-CNS^{H-}_{R}$]{
		\begin{minipage}[b]{0.185\linewidth}
			\includegraphics[width=1\linewidth]{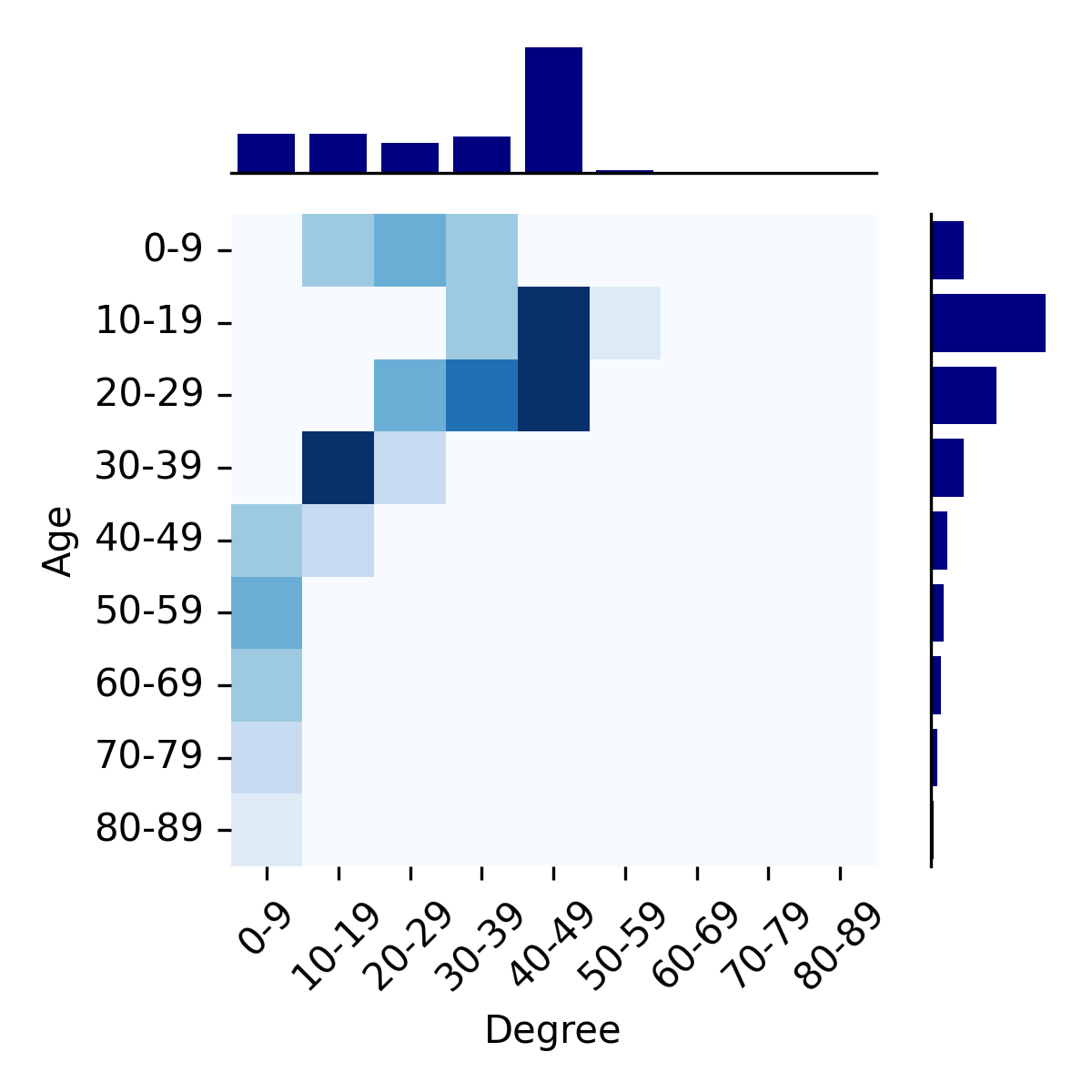}
	\end{minipage}}	\\
 	\subfigure[$DT-CNS^{PH}_{U}$]{
		\begin{minipage}[b]{0.185\linewidth}
			\includegraphics[width=1\linewidth]{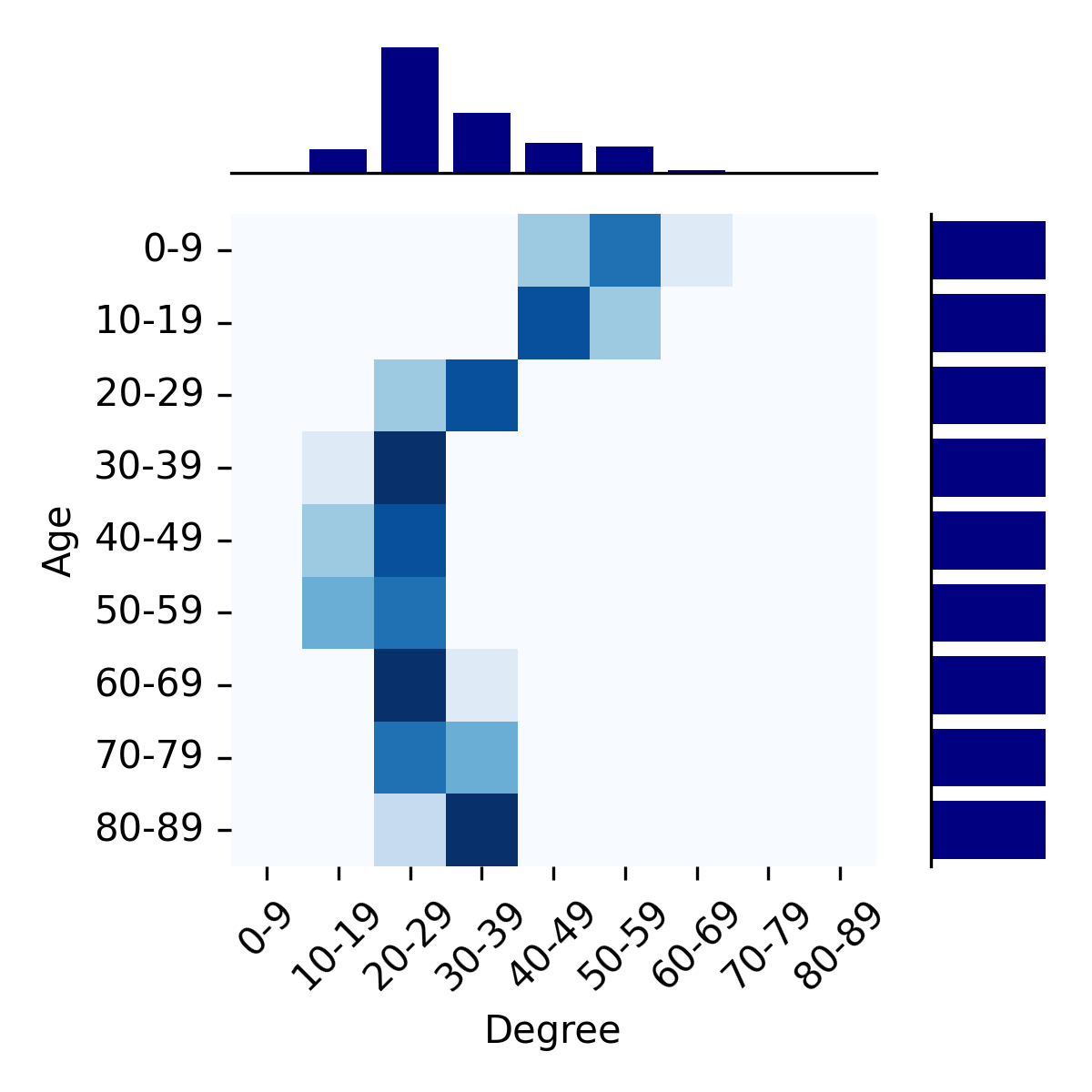}
	\end{minipage}}	
 	\subfigure[$DT-CNS^{PH}_{B}$]{
		\begin{minipage}[b]{0.185\linewidth}
			\includegraphics[width=1\linewidth]{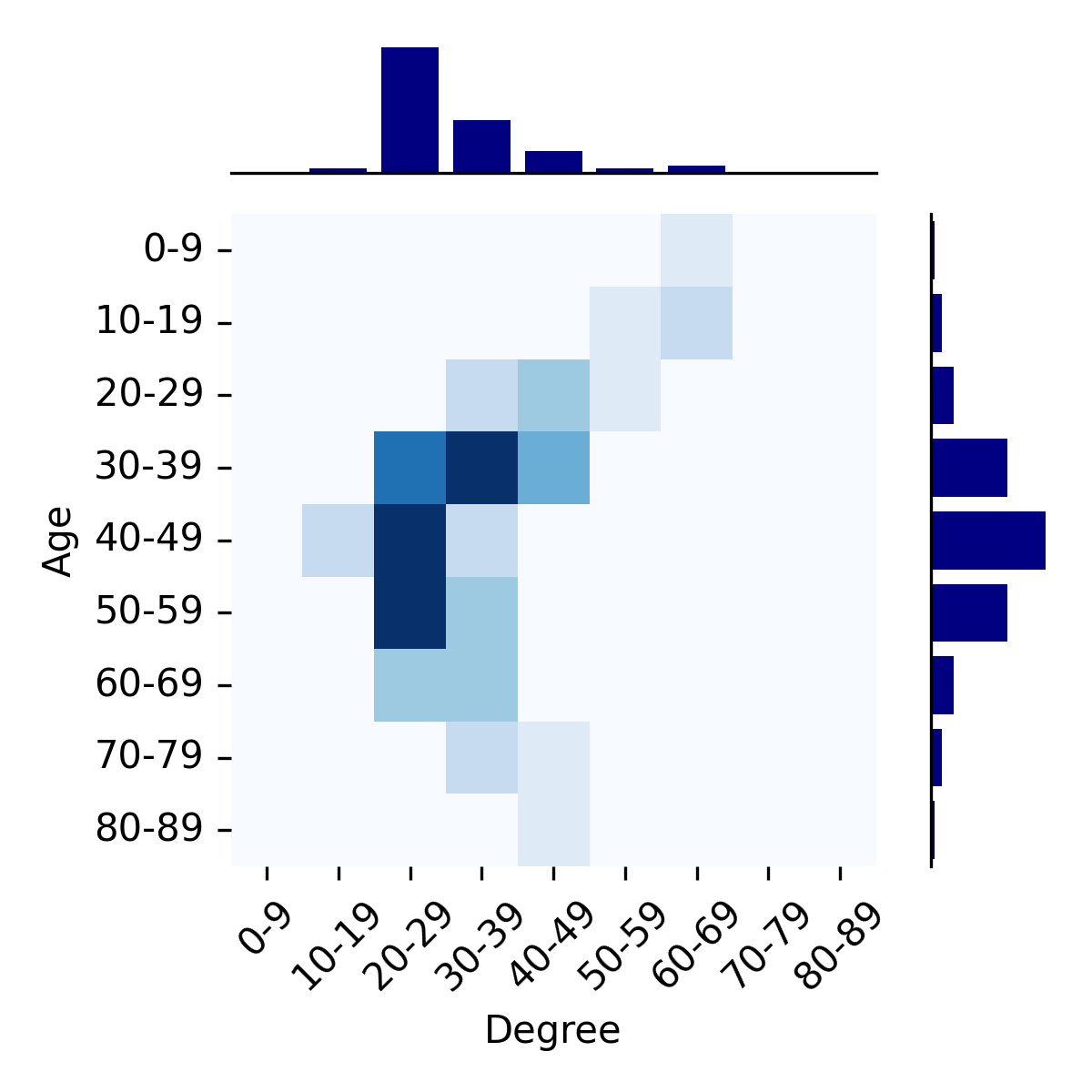}
	\end{minipage}}	
 	\subfigure[$DT-CNS^{PH}_{I}$]{
		\begin{minipage}[b]{0.185\linewidth}
			\includegraphics[width=1\linewidth]{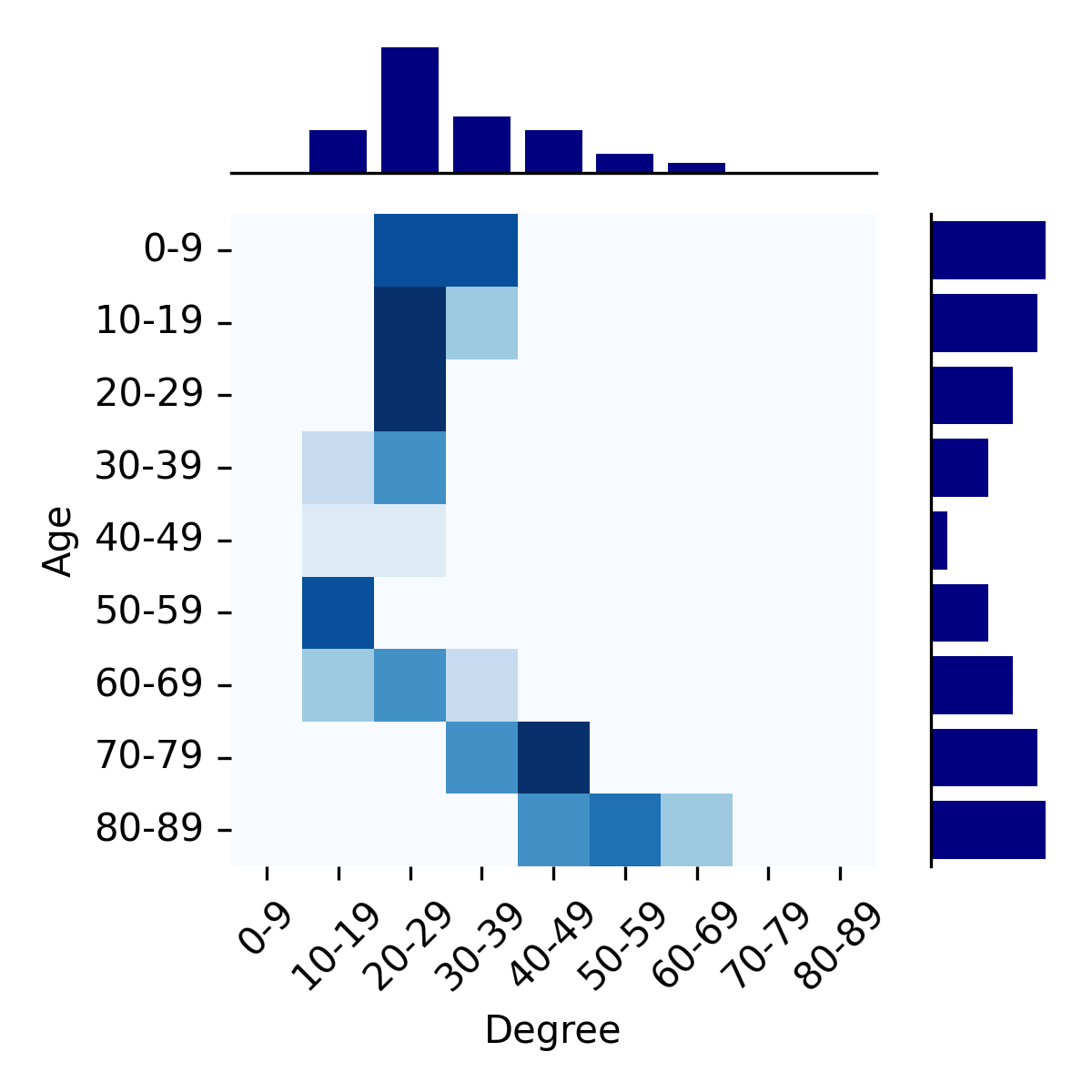}
	\end{minipage}}	
  	\subfigure[$DT-CNS^{PH}_{L}$]{
		\begin{minipage}[b]{0.185\linewidth}
			\includegraphics[width=1\linewidth]{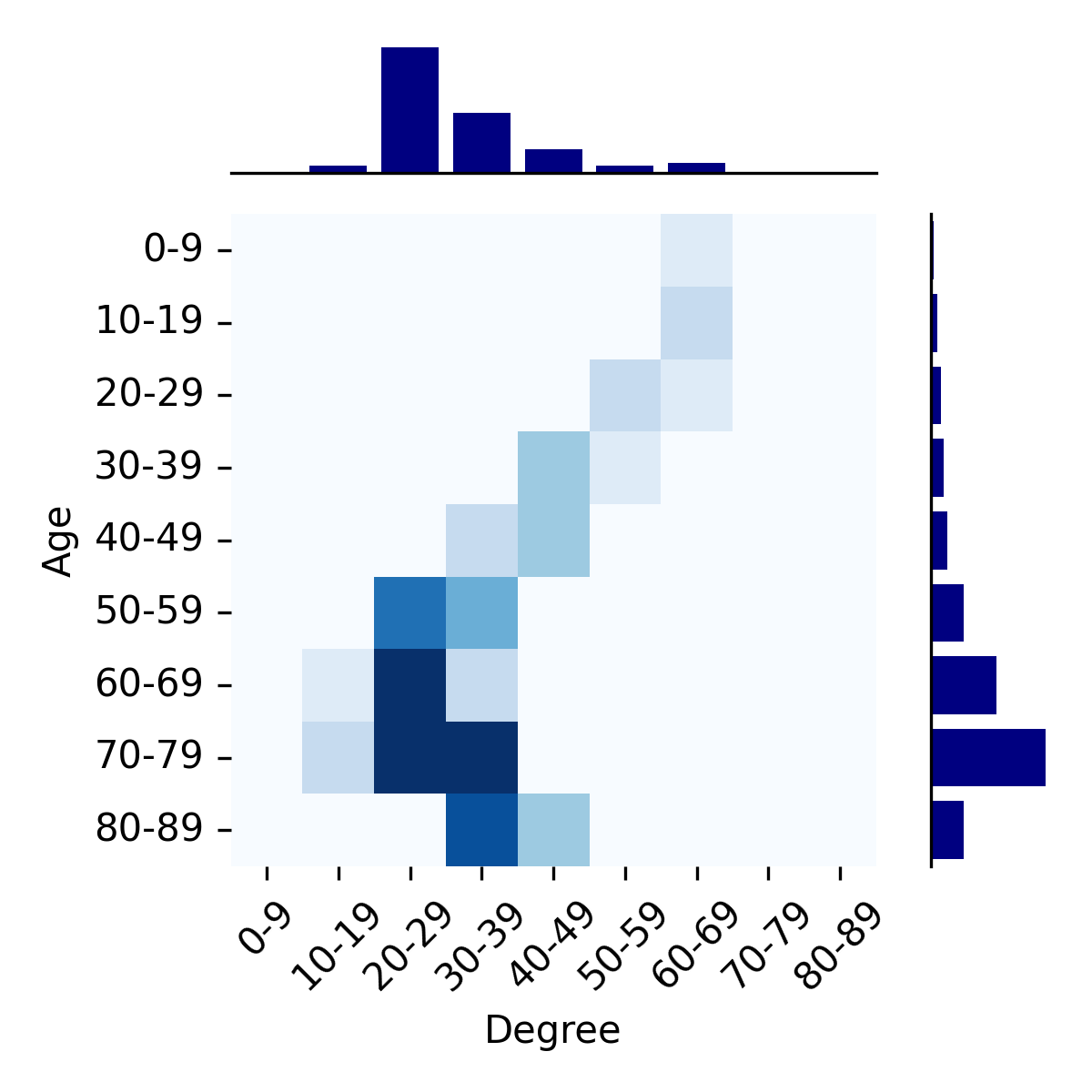}
	\end{minipage}}	
  	\subfigure[$DT-CNS^{PH}_{R}$]{
		\begin{minipage}[b]{0.185\linewidth}
			\includegraphics[width=1\linewidth]{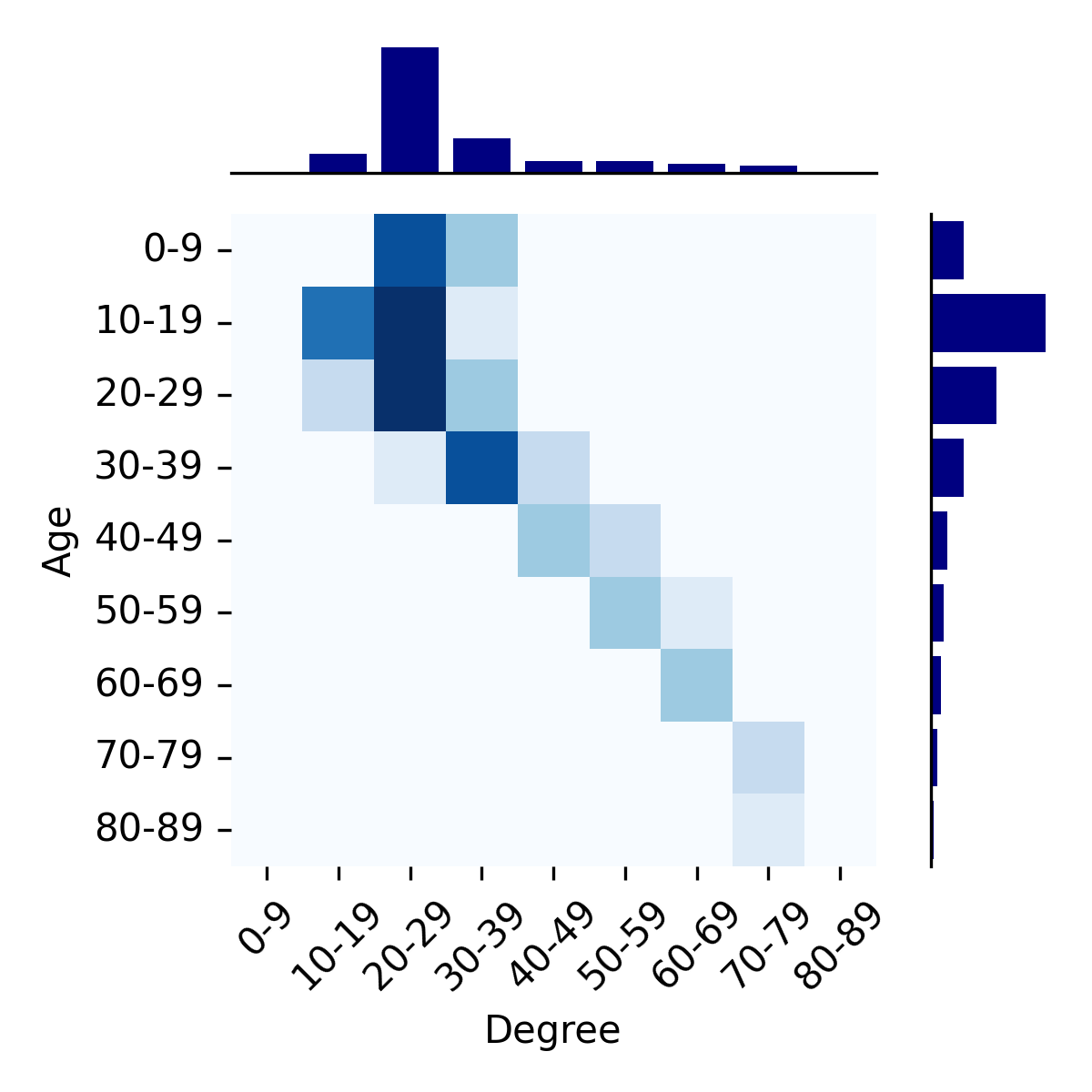}
	\end{minipage}}	
	\caption{The age and degree distributions of social networks generated by DT-CNSs based different features and rules.}
\label{AgeAndDeg}
\end{figure*}

\begin{figure*}[htp] 
	\centering
	\subfigure[$DT-CNS^{P+}_{U}$]{
		\begin{minipage}[b]{0.185\linewidth}
			\includegraphics[width=1\linewidth]{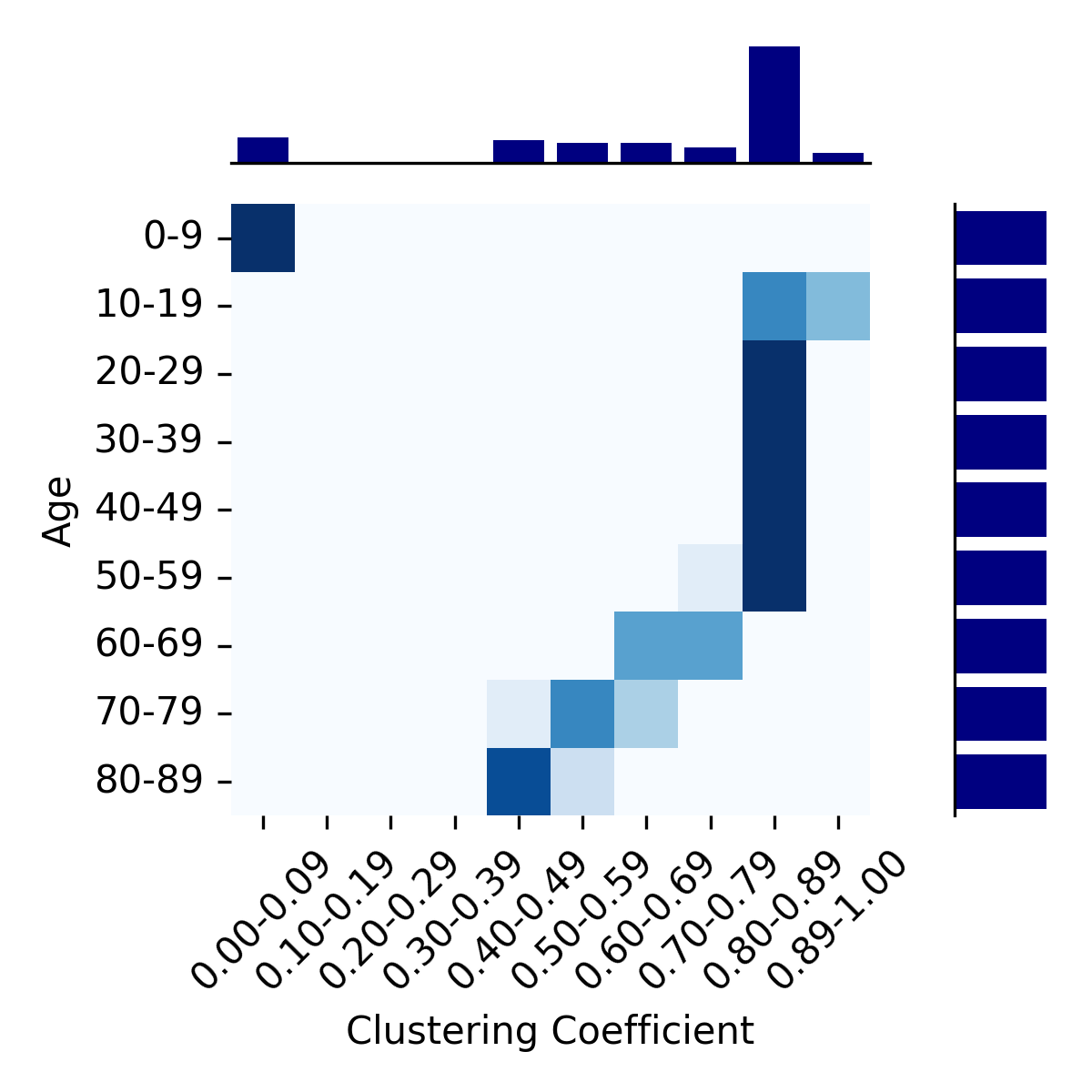}
	\end{minipage}}	
 	\subfigure[$DT-CNS^{P+}_{B}$]{
		\begin{minipage}[b]{0.185\linewidth}
			\includegraphics[width=1\linewidth]{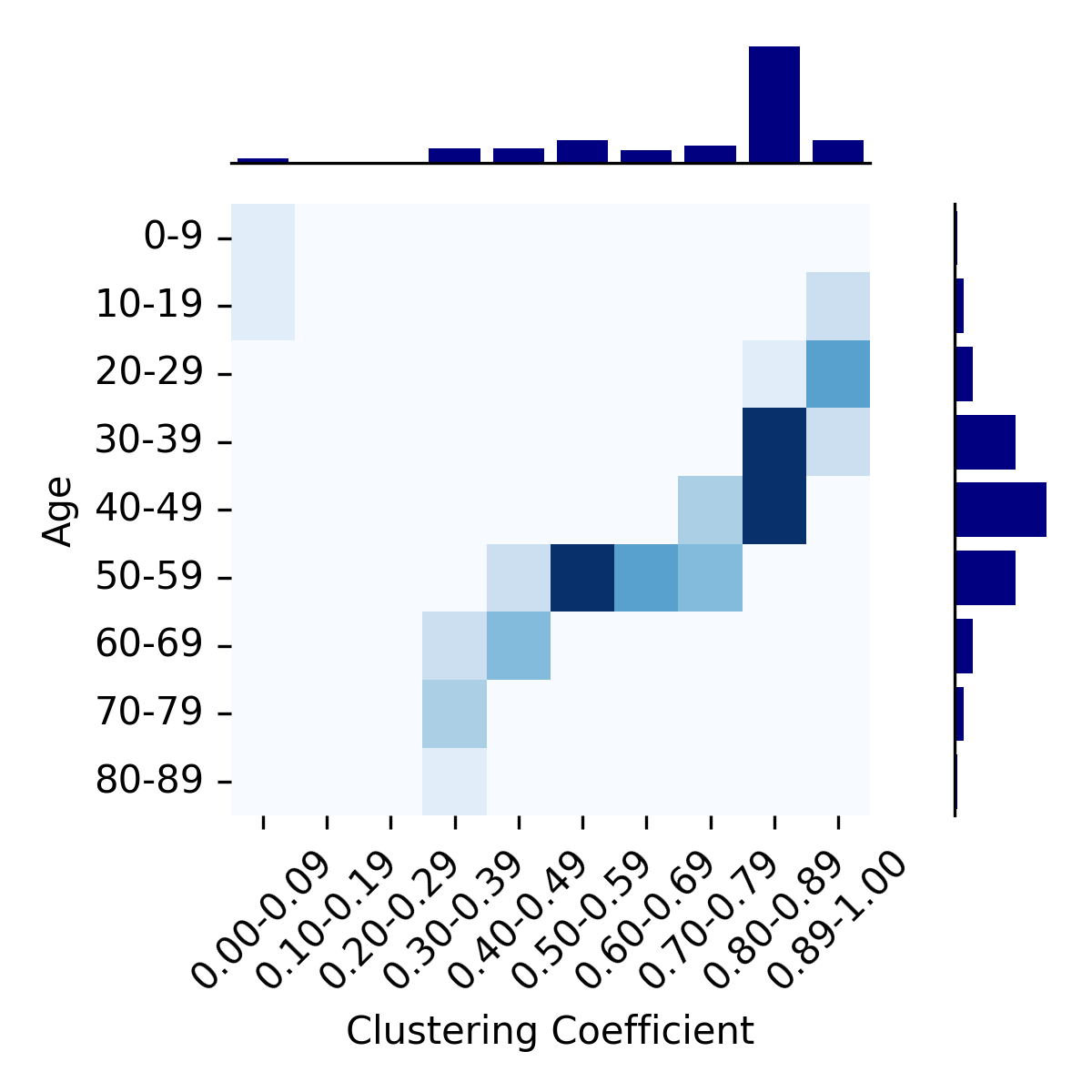}
	\end{minipage}}	
 	\subfigure[$DT-CNS^{P+}_{I}$]{
		\begin{minipage}[b]{0.185\linewidth}
			\includegraphics[width=1\linewidth]{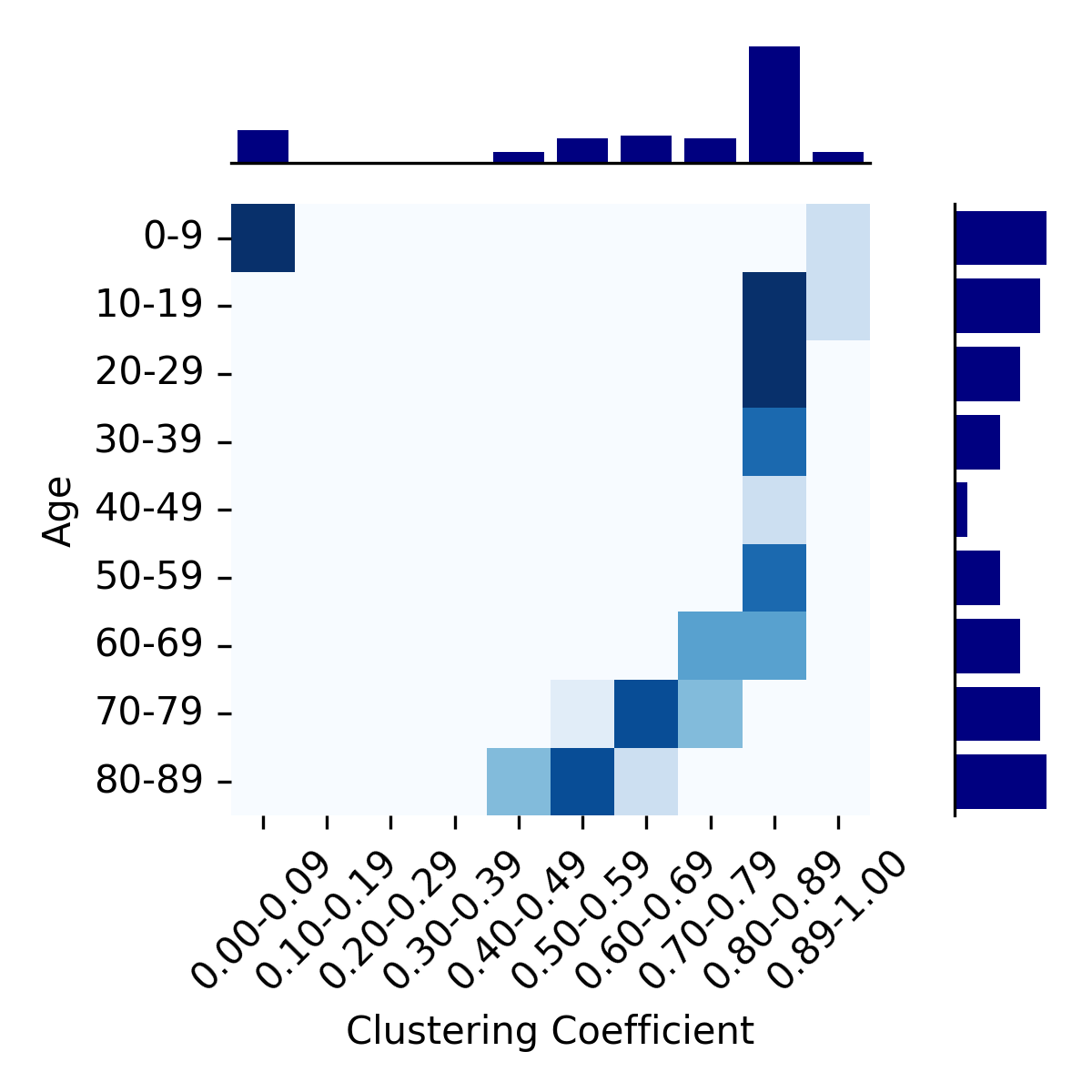}
	\end{minipage}}	
  	\subfigure[$DT-CNS^{P+}_{L}$]{
		\begin{minipage}[b]{0.185\linewidth}
			\includegraphics[width=1\linewidth]{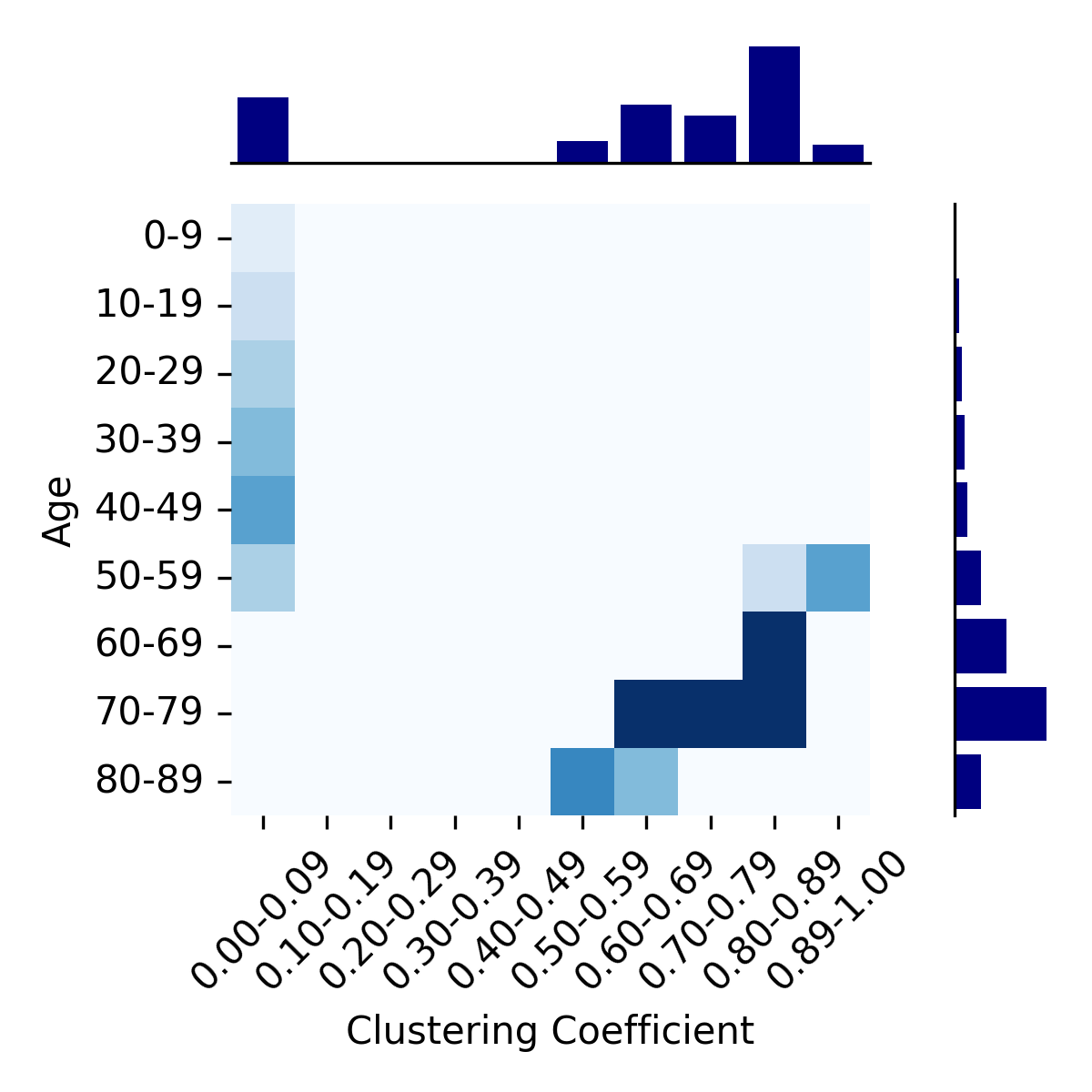}
	\end{minipage}}	
  	\subfigure[$DT-CNS^{P+}_{R}$]{
		\begin{minipage}[b]{0.185\linewidth}
			\includegraphics[width=1\linewidth]{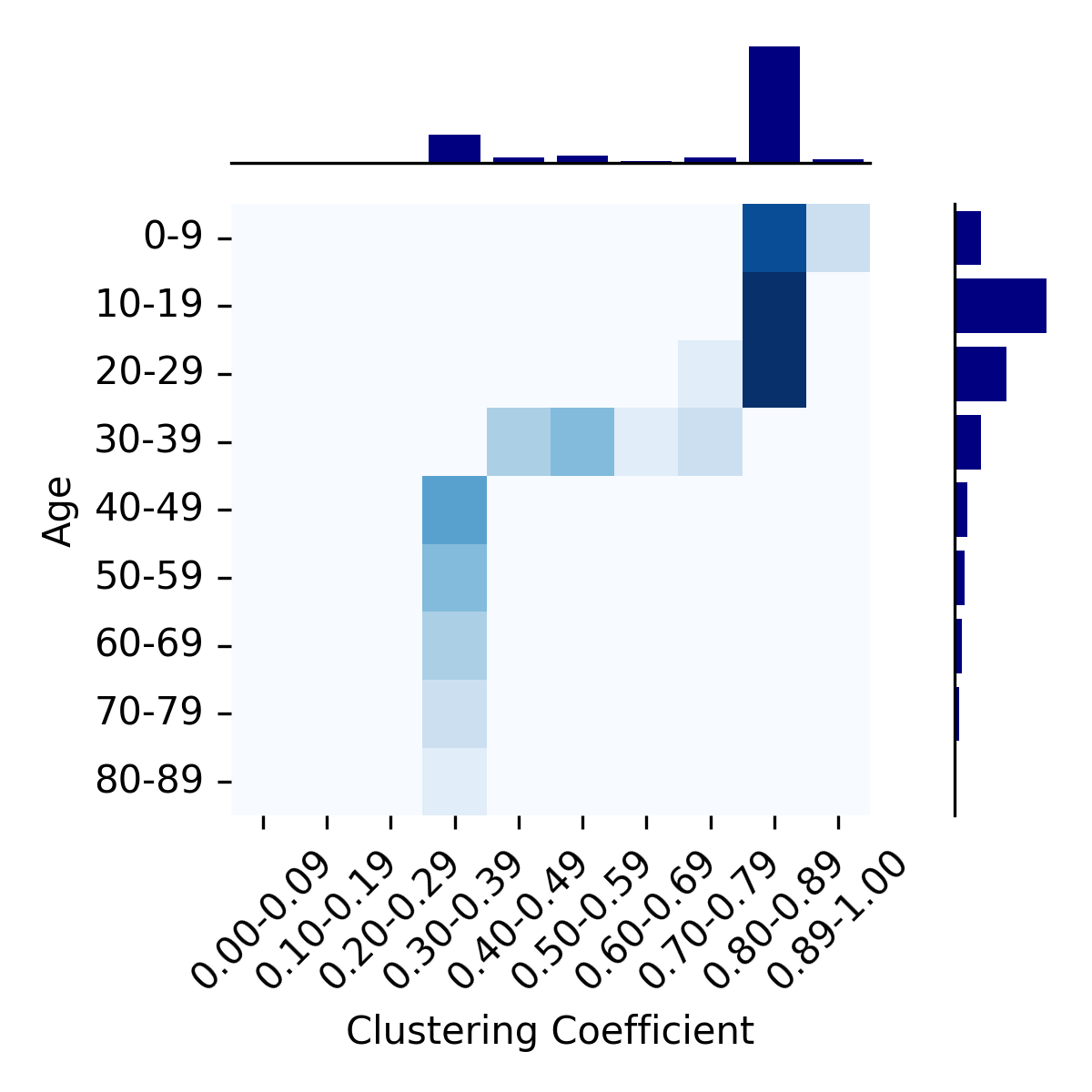}
	\end{minipage}}	\\
 	\subfigure[$DT-CNS^{P-}_{U}$]{
		\begin{minipage}[b]{0.185\linewidth}
			\includegraphics[width=1\linewidth]{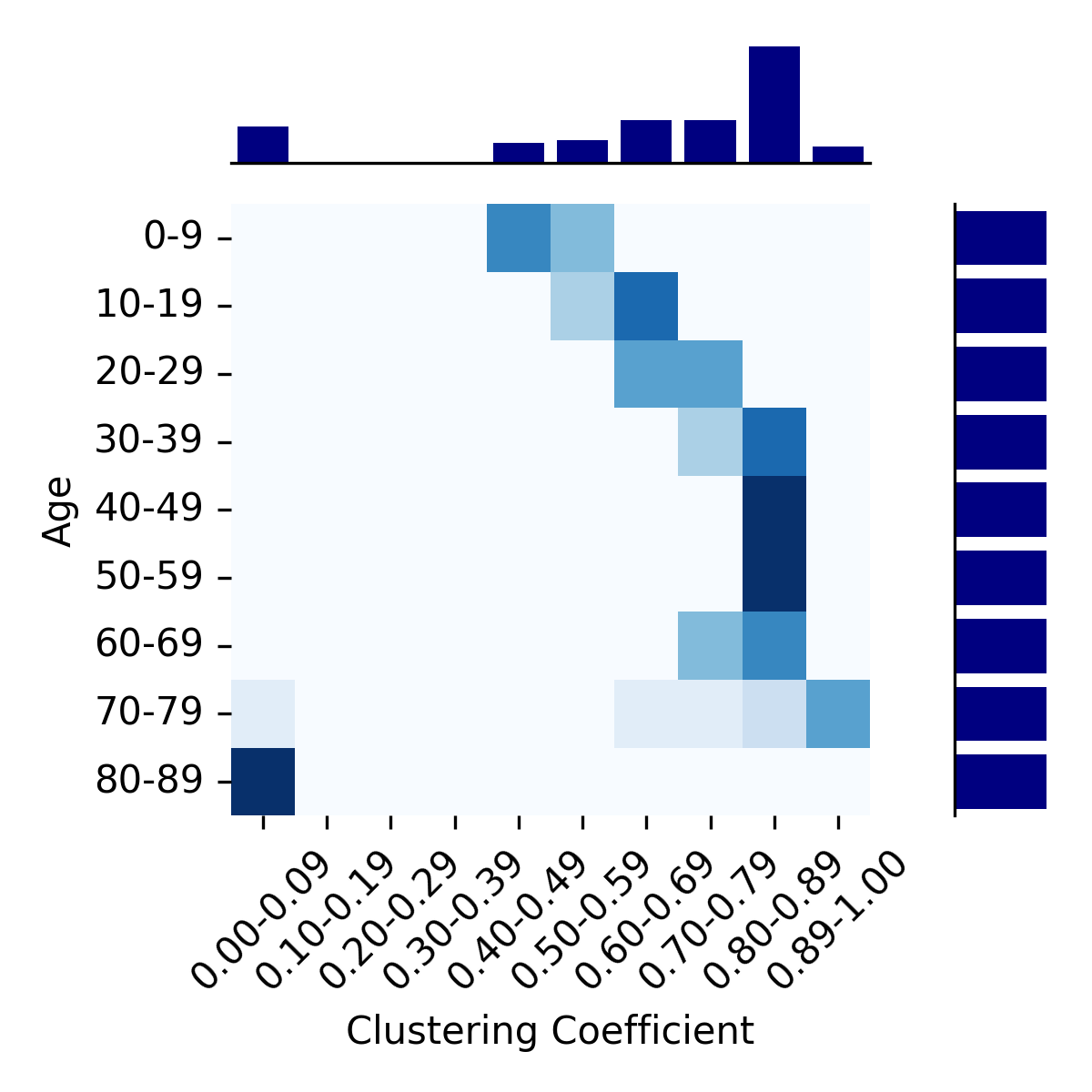}
	\end{minipage}}	
 	\subfigure[$DT-CNS^{P-}_{B}$]{
		\begin{minipage}[b]{0.185\linewidth}
			\includegraphics[width=1\linewidth]{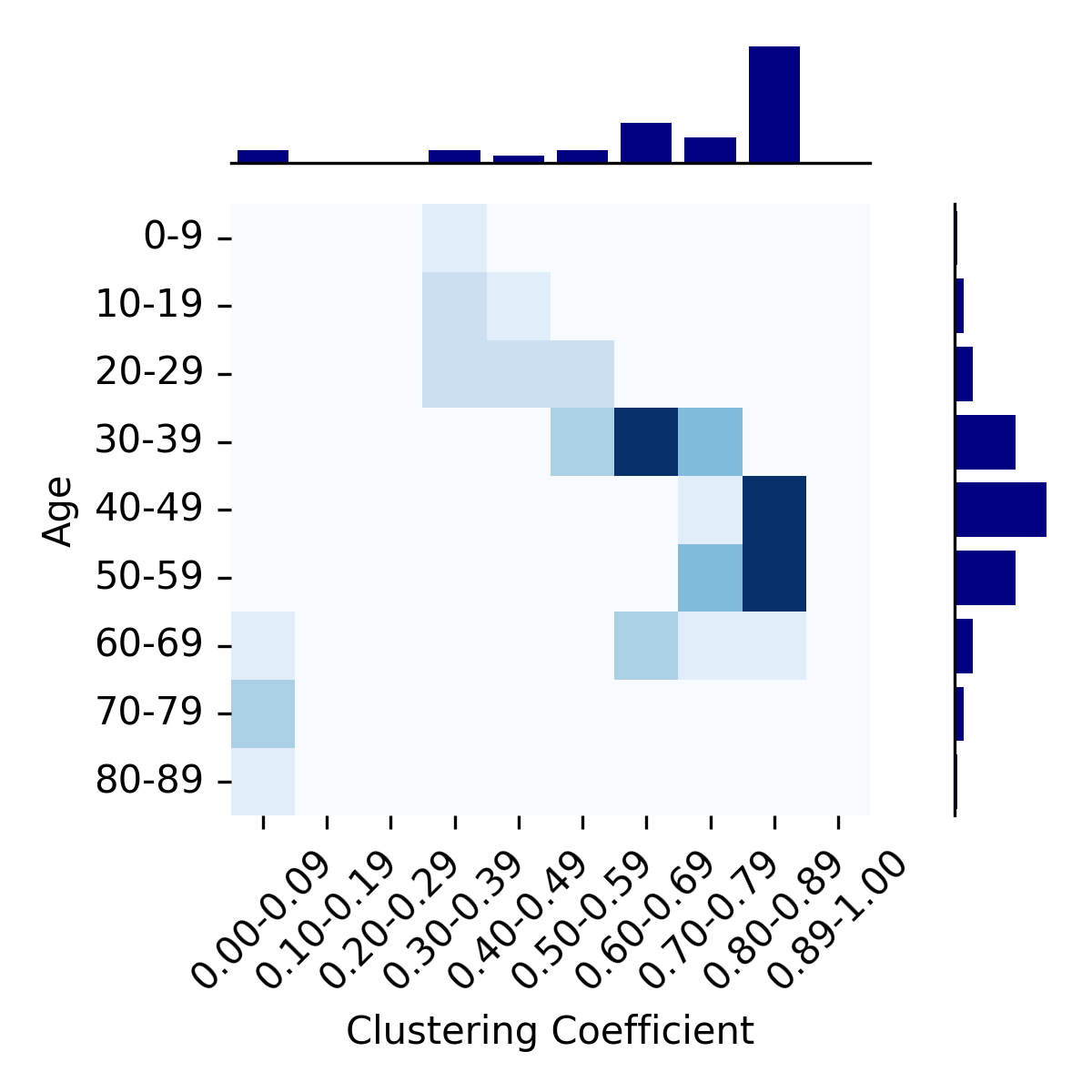}
	\end{minipage}}	
 	\subfigure[$DT-CNS^{P-}_{I}$]{
		\begin{minipage}[b]{0.185\linewidth}
			\includegraphics[width=1\linewidth]{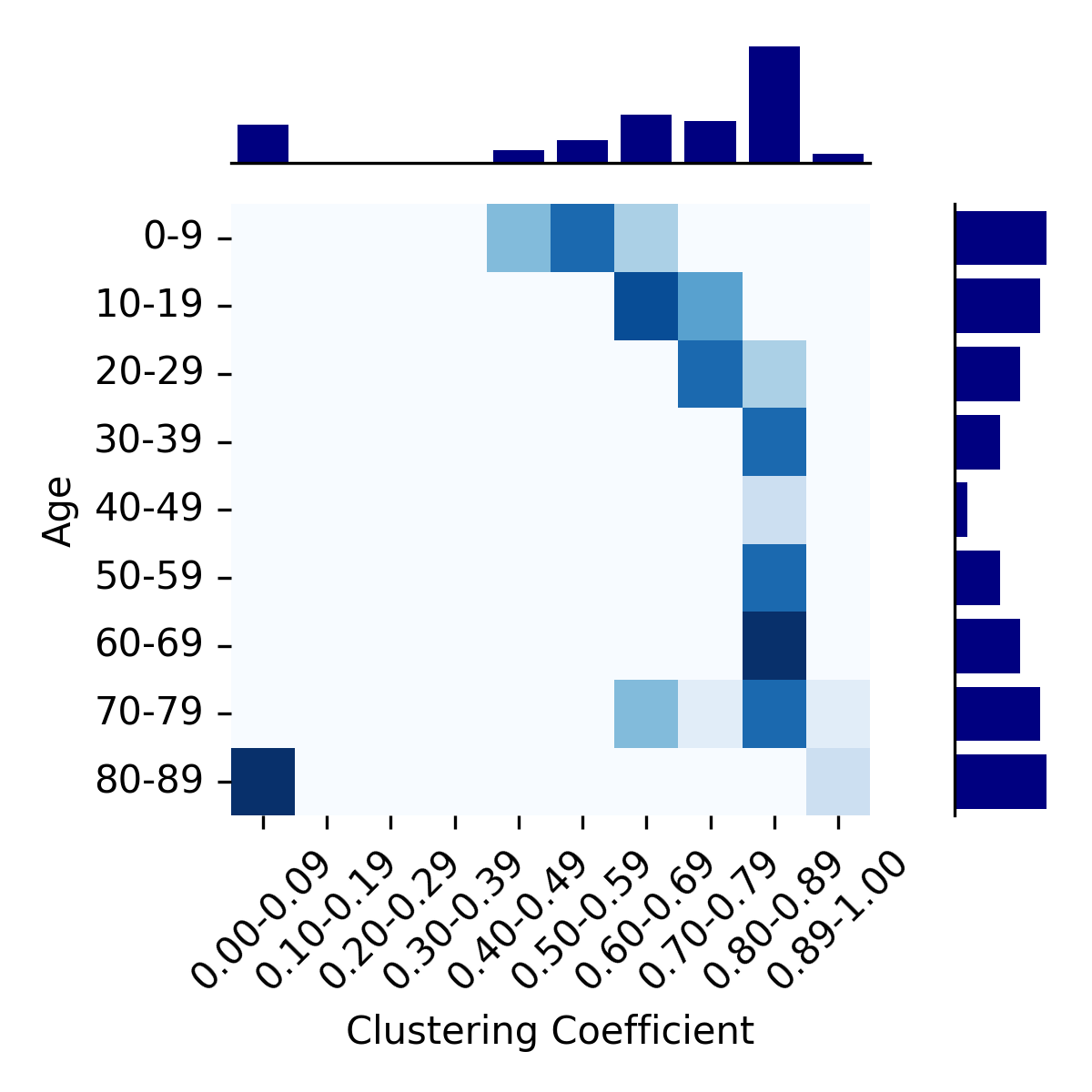}
	\end{minipage}}	
  	\subfigure[$DT-CNS^{P-}_{L}$]{
		\begin{minipage}[b]{0.185\linewidth}
			\includegraphics[width=1\linewidth]{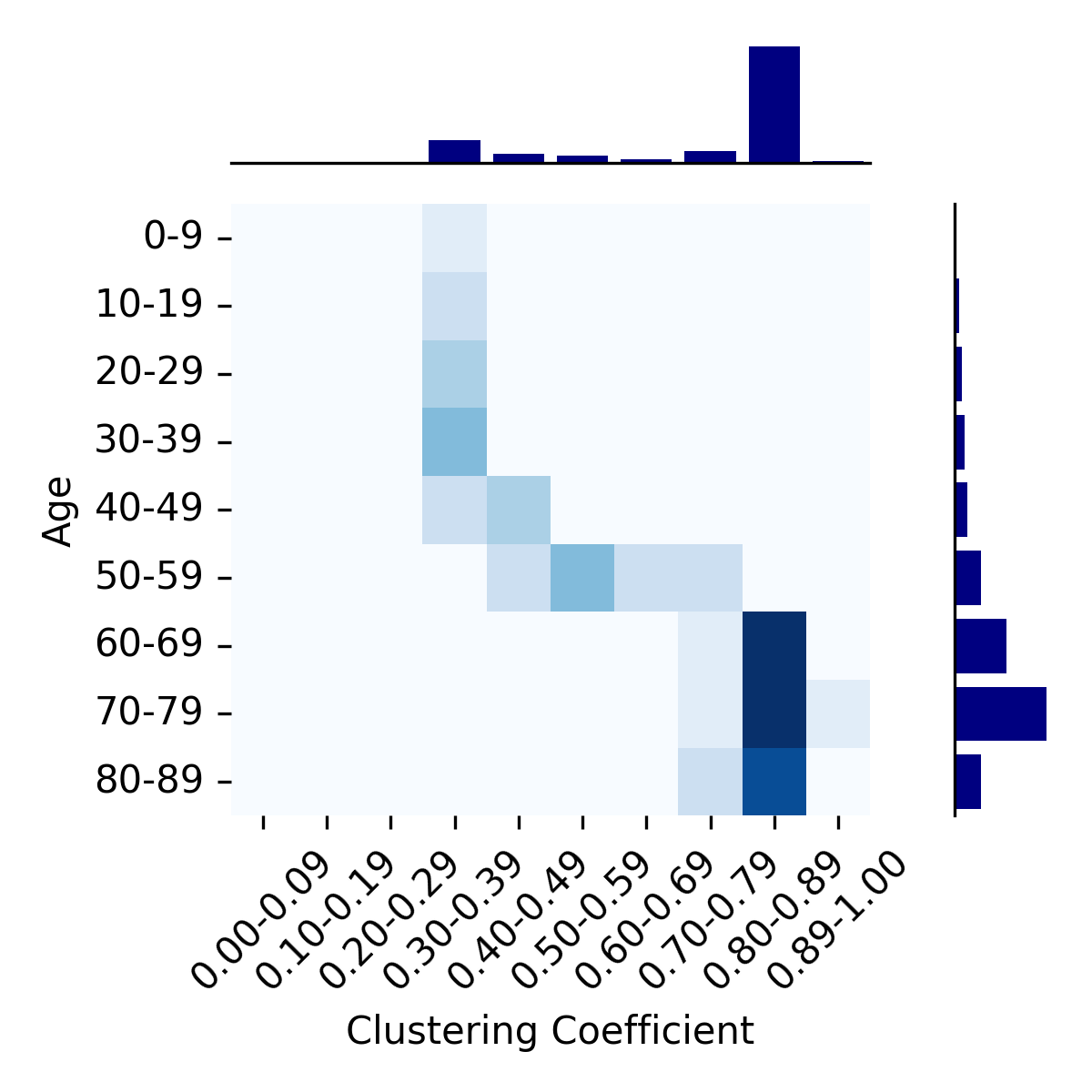}
	\end{minipage}}	
  	\subfigure[$DT-CNS^{P-}_{R}$]{
		\begin{minipage}[b]{0.185\linewidth}
			\includegraphics[width=1\linewidth]{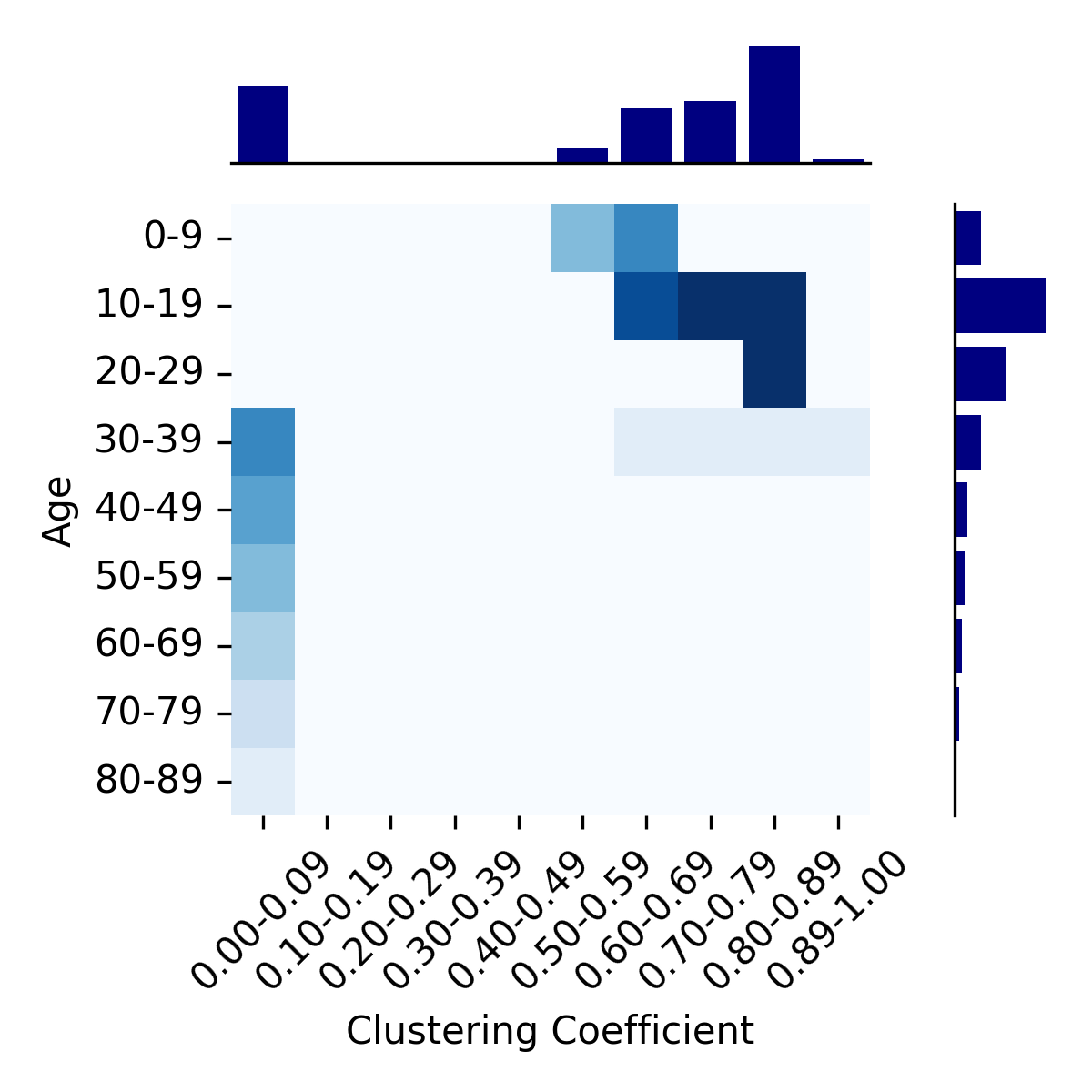}
	\end{minipage}}	\\
 	\subfigure[$DT-CNS^{H+}_{U}$]{
		\begin{minipage}[b]{0.185\linewidth}
			\includegraphics[width=1\linewidth]{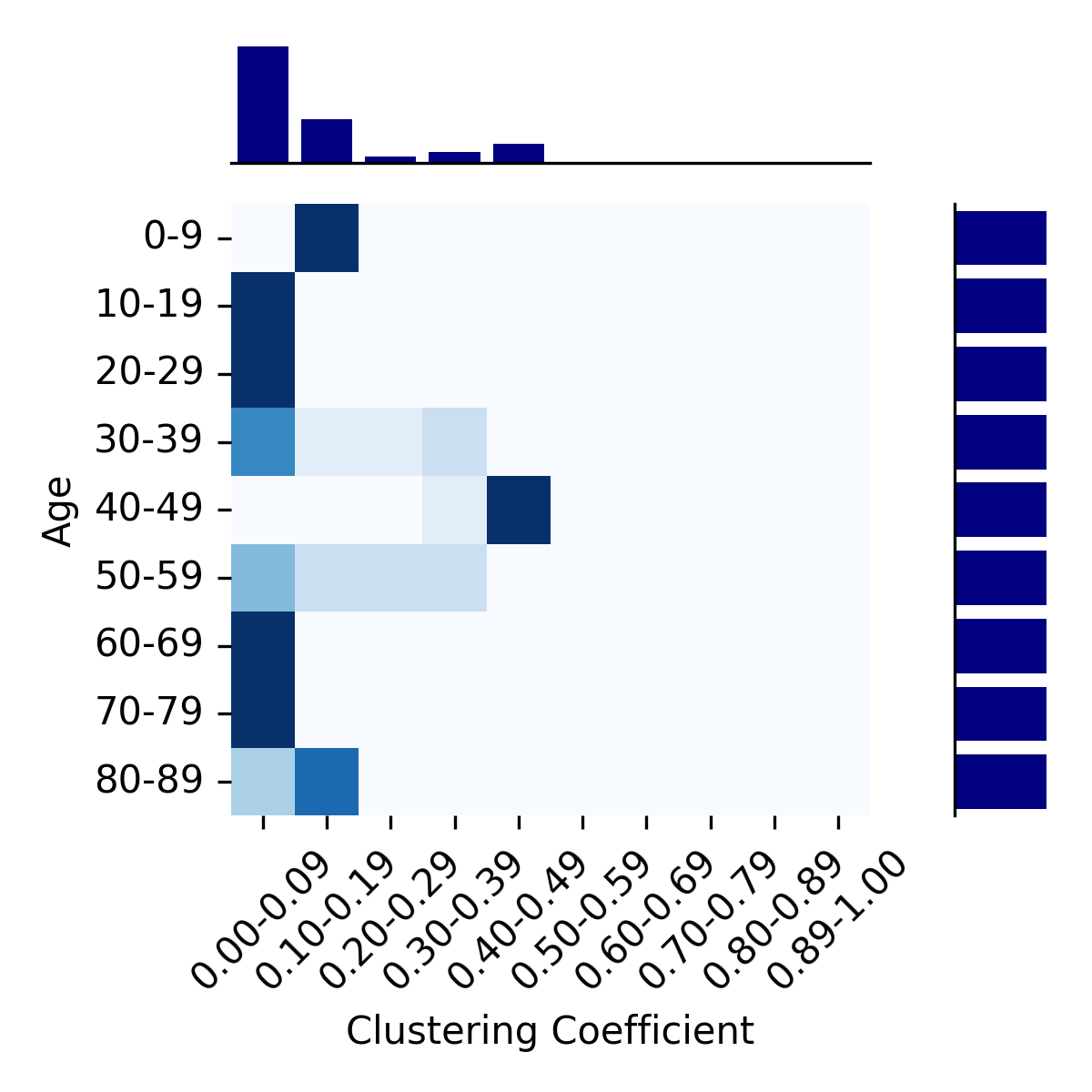}
	\end{minipage}}	
 	\subfigure[$DT-CNS^{H+}_{B}$]{
		\begin{minipage}[b]{0.185\linewidth}
			\includegraphics[width=1\linewidth]{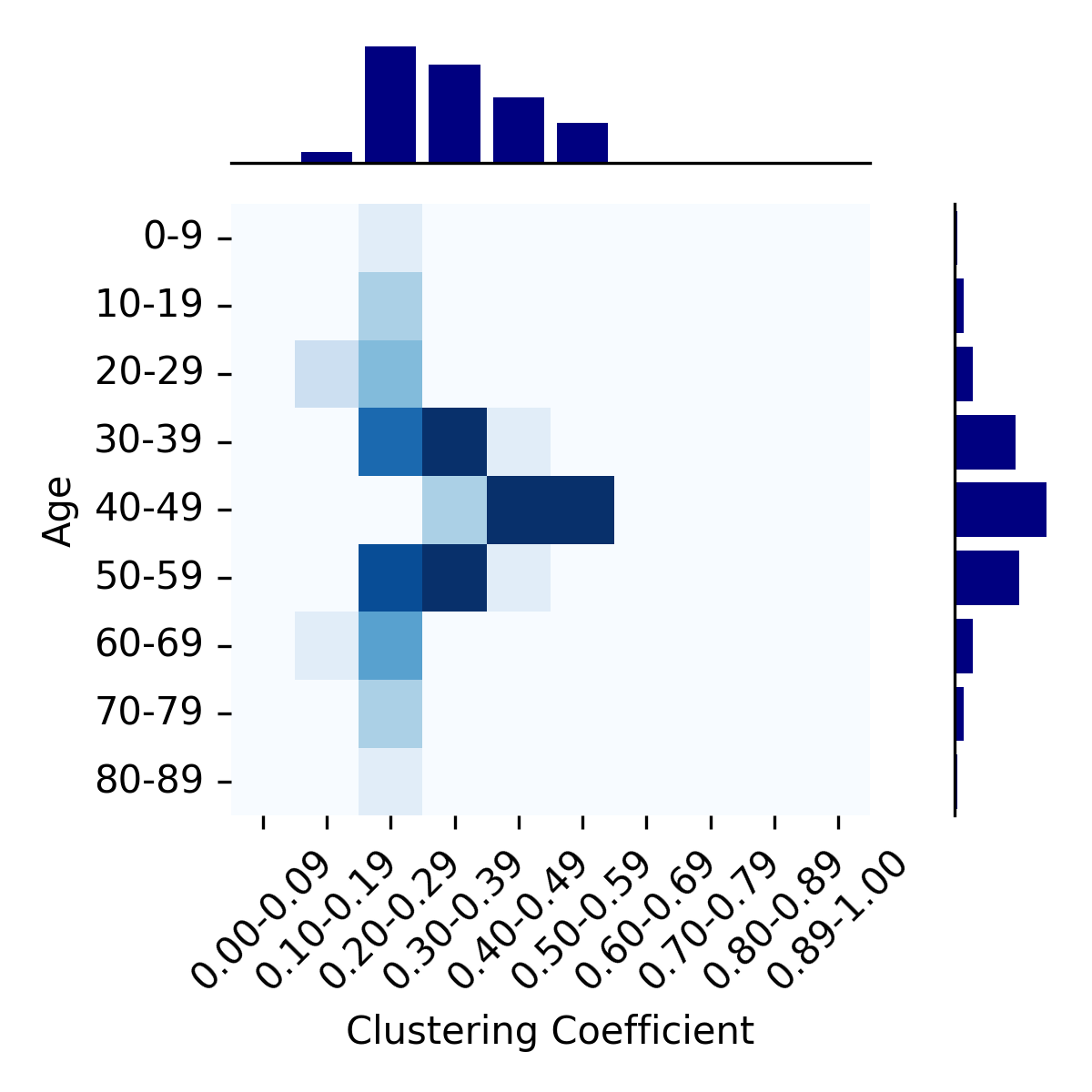}
	\end{minipage}}	
 	\subfigure[$DT-CNS^{H+}_{I}$]{
		\begin{minipage}[b]{0.185\linewidth}
			\includegraphics[width=1\linewidth]{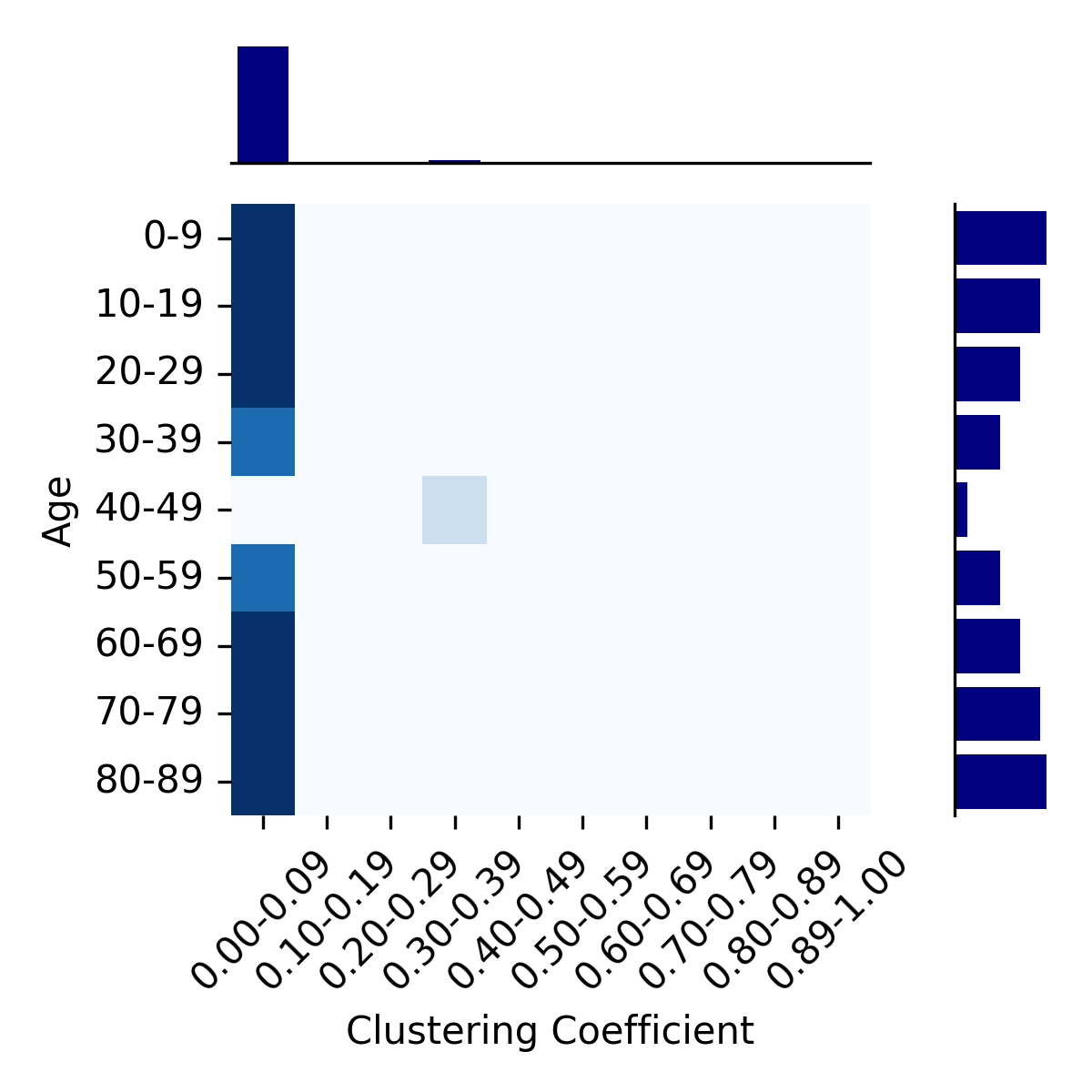}
	\end{minipage}}	
  	\subfigure[$DT-CNS^{H+}_{L}$]{
		\begin{minipage}[b]{0.185\linewidth}
			\includegraphics[width=1\linewidth]{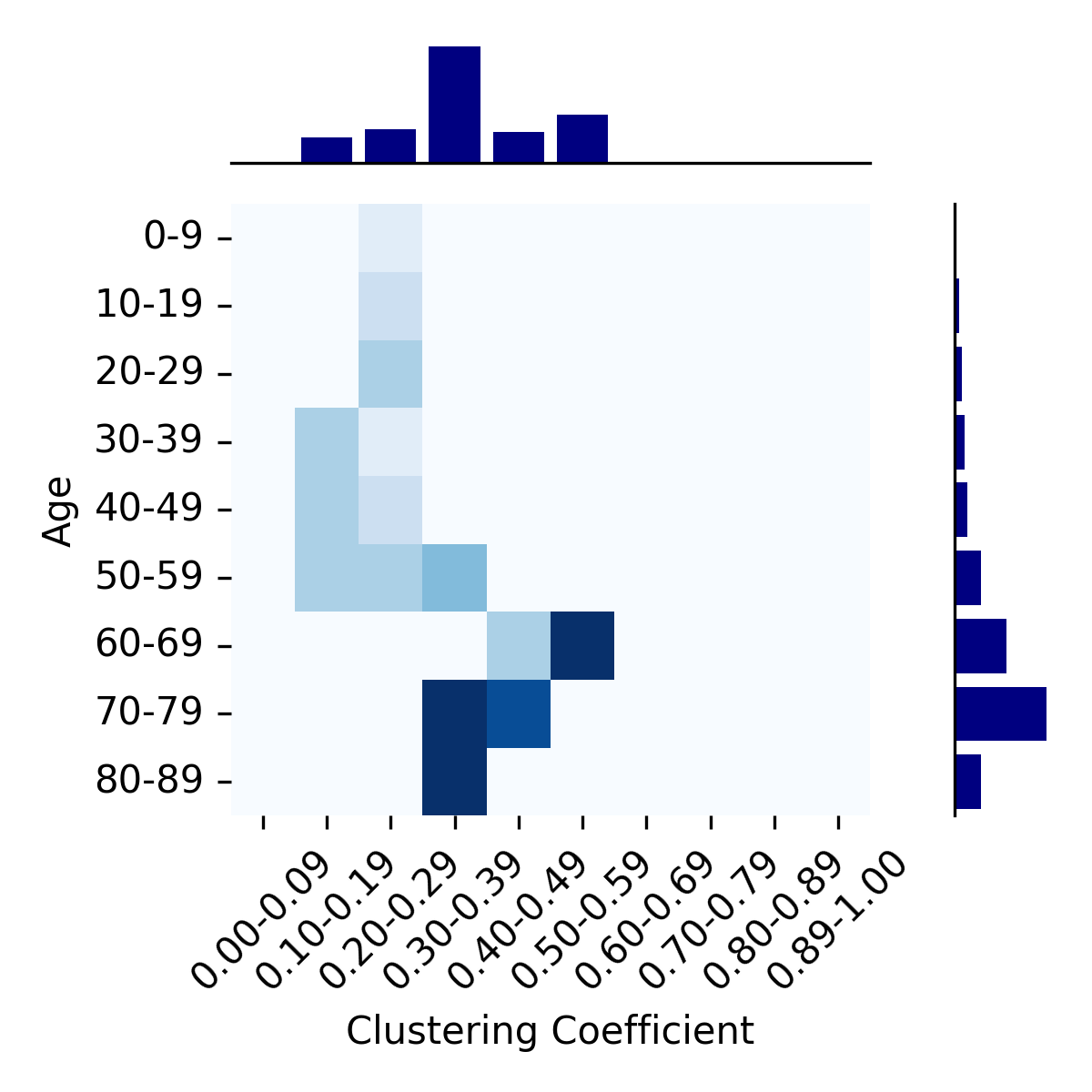}
	\end{minipage}}	
  	\subfigure[$DT-CNS^{H+}_{R}$]{
		\begin{minipage}[b]{0.185\linewidth}
			\includegraphics[width=1\linewidth]{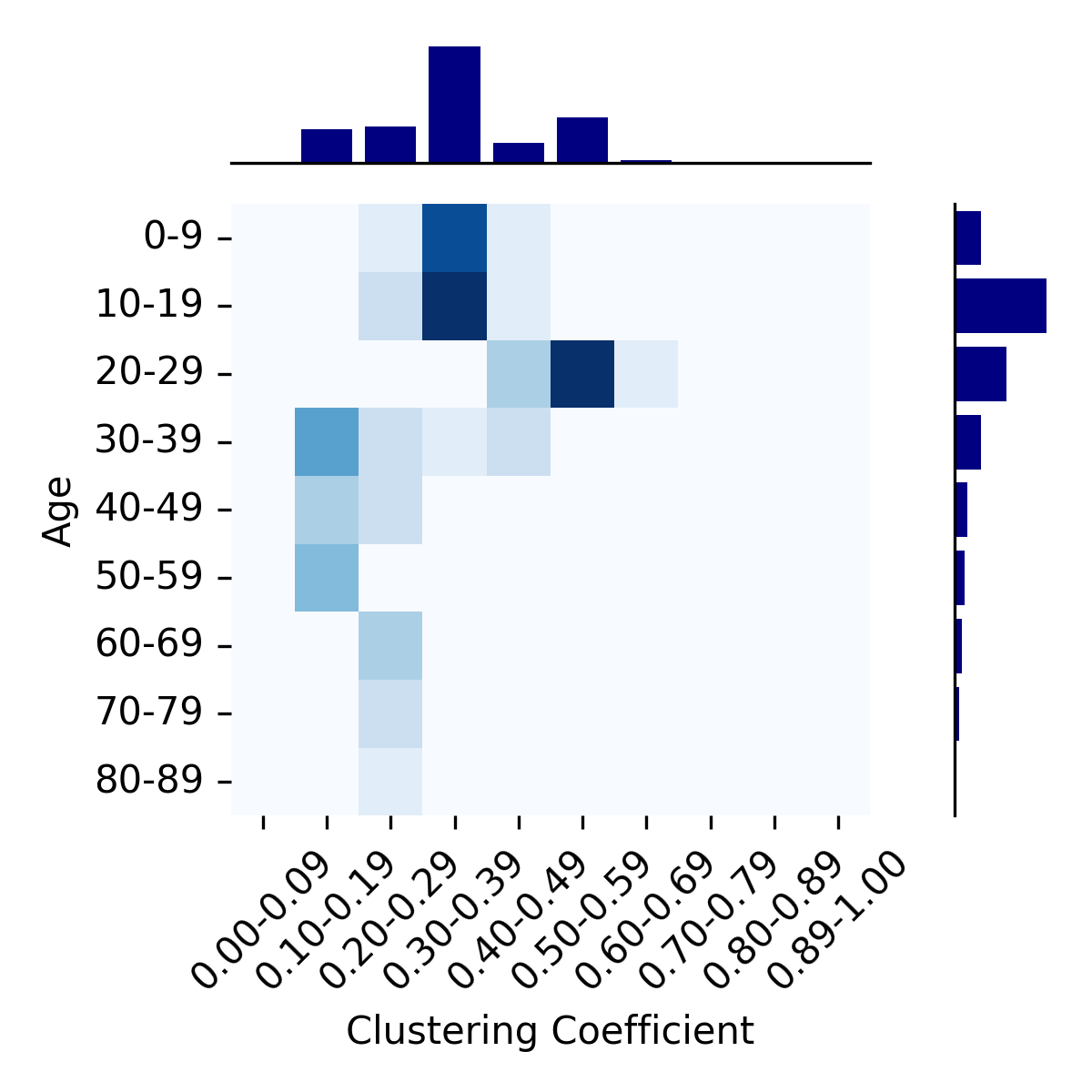}
	\end{minipage}}	\\
 	\subfigure[$DT-CNS^{H-}_{U}$]{
		\begin{minipage}[b]{0.185\linewidth}
			\includegraphics[width=1\linewidth]{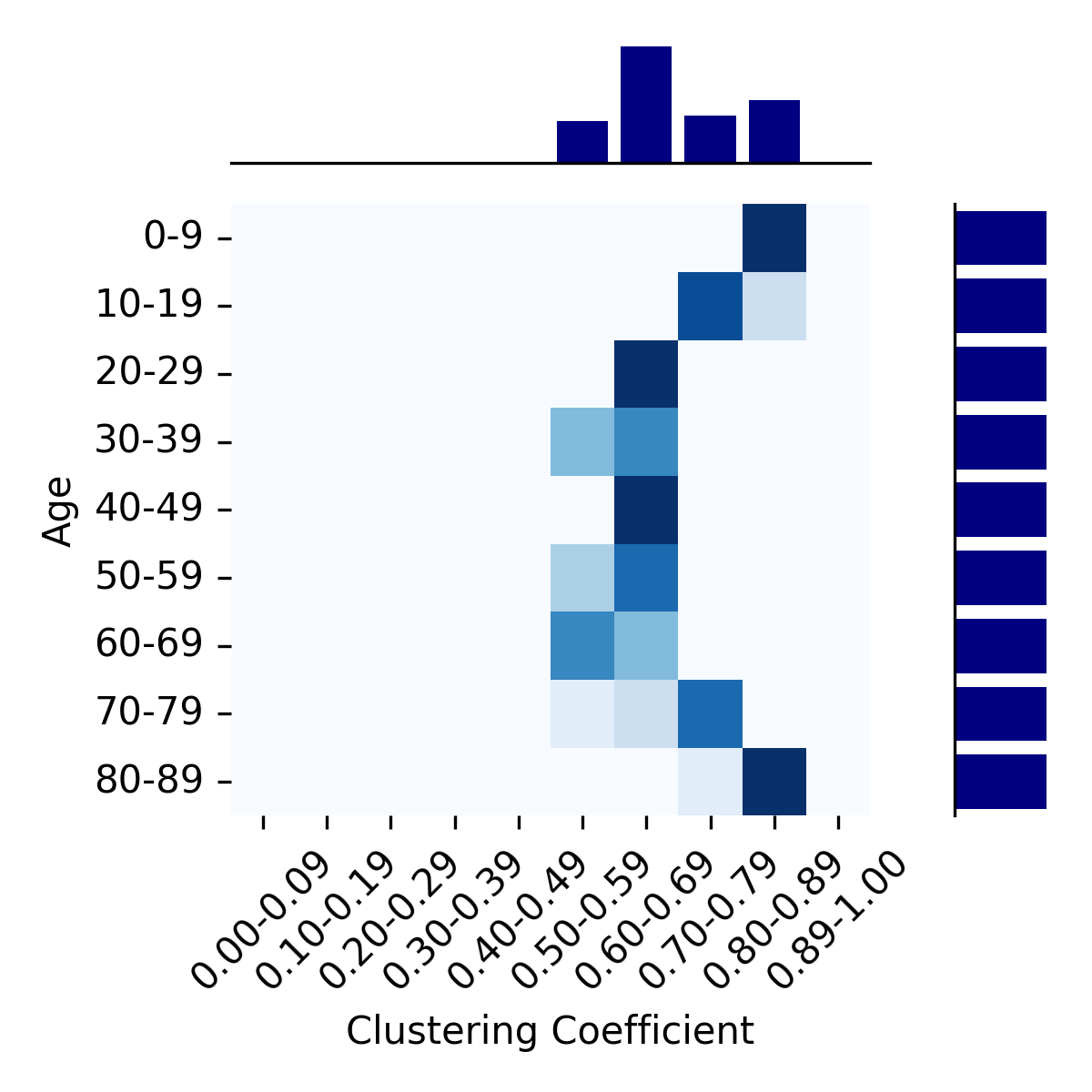}
	\end{minipage}}	
 	\subfigure[$DT-CNS^{H-}_{B}$]{
		\begin{minipage}[b]{0.185\linewidth}
			\includegraphics[width=1\linewidth]{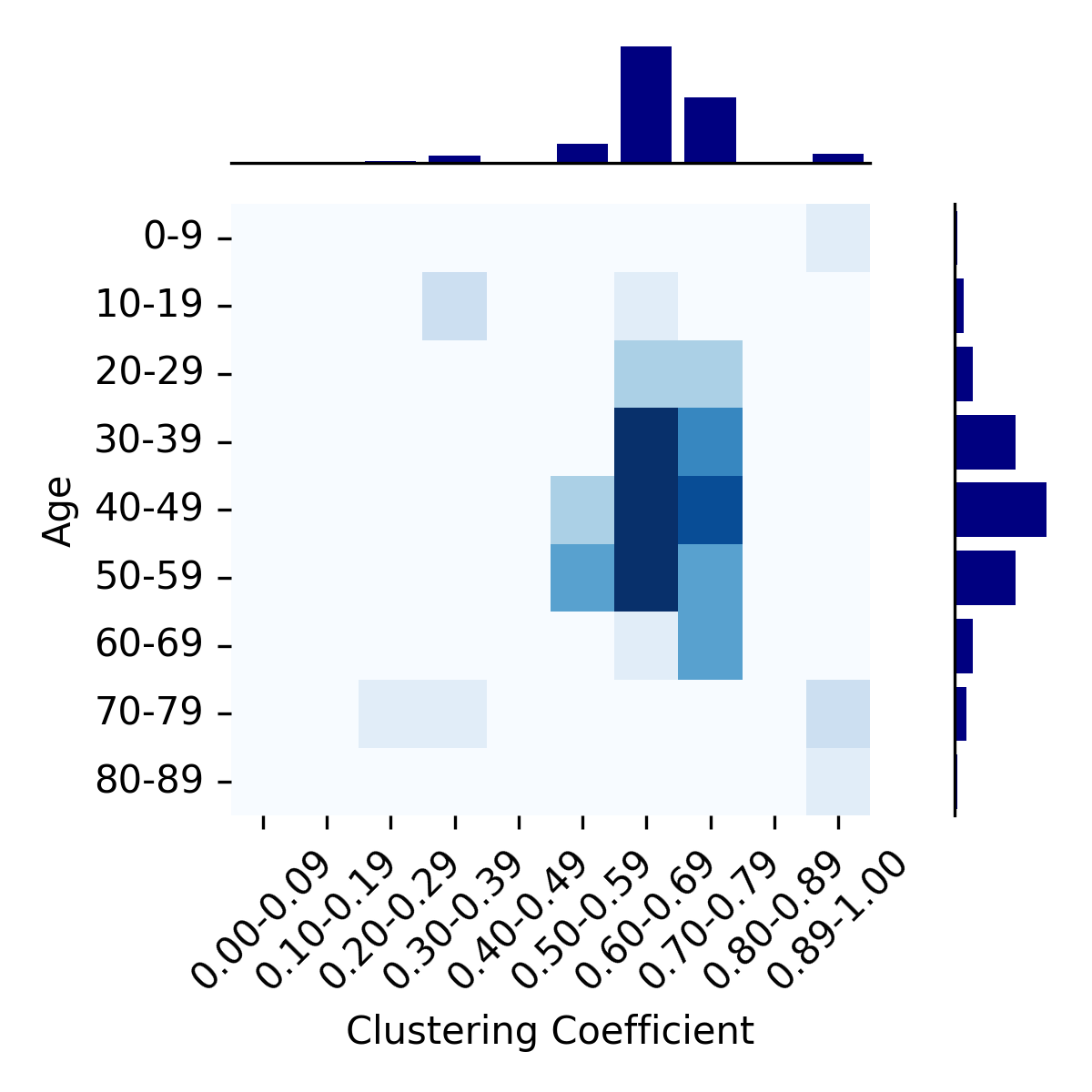}
	\end{minipage}}	
 	\subfigure[$DT-CNS^{H-}_{I}$]{
		\begin{minipage}[b]{0.185\linewidth}
			\includegraphics[width=1\linewidth]{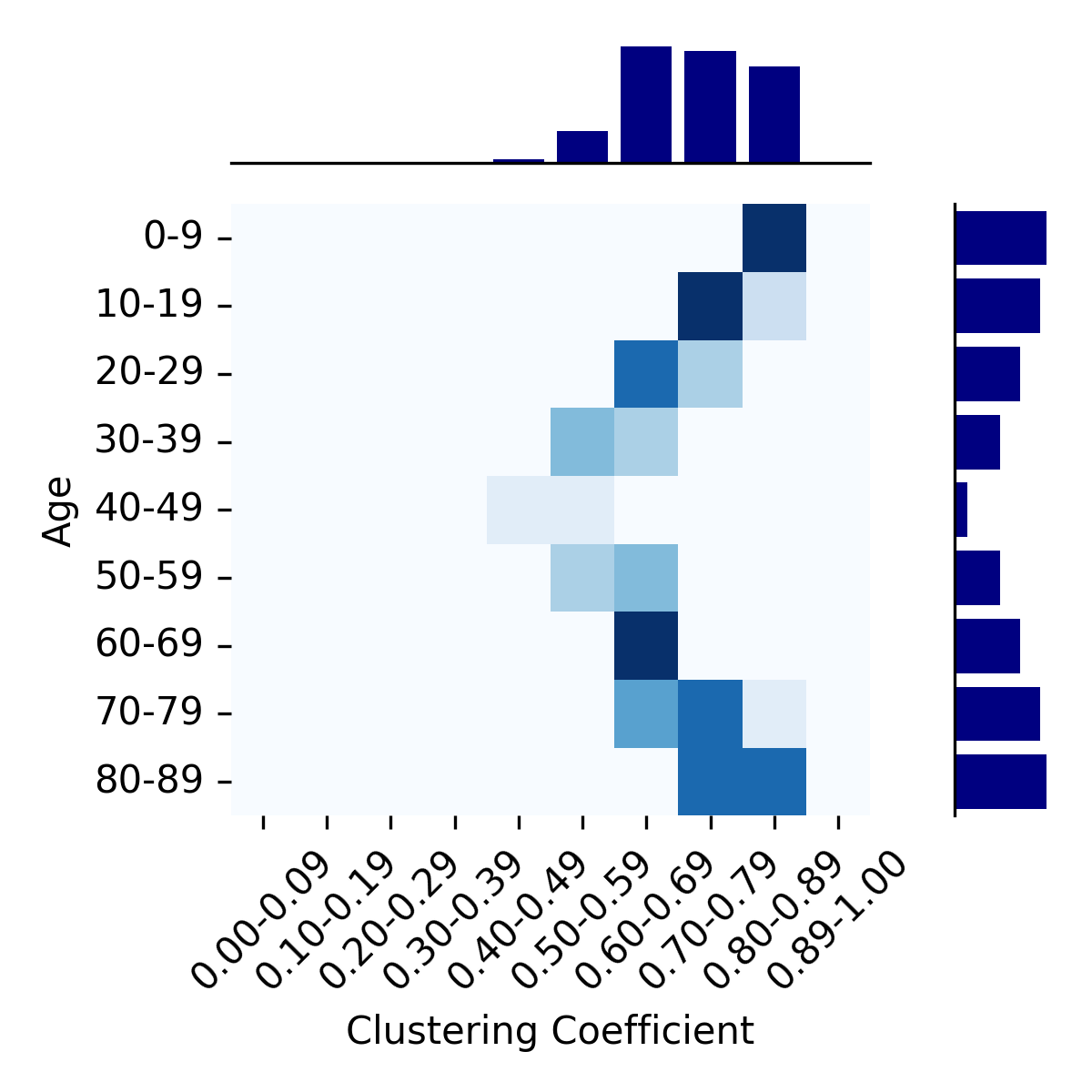}
	\end{minipage}}	
  	\subfigure[$DT-CNS^{H-}_{L}$]{
		\begin{minipage}[b]{0.185\linewidth}
			\includegraphics[width=1\linewidth]{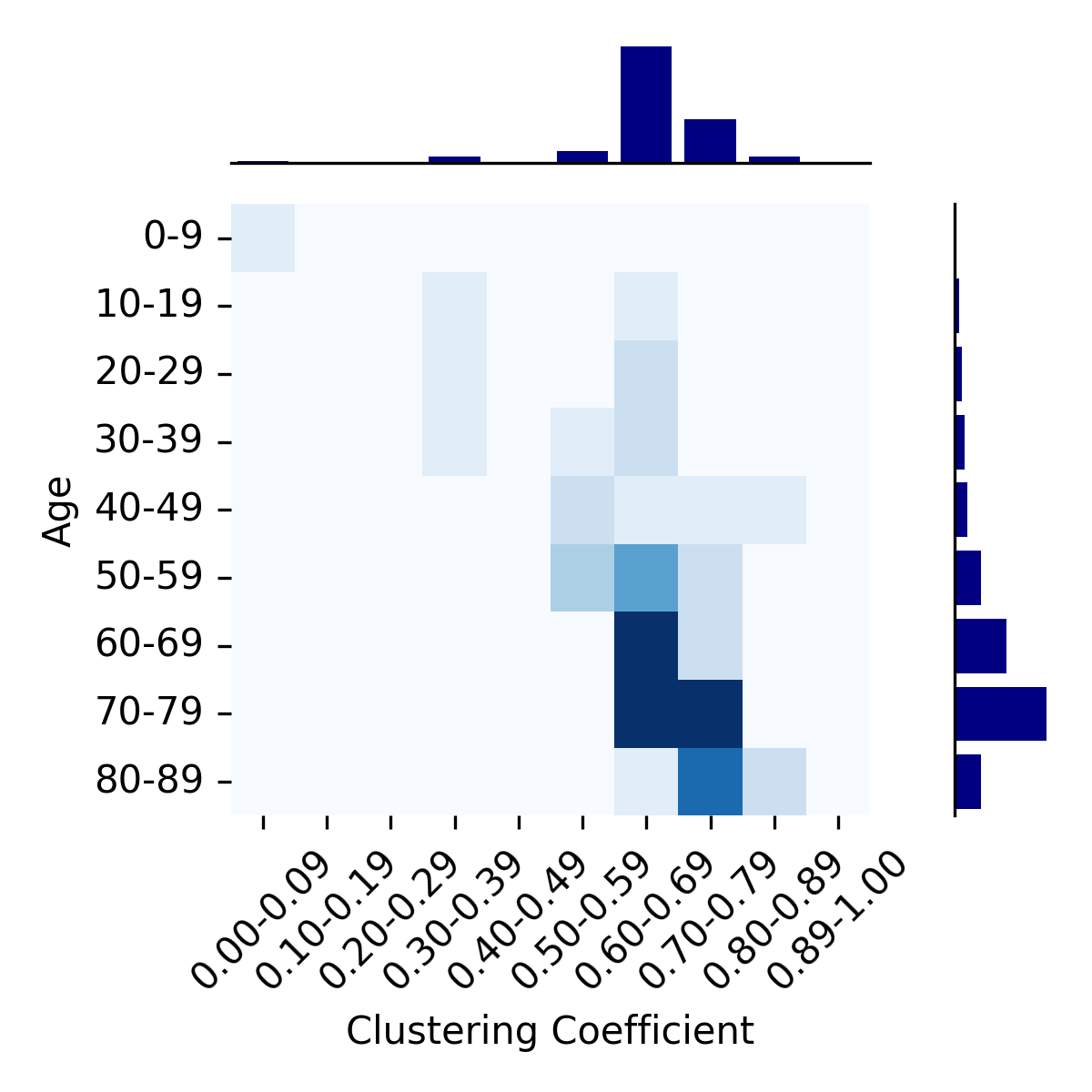}
	\end{minipage}}	
  	\subfigure[$DT-CNS^{H-}_{R}$]{
		\begin{minipage}[b]{0.185\linewidth}
			\includegraphics[width=1\linewidth]{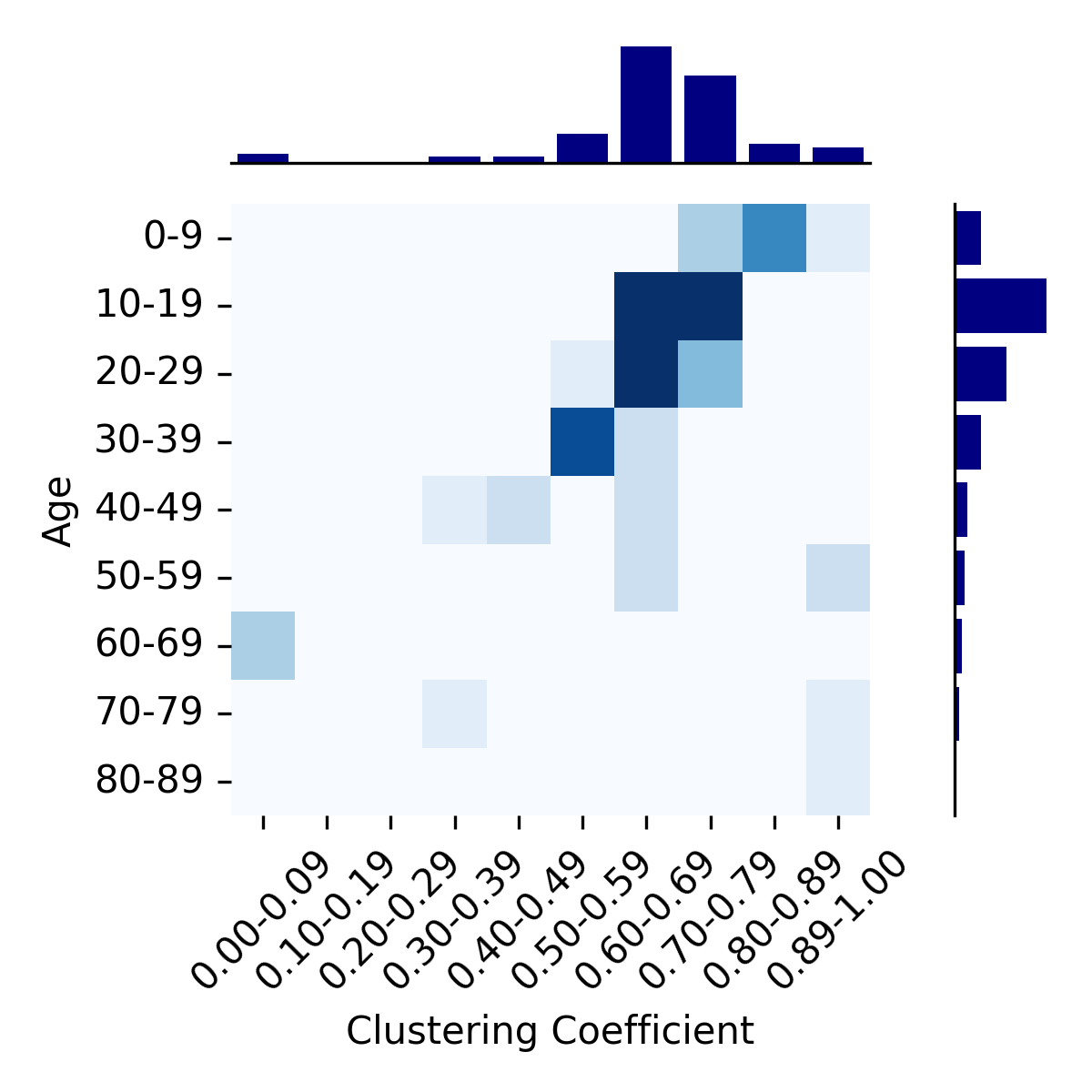}
	\end{minipage}}	\\
 	\subfigure[$DT-CNS^{PH}_{U}$]{
		\begin{minipage}[b]{0.185\linewidth}
			\includegraphics[width=1\linewidth]{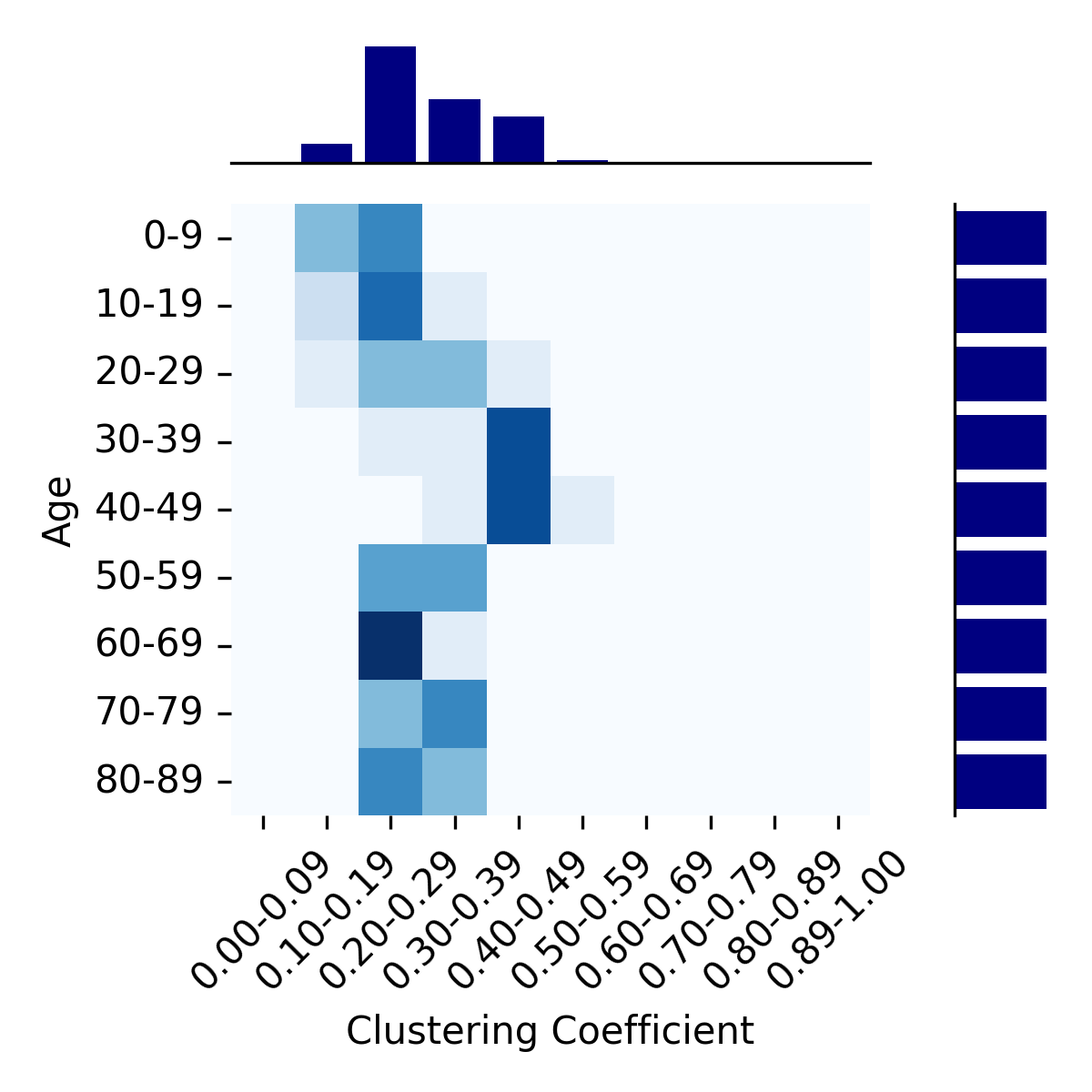}
	\end{minipage}}	
 	\subfigure[$DT-CNS^{PH}_{B}$]{
		\begin{minipage}[b]{0.185\linewidth}
			\includegraphics[width=1\linewidth]{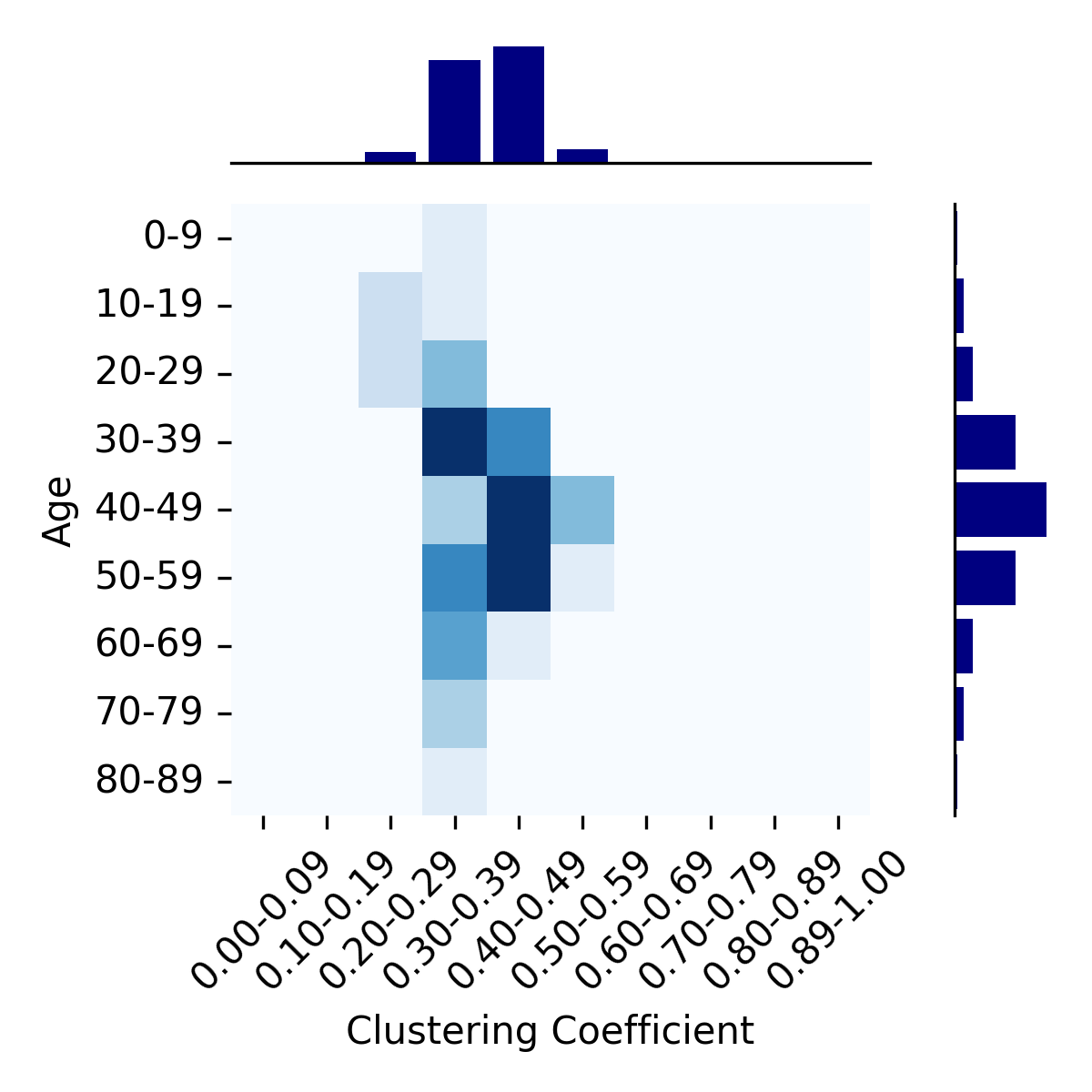}
	\end{minipage}}	
 	\subfigure[$DT-CNS^{PH}_{I}$]{
		\begin{minipage}[b]{0.185\linewidth}
			\includegraphics[width=1\linewidth]{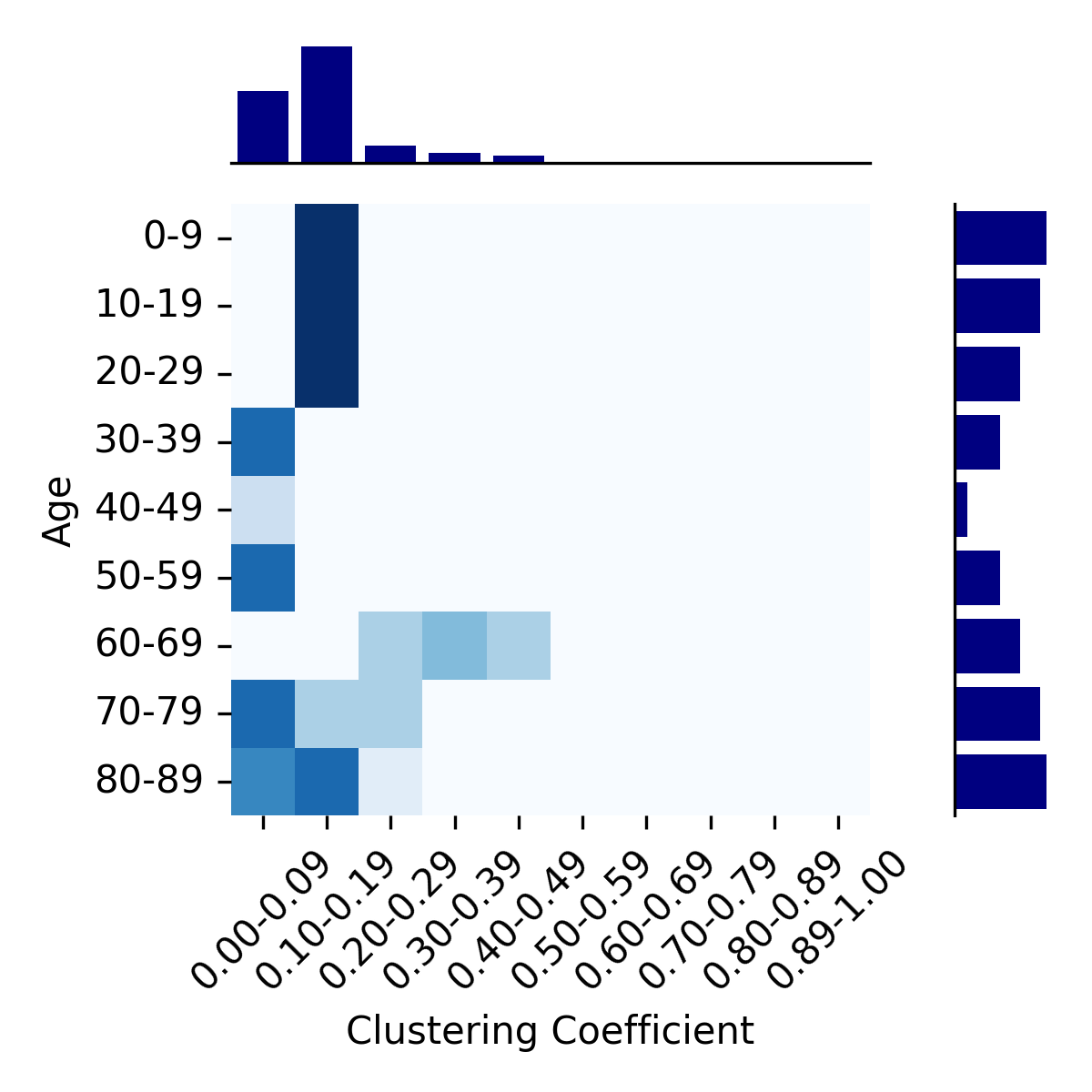}
	\end{minipage}}	
  	\subfigure[$DT-CNS^{PH}_{L}$]{
		\begin{minipage}[b]{0.185\linewidth}
			\includegraphics[width=1\linewidth]{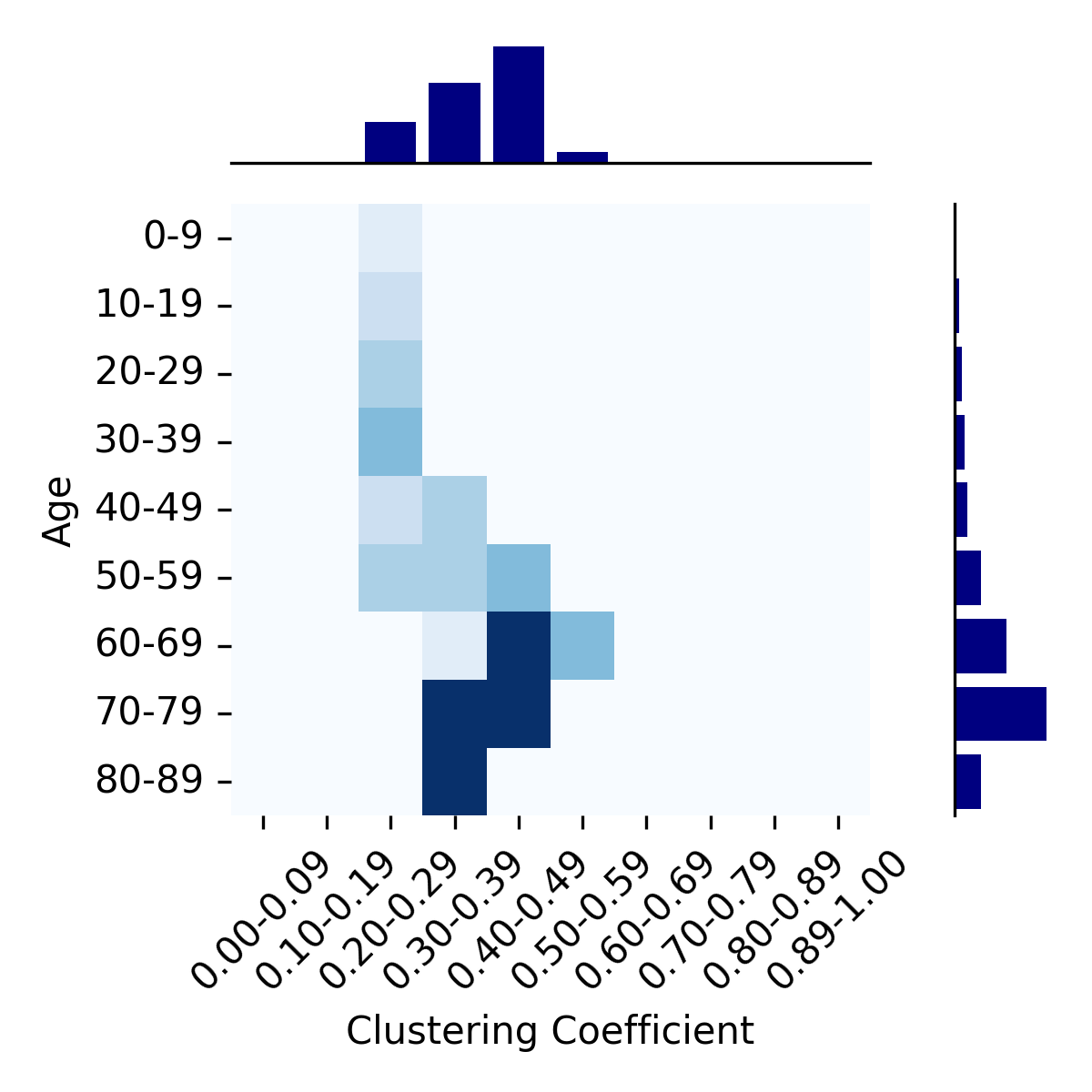}
	\end{minipage}}	
  	\subfigure[$DT-CNS^{PH}_{R}$]{
		\begin{minipage}[b]{0.185\linewidth}
			\includegraphics[width=1\linewidth]{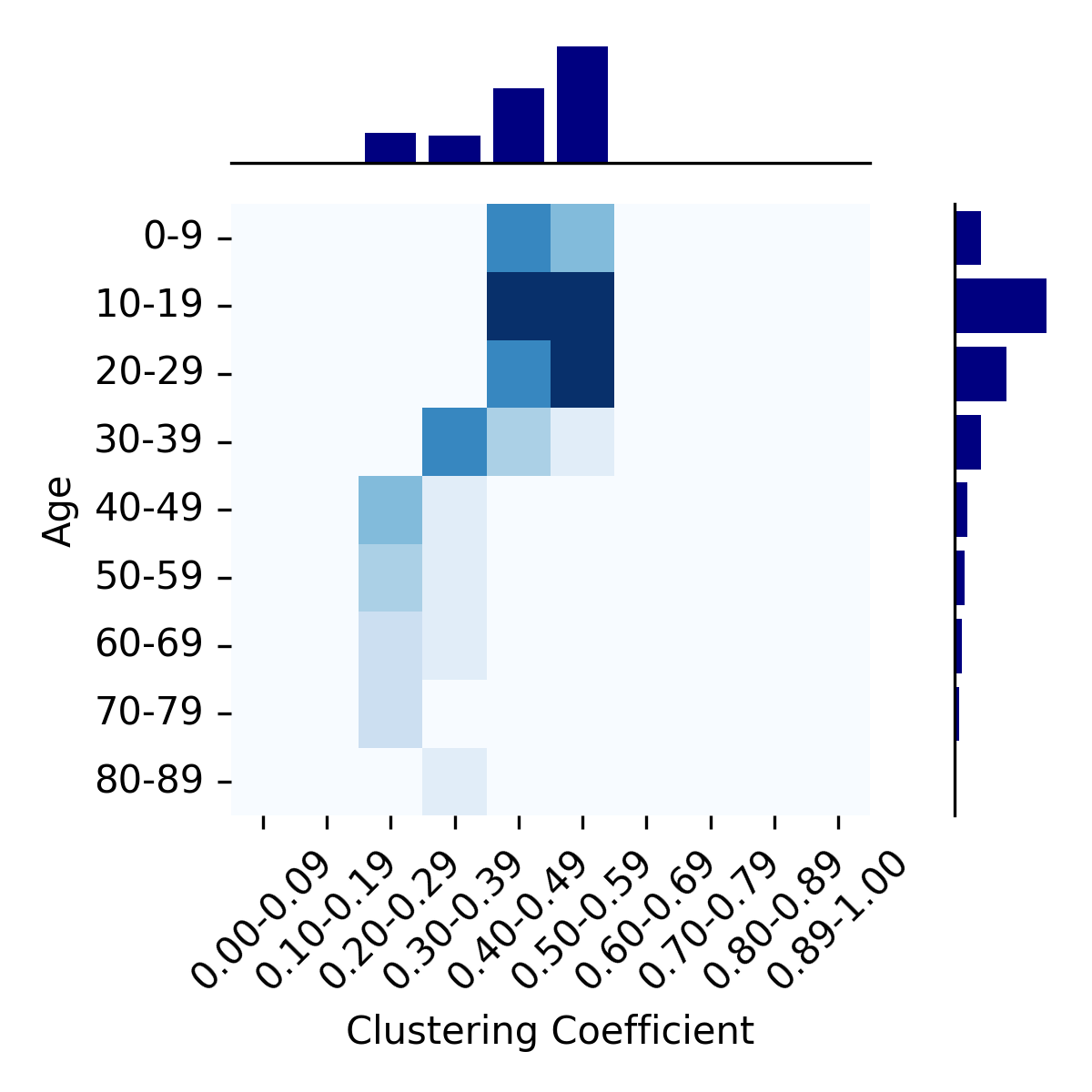}
	\end{minipage}}	
	\caption{The age and clustering coefficient distributions of social networks generated by DT-CNSs based different features and rules}
\label{age and clus}
\end{figure*}

\begin{figure*}[htp] 
	\centering
	\subfigure[$DT-CNS^{P+}_{U}$]{
		\begin{minipage}[b]{0.185\linewidth}
			\includegraphics[width=1\linewidth]{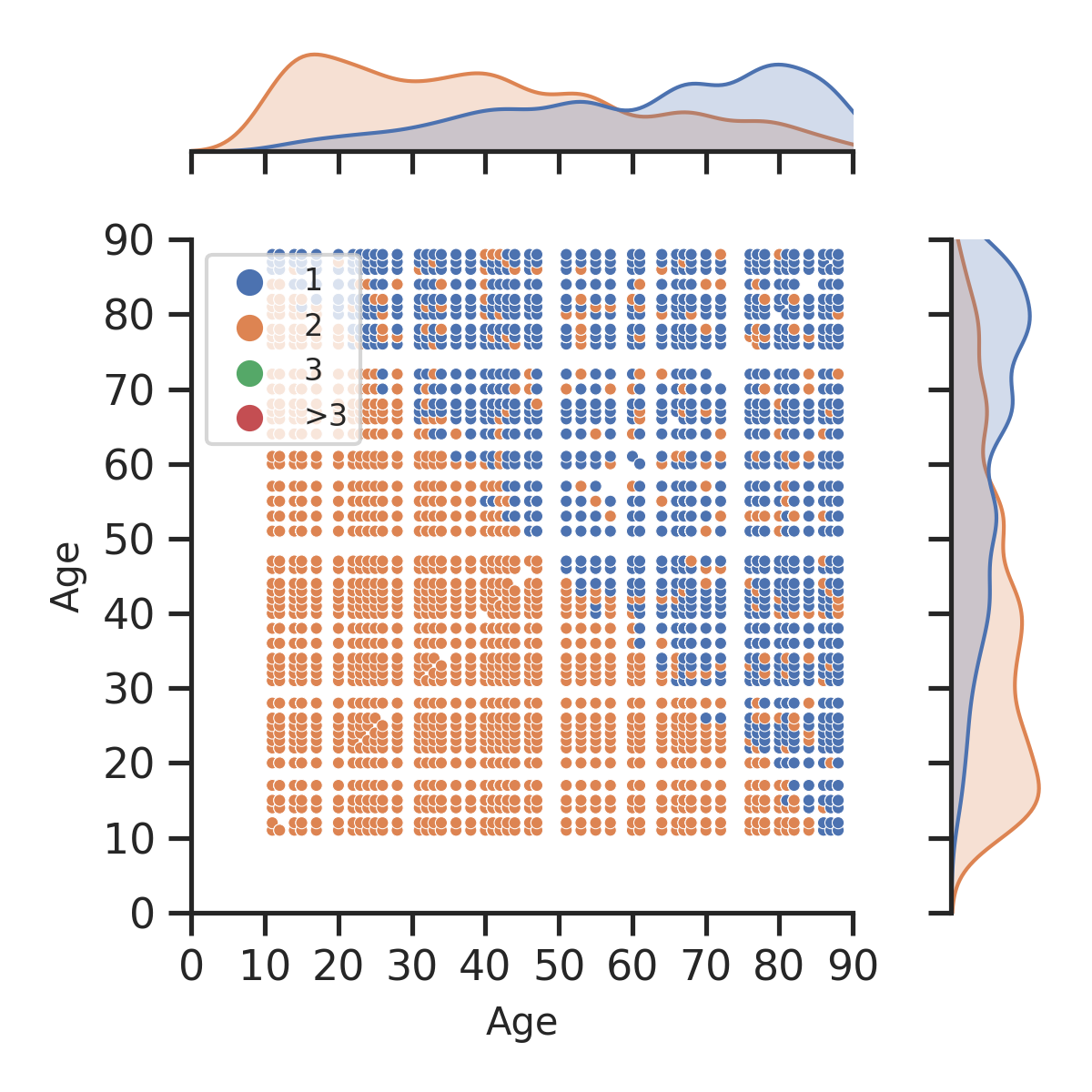}
	\end{minipage}}	
 	\subfigure[$DT-CNS^{P+}_{B}$]{
		\begin{minipage}[b]{0.185\linewidth}
			\includegraphics[width=1\linewidth]{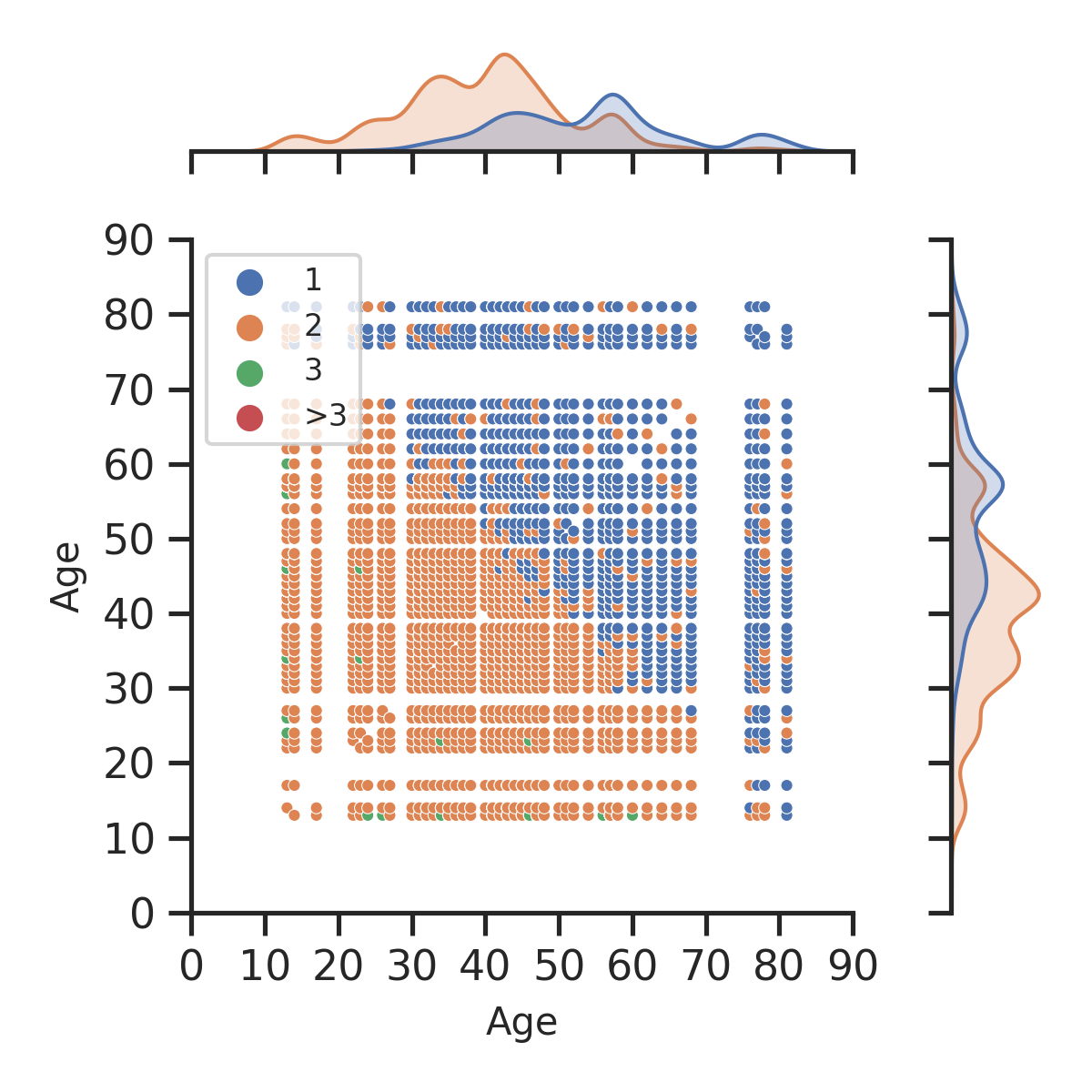}
	\end{minipage}}	
 	\subfigure[$DT-CNS^{P+}_{I}$]{
		\begin{minipage}[b]{0.185\linewidth}
			\includegraphics[width=1\linewidth]{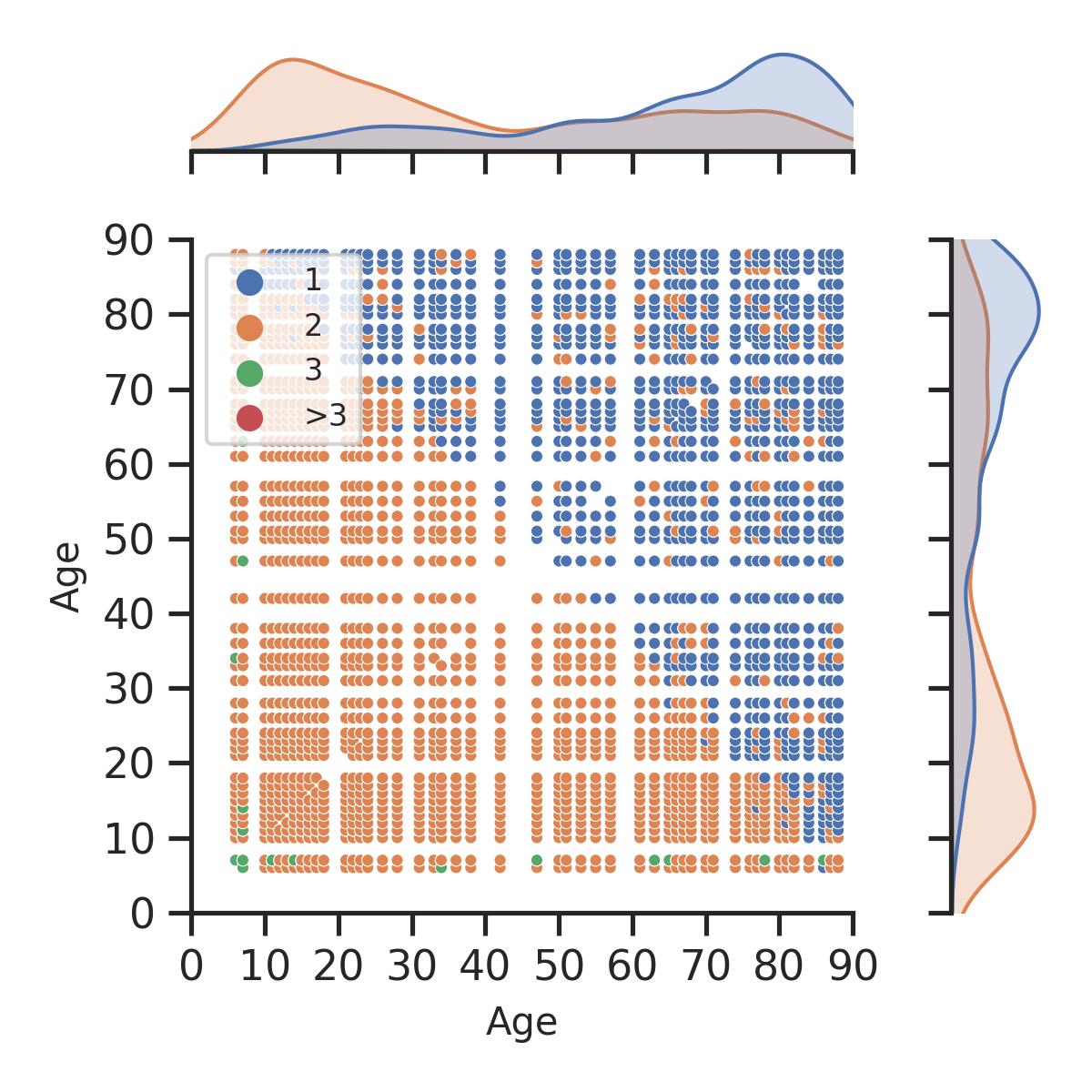}
	\end{minipage}}	
  	\subfigure[$DT-CNS^{P+}_{L}$]{
		\begin{minipage}[b]{0.185\linewidth}
			\includegraphics[width=1\linewidth]{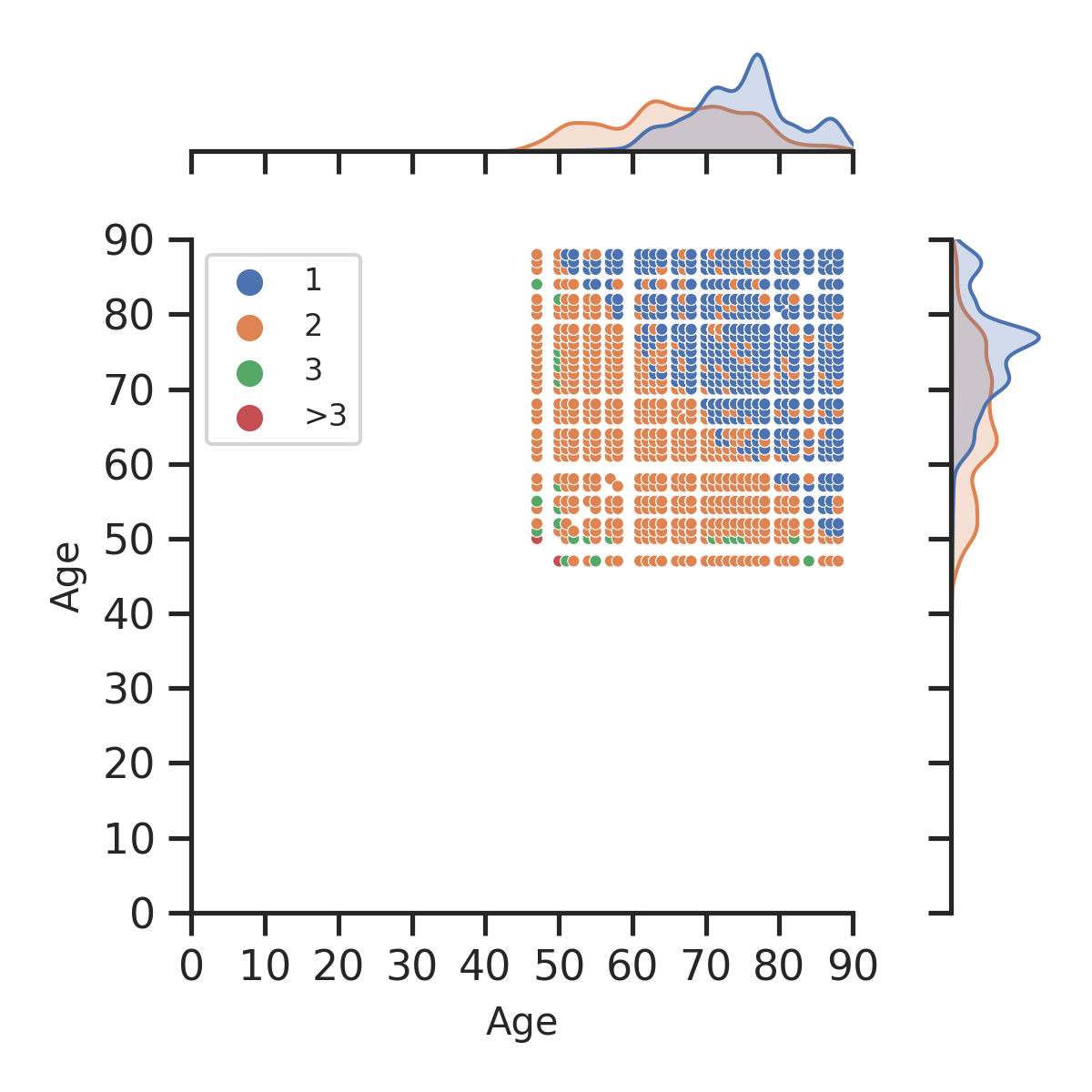}
	\end{minipage}}	
  	\subfigure[$DT-CNS^{P+}_{R}$]{
		\begin{minipage}[b]{0.185\linewidth}
			\includegraphics[width=1\linewidth]{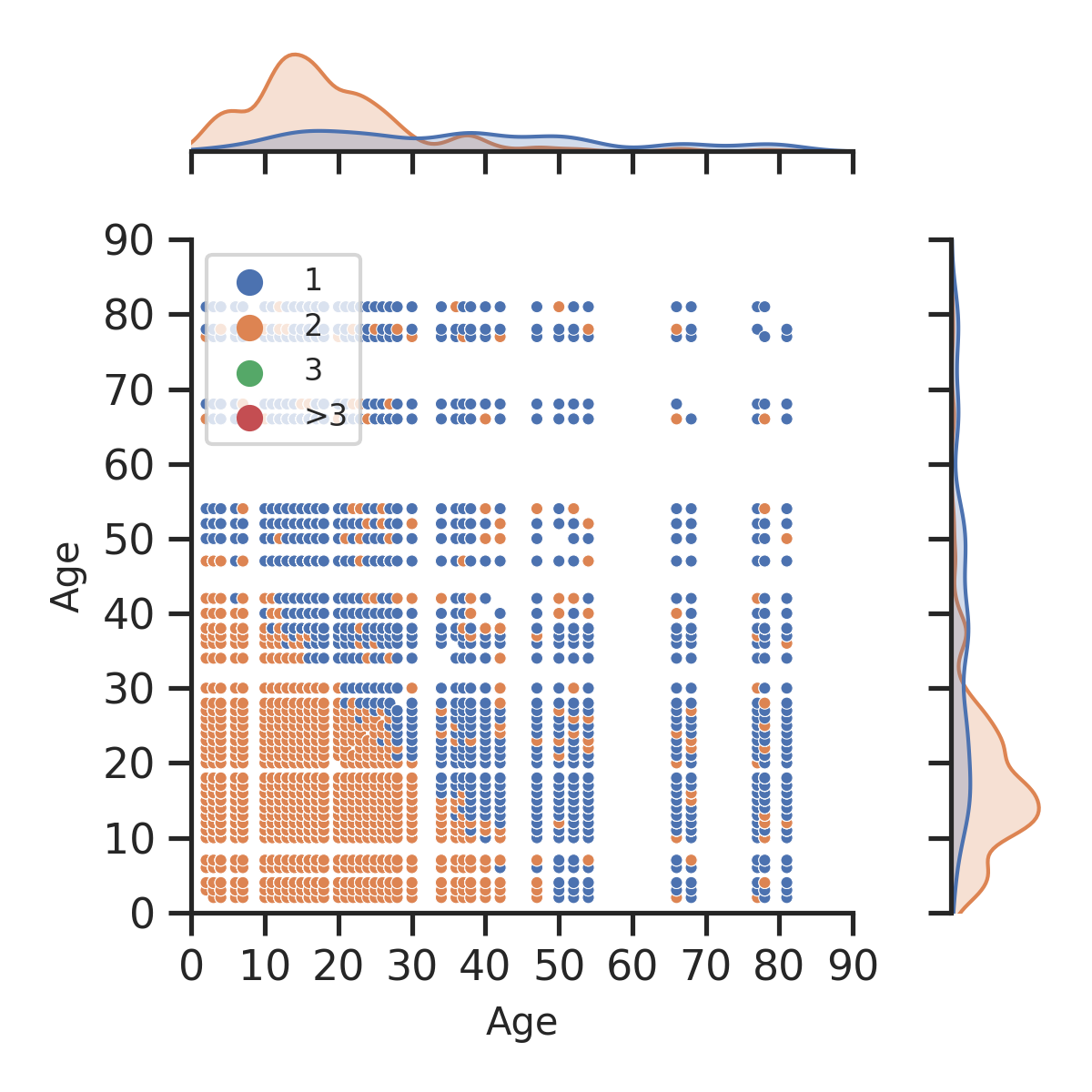}
	\end{minipage}}	\\
 	\subfigure[$DT-CNS^{P-}_{U}$]{
		\begin{minipage}[b]{0.185\linewidth}
			\includegraphics[width=1\linewidth]{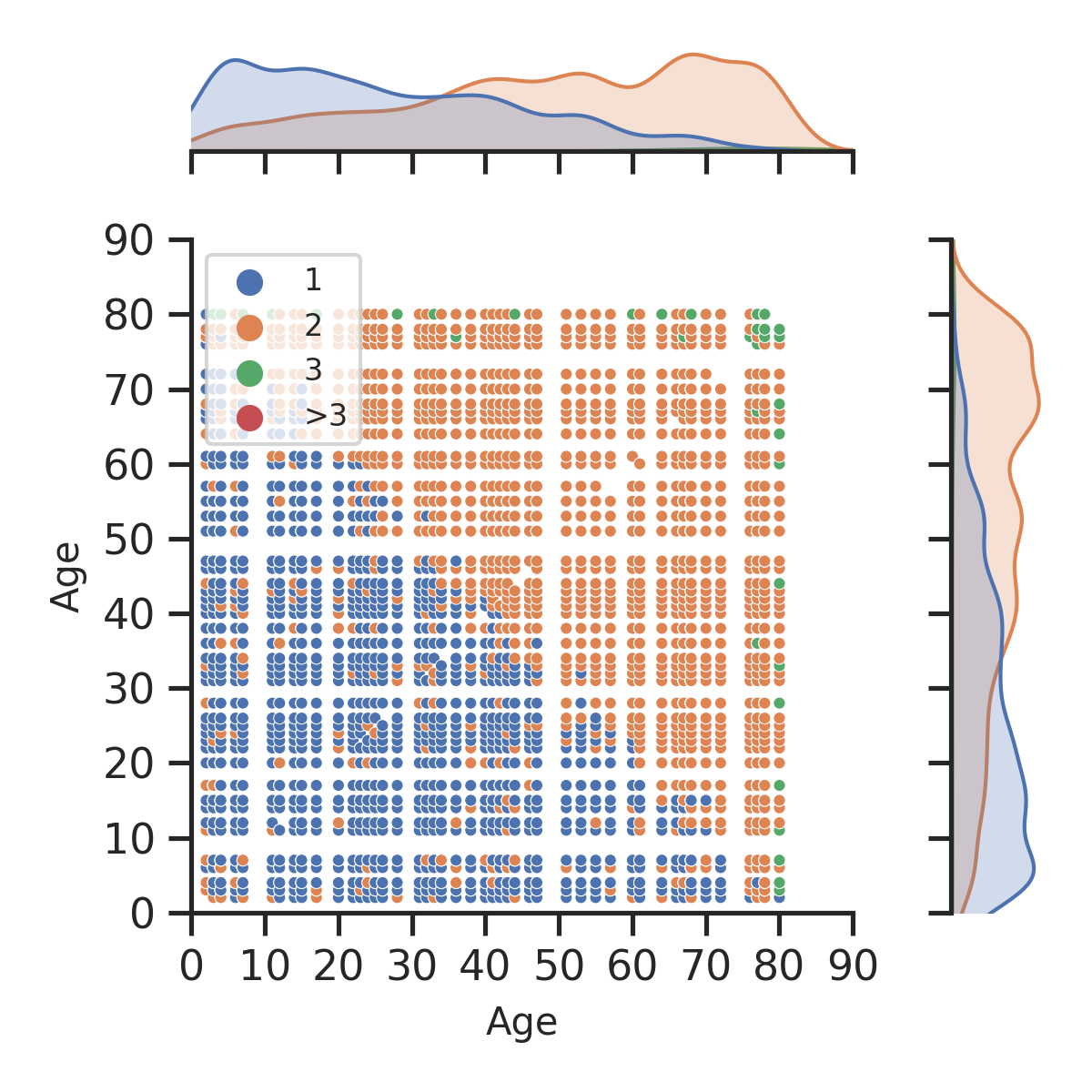}
	\end{minipage}}	
 	\subfigure[$DT-CNS^{P-}_{B}$]{
		\begin{minipage}[b]{0.185\linewidth}
			\includegraphics[width=1\linewidth]{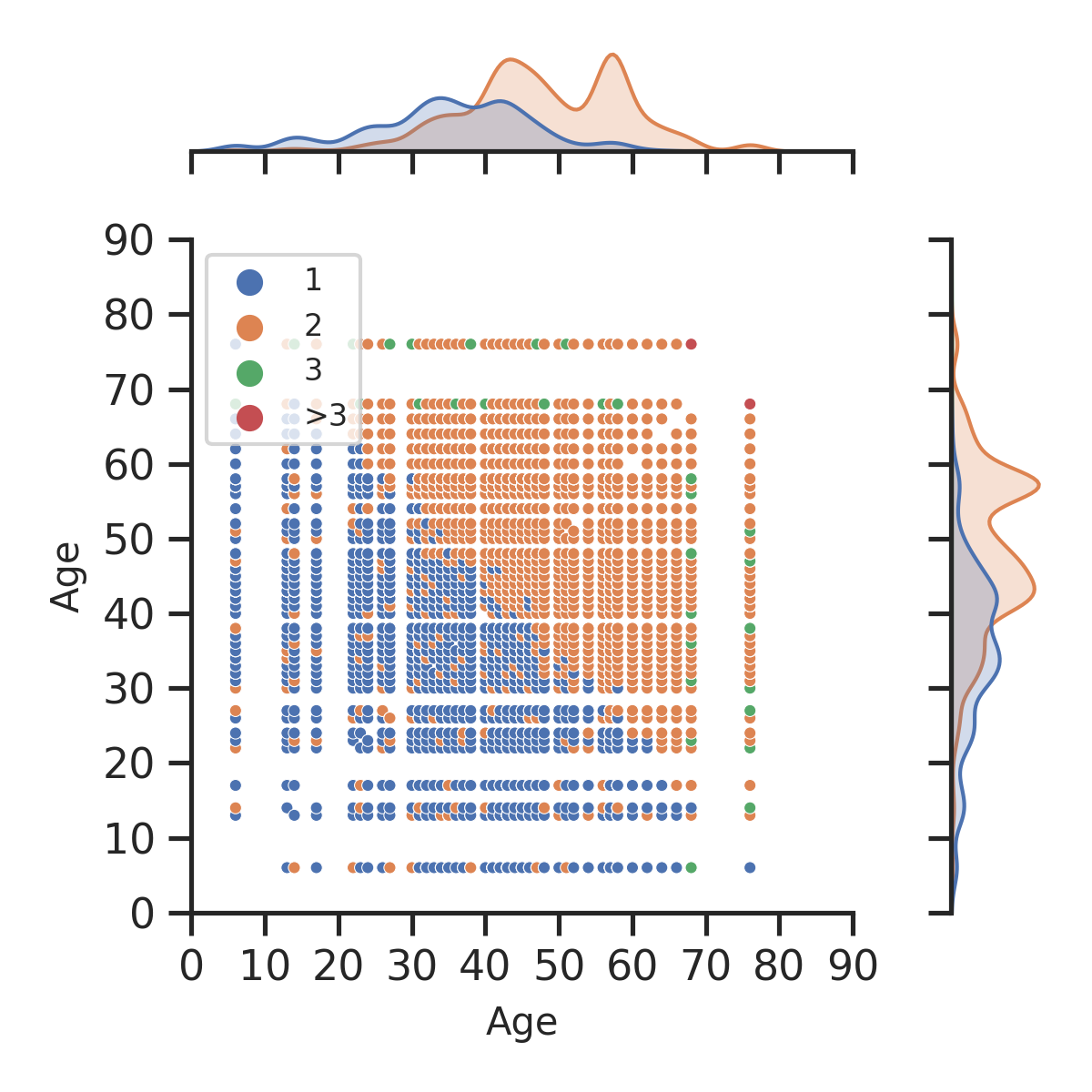}
	\end{minipage}}	
 	\subfigure[$DT-CNS^{P-}_{I}$]{
		\begin{minipage}[b]{0.185\linewidth}
			\includegraphics[width=1\linewidth]{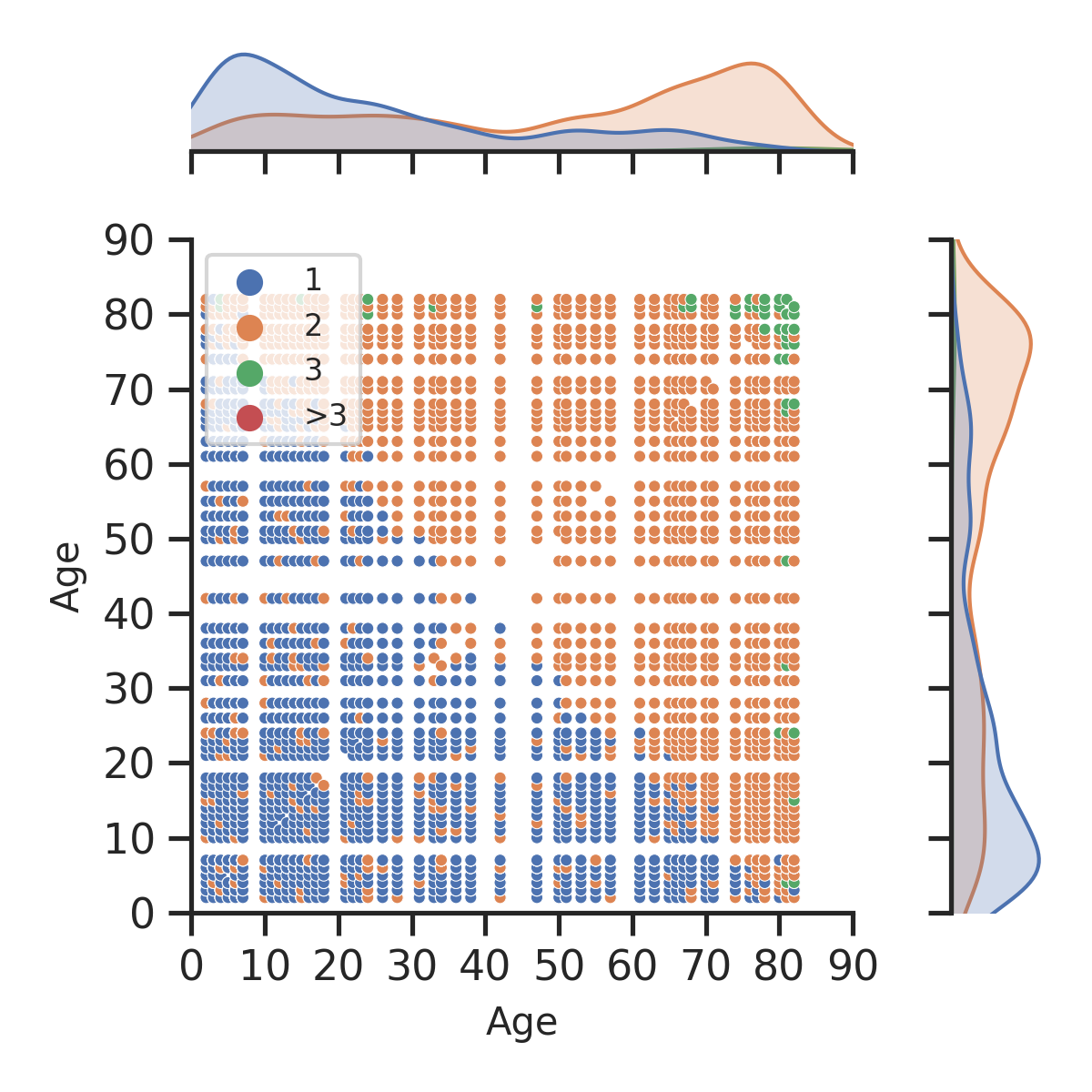}
	\end{minipage}}	
  	\subfigure[$DT-CNS^{P-}_{L}$]{
		\begin{minipage}[b]{0.185\linewidth}
			\includegraphics[width=1\linewidth]{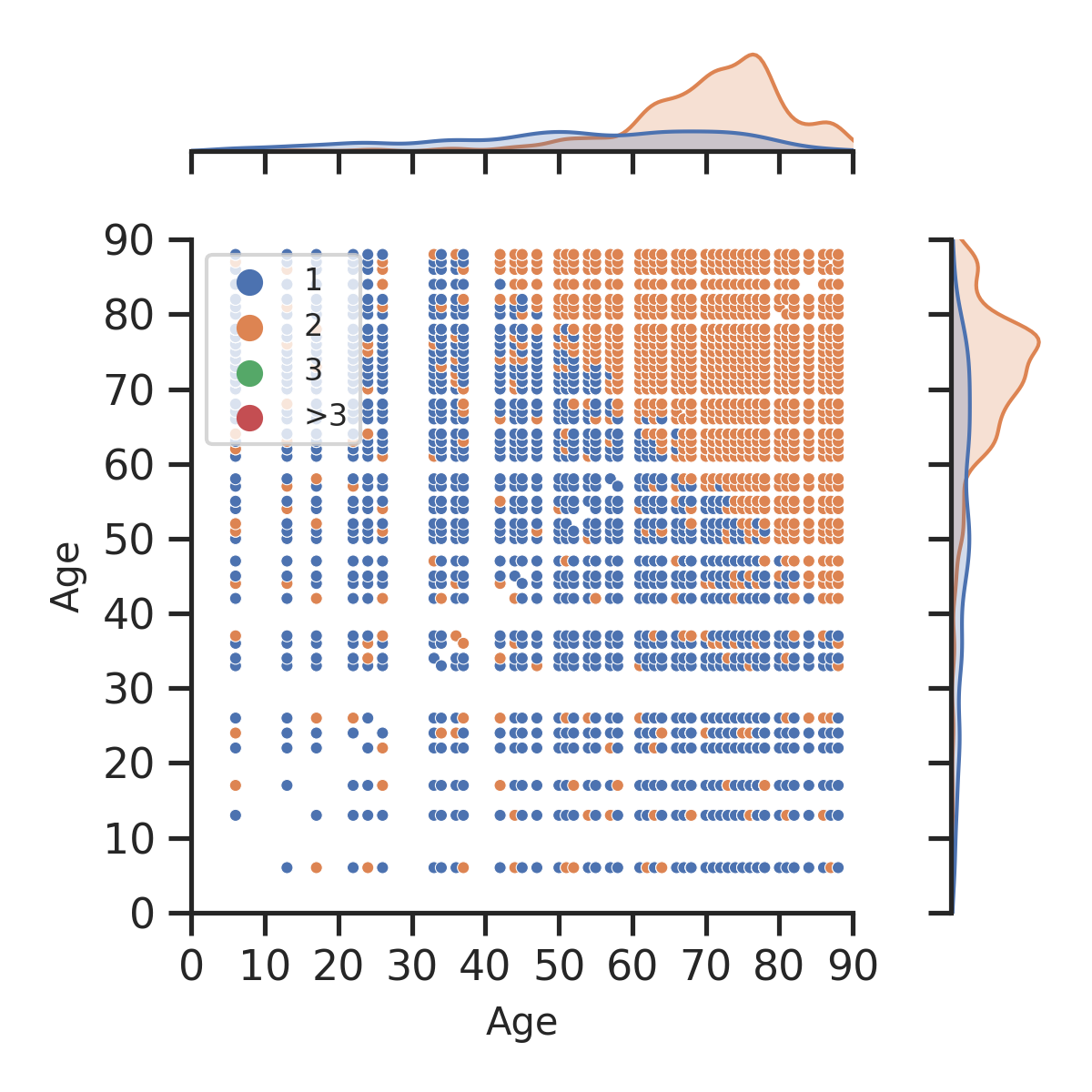}
	\end{minipage}}	
  	\subfigure[$DT-CNS^{P-}_{R}$]{
		\begin{minipage}[b]{0.185\linewidth}
			\includegraphics[width=1\linewidth]{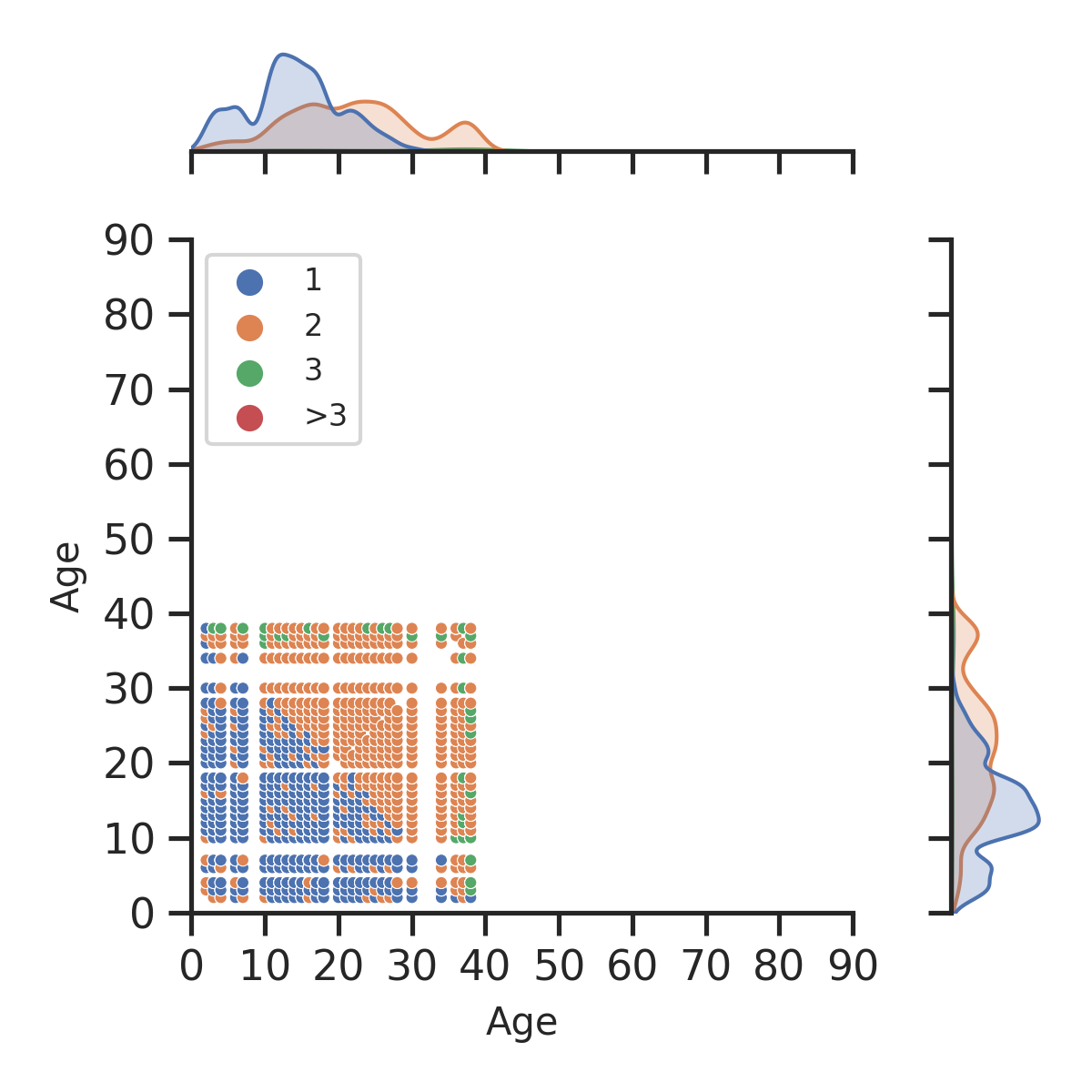}
	\end{minipage}}	\\
 	\subfigure[$DT-CNS^{H+}_{U}$]{
		\begin{minipage}[b]{0.185\linewidth}
			\includegraphics[width=1\linewidth]{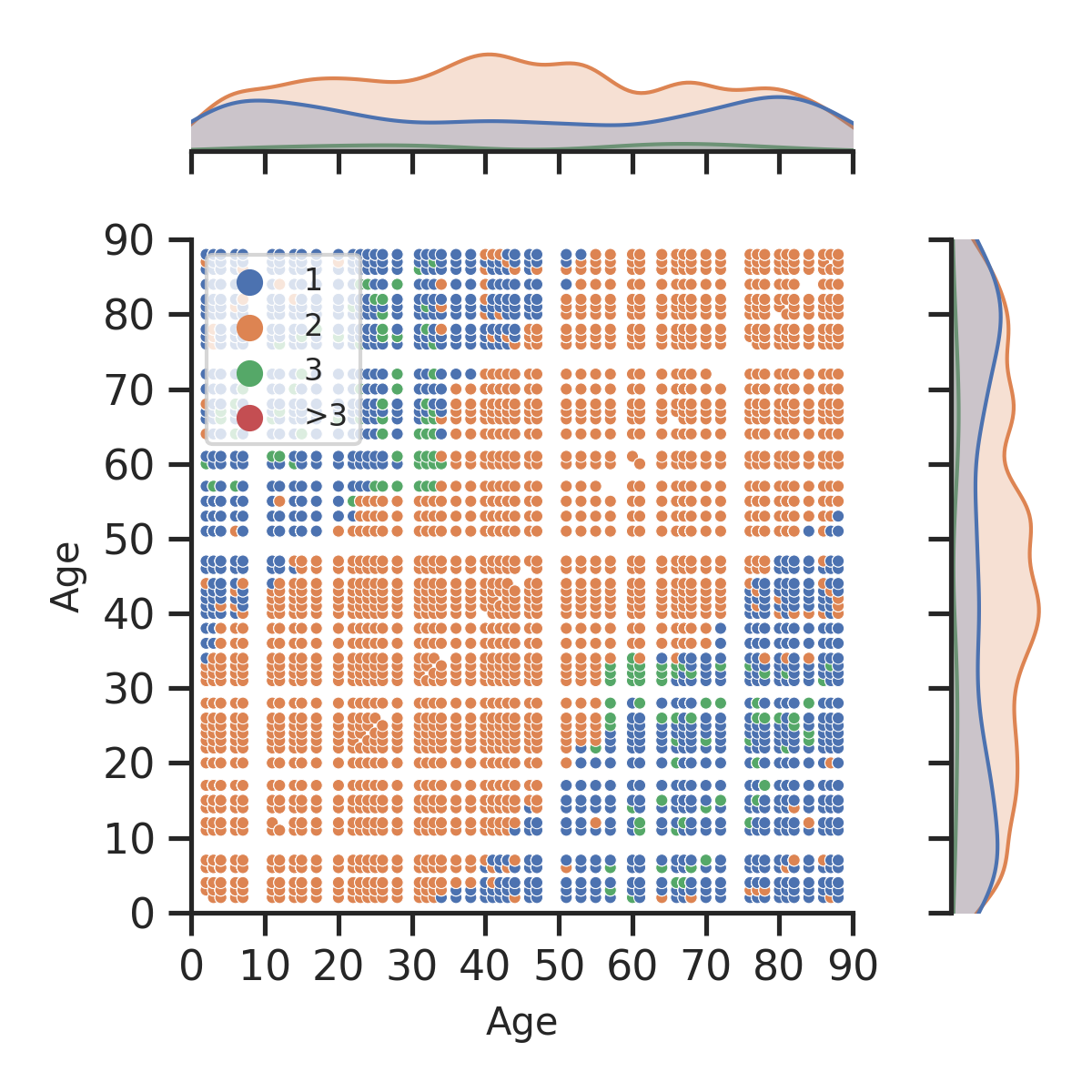}
	\end{minipage}}	
 	\subfigure[$DT-CNS^{H+}_{B}$]{
		\begin{minipage}[b]{0.185\linewidth}
			\includegraphics[width=1\linewidth]{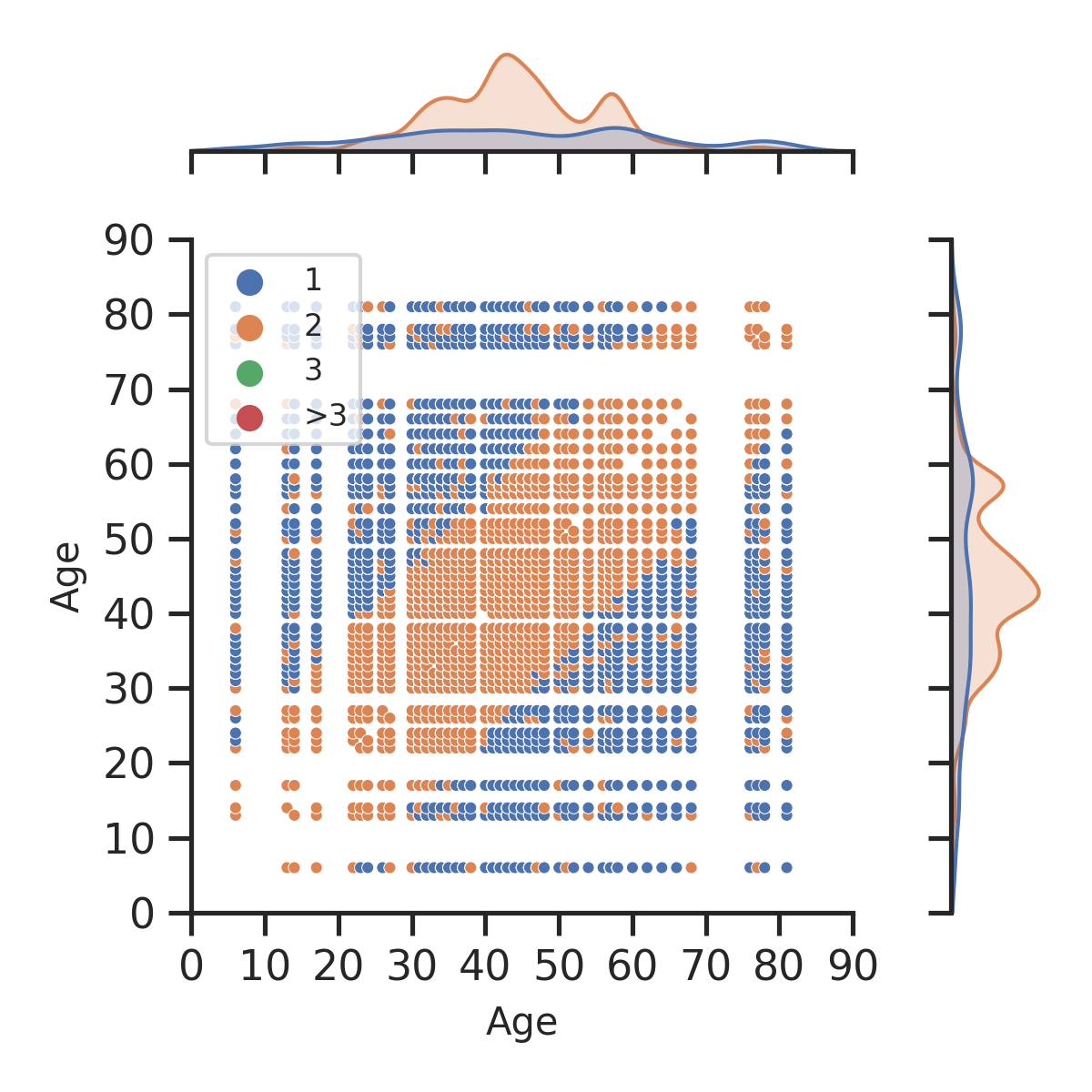}
	\end{minipage}}	
 	\subfigure[$DT-CNS^{H+}_{I}$]{
		\begin{minipage}[b]{0.185\linewidth}
			\includegraphics[width=1\linewidth]{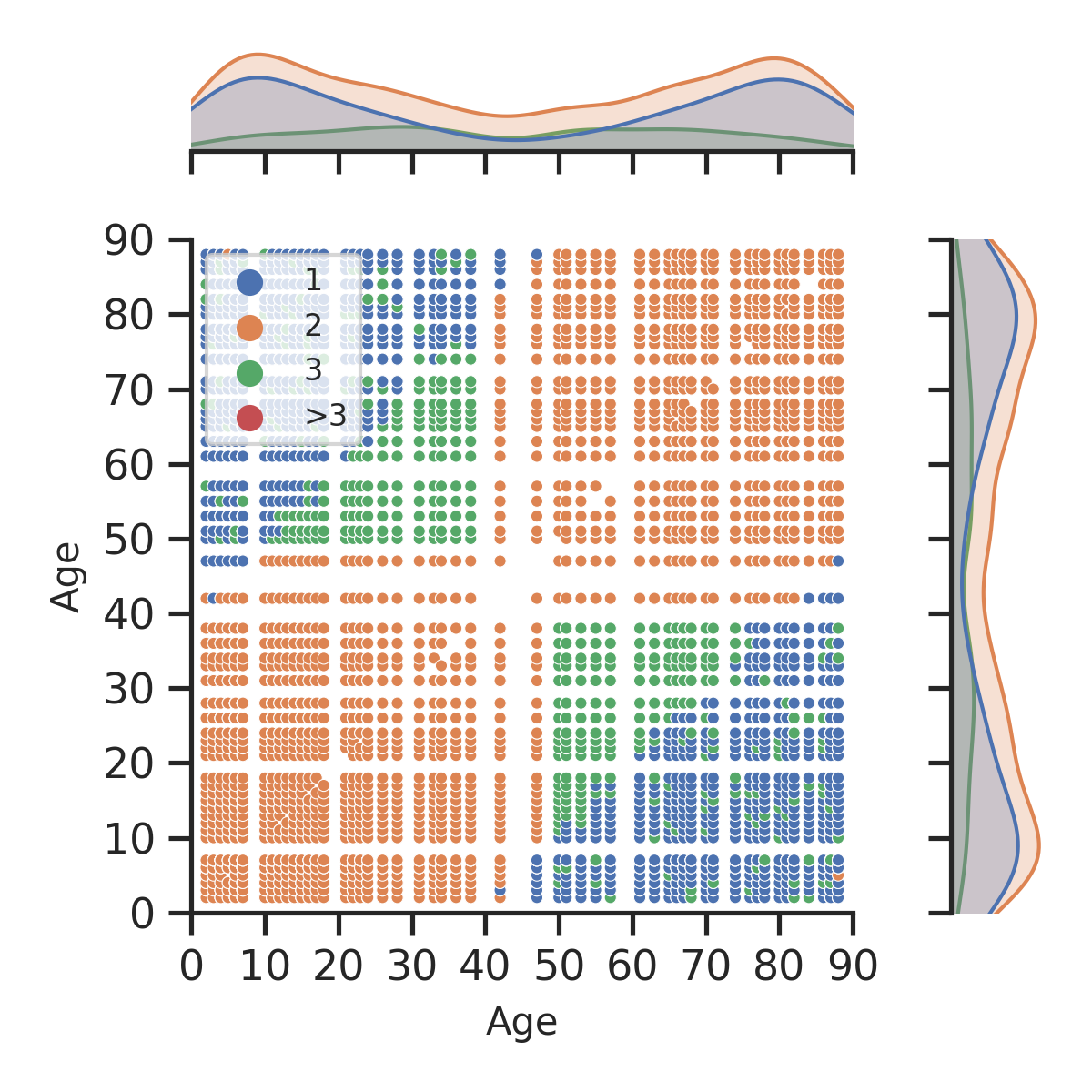}
	\end{minipage}}	
  	\subfigure[$DT-CNS^{H+}_{L}$]{
		\begin{minipage}[b]{0.185\linewidth}
			\includegraphics[width=1\linewidth]{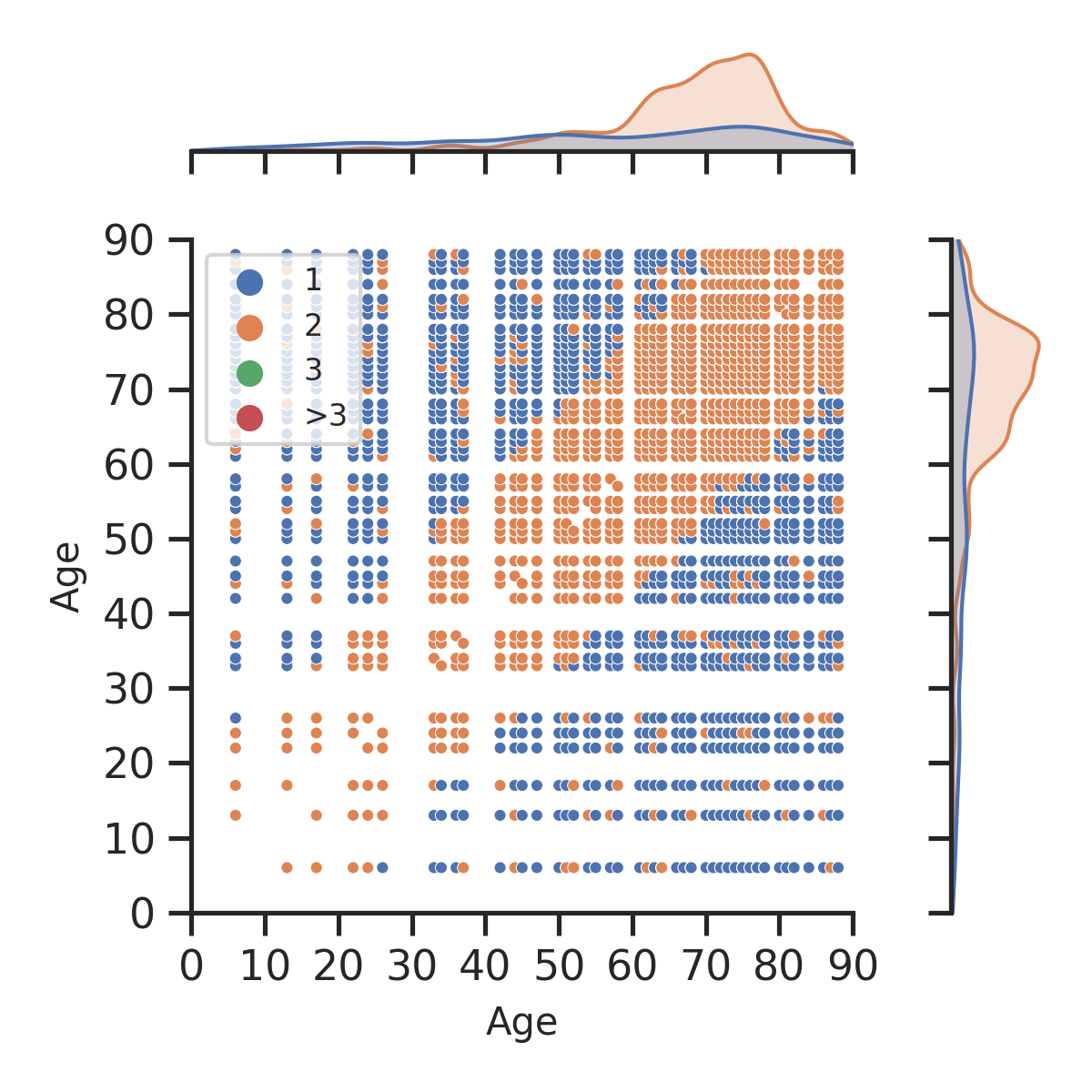}
	\end{minipage}}	
  	\subfigure[$DT-CNS^{H+}_{R}$]{
		\begin{minipage}[b]{0.185\linewidth}
			\includegraphics[width=1\linewidth]{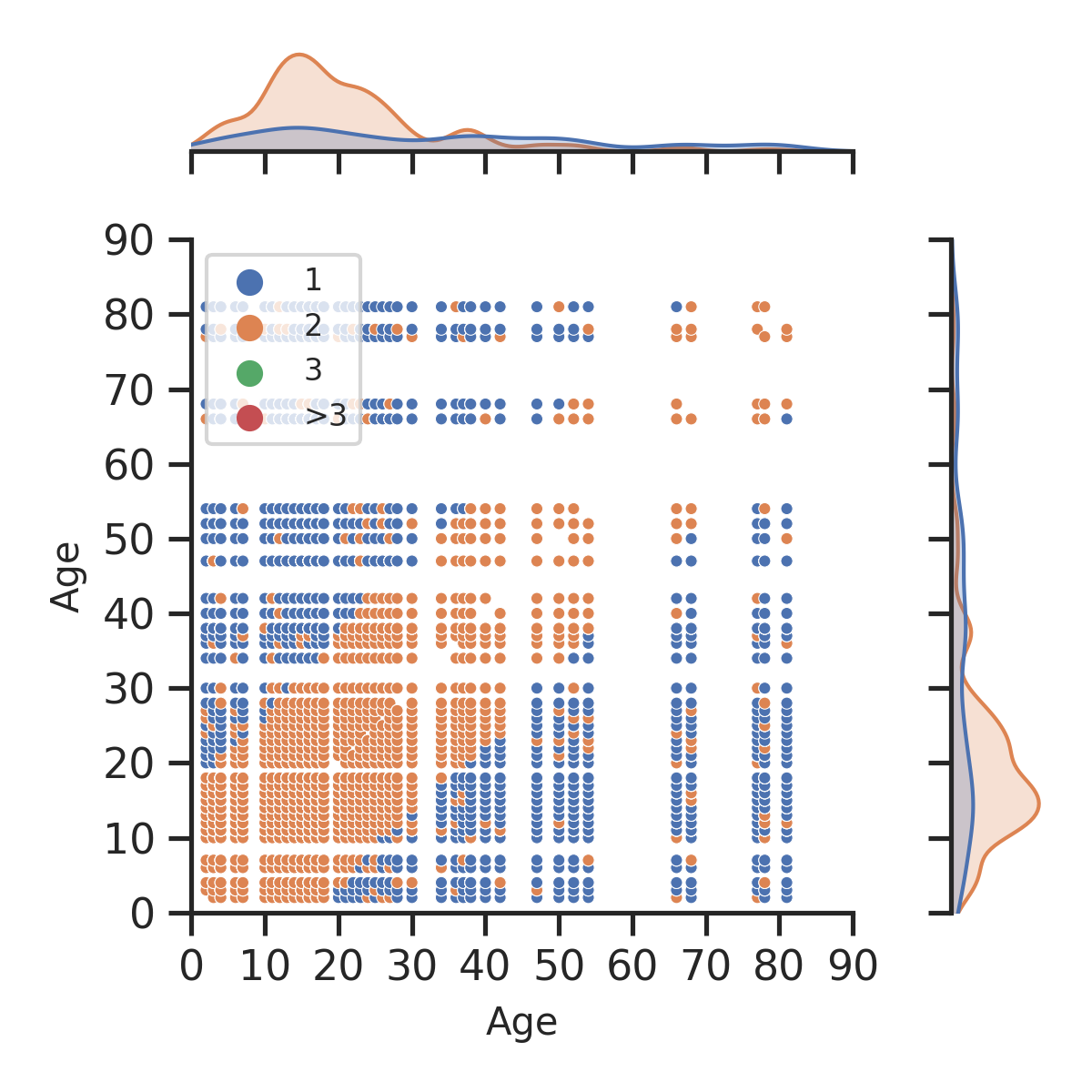}
	\end{minipage}}	\\
 	\subfigure[$DT-CNS^{H-}_{U}$]{
		\begin{minipage}[b]{0.185\linewidth}
			\includegraphics[width=1\linewidth]{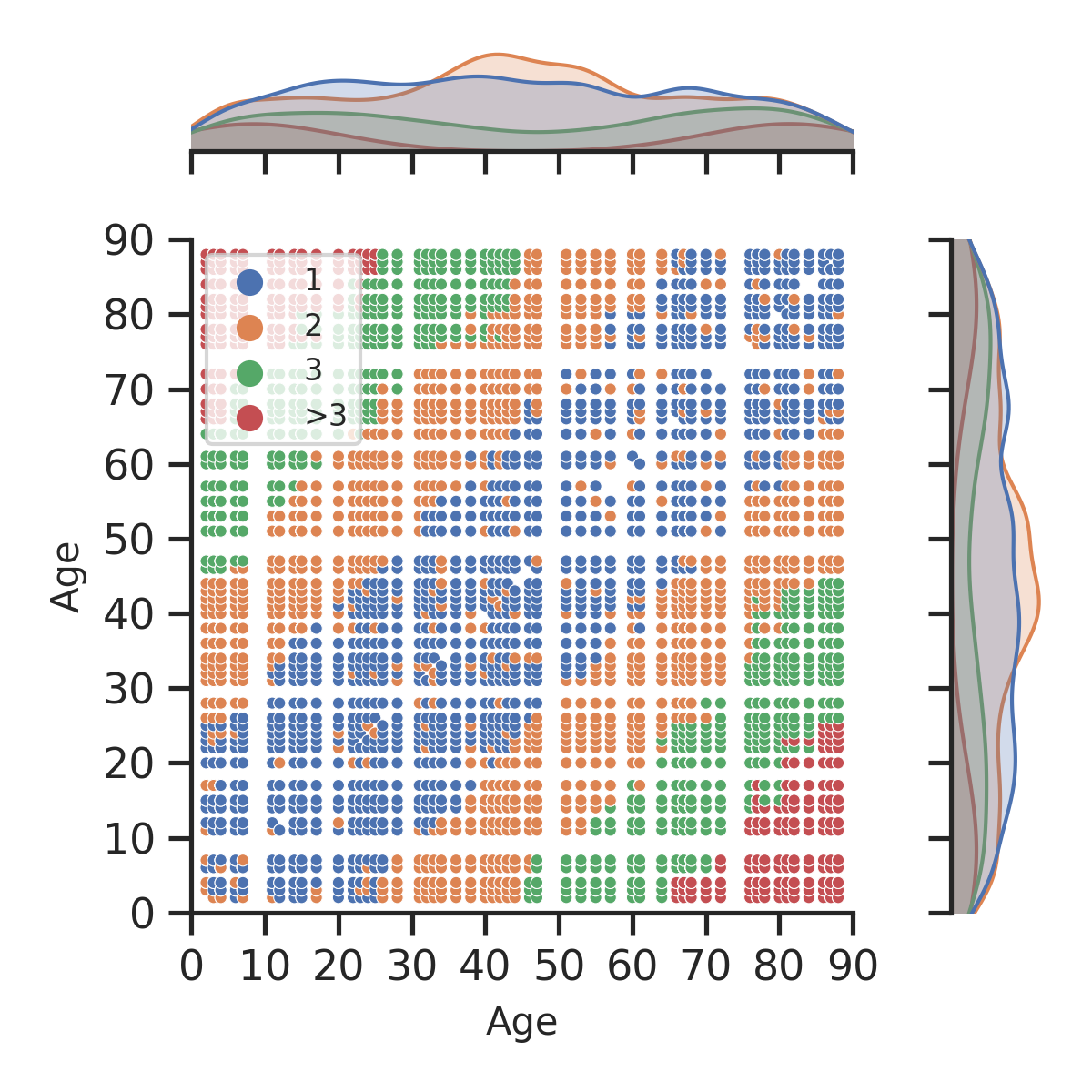}
	\end{minipage}}	
 	\subfigure[$DT-CNS^{H-}_{B}$]{
		\begin{minipage}[b]{0.185\linewidth}
			\includegraphics[width=1\linewidth]{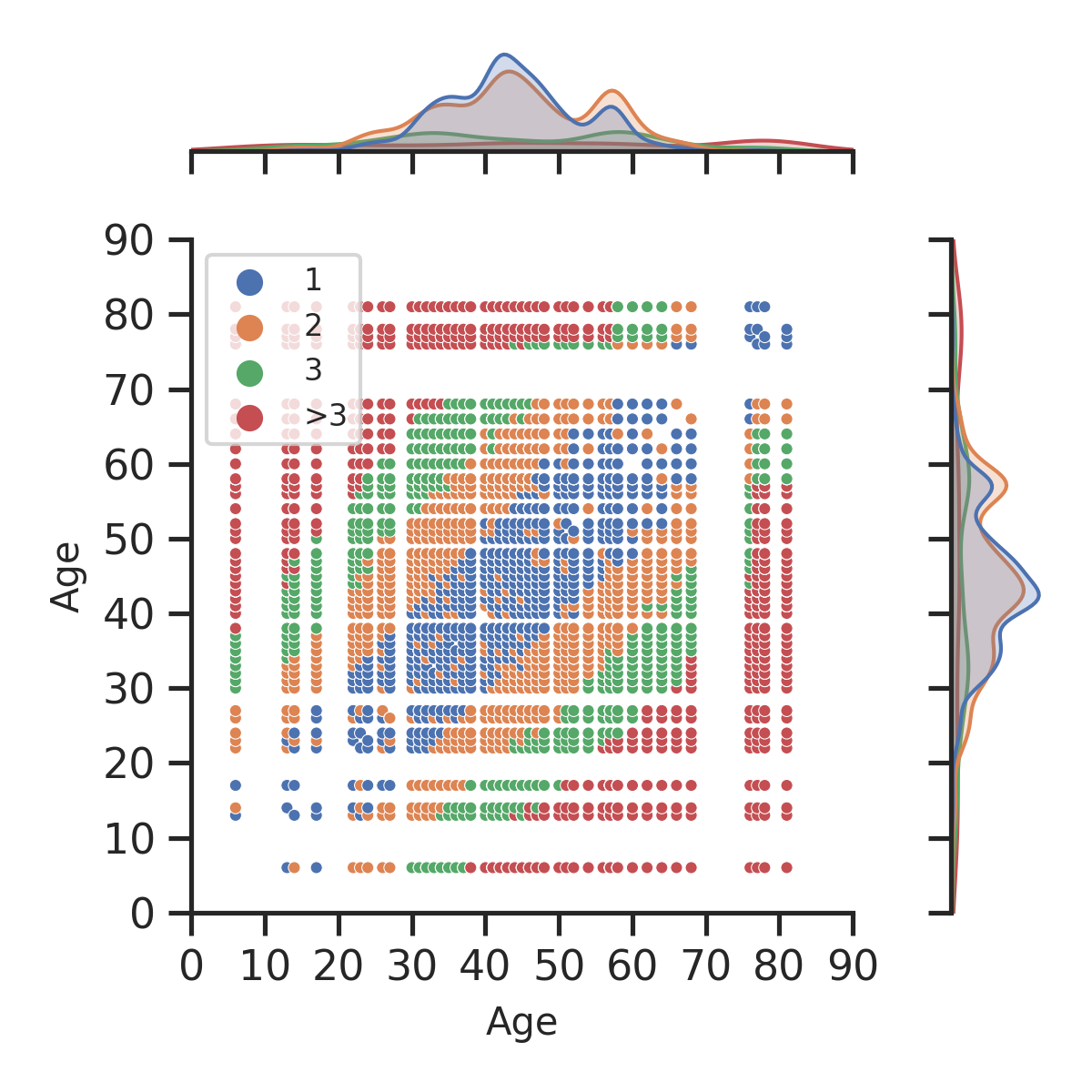}
	\end{minipage}}	
 	\subfigure[$DT-CNS^{H-}_{I}$]{
		\begin{minipage}[b]{0.185\linewidth}
			\includegraphics[width=1\linewidth]{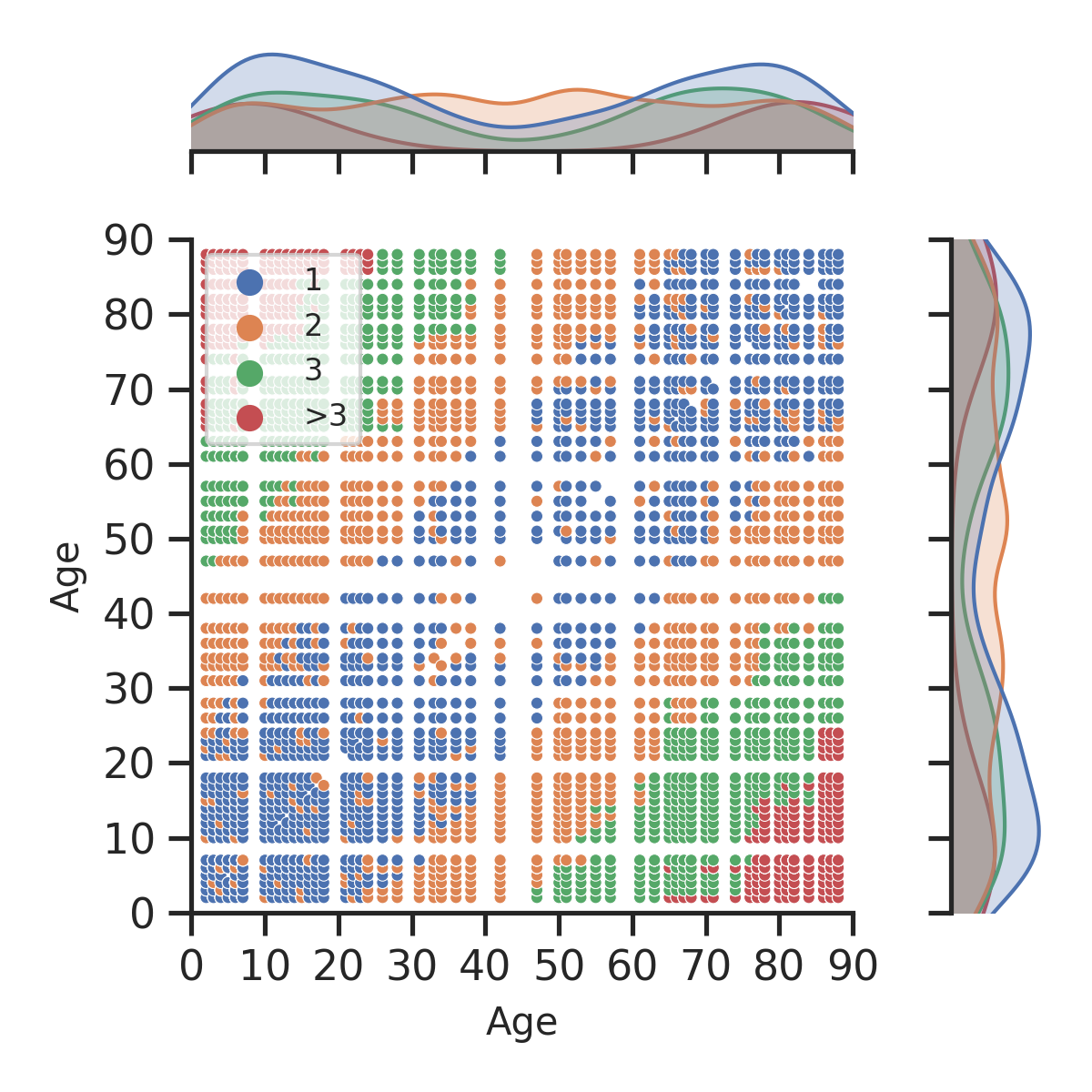}
	\end{minipage}}	
  	\subfigure[$DT-CNS^{H-}_{L}$]{
		\begin{minipage}[b]{0.185\linewidth}
			\includegraphics[width=1\linewidth]{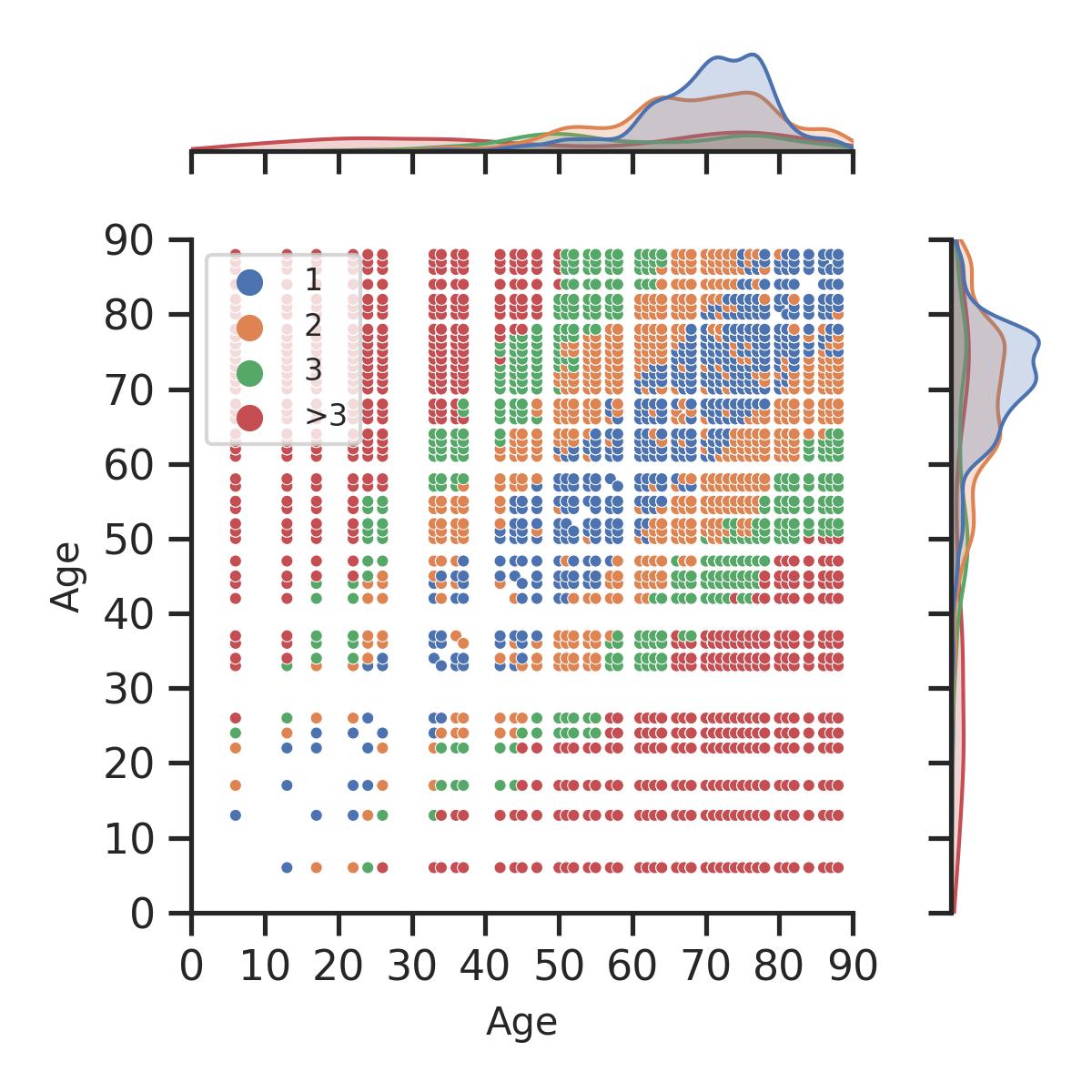}
	\end{minipage}}	
  	\subfigure[$DT-CNS^{H-}_{R}$]{
		\begin{minipage}[b]{0.185\linewidth}
			\includegraphics[width=1\linewidth]{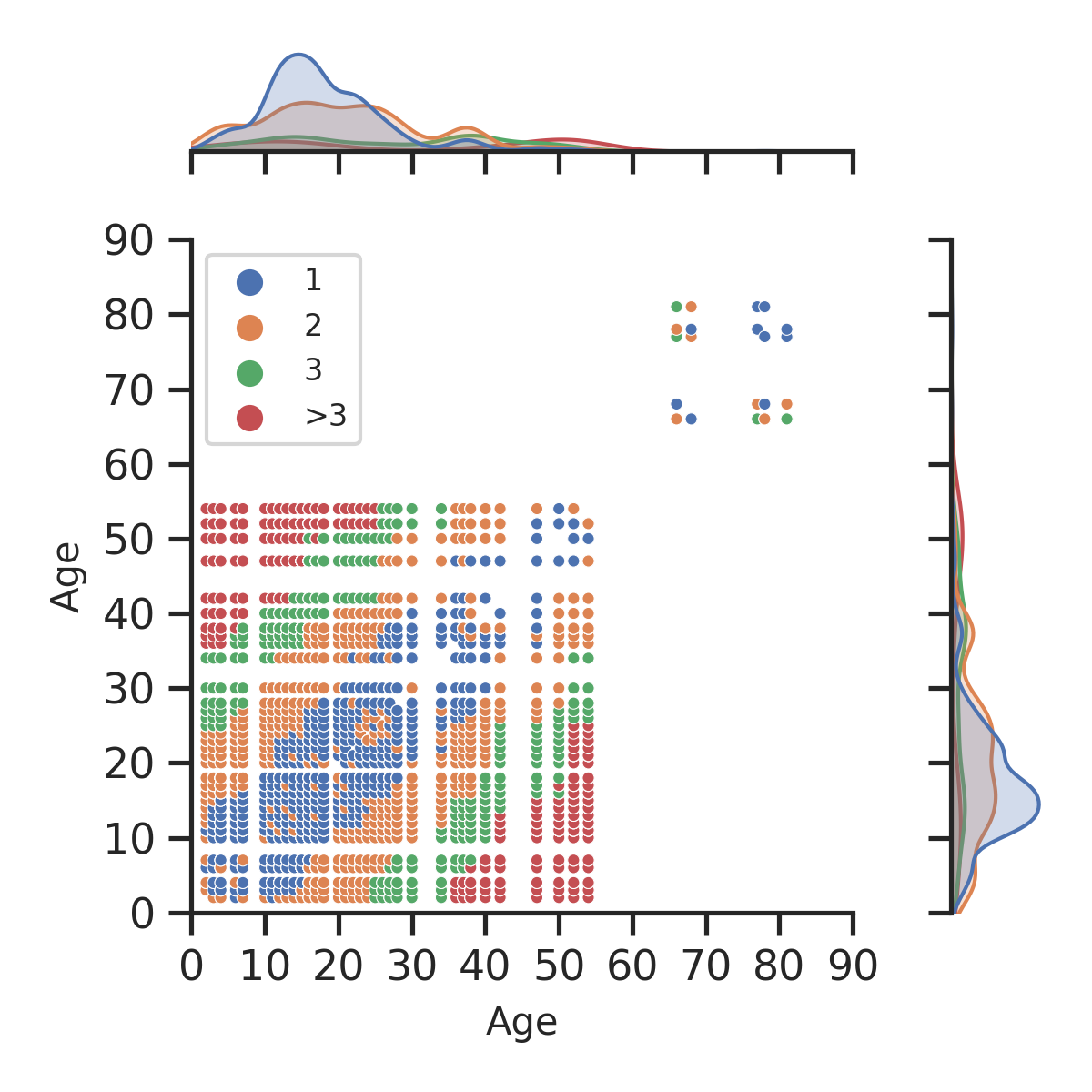}
	\end{minipage}}	\\
 	\subfigure[$DT-CNS^{PH}_{U}$]{
		\begin{minipage}[b]{0.185\linewidth}
			\includegraphics[width=1\linewidth]{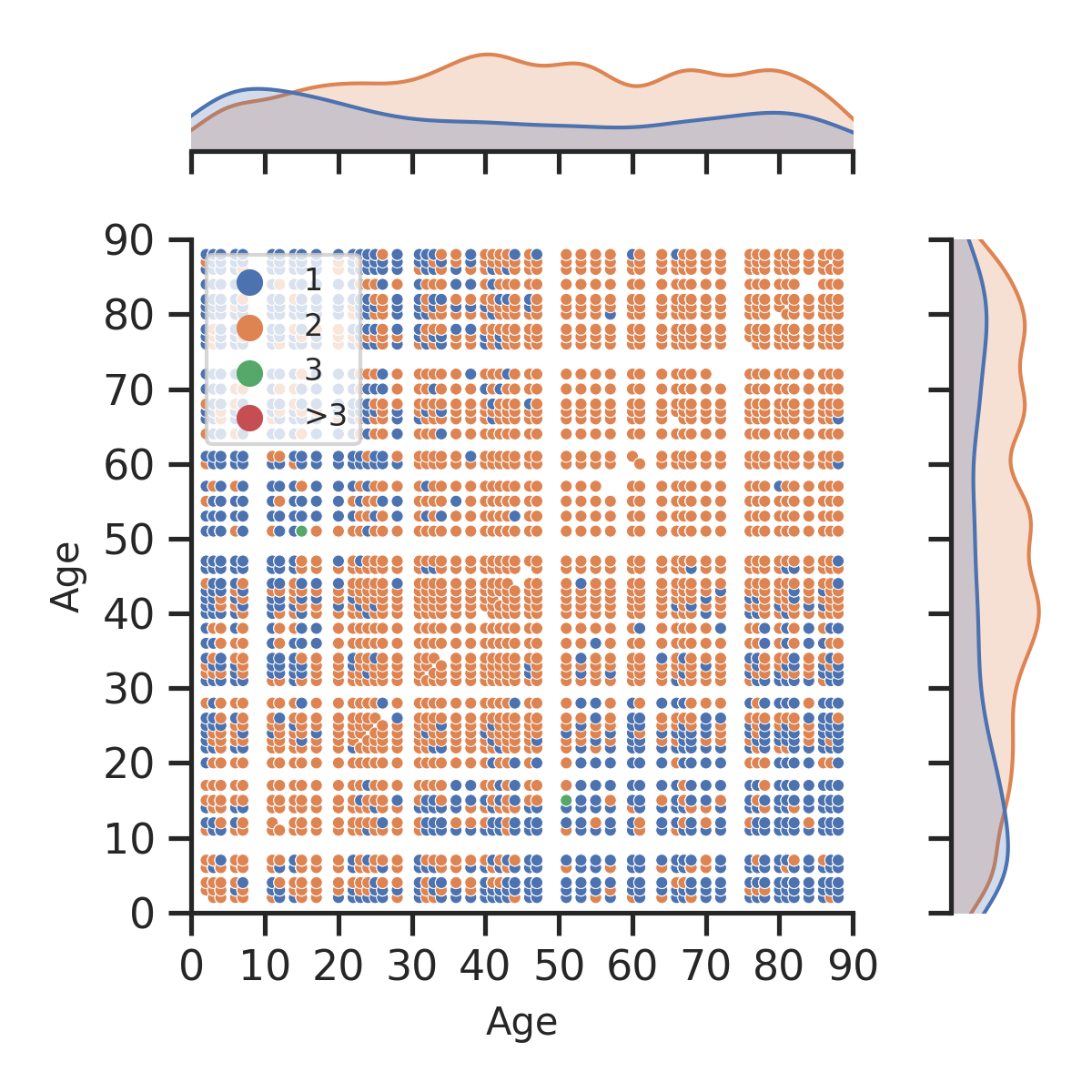}
	\end{minipage}}	
 	\subfigure[$DT-CNS^{PH}_{B}$]{
		\begin{minipage}[b]{0.185\linewidth}
			\includegraphics[width=1\linewidth]{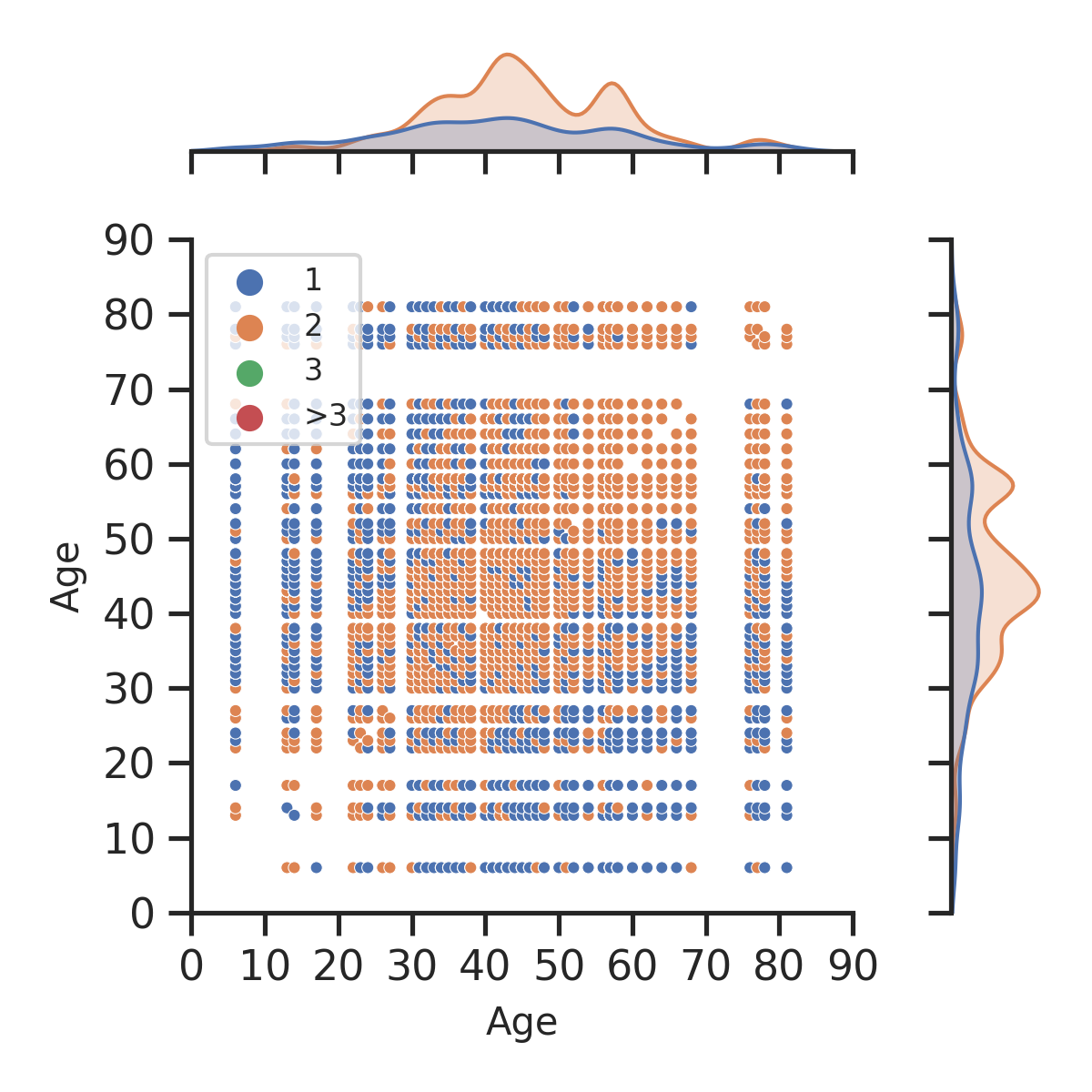}
	\end{minipage}}	
 	\subfigure[$DT-CNS^{PH}_{I}$]{
		\begin{minipage}[b]{0.185\linewidth}
			\includegraphics[width=1\linewidth]{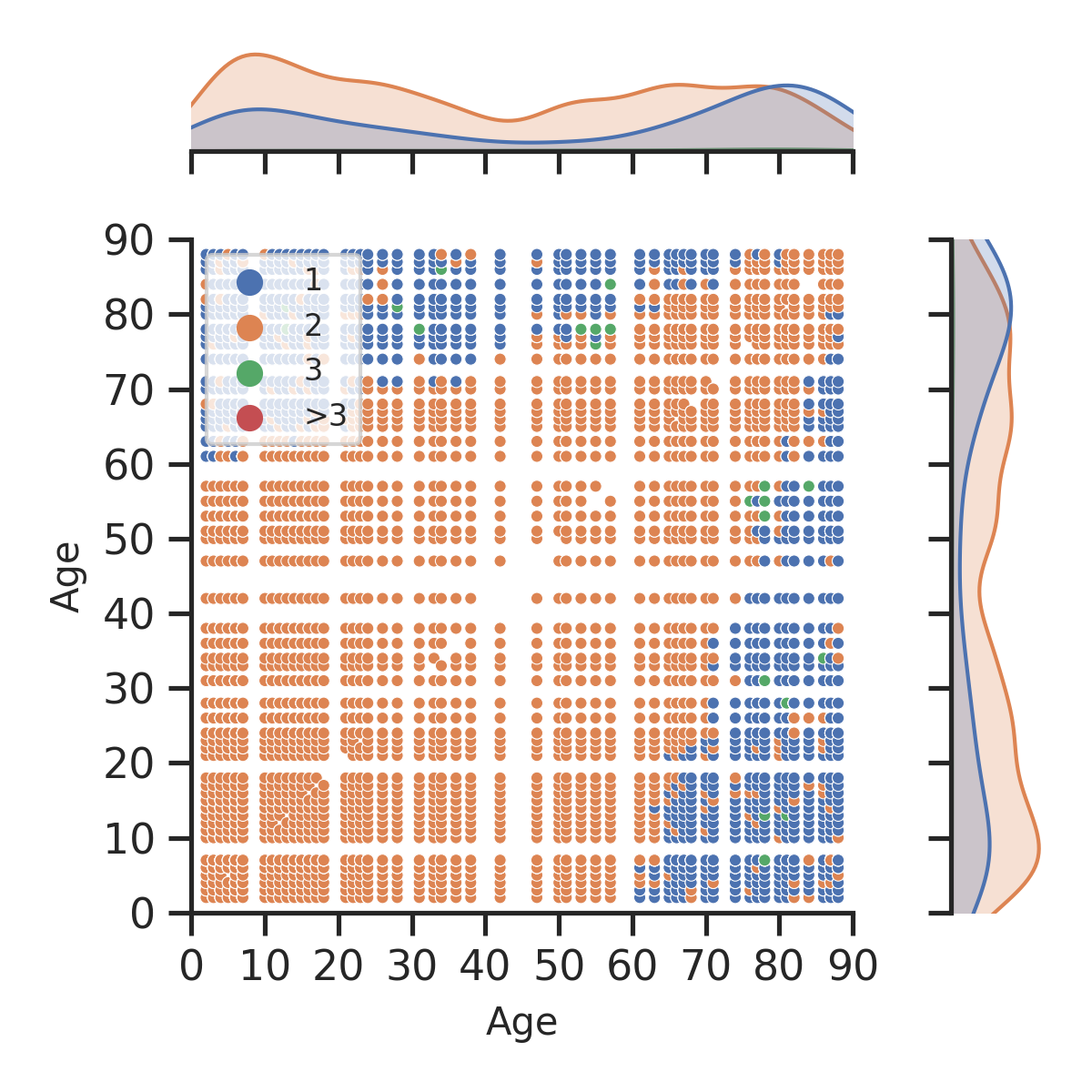}
	\end{minipage}}	
  	\subfigure[$DT-CNS^{PH}_{L}$]{
		\begin{minipage}[b]{0.185\linewidth}
			\includegraphics[width=1\linewidth]{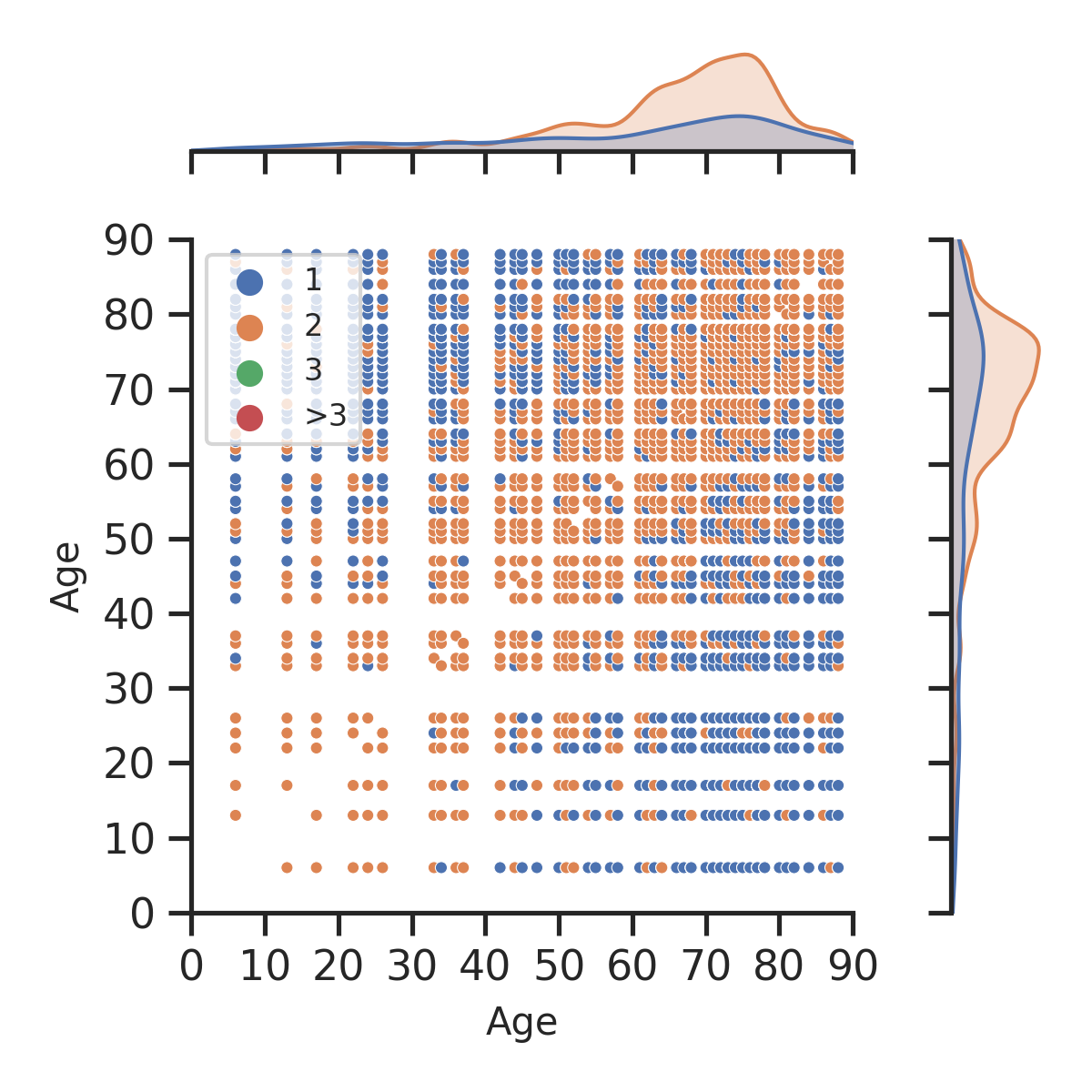}
	\end{minipage}}	
  	\subfigure[$DT-CNS^{PH}_{R}$]{
		\begin{minipage}[b]{0.185\linewidth}
			\includegraphics[width=1\linewidth]{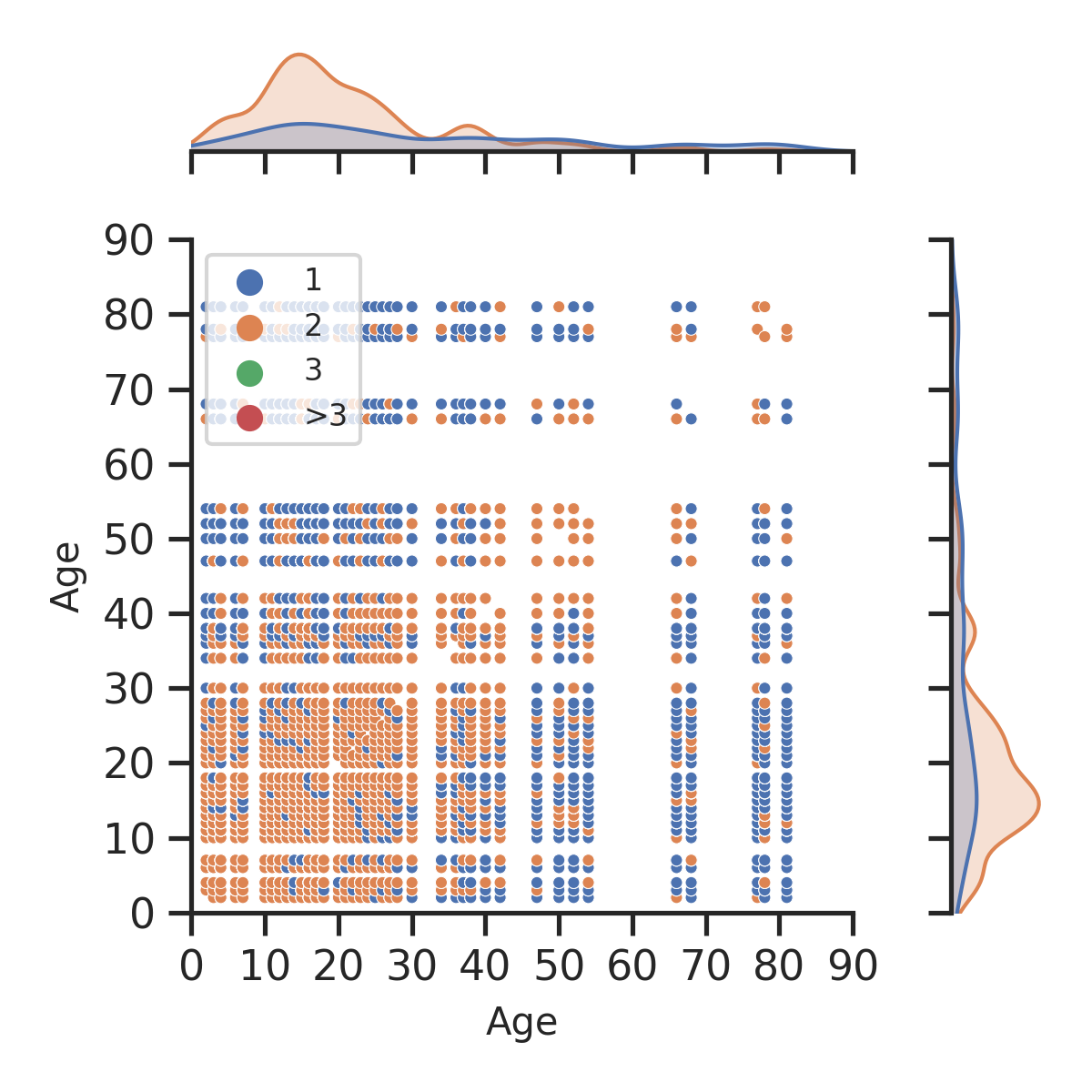}
	\end{minipage}}	
	\caption{The age and shortest path length distributions between connected nodes of social networks based on different features and rules}
\label{AgeSP}
\end{figure*}

Given the same threshold factor $\lambda_e = 1400$, that determines the number of edges, the simulated networks are correspondingly characterised with varying network topology (See Fig.~\ref{AgeAndDeg}, Fig.~\ref{age and clus} and Fig.~\ref{AgeSP}).

\paragraph{Degree Distribution} is a distribution of node degrees in a given network~\cite{musial2013kind}. As shown in Table~\ref{networkinfo}, the node degree of the target network fluctuates around the average value of $31.11$ with a standard deviation of $12.90$, ranging from $6.00$ to $70.00$. 

Fig~\ref{AgeAndDeg} shows age and degree distributions with a heatmap and two histogram plots for age (See Fig.~\ref{featfig}) and node degree (See Tab.~\ref{networkinfo}). We find that the node degrees of each age group varies with the interaction rules related to preferences for age. For example, the $DT-CNS^{P+}$ models, built on positive preferential attachment to age, tend to have more older nodes connected with each other and those older nodes have higher node degrees than in case of other models. Similarly, the $DT-CNS^{H-}$ models built on homophily when it comes to age wire more nodes within the same age groups, enabling the nodes in age groups with more members to have higher node degrees. 

Given the same interaction rules, the node degrees of respective age groups also vary with the shapes of age distributions. The symmetric and asymmetric shape distributions can lead to a different trend of node degree changes with age change. For example, for $DT-CNS_{B}^{H+}$ based on the $H+$ (heterophily) rule and the bell-shape age distribution, the node degree first decreases from $[70-79]$ to $[10-19]$ when the node age transits from $[0-9]$ to $[40-49]$, which then increases from $[10-19]$ to $[70-79]$ as the node age transits from $[50-59]$ to $[60-89]$. This contrasts with the case of $DT-CNS_{L}^{H+}$, where the node degree first decreases from $[70-79]$ to $[10-19]$ when the node age transits from $[0-9]$ to $[60-69]$, which then increases from $[10-19]$ to $[40-49]$ as the node age transits from $[70-79]$ to $[80-89]$. This phenomenon can be caused by the number of nodes in each age group, as nodes in sparse age groups are preferred and connected with those dissimilar others in dense age groups. This inevitably leads to higher node degrees for these sparse age groups. In contrast, the dense age groups have lower node degrees as they prefer the sparse age groups and have limit connections with these preferred nodes.
In addition, when we optimise the combined preferences considering both preferential attachment and homophily ($DT-CNS^{PH}$ models), the shapes of degree distributions approach the shape of a power-law distribution. This indicates that the introduction of features and the optimisation of corresponding preferences enable better achievement of the target states of networks. 

\paragraph{Clustering Coefficient} describes the probability of a node's neighbours to be connected. Its value is between $0$ and $1$~\cite{musial2013kind}.  As shown in Table~\ref{networkinfo}, in the target network, clustering coefficient fluctuates around an average value of $0.47$ with a standard deviation of $0.08$, ranging from $0.79$ to $0.33$. 

Fig~\ref{age and clus} shows age and clustering coefficient distributions with a heatmap and two histogram plots for age (See Fig.~\ref{featfig}) and clustering coefficient (See Tab.~\ref{networkinfo}). We find that nodes in age groups with bigger number of people with connection preferences for a similar group of nodes tend to cluster with a higher clustering coefficient. For example, the young and middle-aged nodes in the $DT-CNS^{P+}$ models, built on positive preferential attachment to age, prefer to be connected with old nodes, where these nodes cluster around the popular old nodes.
Similarly, the nodes in $DT-CNS^{H-}$ models, where interactions are based on homophily phenomenon, like to be connected with similar others. This generally leads to a higher clustering coefficient than in the case of other DT-CNSs. However, nodes in $DT-CNS^{H+}$ models, driven by heterophily phenomenon, prefer to be connected with dissimilar others. They generally have lower clustering coefficients than in the case of other DT-CNSs because they select different nodes to connect and do not cluster around the same node. When we optimise the combined preferences considering both preferential attachment and homophily, the shapes of clustering coefficient distributions vary with age distributions. As nodes tend to connect with similar and old nodes (referring to the optimised preferences for both preferential attachment and homophily in Table \ref{params} in the appendix \ref{app}), the old and dense age groups tend to have higher clustering coefficient. Among them, the $DT-CNS^{PH}_{U}$ and the $DT-CNS^{PH}_I$ models characterised with more even age group allocations generate a clustering coefficient with a power-law shape because less nodes fulfil one of the requirements for larger age values and denser age groups, which 
induces a lower clustering coefficient in the respective cases.

\paragraph{Shortest Path Length} between two nodes describes the number of edges along the shortest path between a pair of nodes \cite{musial2013kind}. As shown in Table~\ref{networkinfo}, the shortest path length in the target network fluctuates around an average value of $1,65$ with a standard deviation of $0.48$, ranging from $1.00$ to $3.00$.

Fig~\ref{AgeSP} shows age and shortest path length distributions with a scatter plot and two same Kernel Density Estimation (KDE) plots for shortest path length referring to the age values (See Fig.~\ref{featfig} for age distributions and Tab.~\ref{networkinfo} for information related to the shortest path length). Specifically, in Fig~\ref{AgeSP}, we present the shortest path length at $1$, $2$, $3$ and over $3$ between the connected nodes with blue, orange, green and red dots. The x-axis and y-axis represent age values and the scattered dots represent the shortest path lengths between given pair of nodes of given ages. We find that the shortest path lengths between the nodes are generated by all the $DT-CNS$ models range between $1$ and $3$ except for the $DT-CNS^{H-}$ models built on the $H-$ (homophily) rule. This is because in $DT-CNS^{H-}$ models, similar nodes cluster within the similar age groups, leading to longer paths to connect the dissimilar nodes. For example, in the $DT-CNS_L^{H-}$ model, which is built with a left-skewed age distribution and the homophily principle, a large number of paths between $0-9$ and $80-89$ age groups have value over $3$ as a high proportion of young nodes in the left-skewed age distribution cluster together without many direct connections with the nodes for which the age range is $89-90$. In addition, the distribution of the shortest path lengths is also highly related to the preference principles. The age groups with preferred features have shorter paths to other nodes. For example, the nodes which are old in $DT-CNS^{P+}$ models have more direct connections with others and the shortest paths of length of $1$ due to the fact that they have preferred value of feacture age. When we optimise the combined preferences considering both preferential attachment and homophily, due to the complex interaction rules, the shapes of shortest path lengths distributions vary greatly with age distributions and become very hard to interpret. This indicates that the increasing dynamics level depreciates the network patterns' interpretability but enables the creation of more complex network and more faithful scenarios.

\paragraph{Summary Statistics,} related to the target network and the simulated networks, are presented in Tab.~\ref{networkinfo}. It incorporates information related to the degree distributions, clustering coefficient distributions and shortest path length distributions. In addition, we use JS divergence to evaluate the distances between these network patterns with that of the target network (See Tab.~\ref{jspatterns} in the appendix \ref{app2}). More specifically, the most similar network patterns are identified with green colour. Compared with the target network, the network simulated by the $DT-CNS_{U}^{PH}$ model generates the most similar degree distribution and shortest path length distribution. $DT-CNS_{L}^{PH}$ model generates the most similar clustering coefficient distribution. This indicates that network models, driven by different features and the optimised interaction rules, approach different characteristics of the target network. These differences challenge the evaluation and development of DT-CNS models, remaining a research gap to be addressed in future studies. However, in our experiments, we only focus on the degree distribution of target network and investigate how heterogeneous features and rules influence the performance of DTCNSs in recreating similar degree distributions.

\begin{table}[htp]
\centering
\scriptsize
\caption{Topological information of the networks generated with the DT-CNSs.}
\label{networkinfo}
\setlength{\tabcolsep}{2pt}
\renewcommand{\arraystretch}{1.5}
\begin{tabular}{|c|c|c|c|c|c|c|c|c|c|c|c|c|c|c|c|c|}
\hline
\multirow{2}{*}{Features}& \multirow{2}{*}{Rules} &\multicolumn{2}{|c|}{Nodes} &\multicolumn{4}{|c|}{Node Degree} &\multicolumn{4}{|c|}{Clustering coefficient} & \multicolumn{5}{|c|}{Shortest path length}  \\
\cline{3-17}
& & Connected & Unconnected & Avg. & Std. & Max. & Min.&Avg. & Std. & Max. & Min.& Fake Paths& Avg. & Std. & Max. & Min.\\
\hline
\multicolumn{2}{|c|}{Target Network} & 90& 0& 31.11&12.90& 70& 6 & 0.47&0.08&0.79& 0.33& 0&1.65& 0.48& 3 & 1\\
\hline
\multirow{5}{*}{Uniform}&$P+$&80& 10& 31.11& 21.20& 70& 0 &0.66&0.27&1.00&0.00&845&20.22&36.09&90&1\\
\cline{2-17}
&$P-$&81& 9& 31.11& 21.48& 68& 0&0.67&0.28&1.00&0.00&765&18.47&34.76&90&1\\
\cline{2-17}
&$H+$&90& 0& 31.11& 9.64& 52& 16 & 0.12&0.14&0.45&0.00&0&1.69&0.54&3&1\\
\cline{2-17}
&$H-$&90& 0& 31.11& 6.16& 43& 19 & 0.69&0.09&0.89&0.57&0&2.05&0.98&5&1\\
\cline{2-17}
&$PH$&90& 0& \textcolor{green}{31.11}& \textcolor{green}{11.56}& \textcolor{green}{63}& \textcolor{green}{13} & 0.31&0.09&0.52&0.16&\textcolor{green}{0}&\textcolor{green}{1.65}&\textcolor{green}{0.48}&\textcolor{green}{3}&\textcolor{green}{1}\\
\hline
\multirow{5}{*}{Bell}&$P+$&89& 1& 31.11& 21.34& 77& 0 & 0.73&0.21&1.00&0.00&89&3.61&13.03&90&1\\
\cline{2-17}
&$P-$&87& 3& 31.11& 21.50& 75& 0 & 0.70&0.22&0.89&0.00&264&7.46&21.93&90&1\\
\cline{2-17}
&$H+$&90& 0& 31.11& 15.70& 73& 13 & 0.36&0.11&0.57&0.16&0.00&1.65&0.48&2&1\\
\cline{2-17}
&$H-$&90& 0& 31.11& 12.85& 47& 2 & 0.68&0.10&1.00&0.30&0&2.12&1.16&8&1\\
\cline{2-17}
&$PH$&90& 0& 31.11& 10.32& 68& 17 & 0.40&0.06&0.52&0.28&0&1.65&0.48&2&1\\
\hline
\multirow{5}{*}{Inverse bell}&$P+$&79& 11& 31.11& 21.24& 66& 0 & 0.66&0.28&1.00&0.00&924&21.96&37.26&90&1\\
\cline{2-17}
&$P-$&80& 10& 31.11& 21.54& 67& 0 & 0.66&0.29&1.00&0.00&845&20.23&36.08&90&1\\
\cline{2-17}
&$H+$&90& 0& 31.11& 8.03& 42& 12 & 0.01&0.05&0.36&0.00&0&1.79&0.67&3&1\\
\cline{2-17}
&$H-$&90& 0& 31.11& 4.71& 40& 22 & 0.72&0.10&0.88&0.48&0&2.18&1.07&5&1\\
\cline{2-17}
&$PH$&90& 0& 31.11& 12.63& 63& 10&0.14&0.10&0.47&0.00&0&1.66&0.5&3&1\\
\hline
\multirow{5}{*}{Left skewed}&$P+$&75& 15& 31.11& 21.34& 64& 0& 0.61&0.32&1.00&0.00&1230&28.69&40.82&90&1\\
\cline{2-17}
&$P-$& 90& 0& 31.11& 22.47& 81& 7 & 0.74&0.19&0.90&0.30&0&1.65&0.48&2&1\\
\cline{2-17}
&$H+$&90& 0& 31.11& 16.80& 73& 14 & 0.37&0.11&0.54&0.15&0&1.65&0.48&2&1\\
\cline{2-17}
&$H-$&90& 0& 31.11& 14.50& 50& 1 & 0.66&0.11&0.83&0.00&0&2.42&1.67&9&1\\
\cline{2-17}
&$PH$&90& 0& 31.11& 11.25& 67& 15 & \textcolor{green}{0.39}&\textcolor{green}{0.08}&\textcolor{green}{0.53}&\textcolor{green}{0.23}&0&1.65&0.48&2&1\\
\hline
\multirow{5}{*}{Right skewed}&$P+$&90& 0& 31.11& 22.41& 79& 7 & 0.73&0.2&0.93&0.31&0&1.65&0.48&2&1\\
\cline{2-17}
&$P-$&73& 17& 31.11& 21.69& 62& 0& 0.59&0.34&1.00&0.00&1377&31.92&42.04&90&1\\
\cline{2-17}
&$H+$&90& 0& 31.11& 17.07& 74& 15 & 0.36&0.12&0.6&0.14&0&1.65&0.48&2&1\\
\cline{2-17}
&$H-$&90& 0& 31.11& 15.48& 51& 1 & 0.67&0.17&1.00&0.00&504&13.06&29.21&90&1\\
\cline{2-17}
&$PH$&90& 0& 31.11& 14.24& 74& 14& 0.46&0.09&0.58&0.25&0&1.65&0.48&2&1\\
\hline
\end{tabular}
\end{table}

Tab.~\ref{networkinfo} shows the number of connected and unconnected nodes and the information about the degree distributions, clustering coefficient distributions and the shortest path length distributions that vary depending on the values of the features and rules employed. Generally, all the nodes get connected with the $DT-CNS^{H+}$, $DT-CNS^{H-}$ and $DT-CNS^{PH}$ paradigms. $DT-CNS^{P+}$ or $DT-CNS^{P-}$ paradigms, driven by preferential attachment to old/young ages, result in some unconnected old/young nodes given any age distribution due to the strong preference for young/old ages. In contrast, $DT-CNS^{H+}$, $DT-CNS^{H-}$ and $DT-CNS^{PH}$ paradigms consider the effect of similar or dissimilar ages, where all nodes can find similar/dissimilar others within/across age groups and connect with them. The node degree distributions for each DT-CNS paradigm share the same mean values but have different standard deviation, maximum and minimum values. Degree distributions, generated by $DT-CNS^{H-}$ and $DT-CNS^{PH}$ paradigms, generally have smaller standard deviations and maximum values. In contrast, $DT-CNS^{P+}$ and $DT-CNS^{P-}$ paradigms have larger ones. The maximum degree of the respective paradigms are similar with the number of connected nodes, which indicates that the most popular node tends to directly connect with most of the connected nodes (resulting in power law node degree distribution). The node clustering coefficient distributions for $DT-CNS^{P+}$ and $DT-CNS^{P-}$ models based on preferential attachment principles and $DT-CNS^{H-}$ models built on homophily generally have larger average values, standard deviations and maximum values than other models, which results from the dense clusters created by preferred nodes (See Fig.~\ref{age and clus}). 
The shortest path length distributions are especially influenced by the number of unconnected nodes and the corresponding non-existing ('fake') paths, which are assumed as the maximum path length plus one (i.e. $90$). For the fully connected networks, the average shortest path length fluctuates around $2$, given different age distributions and preferences. It increases significantly when there are fake paths for unconnected node pairs.

\subsection{Simulation-based Processes}
In this section, we build simulation-based processes to analyse disaster resilience given different network structures under the influence of various age distributions. The susceptible nodes get infected given exposure with fixed transmissibility rate, while infected nodes get infectious over time without recovery. The epidemic spread on the networks propagates one step (edge) away for each time step. 

\subsubsection{Seed Selection}

We investigate the most severe case of epidemic explosion by assuming a single seed selection in the initial stage based on the largest node degree (See Tab.~\ref{seedselection}) and allow the epidemic propagation for $5$ time steps within $6$ steps (edges) away from the seed (first infection). Within this time and distance range, most of the connected nodes finally get infected as they are directly/indirectly connected with the seed. As shown in Tab.~\ref{networkinfo} and Tab.~\ref{seedselection}, almost all the connected nodes are connected with the seed node, resulting in a higher infection risk and the full infection.

\begin{table}[htp]
\centering
\small
\caption{The single seed selection based on the largest node degree.}
\label{seedselection}
\setlength{\tabcolsep}{5pt}
\renewcommand{\arraystretch}{1.5}
\begin{tabular}{|c|c|c|c|}
\hline
Features& Rules&\multicolumn{2}{|c|}{Seed}\\
\cline{3-4}
& &  Node degree & Age group \\
\hline
\multirow{5}{*}{Uniform}&$P+$& 70 & 80-89\\
\cline{2-4}
&$P-$& 68 &0-9\\
\cline{2-4}
&$H+$& 52&80-89\\ 
\cline{2-4}
&$H-$& 43& 20-29\\
\cline{2-4}
&$PH$& 63& 0-9\\
\hline
\multirow{5}{*}{Bell}&$P+$& 77&70-79\\
\cline{2-4}
&$P-$& 75 & 10-19\\
\cline{2-4}
&$H+$& 73& 70-79\\
\cline{2-4}
&$H-$& 47& 40-49\\
\cline{2-4}
&$PH$& 68& 0-9\\
\hline
\multirow{5}{*}{Inverse bell}&$P+$& 66&80-89\\
\cline{2-4}
&$P-$& 67&0-9\\
\cline{2-4}
&$H+$& 42&80-89\\
\cline{2-4}
&$H-$& 40&20-29\\
\cline{2-4}
&$PH$& 63&80-89\\
\hline
\multirow{5}{*}{Left skewed}&$P+$& 64&80-89\\
\cline{2-4}
&$P-$& 81& 20-29\\
\cline{2-4}
&$H+$& 73&20-29\\
\cline{2-4}
&$H-$& 50&70-79\\
\cline{2-4}
&$PH$& 67&0-9\\ 
\hline
\multirow{5}{*}{Right skewed}&$P+$&79&50-59\\
\cline{2-4}
&$P-$& 62&0-9\\
\cline{2-4}
&$H+$& 74&70-79\\
\cline{2-4}
&$H-$& 51&10-19\\
\cline{2-4}
&$PH$& 74&70-79\\
\hline
\end{tabular}
\end{table}

As shown in Tab.~\ref{seedselection}, the seeds selected for all the modelling paradigms generally fall in an age group which are preferred by much denser age groups. For example, the old (young) nodes are preferred by all other nodes in $DT-CNS^{P+}$ paradigms ($DT-CNS^{P-}$ paradigms) and thus have the largest number of connections, which almost directly connect all the connected nodes in the respective model (See Tab.~\ref{networkinfo}).   

\subsubsection{Infection Occurrence}
The infection occurrence refers to the proportion of infected nodes to the entire population within a specific distance from the seed. Given the infectious seed, we simulate an epidemic spreading process based on different transmissibilities (ranging in $[0.2,0.4,0.6,0.8]$) and investigate their impact on the infection occurrence.

\begin{figure*}[htp] 
	\centering
	\subfigure[$1.0$ transmissibility given an exposure]{
		\begin{minipage}[b]{0.95\linewidth}
			\includegraphics[width=1\linewidth]{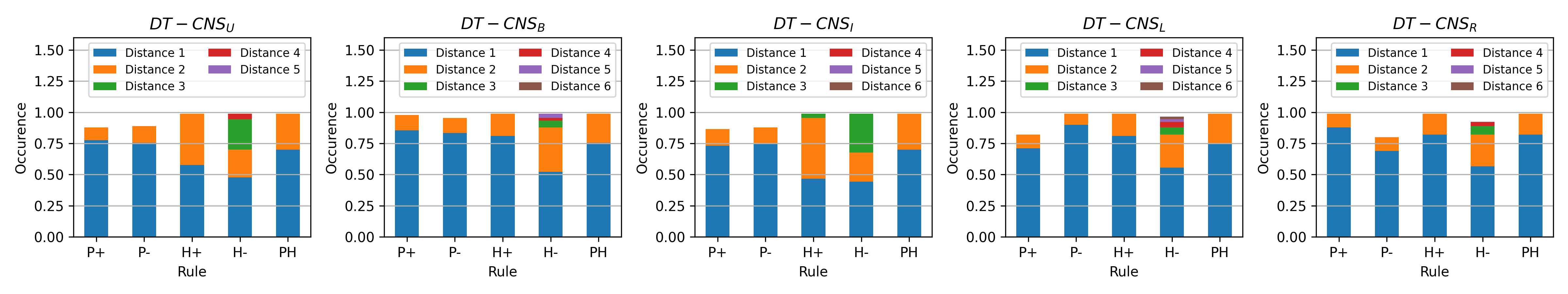}
	\end{minipage}}\\
	\subfigure[$0.8$ transmissibility given an exposure]{
		\begin{minipage}[b]{0.95\linewidth}
			\includegraphics[width=1\linewidth]{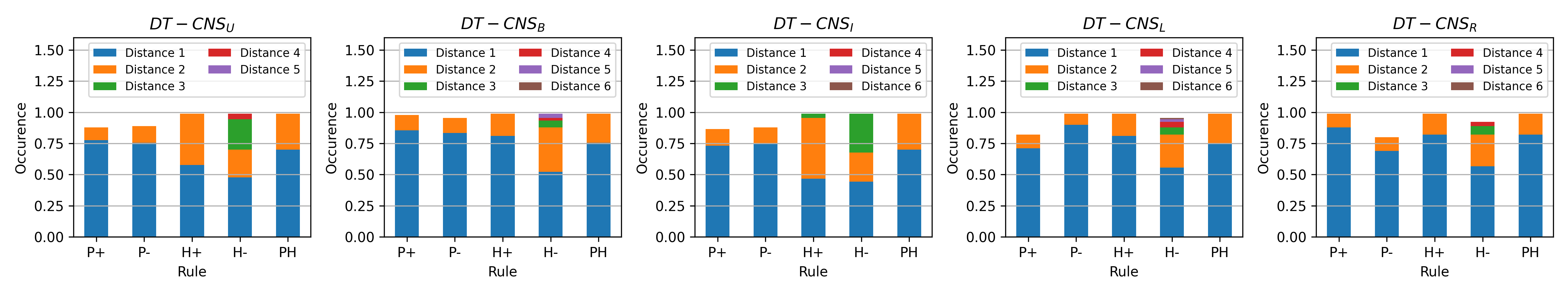}
	\end{minipage}}\\
	\subfigure[$0.6$ transmissibility given an exposure]{
		\begin{minipage}[b]{0.95\linewidth}
			\includegraphics[width=1\linewidth]{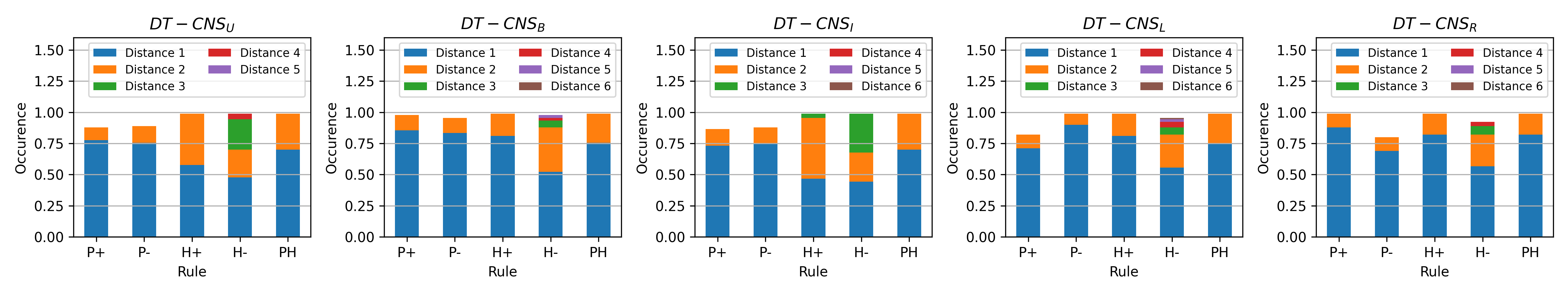}
	\end{minipage}}\\
	\subfigure[$0.4$ transmissibility given an exposure]{
		\begin{minipage}[b]{0.95\linewidth}
			\includegraphics[width=1\linewidth]{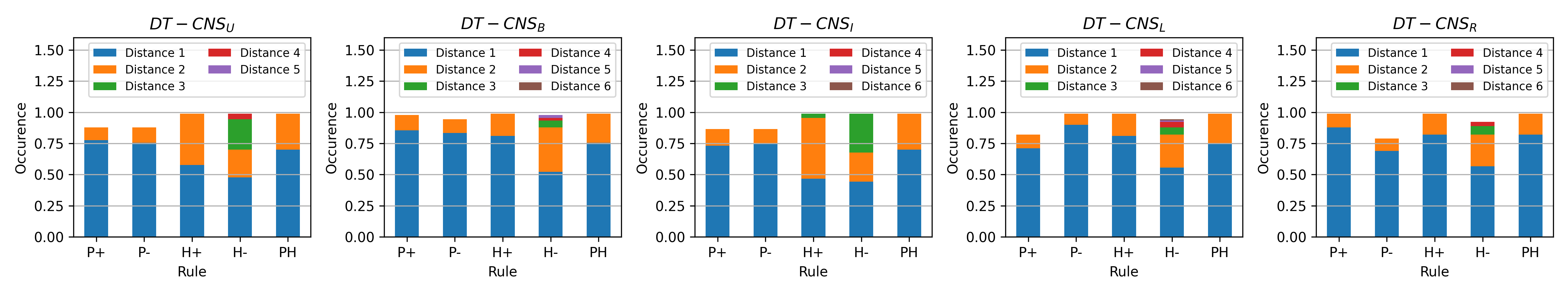}
	\end{minipage}}\\
	\subfigure[$0.2$ transmissibility given an exposure]{
		\begin{minipage}[b]{0.95\linewidth}
			\includegraphics[width=1\linewidth]{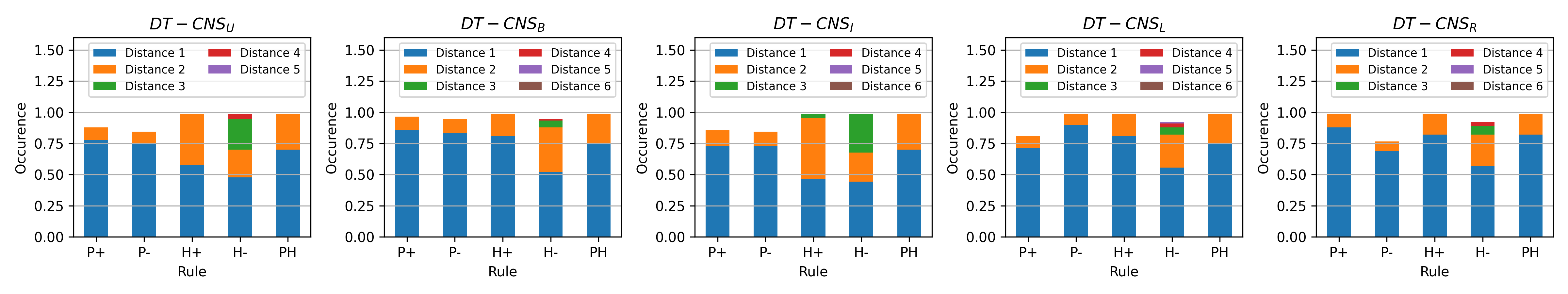}
			
	\end{minipage}}\\
	\caption{The infection occurrence within certain distance to the seed given different transmissibilities.}
\label{seeddistance}
\end{figure*}

In Fig.~\ref{seeddistance}, we represent the number of infection occurrences within specific number of steps (edges away) from the first infection. The vertical axis and the horizontal axis each represents the rules of network formation and the infection occurrence among the population (90 nodes in total). As we simulate the epidemic spread for $6$ time steps and allow the epidemic propagation by one step (edge) for each time step, nodes within $6$ steps away from the first infection are likely to be infected and thus have infection risks in the epidemic simulations. 
Nodes out of this range will not be exposed to the infection risks and thus some of the nodes stay uninfected despite the increase of transmissibility. The infection occurrence 
generally stays the same given any transmissibility between $0.4$ and $1.0$. 
Given different rules, the $DT-CNS^{P+}$ and the $DT-CNS^{P-}$ paradigms can achieve the biggest number of infections within a distance of $1$ (one edge away) from the seed; $DT-CNS^{H+}$ and $DT-CNS^{PH}$ reach the upper limit of $90$ within a distance of $2$; $DT-CNS^{H-}$ gets the biggest number of infections within a distance of around $6$. This indicates that the networks, driven by $P+$ or $P-$ rule (preferential attachment to old/young ages), reach the maximum infection occurrence within the first time step as most nodes are directly connected with the seed and can get infected. Given various age distributions, $DT-CNS_I$ has the fewest infections within the distance of $1$, indicating the less explosive effect of the epidemic outbreak from the seed. The above mentioned phenomenon suggests that the isolation policies imposed on nodes $1$ or $2$ steps away from the seed can efficiently reduce infection occurrence especially in networks driven by the $H-$ rule, which describes the nodes' preferences for similar features. 

\begin{figure*}[htp] 
	\centering
	\subfigure[$1.0$ transmissibility given an exposure]{
		\begin{minipage}[b]{0.95\linewidth}
			\includegraphics[width=1\linewidth]{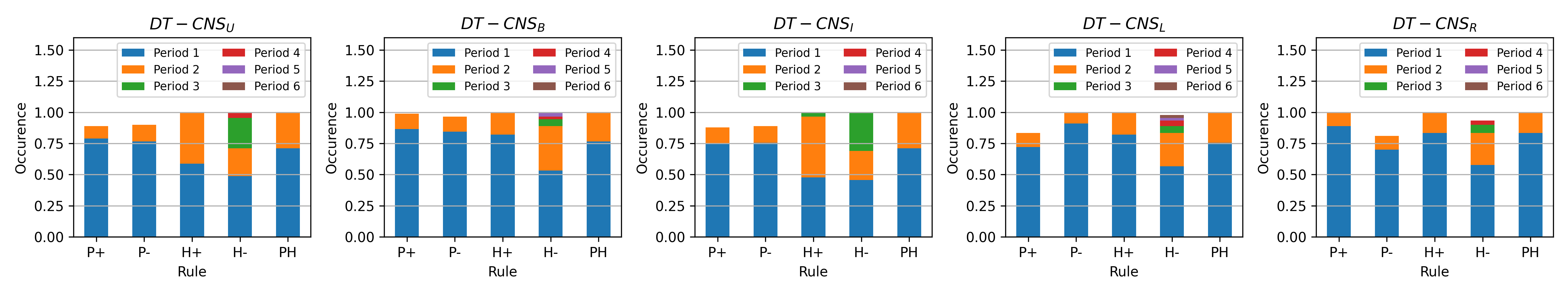}
	\end{minipage}}\\
	\subfigure[$0.8$ transmissibility given an exposure]{
		\begin{minipage}[b]{0.95\linewidth}
			\includegraphics[width=1\linewidth]{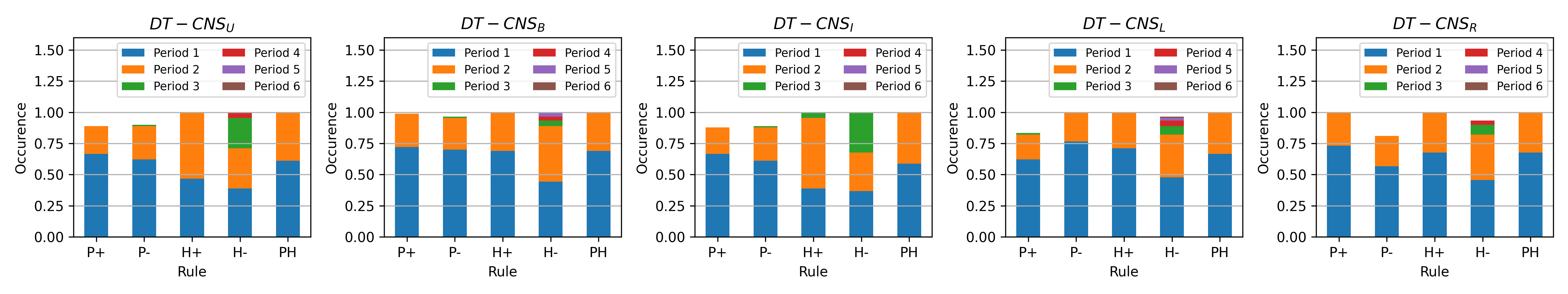}
	\end{minipage}}\\
	\subfigure[$0.6$ transmissibility given an exposure]{
		\begin{minipage}[b]{0.95\linewidth}
			\includegraphics[width=1\linewidth]{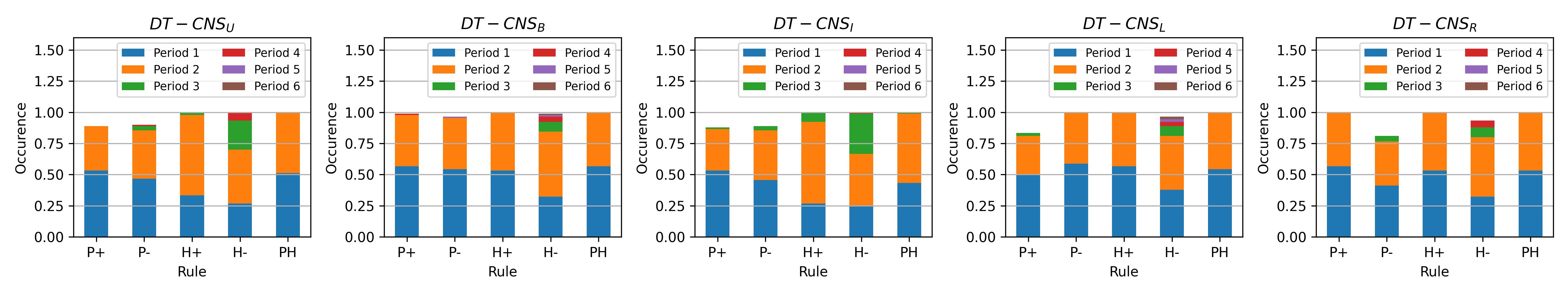}
	\end{minipage}}\\
	\subfigure[$0.4$ transmissibility given an exposure]{
		\begin{minipage}[b]{0.95\linewidth}
			\includegraphics[width=1\linewidth]{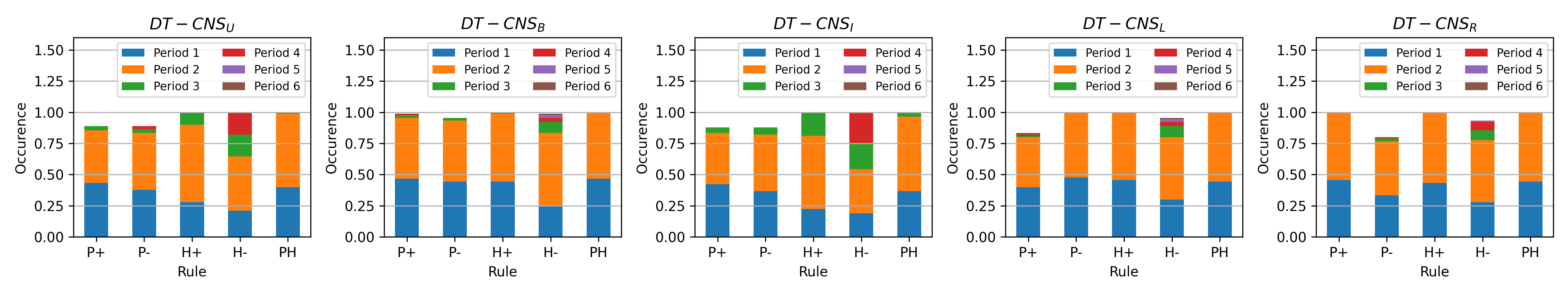}
	\end{minipage}}\\
	\subfigure[$0.2$ transmissibility given an exposure]{
		\begin{minipage}[b]{0.95\linewidth}
			\includegraphics[width=1\linewidth]{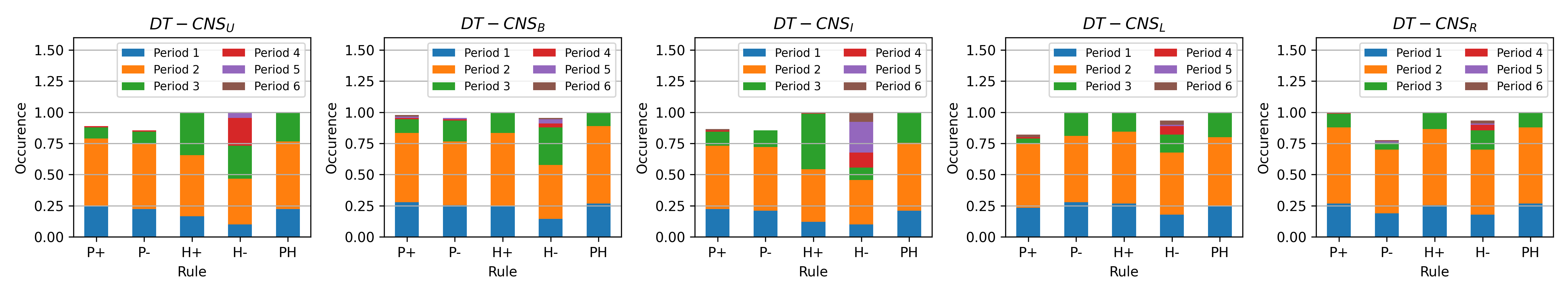}
	\end{minipage}}\\
	\caption{The infection occurrence over time with different transmissibilities given an exposure.}
\label{trans}
\end{figure*}

With decreasing transmissibility of the epidemic spread given the exposure, it takes more time steps to achieve the infection status shown in Fig.~\ref{seeddistance}. In Fig.~\ref{trans}, the number of infected nodes within the first time step (represented as Period 1) gets smaller when the transmissibility decreases from $1.0$ to $0.2$, while the infections increase within Period 2 and Period 3. Given a smaller transmissibility, the nodes are less likely to get infected in Period 1 and thus stay healthy. However, these healthy nodes in Period 2 and Period 3 get exposed to more infected neighbours and finally get infected, leading to a delayed increase in infection numbers. The decrease of transmissibility results in increase of time needed to infect all the nodes and in the same time gives more time to react to the disaster. Compared with the other models, $DT-CNS^{H-}$ models, driven by the homophily principle, have smaller infection occurrences despite different transmissibilities. In addition, the infection occurrences for $DT-CNS^{H-}$ models also increase slowly compared to other models. This is because the seed node clusters around a limited number of similar nodes (See Tab.~\ref{seedselection}). This leads to a social distancing between the infectious seed and dissimilar nodes. The increase in transmissibility does not bridge the social distance between the seed node and the dissimilar nodes. Therefore, for $DT-CNS^{H-}$ models, it still takes more time for the epidemic spread to reach and infect these nodes. For infection occurrences in various age distributions over different time steps, $DT-CNS_I$ generally has the fewest infections within the first time step. This also results from limited connections between the seed node and others (See Tab.~\ref{seedselection}). The abovementioned phenomenon indicates that the increase in transmissibility can have a greater impact on networks that densely clusters around the seed node. 

\subsubsection{People at Risk}

We introduce the People at Risk ($PaR$) measure to have an intuitive representation of the proportion of people at risk of a disaster:
\begin{equation}
    PaR(T,D) = \frac{\sum\limits_{v_{i,t}\in V_t} \mathbf{r}(v_{i,t})\delta_{\mathrm{l}(v_{i,t},\mathbf{s}_{t}),D}\delta_{t,T}}{N}, \quad D\leq T
\end{equation}
which is defined as $PaR(T,D)$ to help identify the easiest-to-be-infected proportion of the population in the limited time $[0,T]$ and the space within the distance $[0,D]$ away from the seed. $\delta_{t,T}$ and $\delta_{\mathrm{l}(v_{i,t},\mathbf{s}_{t}),D}$ each represents the Kronecker functions related to time $t$ and the distance $\mathrm{l}(v_{i,t},\mathbf{s}_{t})$ between node $v_{i,t}$ and seed $\mathbf{s}_t$. $N$ represents the number of nodes.

\begin{figure*}[htp] 
	\centering
 	\subfigure[$PaR(1,1)$]{
		\begin{minipage}[b]{0.95\linewidth}
			\includegraphics[width=1\linewidth]{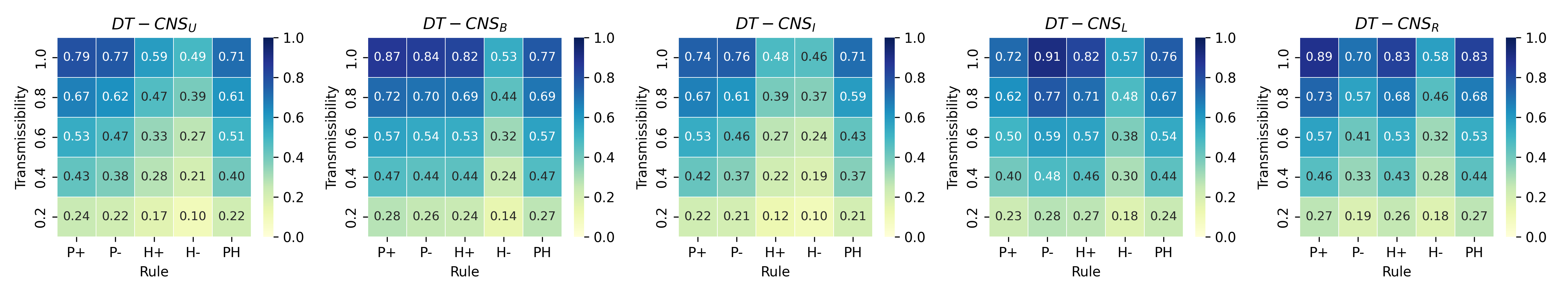}
	\end{minipage}}\\
  	\subfigure[$PaR(2,2)$]{
		\begin{minipage}[b]{0.95\linewidth}
			\includegraphics[width=1\linewidth]{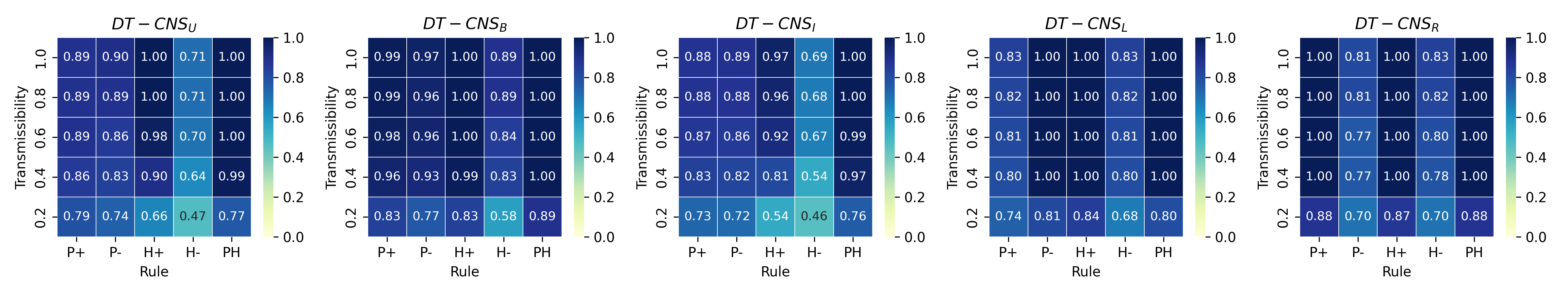}
	\end{minipage}}\\
	\caption{The $PaR(1,1)$ and $PaR(2,2)$ values with different transmissibilities, age distributions and rules of network formation.}
\label{PaR}
\end{figure*}

Fig.~\ref{PaR} shows the $PaR(1,1)$ and the $PaR(2,2)$ values that vary with transmissibilities, rules of network formation and the age feature distributions. First, the $PaR(1,1)$ decreases as transmissibility decreases. $DT-CNS^{P+}$, $DT-CNS^{P-}$, $DT-CNS^{H+}$ and $DT-CNS^{PH}$ generally have a higher $PaR(1,1)$, and this is due to the higher number of nodes connected to the seed (See Fig.~\ref{seeddistance}, where there are more occurrences of infections within the distance 1). In addition, $DT-CNS^{U}$ and $DT-CNS^{I}$, due to a higher age diversity, have smaller $PaR(1,1)$ values. The $PaR(2,2)$ values increase significantly as the transmission propagates one step further. Almost all the connected nodes get infected within two time steps, despite the differences in transmissibilities (See Fig.~\ref{trans}). This can be caused by the significant number of direct connections between the popular seed node and others (See Tab.~\ref{networkinfo}). Similar with the case of $PaR(1,1)$ values, the $PaR(2,2)$ values for the $DT-CNS^{H-}$ models are also smaller than for the other models, resulting from the limited number of connections with the seed node 
(See Tab.~\ref{networkinfo}). The above-mentioned differences between $PaR(1,1)$ and $PaR(22)$ values indicate the necessity of epidemic control in primary time and distance to the seed node. 

\begin{figure*}[htp] 
	\centering
	\subfigure[$1.0$ transmissibility given an exposure]{
		\begin{minipage}[b]{0.95\linewidth}
			\includegraphics[width=1\linewidth]{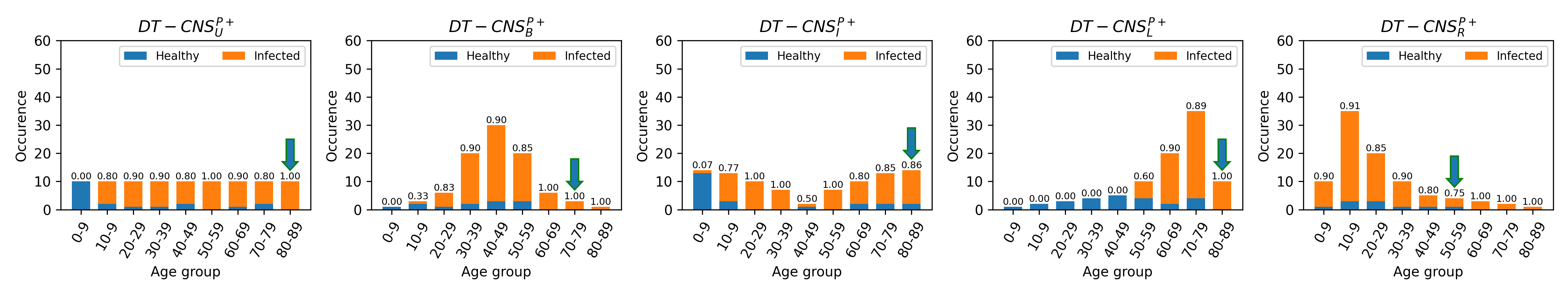}
	\end{minipage}}\\
	\subfigure[$0.8$ transmissibility given an exposure]{
		\begin{minipage}[b]{0.95\linewidth}
			\includegraphics[width=1\linewidth]{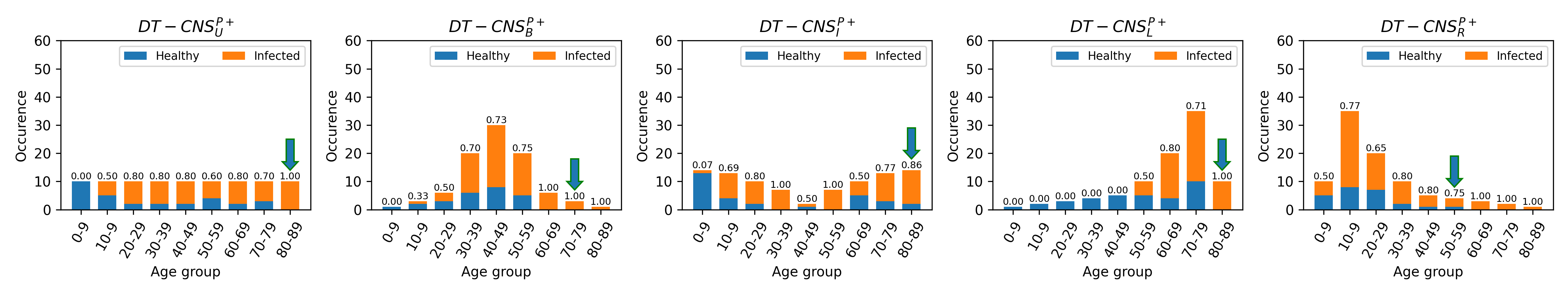}
	\end{minipage}}\\
	\subfigure[$0.6$ transmissibility given an exposure]{
		\begin{minipage}[b]{0.95\linewidth}
			\includegraphics[width=1\linewidth]{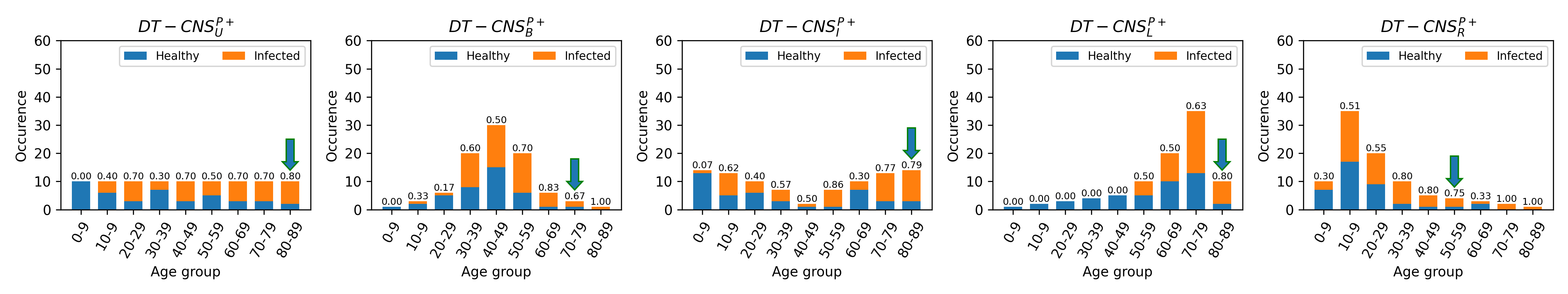}
	\end{minipage}}\\
	\subfigure[$0.4$ transmissibility given an exposure]{
		\begin{minipage}[b]{0.95\linewidth}
			\includegraphics[width=1\linewidth]{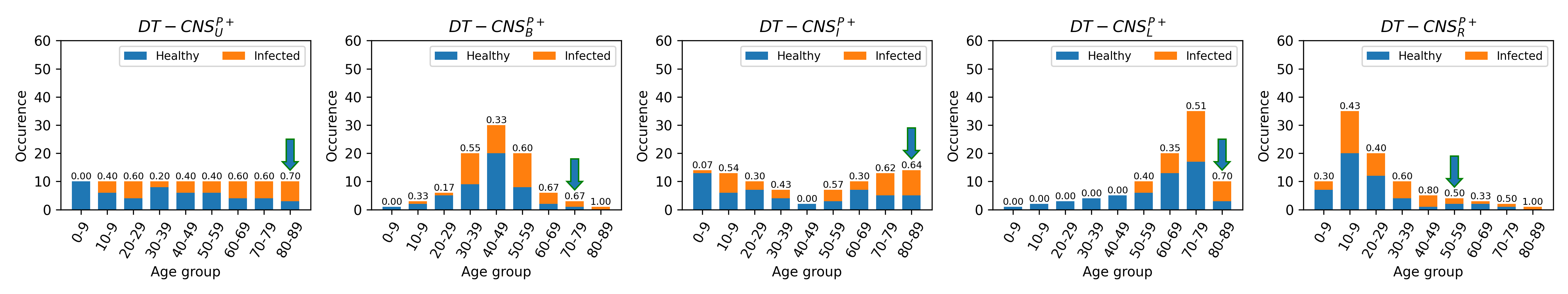}
	\end{minipage}}\\
	\subfigure[$0.2$ transmissibility given an exposure]{
		\begin{minipage}[b]{0.95\linewidth}
			\includegraphics[width=1\linewidth]{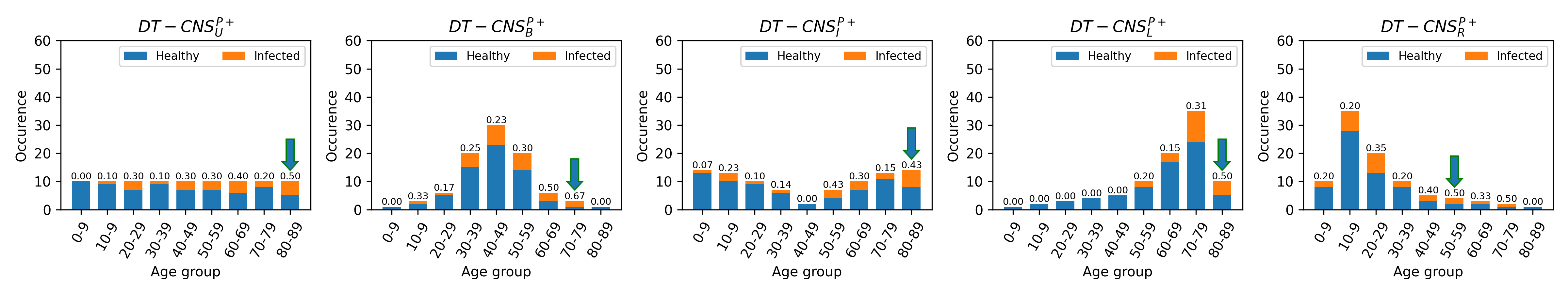}
	\end{minipage}}\\
	\caption{The PaR(1,1) for each age group given different transmissibilities, age distributions and the $P+$ rule.}
\label{GaRp}
\end{figure*}

\begin{figure*}[htp] 
	\centering
	\subfigure[$1.0$ transmissibility given an exposure]{
		\begin{minipage}[b]{0.95\linewidth}
			\includegraphics[width=1\linewidth]{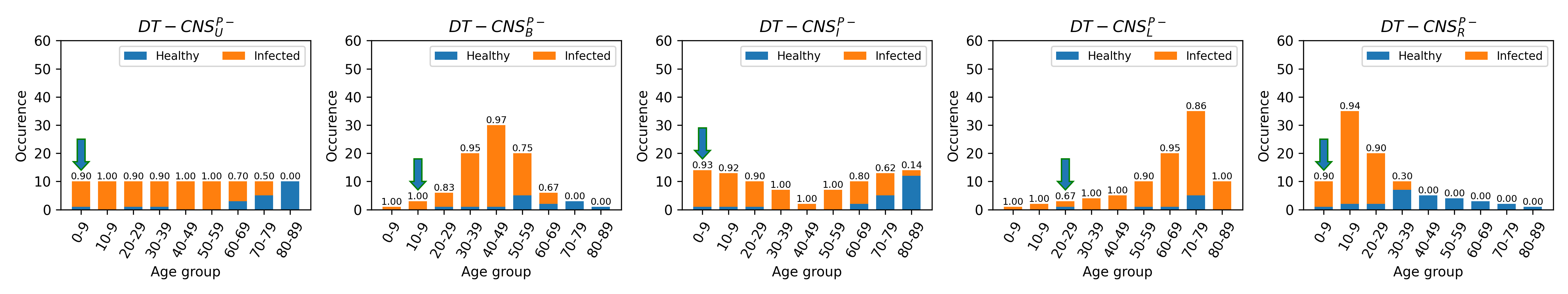}
	\end{minipage}}\\
	\subfigure[$0.8$ transmissibility given an exposure]{
		\begin{minipage}[b]{0.95\linewidth}
			\includegraphics[width=1\linewidth]{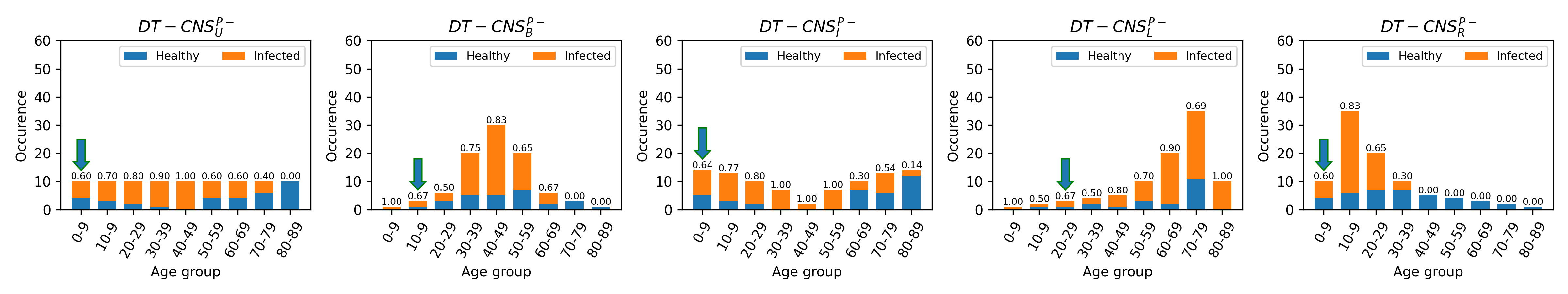}
	\end{minipage}}\\
	\subfigure[$0.6$ transmissibility given an exposure]{
		\begin{minipage}[b]{0.95\linewidth}
			\includegraphics[width=1\linewidth]{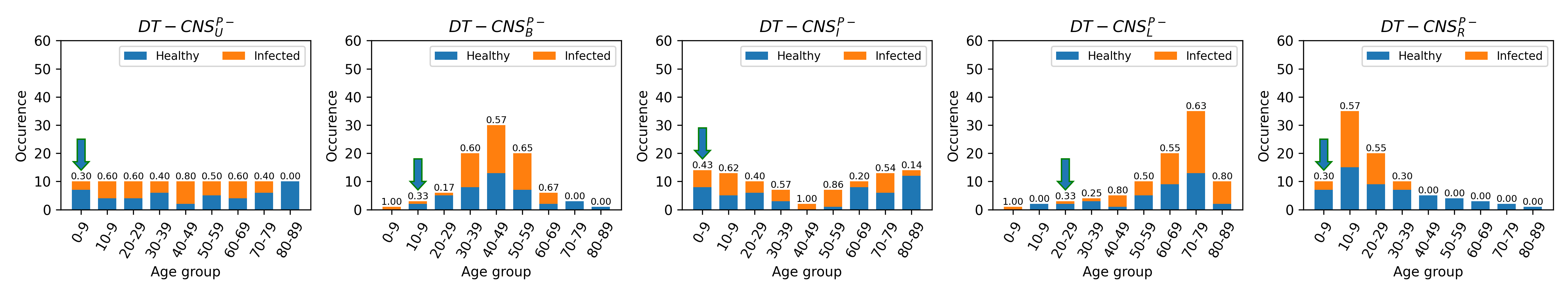}
	\end{minipage}}\\
	\subfigure[$0.4$ transmissibility given an exposure]{
		\begin{minipage}[b]{0.95\linewidth}
			\includegraphics[width=1\linewidth]{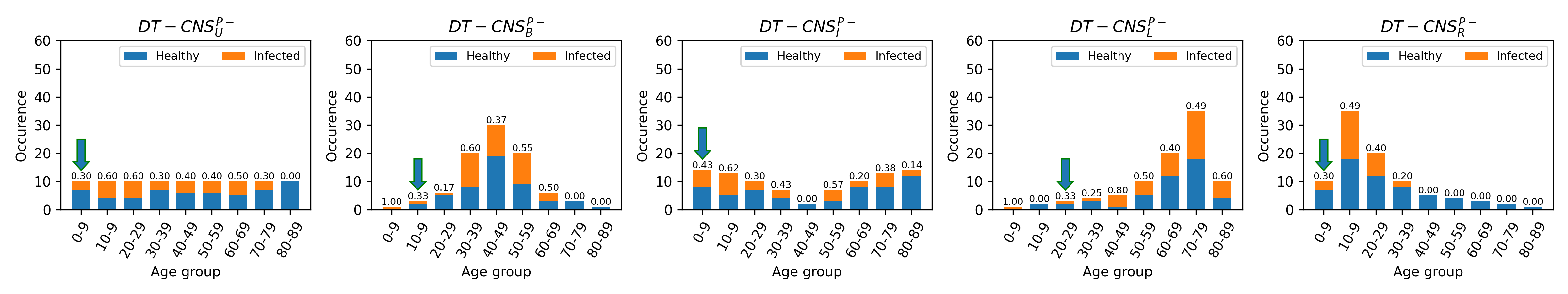}
	\end{minipage}}\\
	\subfigure[$0.2$ transmissibility given an exposure]{
		\begin{minipage}[b]{0.95\linewidth}
			\includegraphics[width=1\linewidth]{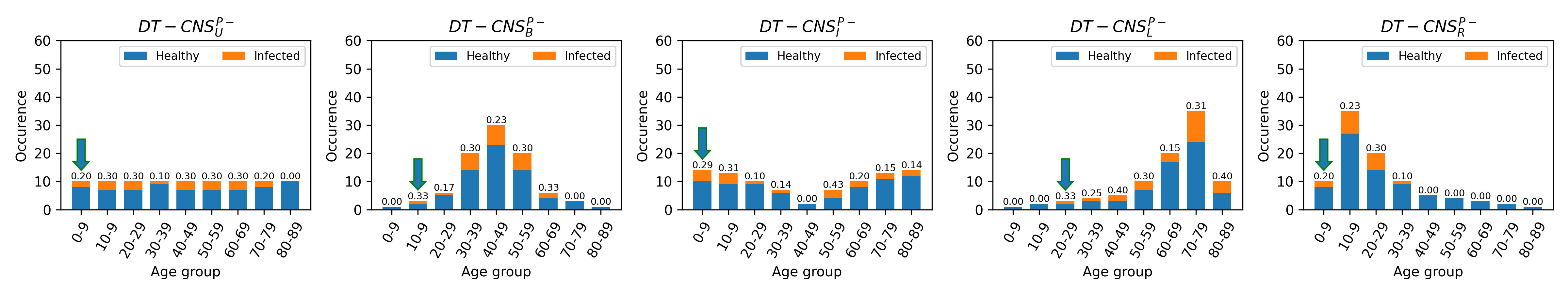}
	\end{minipage}}\\
	\caption{The PaR(1,1) for each age group given different transmissibilities, age distributions and the $P-$ rule.}
\label{GaRp1}
\end{figure*}

\begin{figure*}[htp] 
	\centering
	\subfigure[$1.0$ transmissibility given an exposure]{
		\begin{minipage}[b]{0.95\linewidth}
			\includegraphics[width=1\linewidth]{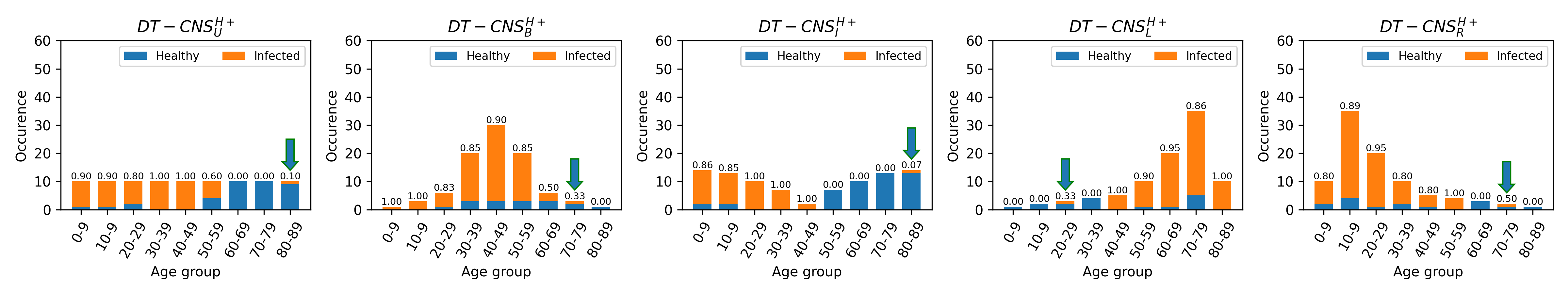}
	\end{minipage}}\\
	\subfigure[$0.8$ transmissibility given an exposure]{
		\begin{minipage}[b]{0.95\linewidth}
			\includegraphics[width=1\linewidth]{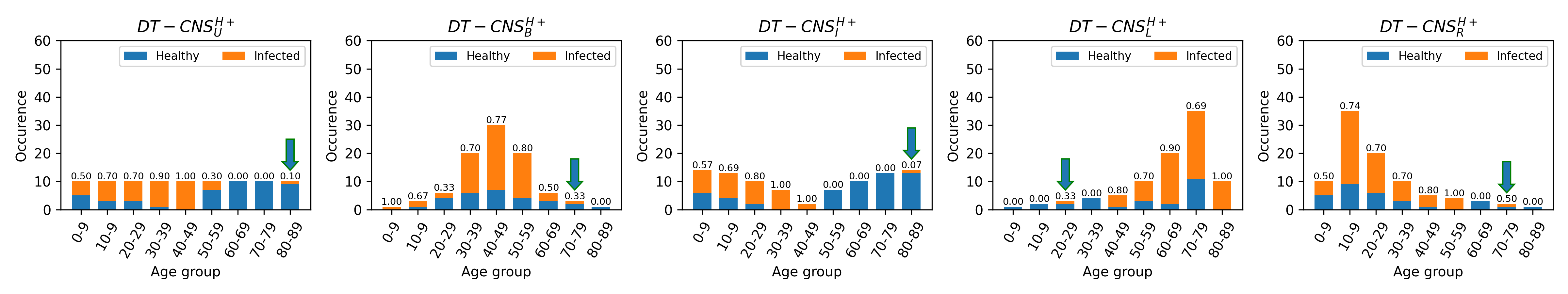}
	\end{minipage}}\\
	\subfigure[$0.6$ transmissibility given an exposure]{
		\begin{minipage}[b]{0.95\linewidth}
			\includegraphics[width=1\linewidth]{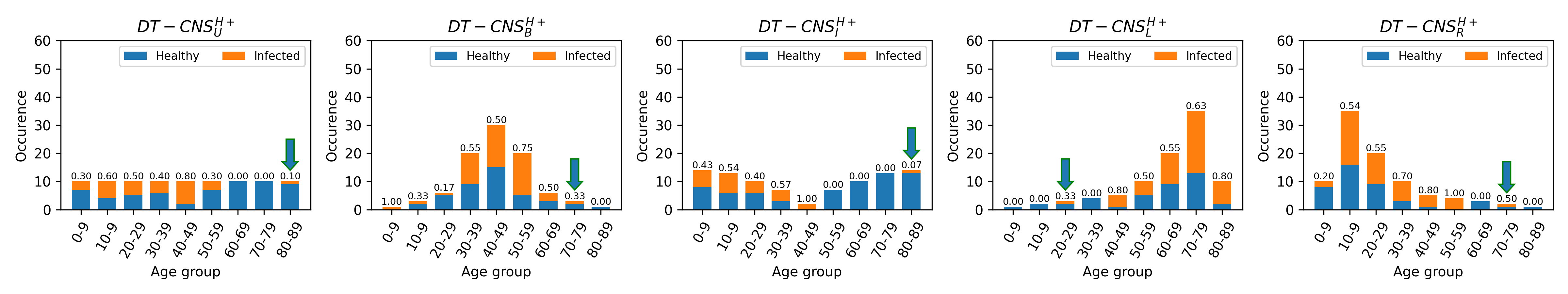}
	\end{minipage}}\\
	\subfigure[$0.4$ transmissibility given an exposure]{
		\begin{minipage}[b]{0.95\linewidth}
			\includegraphics[width=1\linewidth]{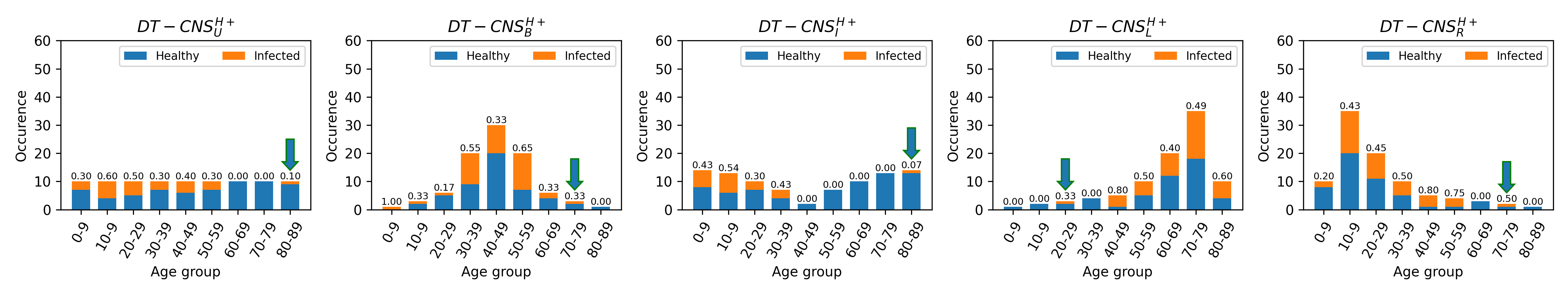}
	\end{minipage}}\\
	\subfigure[$0.2$ transmissibility given an exposure]{
		\begin{minipage}[b]{0.95\linewidth}
			\includegraphics[width=1\linewidth]{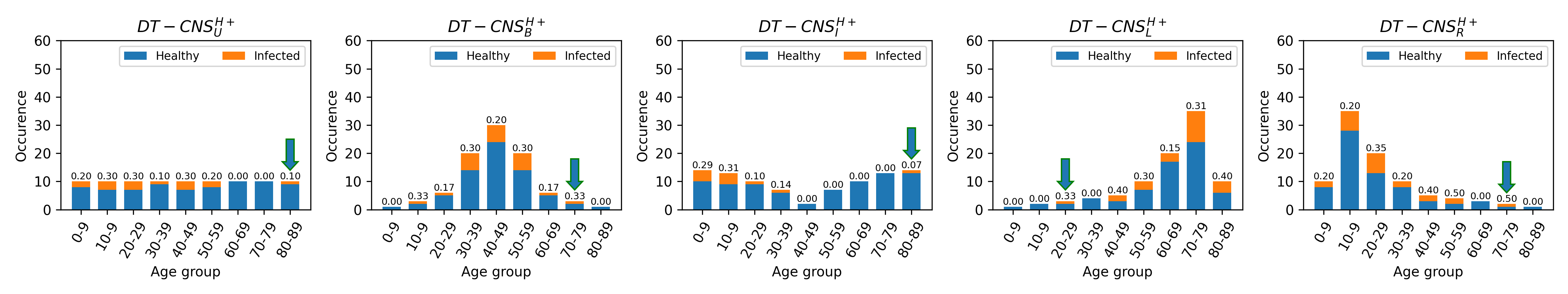}
	\end{minipage}}\\
	\caption{The PaR(1,1) for each age group given different transmissibilities, age distributions and the $H+$ rule.}
\label{GaRh}
\end{figure*}

\begin{figure*}[htp] 
	\centering
	\subfigure[$1.0$ transmissibility given an exposure]{
		\begin{minipage}[b]{0.95\linewidth}
			\includegraphics[width=1\linewidth]{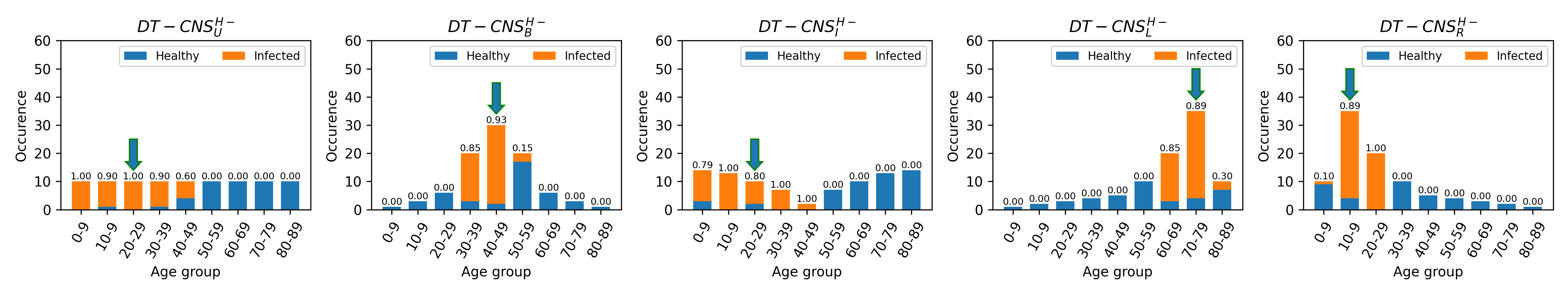}
	\end{minipage}}\\
	\subfigure[$0.8$ transmissibility given an exposure]{
		\begin{minipage}[b]{0.95\linewidth}
			\includegraphics[width=1\linewidth]{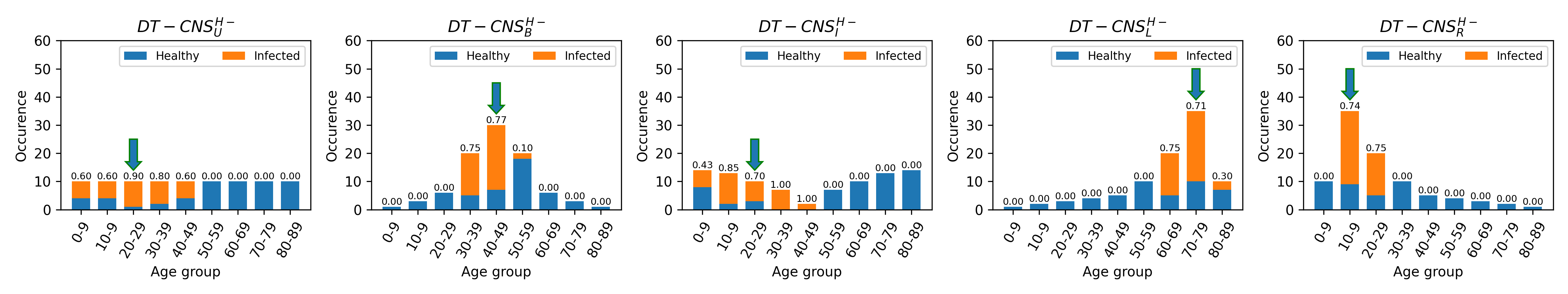}
	\end{minipage}}\\
	\subfigure[$0.6$ transmissibility given an exposure]{
		\begin{minipage}[b]{0.95\linewidth}
			\includegraphics[width=1\linewidth]{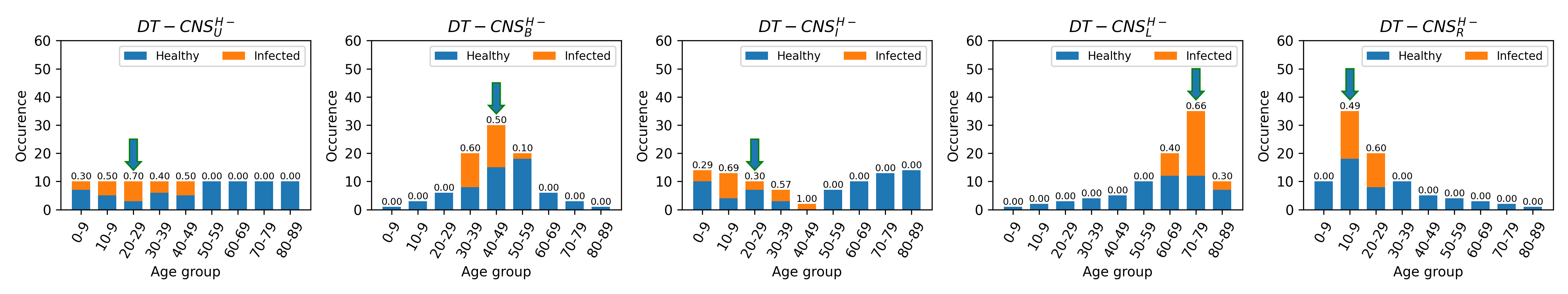}
	\end{minipage}}\\
	\subfigure[$0.4$ transmissibility given an exposure]{
		\begin{minipage}[b]{0.95\linewidth}
			\includegraphics[width=1\linewidth]{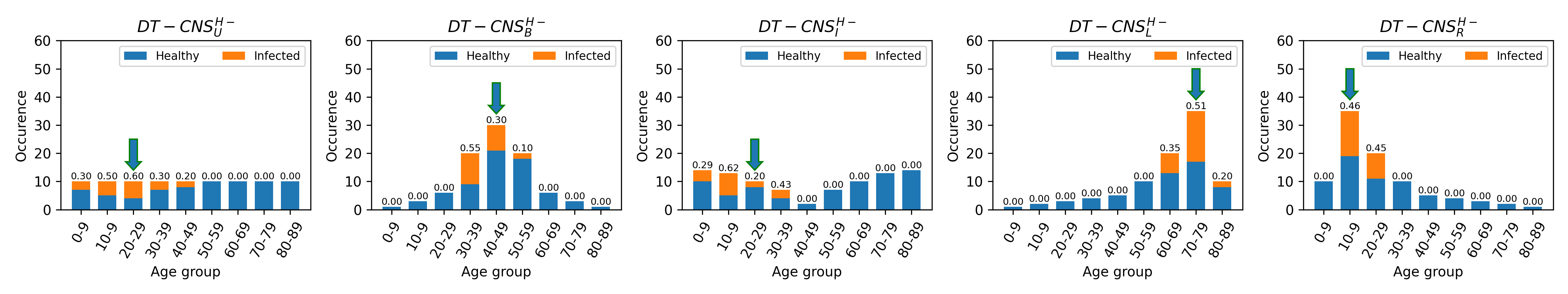}
	\end{minipage}}\\
	\subfigure[$0.2$ transmissibility given an exposure]{
		\begin{minipage}[b]{0.95\linewidth}
			\includegraphics[width=1\linewidth]{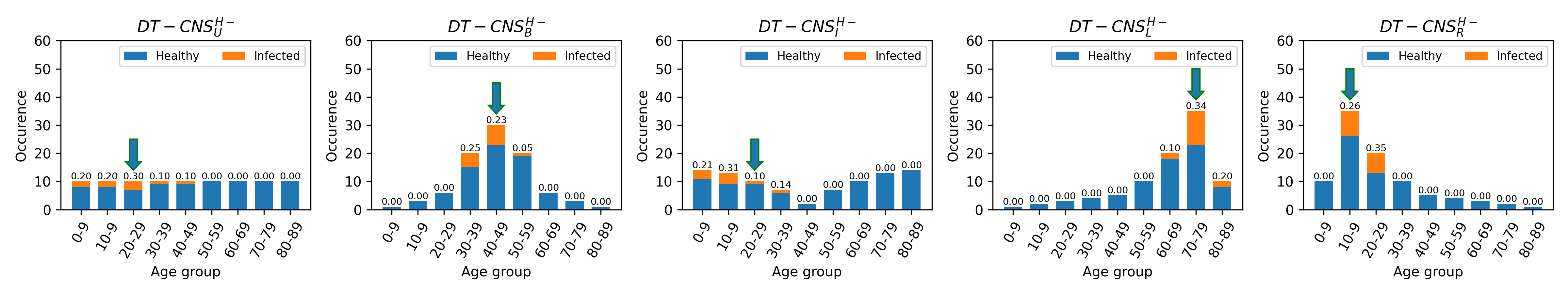}
	\end{minipage}}\\
	\caption{The PaR(1,1) for each age group given different transmissibilities, age distributions and the $H-$ rule.}
\label{GaRh1}
\end{figure*}

\begin{figure*}[htp] 
	\centering
	\subfigure[$1.0$ transmissibility given an exposure]{
		\begin{minipage}[b]{0.95\linewidth}
			\includegraphics[width=1\linewidth]{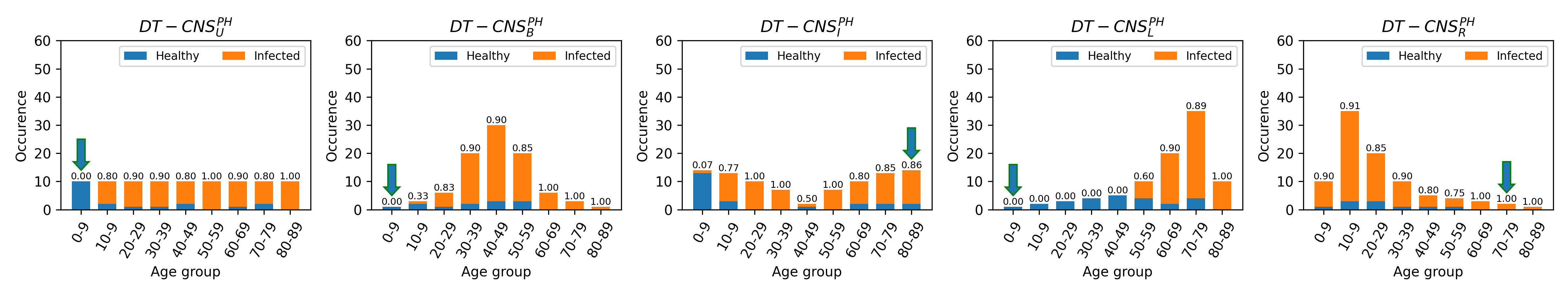}
	\end{minipage}}\\
	\subfigure[$0.8$ transmissibility given an exposure]{
		\begin{minipage}[b]{0.95\linewidth}
			\includegraphics[width=1\linewidth]{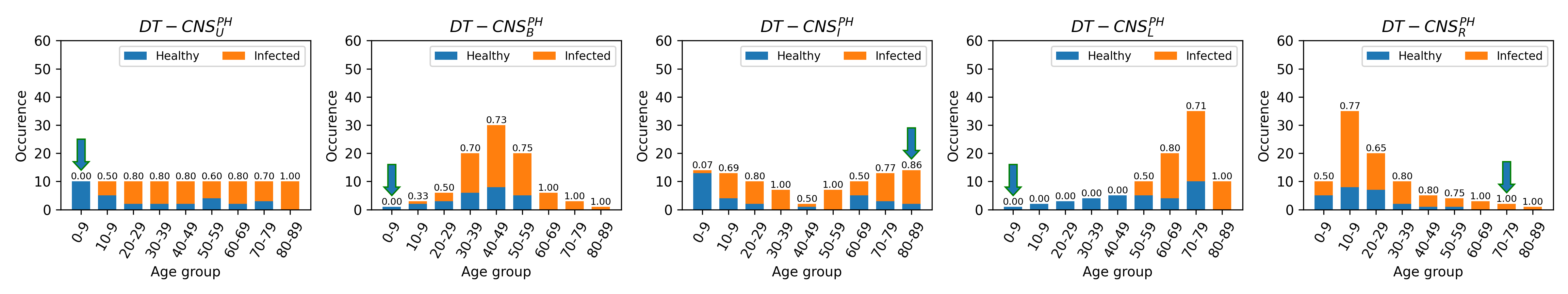}
	\end{minipage}}\\
	\subfigure[$0.6$ transmissibility given an exposure]{
		\begin{minipage}[b]{0.95\linewidth}
			\includegraphics[width=1\linewidth]{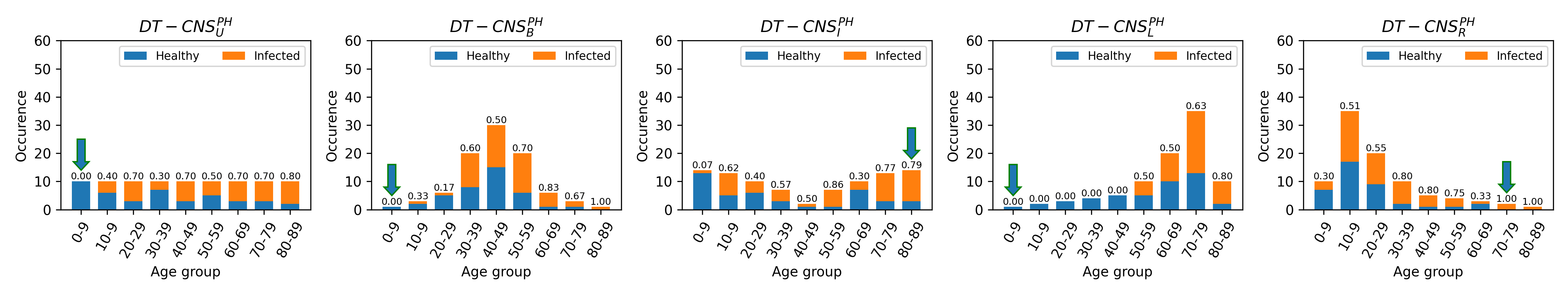}
	\end{minipage}}\\
	\subfigure[$0.4$ transmissibility given an exposure]{
		\begin{minipage}[b]{0.95\linewidth}
			\includegraphics[width=1\linewidth]{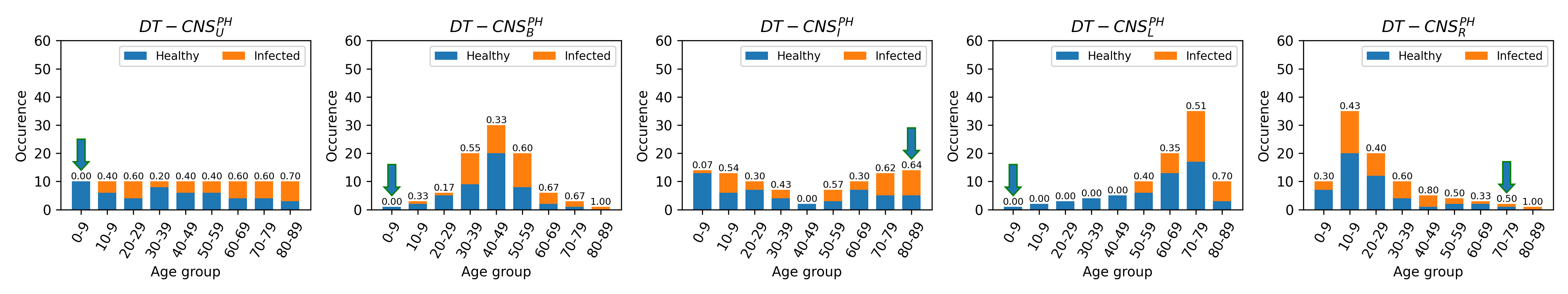}
	\end{minipage}}\\
	\subfigure[$0.2$ transmissibility given an exposure]{
		\begin{minipage}[b]{0.95\linewidth}
			\includegraphics[width=1\linewidth]{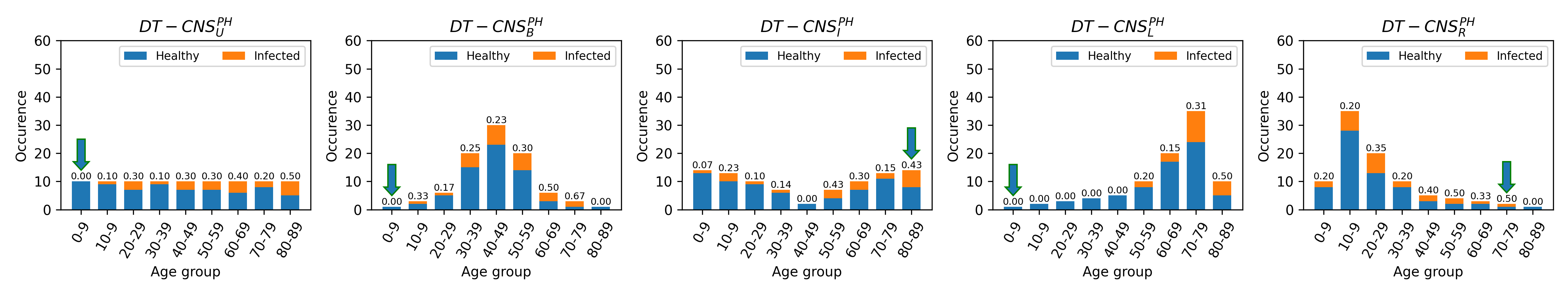}
	\end{minipage}}\\
	\caption{The PaR(1,1) for each age group given different transmissibilities, age distributions and the $PH$ rule.}
\label{GaRph}
\end{figure*}

To better understand the diversity of age features and their influence on the infection status, we respectively calculate the $PaR(1,1)$ of each age group based on different transmissibilities, age groups and rules (See Fig.~\ref{GaRp}, Fig.~\ref{GaRp1}, Fig.~\ref{GaRh}, Fig.~\ref{GaRh1} and Fig.~\ref{GaRph}). We also identify the age group where the epidemic starts with a green arrow in the corresponding $PaR(1,1)$ figures.

In Fig.~\ref{GaRp}, the $PaR(1,1)$ increases with the increasing transmissibility. In addition, there is a significant upward trend of $PaR(1,1)$ when the transmissibility changes from $0.2$ to $0.4$. This indicates an increasing level of difficulty to resist an epidemic outbreak given an increased transmissibility, because a higher $PaR(1,1)$ value represents a higher proportion of infected nodes and more resources that are needed to keep the epidemic under control. Given the $P+$ rule (preferential attachment to old ages) of network formation but different age distributions, older age groups and the denser age groups, each characterised with higher node degrees and node clustering coefficient, tend to have higher $PaR(1,1)$ values (See Fig.~\ref{AgeAndDeg}, Fig.~\ref{age and clus} and Tab.~\ref{networkinfo}). Due to the features they prefer and the similar preferences within the same age group, these nodes get directly connected with the seed node, exposing the corresponding age group to much higher infection risk. For example, with the $DT-CNS_R$ paradigms and the epidemic transmission from older age groups around the age of $50-59$, the younger and denser age groups around the age of $0-39$ have higher $PaR(1,1)$ values due to the significant number of direct connections with the seed node.
Therefore, the preference for old ages with the $P+$ rule (positive preferential attachment to age) may pose the old at higher risk. Similarly, the density of age groups can also lead to higher infection risks. However, this is not always the case in reality and is constrained by our assumptions related to features and the corresponding preferences.

As is shown in Fig.~\ref{GaRp1}, given $P-$ rule (preferential attachment to young ages) of network formation, young and dense age groups have higher $PaR(1,1)$ values due to the similarly preferred features within the corresponding age group. Most of these nodes get directly connected with the seed node and have higher infection risks. This indicates a relatively lower resistance level to such an epidemic outbreak as a higher proportion of the respective age groups tend to be infected and treated in epidemic control. The higher $PaR(1,1)$ values in young-age groups given the $P+$ rule contrasts with the lower ones given the $P-$ rule (See Fig.~\ref{GaRp}). However, the preferential attachment principles, either $P+$ or $P-$, induces higher infection risks for dense age groups.

As is shown in Fig.~\ref{GaRh}, given the $H+$ rule (preferences for dissimilar ages) of network formation, the age groups, which are denser and have larger age differences with the seed node, have higher $PaR(1,1)$ values due to their similar preferences for dissimilar ages and thus, the significant number of direct connections across the age groups (between the age group of the seed node and the other age groups). Compared with the $DT-CNS^{P-}$ models built on negative preferential attachment in age values, the $DT-CNS^{H+}$ models have a different seed node but similar $PaR(1,1)$ value distributions in the respective age groups. This is because, in both these cases, the young and dense age groups prefer to be connected with the seed node. Therefore, in the shortest time and distance to the first infection, the differences in $P-$ and $H+$ rules can not lead to significant differences in infection risks. This also indicates that the preference for the seed node for each age group influences their $PaR(1,1)$ values directly.

In Fig.~\ref{GaRh1}, given the $H-$ rule (preferences for similar ages) of network formation, fewer age groups get involved in the epidemic spreading process since the homophily effect limits the interactions with the seed node to similar age groups 
(See Tab.~\ref{networkinfo}). The $PaR(1,1)$ values are higher for age groups which are dense and in a similar age with the seed node due to the similar preferences for similar features ($H-$ rule). In contrast with the other rules (See Fig.\ref{GaRp}, Fig.~\ref{GaRp1}, Fig.~\ref{GaRh} and Fig.~\ref{GaRh1}), the lower infection risk, as indicated by smaller $PaR(1,1)$ values, represents a higher resistance level to the epidemic outbreak in the early stages of epidemics and distance to the first infection.

In Fig.~\ref{GaRph}, the $PH$ rule (optimised preferences for ages and age difference) of network formation optimally combines the effect of preferential attachment and homophily principles to approach the target degree distributions of the scale-free networks. The $DT-CNS_{U}^{PH}$ and the $DT-CNS_{B}^{PH}$ models combine the negative preferential attachment in ages and the heterophily effect related to the preferences for different ages. For these two models, the old age groups, dissimilar to the seed node and sharing similar preferences for young nodes ($0-9$), have high $PaR(1,1)$ values. In contrast, the $DT-CNS_{I}^{PH}$, $DT-CNS_{L}^{PH}$ and the $DT-CNS_{R}^{PH}$ models combine the positive preferential attachment in ages and the heterophily effect related to the preferences for different ages. For these models, the age groups, which are younger (younger than $60$) and denser, have high $PaR(1,1)$ values. The phenomena mentioned above indicate the influence of density in age groups on the corresponding $PaR(1,1)$ values. Denser age groups, despite the differences and complexities in preferences, generally have higher infection risks and correspondingly lower resistance levels to the epidemic outbreak.

\subsection{Faithfulness}
\begin{figure*}[htp] 
	\centering
 	\subfigure[$DT-CNS_{U}$]{
		\begin{minipage}[b]{0.17\linewidth}
			\includegraphics[width=1\linewidth]{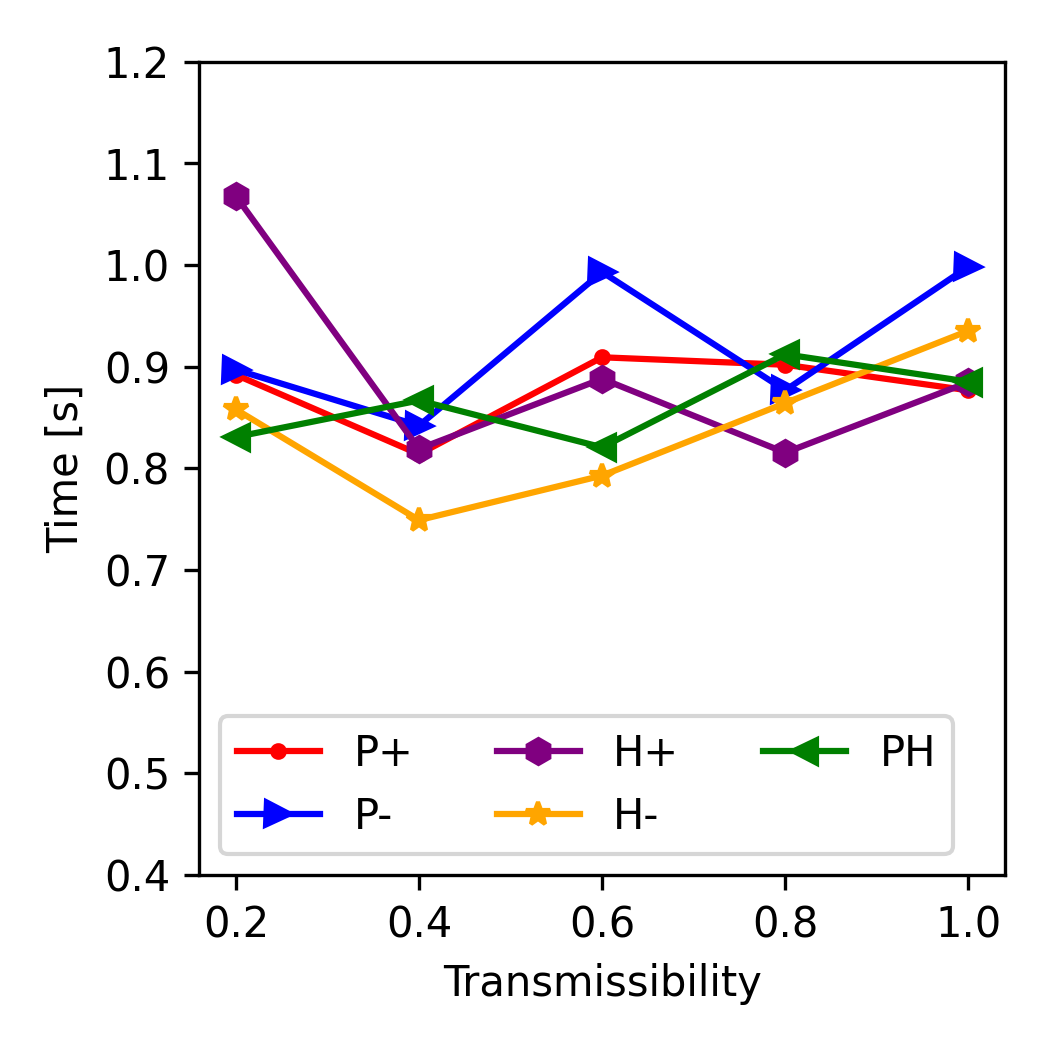}
	\end{minipage}}
  	\subfigure[$DT-CNS_B$]{
		\begin{minipage}[b]{0.17\linewidth}
			\includegraphics[width=1\linewidth]{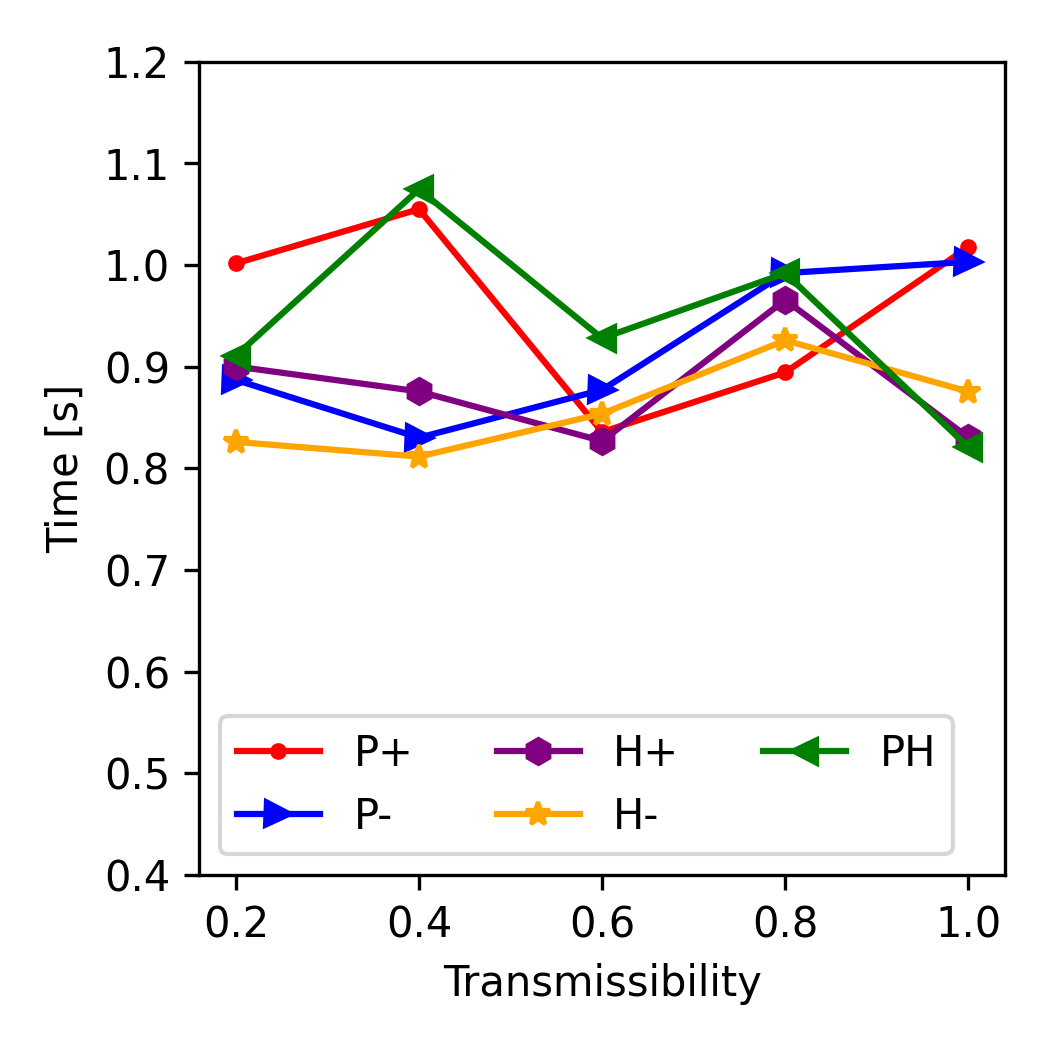}
	\end{minipage}}
   	\subfigure[$DT-CNS_I$]{
		\begin{minipage}[b]{0.17\linewidth}
			\includegraphics[width=1\linewidth]{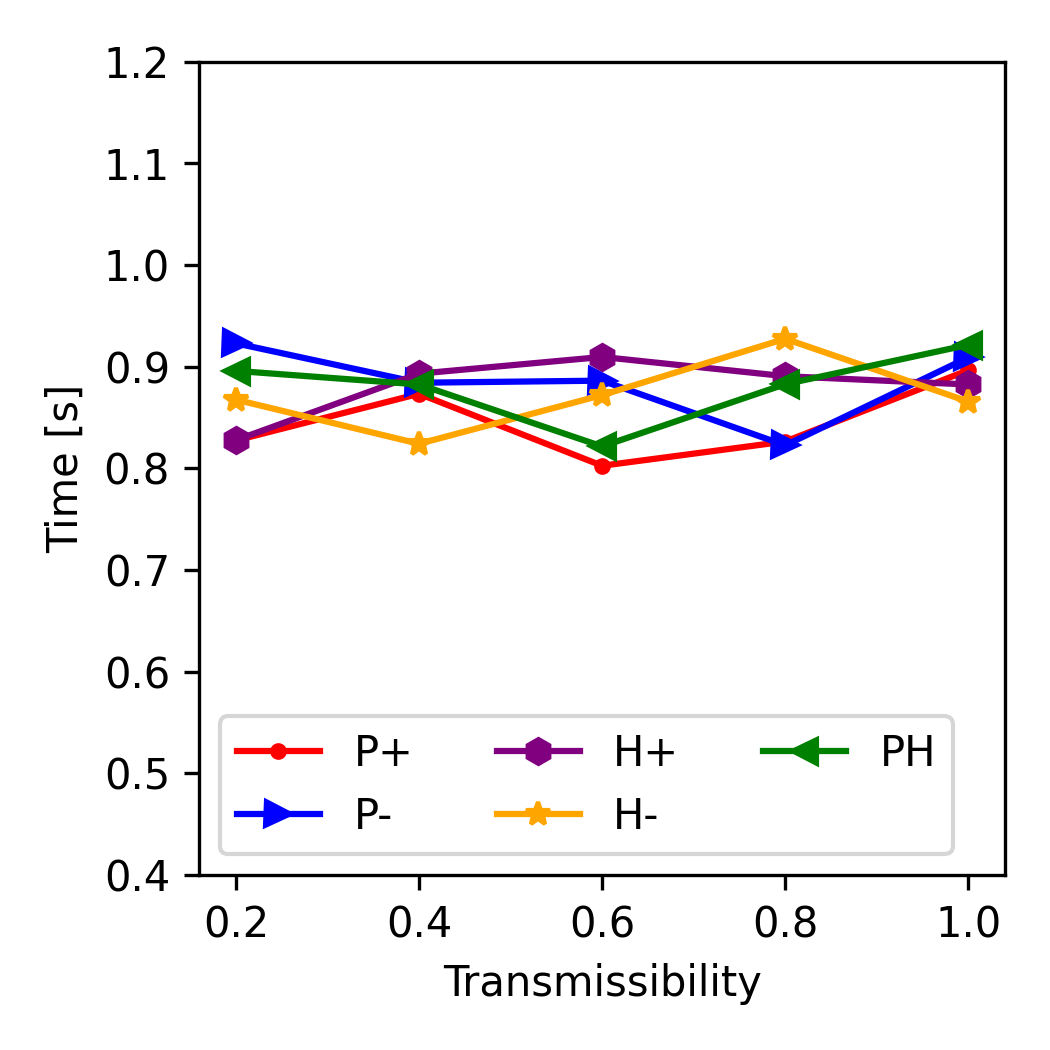}
	\end{minipage}}
   	\subfigure[$DT-CNS_L$]{
		\begin{minipage}[b]{0.17\linewidth}
			\includegraphics[width=1\linewidth]{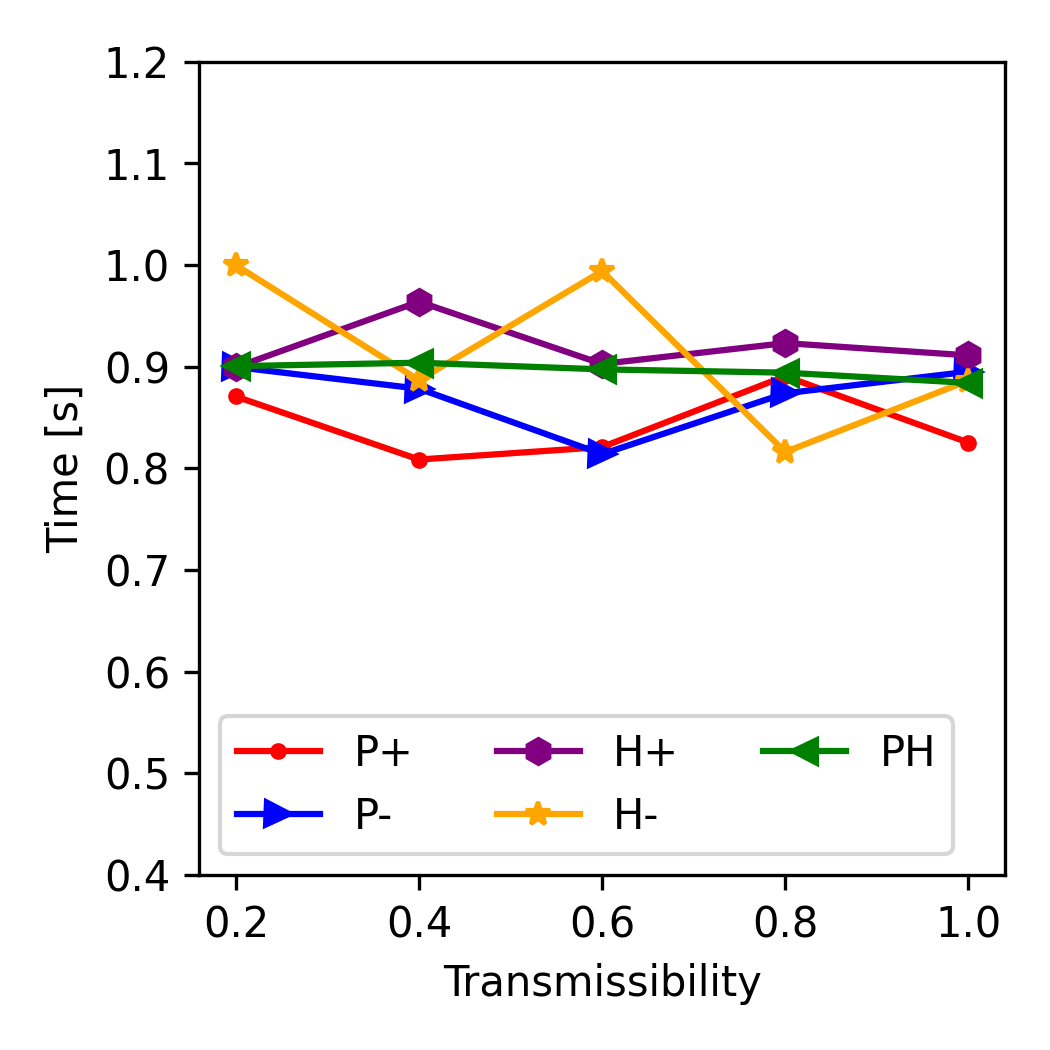}
	\end{minipage}}
   	\subfigure[$DT-CNS_R$]{
		\begin{minipage}[b]{0.17\linewidth}
			\includegraphics[width=1\linewidth]{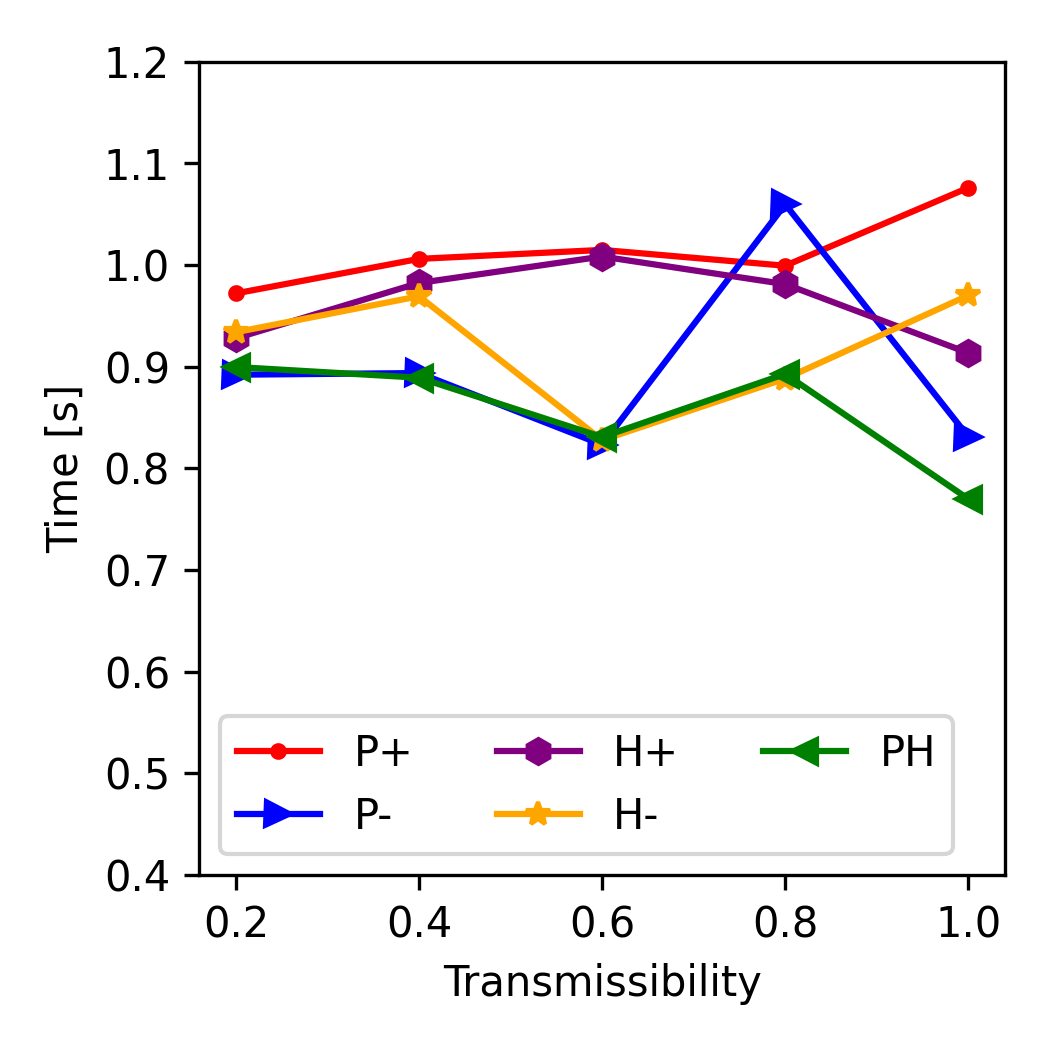}
	\end{minipage}}
	\caption{The runtime of simulation-based DT-CNS given different features and rules.}
\label{time}
\end{figure*}

In this section, we evaluate the simulation-based DT-CNSs based on their network representations' faithfulness. This involves the similarity of simulated network with the target network (see Tab.~\ref{JS}), the simulation runtime (see Fig.~\ref{time}) and the reproducible network statistics.

\begin{table}[h]
\centering
\small
\caption{The similarity of degree distributions with the target degree distribution, as measured by the Jensen–Shannon divergence.}
\label{JS}
\setlength{\tabcolsep}{3pt}
\renewcommand{\arraystretch}{1.5}
\begin{tabular}{|c|c|c|c|c|c|}
\hline
 Paradigms & $P+$ & $P-$ & $H+$ & $H-$ & $PH$ \\
\hline
$DT-CNS_{U}$ &0.51&0.53&\textcolor{blue}{0.36}&0.43&\textcolor{green}{0.26}\\
\hline
$DT-CNS_{B}$ & \textcolor{blue}{0.50}&\textcolor{blue}{0.50}&0.45&0.50&\textcolor{red}{0.27}\\
\hline
$DT-CNS_{I}$ & 0.56&0.54&0.46&\textcolor{blue}{0.42}&\textcolor{red}{0.33}\\
\hline
$DT-CNS_{L}$ &0.56&0.53&0.49&0.53&\textcolor{red}{0.28}\\
\hline
$DT-CNS_{R}$ &0.51&0.59&0.47&0.55&\textcolor{red}{0.28}\\
\hline
\end{tabular}
\end{table}

We evaluate the similarity of simulated networks with the target network from a global perspective using the degree distribution. In addition, we apply the Jensen–Shannon divergence (JS divergence) \cite{della1997inducing} to measure the similarity of degree distributions. A smaller JS divergence indicates a higher similarity level. In Tab.~\ref{JS}, $DT-CNS_{U}$, $DT-CNS_{B}$ and $DT-CNS_{I}$ paradigms, with more diverse age distributions where nodes are more evenly distributed in the age groups with a similar density, tend to generate more similar degree distributions than the paradigms built on other distributions (See the blue texts in Tab.~\ref{JS}). $DT-CNS^{PH}$ paradigms, with an optimised preference principles have smaller JS values (See the red texts in Tab.~\ref{JS}). $DT-CNS_{U}^{PH}$ paradigm achieves the most similar degree distribution with the target. Therefore, the scale-free network pattern, described by its degree distribution from a global perspective, can be approached based on node features and the nodes' preferences for connecting with others. The consideration of node features is a good way to built up the complexity of a model and approach the reality.

We employ the runtime of simulations to evaluate the efficiency of the $DT-CNS$ paradigms, where, as is shown in Fig.~\ref{time}, the time spent on simulation fluctuates with varying transmissibility. The $DT-CNS^{H-}$ paradigms generally have the lowest runtime due to the relatively lower node degrees and the resulting inactivity of epidemic spread. Overall and as expected, the diversity of age distributions and the different rules of network formation, as well as the resulting patterns from the epidemic spreading process, do not make much difference, indicating a similar efficiency level for all the involved $DT-CNS$ paradigms.

Regarding reproducibility, we ensure the repetition of the same resulting patterns with the involved DT-CNSs by setting up a random seed to generate the same random distributions in the process of feature simulation, network formation and process simulation. These random distributions include (i) the uniform distribution that generates age values within an age group (e.g. $\mathcal{U}(20,29)$ for the [20-29] age group), (ii) the Bernoulli distribution $\mathcal{B}(0.08)$ that generates binary values to represent encounters of node pairs based on the $0.08$ encounter rate in network simulation, (iii) the normal distribution $\mathcal{N}(0,0.005^2)$ that generates the random interference in the mutual evaluation between node pairs in the network simulation and (iv) the Bernoulli distribution that generates binary values to represent infections of nodes based on the aggregated infection risks from the exposures in process simulation.

\subsection{Summary of results}

In this section, we summarise the experiment results on the simulation-based DT-CNSs in disaster resilience scenarios connected with the emerging network patterns and the infection status.

In the network dimension, the diversity of the population increases as the age distribution gets more even. 
Nodes at an older/younger age are more popular than others given $P+$ or $P-$ rules. In contrast, nodes closer/further away from the average age are more popular with $H+$ or $H-$ rules. The $DT-CNS^{PH}_U$ paradigm, based on the most diverse age distribution and driven by an optimised $PH$ rule that prefers a younger and dissimilar age, better approaches the target scale-free network patterns. This indicates that introducing features and combining corresponding preferences enable the DT-CNSs to approach the target states better with a higher complexity level concerning the network attributes and dynamics. Our proposed DT-CNS modelling framework enables the extension of DT-CNSs with a flexible complexity levels and the evaluation of their respective distances to the target.

In the process dimension, we study the infection status in the disaster resilience scenario under the influence of network structures driven by various age features and interaction rules. DT-CNS driven by $P+$, $P-$ and the $H+$ rules can generally achieve the most infections within two steps (edges) away from the seed. The decrease of transmissibility means that more time is needed to reach an infection peak, preserving time to react to the epidemic outbreak. DT-CNS paradigms driven by $P+$, $P-$ or $H+$ rules tend to take less time for the final infection status. This is because the seed node has more direct connections with other nodes, in contrast with the nodes who connect with limited number of similar others under the influence of homophily effect. The heterogeneous features and respective preferences lead to complex network topologies, inducing different infection patterns that occur sooner or later due to various transmissibilities. Therefore, social networks have different infection risks/occurrences and, correspondingly, different resistance levels to an epidemic outbreak within a specific time, dependent on transmissibilities and network topologies driven by features and preferences. As a result, nodes with different features and preferences also suffer different infection risks, present different resistance levels to the epidemic and thus can be treated with different mitigation policies. 

To promote disaster resilience, the diversity of the population and the interaction rules are important drivers of case-dependent policy-making. We employ the people at risk $PaR(1,1)$ to represent the proportion of the infected people to be treated by the policymakers with the highest priority. The experiment results suggest that larger transmissibility, a less diverse population, and the $P+$ (positive preferential attachment in age), $P-$ (negative preferential attachment in age) and $H+$ (heterophily effect) rules contribute to a higher $PaR(1,1)$. Among each age group, the $P+$ and $P-$ rules each lead to higher $PaR(1,1)$ for the older and the younger, while $H+$ and $H-$ lead to clustering infection among several age groups due to their connections to similar and dissimilar others respectively. Therefore, the control of epidemic outbreak requires mitigation policies targeted at heterogeneous population. The age groups with higher $PaR$ values given an upcoming epidemic outbreak should be vaccinated in priority to contain the spread of disease. The social networks characterised with higher $PaR$ values within specific time and space require more stringent and urgent isolation policies when the epidemic spreads.  



\section{Conclusion}
\label{Rep1-1section5}

This study proposes a modelling framework for DT-Oriented CNSs based on heterogeneous node features and interaction rules. We also create an evaluation protocol for a faithful representation of reality. Under the modelling framework and evaluation protocol, we conduct a case study on disaster resilience, where we build and compare the simulation-based DT-CNSs given various features and rules.

We build the DT-CNS modelling framework from a top-down perspective concerned with the network and the process dynamics (see section \ref{Rep1-1section31}). We first employ an inner-rule-based network model to represent the network faithfully and mimic its growth. This network model is built on heterogeneous features (ascribed and topological features -- see Table~\ref{feat}) and various feature preference principles (homophily and preferential attachment principles -- see section \ref{Rep1-1section33}). We also devise the (epidemic) process model based on a seed selection strategy and node adoptability, which all vary with specific conditions concerned with node features and the infection status. Based on these features and rules, the network growth and dynamic process are parameterised with sDNA and pDNA vectors, each incorporating the feature preferences and the consideration of conditions related to transmissibility. 

We also propose an evaluation protocol on faithfulness concerned with similarity, efficiency and reproducibility (see section \ref{Rep1-1section32}). Under this protocol, the definition of faithfulness varies with the simulation-based DT-CNSs, hybrid DT-CNSs and data-driven DT-CNSs due to the ground-truth availability. There are pinpoint-, local- and global-level similarity employed, each focusing on the network components, networks local structures, and the global network statistics. We also use the runtime of the simulation and/or modelling to measure the efficiency and evaluate the reproducibility of the same data, statistics or phenomena.

We finally conduct a case study on disaster resilience and conclude with three findings. First, the diversity of the population, people preferences with who they interact and a change of transmissibility can significantly influence infection patterns and serve as important indicators for policymakers in a disaster resilience scenario. Second, the complexity of network dynamics increases when more feature preference representation principles (e.g. preferential attachment and homophily) are introduced. Such an increase in complexity can improve the faithfulness of network representation by preserving necessary heterogeneous patterns observed in reality. Third, age diversity influences the network structures and induces the epidemic outbreak among specific people. This implies the necessity of targeting the easily infected populations with heterogeneous mitigation policies.

Our proposed modelling framework and evaluation protocol enable the extension of DT-CNSs with a flexible complexity level and the evaluation of their respective distances to the target. This modelling framework can also be employed in our future study to generate more realistic social networks by incorporating more complex and real-world information. Our simulation-based experiments on age features and related interaction rules indicate the complexity of real-world interactions and reveal the challenge of approaching reality where people's 
features and preferences can be uncertain and, thus, even more complex. This poses a research question to be addressed in our future study: how to improve the expressive power of DT-CNSs for achieving more real-world like network representation?


\appendix 

\section{Parameter set-ups for preferences and weights of preferences of the DT-CNSs in this study}
\label{app}

In our experiments, we assume an encounter rate at $0.80$ and a random interference that follows a random normal distribution $\mathbf{N}(0,0.005^2)$. We optimise other parameters of the social network simulators, including preferences and the weights of preferences (See Table \ref{params}).
\begin{table}[h]
\centering
\scriptsize
\caption{The parameters of the DT-CNSs.}
\label{params}
\setlength{\tabcolsep}{3pt}
\renewcommand{\arraystretch}{1.5}
\begin{tabular}{|c|c|c|c|c|}
\hline
Parameters & $p$ & $w^p$ & $h$ & $w^h$ \\
\hline
$DT-CNS_{U}^{P+}$ &1&1.00 &1&0.00\\
\hline
$DT-CNS_{U}^{P-}$ &-1&1.00 &1&0.00 \\
\hline
$DT-CNS_{U}^{H+}$ &1&0.00 &1&1.00 \\
\hline
$DT-CNS_{U}^{H-}$ &1&0.00 &-1&1.00 \\
\hline
$DT-CNS_{U}^{PH}$ &-1&0.05 &1&0.08 \\
\hline
$DT-CNS_{B}^{P+}$ &1&1.00 &1&0.00\\
\hline
$DT-CNS_{B}^{P-}$ &-1&1.00 &1&0.00 \\
\hline
$DT-CNS_{B}^{H+}$ &1&0.00 &1&1.00 \\
\hline
$DT-CNS_{B}^{H-}$ &1&0.00 &-1&1.00 \\
\hline
$DT-CNS_{B}^{PH}$ &-1&0.03 &1&0.06 \\
\hline
$DT-CNS_{I}^{P+}$ &1&1.00 &1&0.00\\
\hline
$DT-CNS_{I}^{P-}$ &-1&1.00 &1&0.00 \\
\hline
$DT-CNS_{I}^{H+}$ &1&0.00 &1&1.00 \\
\hline
$DT-CNS_{I}^{H-}$ &1&0.00 &-1&1.00 \\
\hline
$DT-CNS_{I}^{PH}$ &1&0.68 &-1&  0.73 \\
\hline
$DT-CNS_{L}^{P+}$ &1&1.00 &1&0.00\\
\hline
$DT-CNS_{L}^{P-}$ &-1&1.00 &1&0.00 \\
\hline
$DT-CNS_{L}^{H+}$ &1&0.00 &1&1.00 \\
\hline
$DT-CNS_{L}^{H-}$ &1&0.00 &-1&1.00 \\
\hline
$DT-CNS_{L}^{PH}$ &1&0.02 &-1& 0.08 \\
\hline
$DT-CNS_{R}^{P+}$ &1&1.00 &1&0.00\\
\hline
$DT-CNS_{R}^{P-}$ &-1&1.00 &1&0.00 \\
\hline
$DT-CNS_{R}^{H+}$ &1&0.00 &1&1.00 \\
\hline
$DT-CNS_{R}^{H-}$ &1&0.00 &-1&1.00 \\
\hline
$DT-CNS_{R}^{PH}$ &1&0.02 &-1&  0.06\\
\hline
\end{tabular}
\end{table}

\section{Distance between the network patterns of simulated networks and the target network}
\label{app2}

In our experiments, we employ JS divergence to evaluate the differences between distribution-based network patterns, including degree distributions, clustering coefficient distributions and shortest path length distributions. Tab.~\ref{jspatterns} presents the JS divergence values between different network patterns.

\begin{table}[h]
\centering
\scriptsize
\caption{The JS divergence between the network patterns of simulated networks and the target network.}
\label{jspatterns}
\setlength{\tabcolsep}{3pt}
\renewcommand{\arraystretch}{1.5}
\begin{tabular}{|c|c|c|c|}
\hline
Network Patterns & Degree Distribution & Clustering Coefficient Distribution & Shortest Path Length Distribution\\
\hline
$DT-CNS_{U}^{P+}$ & 0.51 & 0.63& 0.29  \\
\hline
$DT-CNS_{U}^{P-}$ & 0.53 & 0.68 &  0.28  \\
\hline
$DT-CNS_{U}^{H+}$ & 0.36 & 0.68& 0.12  \\
\hline
$DT-CNS_{U}^{H-}$ & 0.43 &0.69&  0.35 \\
\hline
$DT-CNS_{U}^{PH}$ & \textcolor{green}{0.26} &0.60& \textcolor{green}{0.00} \\ 
\hline
$DT-CNS_{B}^{P+}$ & 0.50 &0.63 & 0.09\\
\hline
$DT-CNS_{B}^{P-}$ & 0.50 &0.66& 0.16\\
\hline
$DT-CNS_{B}^{H+}$ &0.45&0.55& 0.01\\
\hline
$DT-CNS_{B}^{H-}$ & 0.50 &0.69& 0.34  \\
\hline
$DT-CNS_{B}^{PH}$ & 0.27 &0.41& 0.01 \\
\hline
$DT-CNS_{I}^{P+}$ & 0.56 & 0.68 &, 0.31 \\
\hline
$DT-CNS_{I}^{P-}$ & 0.54 &0.68&0.30\\
\hline
$DT-CNS_{I}^{H+}$ & 0.46 &0.81&0.22  \\
\hline
$DT-CNS_{I}^{H-}$ & 0.42 &0.67&0.41   \\
\hline
$DT-CNS_{I}^{PH}$ &0.33 &0.70&0.06\\
\hline
$DT-CNS_{L}^{P+}$ & 0.56 & 0.70 &0.37 \\
\hline
$DT-CNS_{L}^{P-}$ & 0.53 &0.68 & 0.01\\
\hline
$DT-CNS_{L}^{H+}$ & 0.49 &0.53& 0.01 \\
\hline
$DT-CNS_{L}^{H-}$ &  0.53 &0.68&  0.37 \\
\hline
$DT-CNS_{L}^{PH}$ & 0.28 &\textcolor{green}{0.45}&0.01\\
\hline
$DT-CNS_{R}^{P+}$ & 0.51 &0.69&0.01 \\
\hline
$DT-CNS_{R}^{P-}$ &0.59 &0.72&  0.40\\
\hline
$DT-CNS_{R}^{H+}$ & 0.47&0.58& 0.01\\
\hline
$DT-CNS_{R}^{H-}$ & 0.55 &0.68& 0.39 \\
\hline
$DT-CNS_{R}^{PH}$ & 0.28 &0.50& 0.01\\
\hline
\end{tabular}
\end{table}

\section*{Acknowledgements}\label{section-acknowledge}

This work was supported by the Australian Research Council, Dynamics and Control of Complex Social Networks under Grant DP190101087.



\begin{thebibliography}{}

\bibitem[Abbasi et~al., 2012]{abbasi2012betweenness}
Abbasi, A., Hossain, L., and Leydesdorff, L. (2012).
\newblock Betweenness centrality as a driver of preferential attachment in the
  evolution of research collaboration networks.
\newblock {\em Journal of Informetrics}, 6(3):403--412.

\bibitem[Abu-El-Haija et~al., 2020]{abu2020n}
Abu-El-Haija, S., Kapoor, A., Perozzi, B., and Lee, J. (2020).
\newblock N-gcn: Multi-scale graph convolution for semi-supervised node
  classification.
\newblock In {\em uncertainty in artificial intelligence}, pages 841--851.
  PMLR.

\bibitem[Aguirre et~al., 2018]{aguirre2018structural}
Aguirre, L.~A., Portes, L.~L., and Letellier, C. (2018).
\newblock Structural, dynamical and symbolic observability: From dynamical
  systems to networks.
\newblock {\em PLoS One}, 13(10):e0206180.

\bibitem[Alberdi and Gilbert, 2019]{alberdi2019guide}
Alberdi, A. and Gilbert, M. T.~P. (2019).
\newblock A guide to the application of hill numbers to dna-based diversity
  analyses.
\newblock {\em Molecular Ecology Resources}, 19(4):804--817.

\bibitem[Ashraf et~al., 2019]{ashraf2019simulation}
Ashraf, A. W.-U., Budka, M., and Musial, K. (2019).
\newblock Simulation and augmentation of social networks for building deep
  learning models.
\newblock {\em arXiv preprint arXiv:1905.09087}.

\bibitem[Barab{\'a}si and Albert, 1999]{barabasi1999emergence}
Barab{\'a}si, A.-L. and Albert, R. (1999).
\newblock Emergence of scaling in random networks.
\newblock {\em science}, 286(5439):509--512.

\bibitem[Boucher, 2015]{boucher2015structural}
Boucher, V. (2015).
\newblock Structural homophily.
\newblock {\em International Economic Review}, 56(1):235--264.

\bibitem[Br{\'o}dka et~al., 2020]{brodka2020interacting}
Br{\'o}dka, P., Musial, K., and Jankowski, J. (2020).
\newblock Interacting spreading processes in multilayer networks: a systematic
  review.
\newblock {\em IEEE Access}, 8:10316--10341.

\bibitem[Bu et~al., 2019]{bu2019link}
Bu, Z., Wang, Y., Li, H.-J., Jiang, J., Wu, Z., and Cao, J. (2019).
\newblock Link prediction in temporal networks: Integrating survival analysis
  and game theory.
\newblock {\em Information Sciences}, 498:41--61.

\bibitem[Budka et~al., 2013]{budka2013molecular}
Budka, M., Juszczyszyn, K., Musial, K., and Musial, A. (2013).
\newblock Molecular model of dynamic social network based on e-mail
  communication.
\newblock {\em Social Network Analysis and Mining}, 3(3):543--563.

\bibitem[Carchiolo et~al., 2021]{carchiolo2021mutual}
Carchiolo, V., Longheu, A., Malgeri, M., Mangioni, G., and Previti, M. (2021).
\newblock Mutual influence of users credibility and news spreading in online
  social networks.
\newblock {\em Future Internet}, 13(5):107.

\bibitem[Chao et~al., 2014a]{chao2014unifying}
Chao, A., Chiu, C.-H., and Jost, L. (2014a).
\newblock Unifying species diversity, phylogenetic diversity, functional
  diversity, and related similarity and differentiation measures through hill
  numbers.
\newblock {\em Annual review of ecology, evolution, and systematics},
  45:297--324.

\bibitem[Chao et~al., 2014b]{chao2014rarefaction}
Chao, A., Gotelli, N.~J., Hsieh, T., Sander, E.~L., Ma, K., Colwell, R.~K., and
  Ellison, A.~M. (2014b).
\newblock Rarefaction and extrapolation with hill numbers: a framework for
  sampling and estimation in species diversity studies.
\newblock {\em Ecological monographs}, 84(1):45--67.

\bibitem[Chen et~al., 2020]{chen2020multi}
Chen, H., Yin, H., Sun, X., Chen, T., Gabrys, B., and Musial, K. (2020).
\newblock Multi-level graph convolutional networks for cross-platform anchor
  link prediction.
\newblock In {\em Proceedings of the 26th ACM SIGKDD International Conference
  on Knowledge Discovery \& Data Mining}, pages 1503--1511.

\bibitem[Chen et~al., 2021]{chen2018gc}
Chen, J., Wang, X., and Xu, X. (2021).
\newblock Gc-lstm: graph convolution embedded lstm for dynamic network link
  prediction.
\newblock {\em Applied Intelligence}, pages 1--16.

\bibitem[Comin and da~Fontoura~Costa, 2011]{comin2011identifying}
Comin, C.~H. and da~Fontoura~Costa, L. (2011).
\newblock Identifying the starting point of a spreading process in complex
  networks.
\newblock {\em Physical Review E}, 84(5):056105.

\bibitem[Courtat et~al., 2011]{courtat2011mathematics}
Courtat, T., Gloaguen, C., and Douady, S. (2011).
\newblock Mathematics and morphogenesis of cities: A geometrical approach.
\newblock {\em Physical Review E}, 83(3):036106.

\bibitem[Della~Pietra et~al., 1997]{della1997inducing}
Della~Pietra, S., Della~Pietra, V., and Lafferty, J. (1997).
\newblock Inducing features of random fields.
\newblock {\em IEEE transactions on pattern analysis and machine intelligence},
  19(4):380--393.

\bibitem[Dong et~al., 2015]{dong2015coupledlp}
Dong, Y., Zhang, J., Tang, J., Chawla, N.~V., and Wang, B. (2015).
\newblock Coupledlp: Link prediction in coupled networks.
\newblock In {\em Proceedings of the 21th ACM SIGKDD International Conference
  on Knowledge Discovery and Data Mining}, pages 199--208.

\bibitem[Ertug et~al., 2022]{ertug2022does}
Ertug, G., Brennecke, J., Kov{\'a}cs, B., and Zou, T. (2022).
\newblock What does homophily do? a review of the consequences of homophily.
\newblock {\em Academy of Management Annals}, 16(1):38--69.

\bibitem[Faisal et~al., 2020]{faisal2020comparative}
Faisal, M., Zamzami, E., et~al. (2020).
\newblock Comparative analysis of inter-centroid k-means performance using
  euclidean distance, canberra distance and manhattan distance.
\newblock In {\em Journal of Physics: Conference Series}, volume 1566, page
  012112. IOP Publishing.

\bibitem[Fortunato et~al., 2006]{fortunato2006scale}
Fortunato, S., Flammini, A., and Menczer, F. (2006).
\newblock Scale-free network growth by ranking.
\newblock {\em Physical review letters}, 96(21):218701.

\bibitem[Ganesh et~al., 2005]{ganesh2005effect}
Ganesh, A., Massouli{\'e}, L., and Towsley, D. (2005).
\newblock The effect of network topology on the spread of epidemics.
\newblock In {\em Proceedings IEEE 24th Annual Joint Conference of the IEEE
  Computer and Communications Societies.}, volume~2, pages 1455--1466. IEEE.

\bibitem[Gao et~al., 2017]{gao2017community}
Gao, F., Musial, K., and Gabrys, B. (2017).
\newblock A community bridge boosting social network link prediction model.
\newblock In {\em Proceedings of the 2017 IEEE/ACM International Conference on
  Advances in Social Networks Analysis and Mining 2017}, pages 683--689.

\bibitem[Gao and Musial-Gabrys, 2016]{gao2016hybrid}
Gao, F. and Musial-Gabrys, K. (2016).
\newblock Hybrid structure-based link prediction model.
\newblock In {\em 2016 IEEE/ACM International Conference on Advances in Social
  Networks Analysis and Mining (ASONAM)}, pages 1221--1228. IEEE.

\bibitem[Hong et~al., 2019]{hong2019deep}
Hong, R., He, Y., Wu, L., Ge, Y., and Wu, X. (2019).
\newblock Deep attributed network embedding by preserving structure and
  attribute information.
\newblock {\em IEEE Transactions on Systems, Man, and Cybernetics: Systems}.

\bibitem[Ivie and Thain, 2018]{ivie2018reproducibility}
Ivie, P. and Thain, D. (2018).
\newblock Reproducibility in scientific computing.
\newblock {\em ACM Computing Surveys (CSUR)}, 51(3):1--36.

\bibitem[Jeong et~al., 2003]{jeong2003measuring}
Jeong, H., N{\'e}da, Z., and Barab{\'a}si, A.-L. (2003).
\newblock Measuring preferential attachment in evolving networks.
\newblock {\em EPL (Europhysics Letters)}, 61(4):567.

\bibitem[Jia et~al., 2021a]{jia2021directed}
Jia, M., Gabrys, B., and Musial, K. (2021a).
\newblock Directed closure coefficient and its patterns.
\newblock {\em Plos one}, 16(6):e0253822.

\bibitem[Jia et~al., 2021b]{jia2021measuring}
Jia, M., Gabrys, B., and Musial, K. (2021b).
\newblock Measuring quadrangle formation in complex networks.
\newblock {\em IEEE Transactions on Network Science and Engineering}.

\bibitem[Jin et~al., 2018]{jin2018robust}
Jin, D., Liu, Z., He, D., Gabrys, B., and Musial, K. (2018).
\newblock Robust detection of communities with multi-semantics in large
  attributed networks.
\newblock In {\em International Conference on Knowledge Science, Engineering
  and Management}, pages 362--376. Springer.

\bibitem[Jin et~al., 2020]{jin2020modmrf}
Jin, D., Zhang, B., Song, Y., He, D., Feng, Z., Chen, S., Li, W., and Musial,
  K. (2020).
\newblock Modmrf: A modularity-based markov random field method for community
  detection.
\newblock {\em Neurocomputing}, 405:218--228.

\bibitem[Jovanovski et~al., 2021]{jovanovski2021modeling}
Jovanovski, P., Tomovski, I., and Kocarev, L. (2021).
\newblock Modeling the spread of multiple contagions on multilayer networks.
\newblock {\em Physica A: Statistical Mechanics and its Applications},
  563:125410.

\bibitem[Junuthula et~al., 2016]{junuthula2016evaluating}
Junuthula, R.~R., Xu, K.~S., and Devabhaktuni, V.~K. (2016).
\newblock Evaluating link prediction accuracy in dynamic networks with added
  and removed edges.
\newblock In {\em 2016 IEEE international conferences on big data and cloud
  computing (BDCloud), social computing and networking (SocialCom), sustainable
  computing and communications (SustainCom)(BDCloud-SocialCom-SustainCom)},
  pages 377--384. IEEE.

\bibitem[Karczmarczyk et~al., 2018]{karczmarczyk2018influencing}
Karczmarczyk, A., Bortko, K., Bartk{\'o}w, P., Pazura, P., and Jankowski, J.
  (2018).
\newblock Influencing information spreading processes in complex networks with
  probability spraying.
\newblock In {\em 2018 IEEE/ACM International Conference on Advances in Social
  Networks Analysis and Mining (ASONAM)}, pages 1038--1046. IEEE.

\bibitem[Kendrick et~al., 2018]{kendrick2018change}
Kendrick, L., Musial, K., and Gabrys, B. (2018).
\newblock Change point detection in social networks—critical review with
  experiments.
\newblock {\em Computer Science Review}, 29:1--13.

\bibitem[Kim et~al., 2019]{kim2019advancing}
Kim, J.-S., Kavak, H., Manzoor, U., and Z{\"u}fle, A. (2019).
\newblock Advancing simulation experimentation capabilities with runtime
  interventions.
\newblock In {\em 2019 Spring Simulation Conference (SpringSim)}, pages 1--11.
  IEEE.

\bibitem[Kossinets and Watts, 2009]{kossinets2009origins}
Kossinets, G. and Watts, D.~J. (2009).
\newblock Origins of homophily in an evolving social network.
\newblock {\em American journal of sociology}, 115(2):405--450.

\bibitem[Koutra et~al., 2013]{koutra2013deltacon}
Koutra, D., Vogelstein, J.~T., and Faloutsos, C. (2013).
\newblock Deltacon: A principled massive-graph similarity function.
\newblock In {\em Proceedings of the 2013 SIAM international conference on data
  mining}, pages 162--170. SIAM.

\bibitem[Kr{\'o}l et~al., 2015]{krol2015propagation}
Kr{\'o}l, D., Fay, D., and Gabry{\'s}, B. (2015).
\newblock {\em Propagation phenomena in real world networks}.
\newblock Springer.

\bibitem[Li et~al., 2014]{li2014deep}
Li, X., Du, N., Li, H., Li, K., Gao, J., and Zhang, A. (2014).
\newblock A deep learning approach to link prediction in dynamic networks.
\newblock In {\em Proceedings of the 2014 SIAM International Conference on Data
  Mining}, pages 289--297. SIAM.

\bibitem[Liao et~al., 2018]{liao2018attributed}
Liao, L., He, X., Zhang, H., and Chua, T.-S. (2018).
\newblock Attributed social network embedding.
\newblock {\em IEEE Transactions on Knowledge and Data Engineering},
  30(12):2257--2270.

\bibitem[Liu et~al., 2020a]{liu2020using}
Liu, F., Li, X., and Zhu, G. (2020a).
\newblock Using the contact network model and metropolis-hastings sampling to
  reconstruct the covid-19 spread on the “diamond princess”.
\newblock {\em Science bulletin}, 65(15):1297--1305.

\bibitem[Liu et~al., 2020b]{liu2020semi}
Liu, X., Song, W., Musial, K., Zhao, X., Zuo, W., and Yang, B. (2020b).
\newblock Semi-supervised stochastic blockmodel for structure analysis of
  signed networks.
\newblock {\em Knowledge-Based Systems}, 195:105714.

\bibitem[Liu et~al., 2021]{liu2021block}
Liu, X., Yang, B., Song, W., Musial, K., Zuo, W., Chen, H., and Yin, H. (2021).
\newblock A block-based generative model for attributed network embedding.
\newblock {\em World Wide Web}, 24(5):1439--1464.

\bibitem[L{\"u} and Zhou, 2011]{lu2011link}
L{\"u}, L. and Zhou, T. (2011).
\newblock Link prediction in complex networks: A survey.
\newblock {\em Physica A: statistical mechanics and its applications},
  390(6):1150--1170.

\bibitem[McPherson et~al., 2001]{mcpherson2001birds}
McPherson, M., Smith-Lovin, L., and Cook, J.~M. (2001).
\newblock Birds of a feather: Homophily in social networks.
\newblock {\em Annual review of sociology}, 27(1):415--444.

\bibitem[Min and San~Miguel, 2018]{min2018competing}
Min, B. and San~Miguel, M. (2018).
\newblock Competing contagion processes: Complex contagion triggered by simple
  contagion.
\newblock {\em Scientific reports}, 8(1):1--8.

\bibitem[Musial et~al., 2013a]{musial2013creation}
Musial, K., Budka, M., and Juszczyszyn, K. (2013a).
\newblock Creation and growth of online social network.
\newblock {\em World Wide Web}, 16(4):421--447.

\bibitem[Musial et~al., 2013b]{musial2013kind}
Musial, K., Gabrys, B., and Buczko, M. (2013b).
\newblock What kind of network are you? using local and global characteristics
  in network categorisation tasks.
\newblock In {\em Proceedings of the 2013 IEEE/ACM International Conference on
  Advances in Social Networks Analysis and Mining}, pages 1366--1373.

\bibitem[Musial et~al., 2012]{musial2012triad}
Musial, K., Juszczyszyn, K., and Budka, M. (2012).
\newblock Triad transition probabilities characterize complex networks.
\newblock {\em Awareness Magazine}.

\bibitem[Nikolentzos et~al., 2017]{nikolentzos2017matching}
Nikolentzos, G., Meladianos, P., and Vazirgiannis, M. (2017).
\newblock Matching node embeddings for graph similarity.
\newblock In {\em Thirty-first AAAI conference on artificial intelligence}.

\bibitem[Pastor-Satorras and Vespignani, 2001]{pastor2001epidemic}
Pastor-Satorras, R. and Vespignani, A. (2001).
\newblock Epidemic spreading in scale-free networks.
\newblock {\em Physical review letters}, 86(14):3200.

\bibitem[Patel and Guo, 2021]{patel2021graph}
Patel, R. and Guo, Y. (2021).
\newblock Graph based link prediction between human phenotypes and genes.
\newblock {\em arXiv preprint arXiv:2105.11989}.

\bibitem[Qin et~al., 2018]{qin2018adaptive}
Qin, M., Jin, D., Lei, K., Gabrys, B., and Musial-Gabrys, K. (2018).
\newblock Adaptive community detection incorporating topology and content in
  social networks.
\newblock {\em Knowledge-based systems}, 161:342--356.

\bibitem[Qiu et~al., 2016]{qiu2016effects}
Qiu, X., Zhao, L., Wang, J., Wang, X., and Wang, Q. (2016).
\newblock Effects of time-dependent diffusion behaviors on the rumor spreading
  in social networks.
\newblock {\em Physics Letters A}, 380(24):2054--2063.

\bibitem[Russell et~al., 2020]{russell2020estimating}
Russell, T.~W., Hellewell, J., Jarvis, C.~I., Van~Zandvoort, K., Abbott, S.,
  Ratnayake, R., Flasche, S., Eggo, R.~M., Edmunds, W.~J., Kucharski, A.~J.,
  et~al. (2020).
\newblock Estimating the infection and case fatality ratio for coronavirus
  disease (covid-19) using age-adjusted data from the outbreak on the diamond
  princess cruise ship, february 2020.
\newblock {\em Eurosurveillance}, 25(12):2000256.

\bibitem[Sadaf et~al., 2022]{sadaf2022maximising}
Sadaf, A., Mathieson, L., Br{\'o}dka, P., and Musial, K. (2022).
\newblock Maximising influence spread in complex networks by utilising
  community-based driver nodes as seeds.
\newblock In {\em Annual International Conference on Information Management and
  Big Data}, pages 126--141. Springer.

\bibitem[Skardinga et~al., 2021]{skarding2020foundations}
Skardinga, J., Gabrys, B., and Musial, K. (2021).
\newblock Foundations and modelling of dynamic networks using dynamic graph
  neural networks: A survey.
\newblock {\em IEEE Access}.

\bibitem[Sohn, 2017]{sohn2017small}
Sohn, I. (2017).
\newblock Small-world and scale-free network models for iot systems.
\newblock {\em Mobile Information Systems}, 2017.

\bibitem[Telesford et~al., 2011]{telesford2011ubiquity}
Telesford, Q.~K., Joyce, K.~E., Hayasaka, S., Burdette, J.~H., and Laurienti,
  P.~J. (2011).
\newblock The ubiquity of small-world networks.
\newblock {\em Brain connectivity}, 1(5):367--375.

\bibitem[Tsiotas, 2020]{tsiotas2020preferential}
Tsiotas, D. (2020).
\newblock Preferential attachment: a multi-attribute growth process generating
  scale-free networks of different topologies.
\newblock {\em arXiv preprint arXiv:2001.05167}.

\bibitem[Wahid-Ul-Ashraf et~al., 2019]{wahid2019predict}
Wahid-Ul-Ashraf, A., Budka, M., and Musial, K. (2019).
\newblock How to predict social relationships—physics-inspired approach to
  link prediction.
\newblock {\em Physica A: Statistical Mechanics and its Applications},
  523:1110--1129.

\bibitem[Wang et~al., 2018a]{wang2018graph}
Wang, T., Lu, G., Liu, J., and Yan, P. (2018a).
\newblock Graph-based change detection for condition monitoring of rotating
  machines: Techniques for graph similarity.
\newblock {\em IEEE Transactions on Reliability}, 68(3):1034--1049.

\bibitem[Wang et~al., 2016]{wang2016autonomous}
Wang, W., Jiao, P., He, D., Jin, D., Pan, L., and Gabrys, B. (2016).
\newblock Autonomous overlapping community detection in temporal networks: A
  dynamic bayesian nonnegative matrix factorization approach.
\newblock {\em Knowledge-Based Systems}, 110:121--134.

\bibitem[Wang et~al., 2019a]{wang2019simulation}
Wang, W., Zhang, J., Zhao, S., and Zhang, Y. (2019a).
\newblock Simulation of asset pricing in information networks.
\newblock {\em Physica A: Statistical Mechanics and its Applications},
  513:620--634.

\bibitem[Wang et~al., 2021]{wang2021multi}
Wang, W., Zhao, S., and Zhang, J. (2021).
\newblock Multi-asset pricing modeling using holding-based networks in energy
  markets.
\newblock {\em Finance Research Letters}, page 102483.

\bibitem[Wang et~al., 2019b]{wang2019community}
Wang, Y., Jin, D., Musial, K., and Dang, J. (2019b).
\newblock Community detection in social networks considering topic
  correlations.
\newblock In {\em Proceedings of the AAAI Conference on Artificial
  Intelligence}, volume~33, pages 321--328.

\bibitem[Wang et~al., 2018b]{wang2018real}
Wang, Y., Meyer, M.~C., and Wang, J. (2018b).
\newblock Real-time delay minimization for data processing in wirelessly
  networked disaster areas.
\newblock {\em IEEE Access}, 7:2928--2937.

\bibitem[Wen et~al., 2022]{wen2022towards}
Wen, J., Gabrys, B., and Musial, K. (2022).
\newblock Toward digital twin oriented modeling of complex networked systems
  and their dynamics: A comprehensive survey.
\newblock {\em IEEE Access}, 10:66886--66923.

\bibitem[Zhang, 2018]{zhang2018influence}
Zhang, J. (2018).
\newblock Influence of individual rationality on continuous double auction
  markets with networked traders.
\newblock {\em Physica A: Statistical Mechanics and its Applications},
  495:353--392.

\bibitem[Zhang et~al., 2018]{zhang2018convergence}
Zhang, J., McBurney, P., and Musial, K. (2018).
\newblock Convergence of trading strategies in continuous double auction
  markets with boundedly-rational networked traders.
\newblock {\em Review of Quantitative Finance and Accounting}, 50(1):301--352.

\bibitem[Zhang et~al., 2021]{zhang2021vulnerability}
Zhang, J., Xu, Y., and Houser, D. (2021).
\newblock Vulnerability of scale-free cryptocurrency networks to
  double-spending attacks.
\newblock {\em The European Journal of Finance}, pages 1--15.

\bibitem[Zhou et~al., 2018]{zhou2018dynamic}
Zhou, L., Yang, Y., Ren, X., Wu, F., and Zhuang, Y. (2018).
\newblock Dynamic network embedding by modeling triadic closure process.
\newblock In {\em Proceedings of the AAAI Conference on Artificial
  Intelligence}, volume~32.

\end{thebibliography}
\end{document}